\documentclass[a4paper,11pt]{report}

\newcommand{\linespacing}{1.5}
\renewcommand{\baselinestretch}{\linespacing}

\usepackage{graphicx}
\usepackage{epstopdf}
\usepackage[british]{babel}
\usepackage{amsmath}
\usepackage{amsfonts}
\usepackage{amssymb}
\usepackage{color}
\usepackage{bm}
\usepackage{enumitem}
\usepackage{booktabs}
\usepackage{array}
\usepackage{mathrsfs}
\usepackage[T1]{fontenc}
\usepackage[latin1]{inputenc}
\usepackage[table]{xcolor}  
\usepackage{verbatim}
\usepackage{rotating}
\usepackage{dcolumn}
\usepackage{float}
\usepackage[toc]{appendix}
\usepackage{lipsum}
\usepackage{feynmp}
\usepackage[labelfont={bf,scriptsize},textfont={scriptsize},justification=RaggedRight,format=hang]{caption,subfig}
\usepackage{csquotes}
\usepackage{newtxtext,newtxmath}
\usepackage{mwe}    

\RequirePackage{ifpdf}
	
\ifpdf
\else 
\fi

\expandafter\let\csname equation*\endcsname\relax
\expandafter\let\csname endequation*\endcsname\relax

\AtBeginDocument{\renewcommand{\d}{\mathrm{d}}}

\renewcommand{\leq}{\leqslant}

\newcommand{\gNL}{g_{\mathrm{NL}}}
\newcommand{\fNL}{f_{\mathrm{NL}}}

\newcommand{\fNLloc}{f_{\rm NL}^{\rm local}}
\newcommand{\fNLeq}{f_{\rm NL}^{\rm equil}}
\newcommand{\fNLorth}{f_{\rm NL}^{\rm orthog}}

\newcommand{\ipleft}{\langle\kern-0.2em\langle}
\newcommand{\ipright}{\rangle\kern-0.2em\rangle}

\newcommand{\bk}{\mathbf{k}}
\newcommand{\bx}{\mathbf{x}}
\newcommand{\skewR}{\mathcal{M}_{3,R}}

\newcolumntype{s}{>{$\displaystyle}l<{$}}
\newcolumntype{t}{>{$\displaystyle}c<{$}}
\newcolumntype{u}{>{$\displaystyle}r<{$}}
\newcolumntype{v}{>{$\displaystyle}m{4cm}<{$}}

\newcolumntype{d}{D{!}{\;\pm\;}{-1}}

\usepackage[calcwidth]{titlesec}
 \titleformat{\chapter}[display]{\Large\rmfamily\scshape\center}{}{0 em}{}
\titleformat{\section}[block]{\large\rmfamily\scshape}{\thesection}{0.5 em}{}[\vspace{-1.4 em}\rule{\titlewidth}{0.5pt}]
\titleformat{\subsection}{\rmfamily\scshape}{\thesubsection}{0.5 em}{}

\titlespacing*{\section}{0 pt}{*8}{*6}
\titlespacing*{\subsection}{0 pt}{*6}{*4}

\usepackage[a4paper,top=2.5cm,bottom=2.5cm,left=4cm,right=2cm,headsep=10pt]{geometry}


\usepackage{fancyhdr}
\begin{document}

\pagenumbering{roman}

\thispagestyle{empty}
\begin{flushright}
\end{flushright}	
\vskip40mm
\begin{center}
\huge{\textsc{Constraining the early universe 
\\
with primordial black holes}}
\vskip6mm
\vskip14mm
\Large \emph{Author:}
\\
\Large Samuel Mark Young
\vskip14mm
\normalsize
\end{center}
\vfill
\begin{center}
\large
Submitted for the degree of Doctor of Philosophy \\
University of Sussex	\\
March 2016
\end{center}		

\chapter*{Declaration}
I hereby declare that this thesis has not been and will not be submitted in whole or in part to another University for the award of any other degree.
\vskip2mm
The work in this thesis has been completed in collaboration with Christian Byrnes, Misao Sasaki, and Donough Regan, and is comprised of the following papers:

\begin{itemize}

\item Sam Young, Christian T. Byrnes and Misao Sasaki. `Calculating the mass fraction of primordial black holes', published n JCAP 1407 (2014), p. 045. arXiv:1405.7023 [gr-qc].

I was responsible for the original concepts presented in this paper, which were then formulated in collaboration with Christian T. Byrnes and Misao Sasaki. The paper and calculations were all completed by myself, with minor adjustments and additions made by Christian T. Byrnes and Misao Sasaki.

\item Sam Young and Christian T. Byrnes. `Primordial black holes in non-Gaussian regimes', published in JCAP 1308 (2013), p. 052. arXiv:1307.4995 [astro-ph.CO].

The original concept for this paper was proposed by Christian T. Byrnes, and follows closely from the work completed by \cite{Byrnes:2012yx}. The calculations were completed by myself under the supervision of Christian T. Byrnes. I also wrote the great majority of the paper, with minor adjustments completed by Christian T. Byrnes.

\item Sam Young and Christian T. Byrnes. `Long-short wavelength mode coupling tightens primordial black hole constraints', published in Phys. Rev. D91.8 (2015), p. 083521. arXiv:1411.4620 [astro-ph.CO].

The concept for this paper originated from both myself and Christian T. Byrnes. I completed all the calculations for this paper, and wrote a short code using Mathematica to calculate the primordial black hole abundances. The paper was written by myself, again with minor adjustments made by Christian T. Byrnes.

\item Sam Young and Christian T. Byrnes. `Signatures of non-Gaussianity in the isocurvature modes of primordial black hole dark matter', published in JCAP 1504 (2015), p. 034. arXiv:1503.01505 [astro-ph.CO].

The original concepts provided in this paper were formulated by myself, and I performed all of the calculations presented, making use of a code written by myself to calculate PBH abundances. The paper was written by myself, with minor adjustments made by Christian T. Byrnes.

\item Sam Young, Donough Regan and Christian T. Byrnes. `Influence of large local and non-local bispectra on primordial black hole abundance', published in JCAP 1602 (2016), p. 029. arXiv:1512.07224 [astro-ph.CO]. (The numerical simulations and plots in this paper were completed by Donough Regan.)

The idea for this paper was proposed jointly by myself and Donough Regan. The plots contained in this paper were completed by Donough Regan, with use of a short code written by myself to calculate theoretical values, and a code to produce non-Gaussian density maps written by himself. Donough Regan also wrote section 2 of the paper (section 6.2 of this thesis). The remainder of the paper was written by myself - and further minor adjustments were completed by all of the authors.

\end{itemize}

\vskip5mm
Signature:
\vskip20mm
Samuel Mark Young

\thispagestyle{empty}
\newpage
\thispagestyle{plain}

\null\vskip1mm
\begin{center}
\underline{UNIVERSITY OF SUSSEX}
\vskip0mm
\textsc{Samuel Mark Young, Doctor of Philosophy}
\vskip13mm
\Large{\textsc{Constraining the early universe with primordial black holes}}
\vskip0mm
\vskip8mm
\Large{\textsc{Abstract}}
\vskip2mm
\end{center}
\renewcommand{\baselinestretch}{1.5}
\small\normalsize
Inflation is the leading candidate to explain the initial conditions for the Universe we see today. It consists of an epoch of accelerated expansion, and elegantly solves many problems with the Big Bang theory. Non-Gaussianity of the primordial curvature perturbation can potentially be used to discriminate between competing models and provide an understanding of the mechanism of inflation.

Whilst inflation is believed to have lasted at least $50-60$ e-folds, constraints from sources such as the \emph{cosmic microwave background} (CMB) or \emph{large-scale structure of the Universe} (LSS) only span the largest $6-10$ e-folds inside today's Hubble horizon, limiting our ability to constrain the early universe. Strong constraints on the non-Gaussianity parameters exist on CMB/LSS-scales, but there are no constraints on non-Gaussianity on smaller scales. Primordial black holes (PBHs) represent a unique probe to study the small-scale early Universe, placing an upper limit on the primordial power spectrum spanning around $40$ e-folds smaller than those visible in the CMB. PBHs are also a viable dark matter candidate. 

In this thesis, the effect of non-Gaussianity upon the abundance of PBHs, and the implications of such an effect are considered. It is shown that even small non-Gaussianity parameters can have a large effect on the constraints that can be placed on the primordial curvature perturbation power spectrum - which can become stronger or weaker by an order of magnitude. The effects of super-horizon curvature perturbation modes at the time of PBH formation are considered, and it is shown that these have little effect on the formation of a PBH, but can have an indirect effect on the abundance of PBHs due to modal coupling to horizon-scale modes in the presence of non-Gaussianity. By taking into account the effect of modal coupling to CMB-scale modes, many models can be ruled out as a mechanism to produce enough PBHs to constitute dark matter.

\chapter*{Acknowledgements}
\thispagestyle{plain}

\renewcommand{\baselinestretch}{\linespacing}
\small\normalsize

First and foremost, I would like to thank my supervisor Christian Byrnes. His patience (especially his apparently infinite patience with my typos) and guidance throughout have been invaluable, and I feel lucky to have such an amazing supervisor. I would also like to thank my second supervisor David Seery for his help and support during my time at the University.

It would not have been possible to complete the research I have without my collaborators, past and present: Misao Sasaki, Donough Regan, Shaun Hotchkiss, Vanessa Smer and Ilia Musco - as well as David Sullivan. It has been a pleasure to work with these people, and I am extremely grateful for their support, as well as the rest of the faculty in the Department of Astronomy and everyone who provided helpful discussions over the years.

My office mates during my time at the University of Sussex have been a great source of enjoyment, support and encouragement, my thanks go to Gemma Anderson, Nelson Lima, Antonio Vazquez Mata, Lucia Fonseca de la Bella, Jose Pedro Pinto Vieira, Kiattisak Devasurya, and Mateja Gosenca - my thanks to all for making my working environment so enjoyable. I would like to extend my thanks to the remainder of the PhD students in the department who are sadly too numerous to name.

My thanks as well to my friends Barry Dillon and all the poker guys for their support and encouragement, as well as for making my time so enjoyable. In addition I would also like to thank everyone at the lacrosse team, the karate dojo, and who played in the weekly football game for providing a great diversion and stress relief! My family has been very supportive over the years and in particular I would like to thank my parents, Sue and Mark, and my brother Ben.

I feel lucky to have been able to study a subject I love and to have had the support of the many people involved.


\newpage
\tableofcontents
\listoffigures
\phantomsection
\addcontentsline{toc}{chapter}{List of Figures}


\pagestyle{fancy}

\newpage
\pagenumbering{arabic}



%
\chapter{Introduction}
\markboth{Introduction}{Introduction}
\label{chap:intro}
\section*{Prelude}

The study of cosmology is fascinating - the study of all that ever was or ever will be (at least on a large scale), and can connect the vast cosmological scales to the miniscule scales of particle physics. Now is a time when new discoveries are being made, new precision data is available, and yet many mysteries still remain - possibly some of the biggest questions in physics: how did the Universe begin, how will it end, and what's it all made of anyway? 

With the recent observations of the cosmic microwave background radiation, from sources such as the \emph{Planck} and WMAP satellites, it is now possible to obtain a detailed picture of the Universe as it existed over 13 billion years ago, `only' several hundred thousand years after it began. This has led to cosmological inflation becoming the accepted model for how the Universe came to be - although still begs the question of what there was before inflation and how inflation began. There are, however, still competing models to explain the origin of the Universe, and a multitude of different models for inflation itself that are all consistent with current observations. 

Primordial black holes represent a probe that can be used as a microscope to peer into the extremely early Universe and provide unique constraints. Black holes themselves are captivating, objects with gravity so strong that not even light can escape, with infinite density at their core, seemingly defying our understanding of physics. What could be more enthralling than a primordial black hole, a black hole formed within the first fraction of a second after the dawn of the Universe? It is the author's hope that any readers of this thesis will find it as interesting to read as he did to write it (although hopefully not as stressful).

\section{Big Bang cosmology}

Prior to the $20^{th}$ century, very little was known about the Universe as a whole. One of the first important observations was the darkness of the night sky - which has a seemingly obvious explanation that the Earth is blocking the light from the Sun. However, Olbers' paradox, also called the dark night sky paradox, tells us that the Universe could not be static, infinite and eternal. If this was the case then the infinite universe would be filled with an infinite number of stars, and any line of sight from the Earth would terminate at the (very bright) surface of a star - meaning that the night sky should be completely bright, in obvious contradiction to the observed dark night sky. Whilst many solutions to this paradox exist, including a steady-state universe (where the expansion of the universe causes a red-shift of light from distant stars, meaning the total flux of light reaching the Earth is finite) and a fractal distribution of stars (such that some regions of the sky contain no stars, even though the number of stars is infinite), the correct (or at least the currently accepted) argument is the finite age of the Universe - meaning light from distant stars has not yet had time to reach us.

It is only relatively recently the next piece of evidence was discovered by Slipher \cite{Slipher:1913,Slipher:1917}, who investigated the radial velocities of galaxies (though at the time they were referred to as nebulae). This was later confirmed by (and the discovery is often attributed to) \cite{Hubble:1929}, who formulated the famous Hubble law, relating the recessional velocity $v$ of galaxies to their distance from the Earth $r$ using the Hubble constant $H_0$,
\begin{equation}
v=H_0 r.
\label{eqn:hubble}
\end{equation}
Hubble's law tells us that distant galaxies are moving away faster, and that as you go further into the past the Universe was denser and hotter - eventually reaching a singularity in the distant past (now believed to be around $13.7$ billion years ago) into which the whole of creation was compressed. This theory came to be known as the \emph{Big Bang theory}, and the initial singularity as the Big Bang.

What follows is a brief summary of the history of the Big Bang universe: shortly following the Big Bang, the Universe was filled with a quark plasma (sometimes referred to as the primordial particle soup), and would then rapidly undergo several phase transitions as the temperature dropped. During baryogenesis, the quarks bound together to form hadrons - protons and neutrons - forming charged plasma. After this, as the Universe cooled further, the protons and neutrons combine to form atomic nuclei during nucleosynthesis, consisting mainly of hydrogen and helium, before the nuclei and electrons combine to form neutral elements in the epoch of recombination. What followed is known as the cosmic dark ages, during which time the Universe was filled by an expanding, neutral gas. Gravitational interactions eventually pulled the diffuse gas together to form stars and galaxies, and once the first stars ignited, in the epoch of re-ionisation, the neutral gas was re-ionised by radiation from those stars. 

\begin{figure}[t]
\centering
	\includegraphics[width=\linewidth]{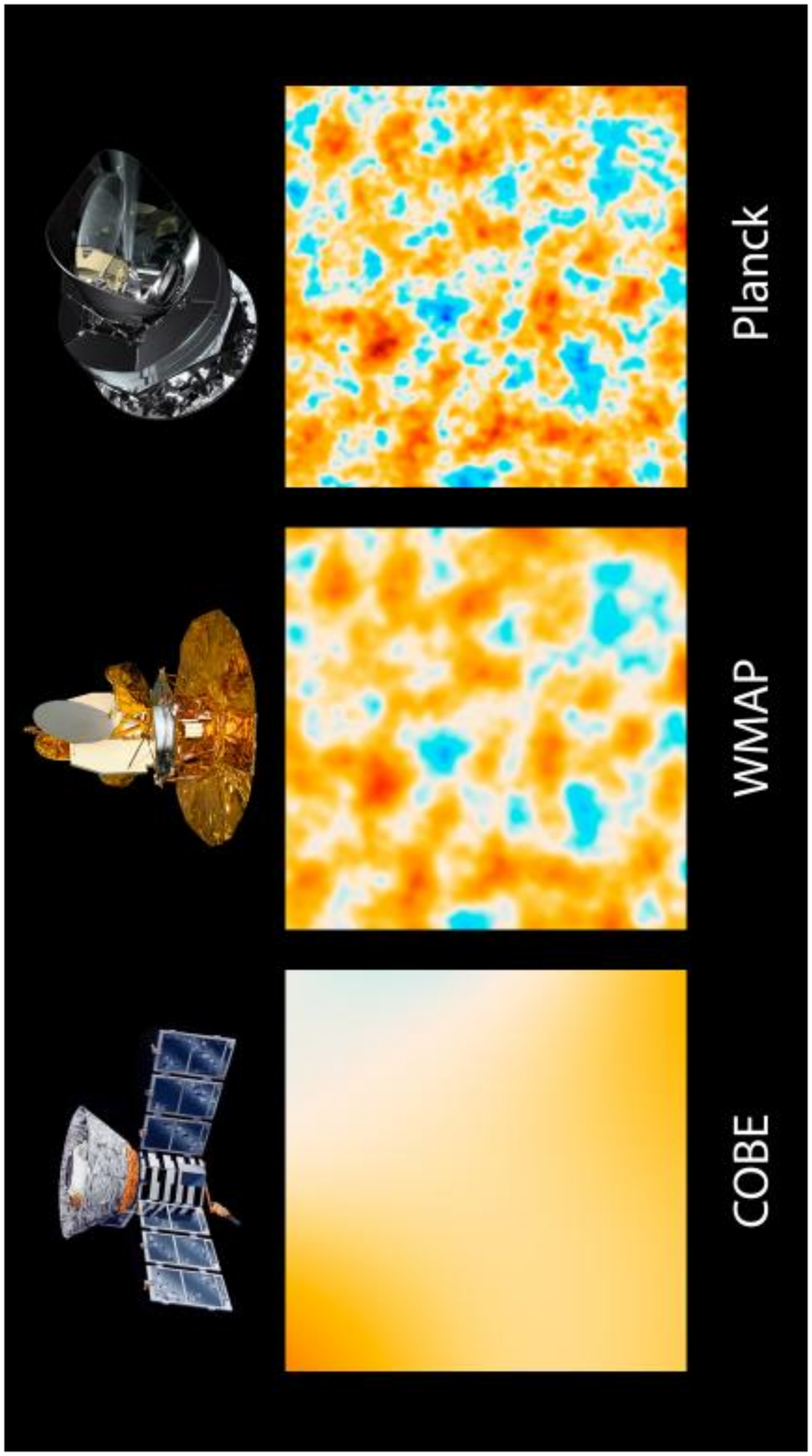}
 \caption[Satellite telescopes of the CMB]{Observations of the CMB by satellite telescopes. Improvements in technology over the years have led to higher resolution images and constraining power. Image credit: NASA/JPL-Caltech/ESA.}
\label{fig:satellites}
\end{figure}

Possibly the most compelling evidence for the Big Bang is the existence of the cosmic microwave background radiation (CMB). When the Universe was very young, it was filled with a hot, (almost) uniform plasma, and was opaque to photons - any free photons were almost immediately absorbed by the plasma before being re-emitted. We are thus unable to see any photons from this time. However, as the Universe cooled, several hundred thousand years after the Big Bang, in the epoch of recombination the charged particles combined to form a neutral gas - photons were now decoupled from the matter content of the Universe and free to propagate without interference. Whilst the Universe was very hot at this time, the expansion of the Universe and cosmic redshift means that the photons are now observed at much lower temperatures than they were when emitted. The photons seen in the CMB are seen to come from a single 2-dimensional shell, known as the surface of last scattering. The CMB represents the oldest light in the Universe and originates from this epoch of recombination, displaying an almost perfect black body spectrum and isotropy. As will be discussed later, the small anisotropies in the CMB are very important for cosmology. The first observation of the CMB was in $1964$ by Penzias and Wilson using the Holmdel Horn Antenna, and was accidental. At the time, they were looking for radio signals bounced off echo balloon satellites, and it was only later that the importance of their discovery became apparent. Since that time, there have been many observations of the CMB, including the Cosmic Background Explorer (COBE) launched in $1989$, the Wilkinson Microwave Anisotropy Probe (WMAP) launched in $2001$, and the \emph{Planck} satellite in $2009$. Figure \ref{fig:satellites} shows the relative resolution of the CMB observed by these 3 satellites.

Another successful prediction of the \emph{Big Bang theory} is the abundance of different elements in the Universe. Before nucleosynthesis, atomic nuclei (consisting of more than a single proton or neutron) were disrupted by high energy photons and were therefore unstable. Knowing how the temperature changed over time, the relative abundances of protons and neutrons at the start of nucleosynthesis can be calculated by simple thermodynamic arguments - with roughly $7$ protons to every neutron. More protons than neutrons are formed due to the lower mass of the proton. The protons and neutrons then combined to form different amounts of elements within the Universe - made up of $74\%$ hydrogen, $24\%$ helium, and trace amounts of heavier elements. The majority of heavier elements we see today were then created by fusion inside stars and spread throughout the Universe during their violent deaths. The abundances of elements predicted by theory matches very closely with the observed abundances of such elements in the Universe.

Some of the formalisms and mathematics that will be used throughout this thesis will now be introduced, as well as a more detailed mathematical discussion of the history of the Universe. Throughout this thesis, natural units will be used, such that the speed of light $c=1$. The Friedmann-Lema\^itre-Robertson-Walker (FLRW) metric is often used to describe a flat expanding Universe,
\begin{equation}
\d s^2=\d t^2+a^2(t)\left(\frac{1}{1-K a^2(t)\chi^2}\d\chi^2+\chi^2 \d\theta^2+\chi^2 \sin^2(\theta)\d\phi^2\right),
\label{eqn:metric}
\end{equation}
where $t$ is the time (sometimes expressed in terms of conformal time, $\mathrm{d}\eta=\mathrm{d}t/a(t)$), $a(t)$ is known as the scale factor, $\chi$ is the comoving distance, $K$ represents the spatial curvature, and $\theta$ and $\phi$ are radial coordinates. The scale factor represents the expansion of the Universe, and is typically defined to have a value of $1$ today - and was smaller in the past. Objects which are moving with the expansion of the Universe, and do not have any \emph{peculiar} velocity, therefore have fixed comoving coordinates. 

The expansion history of the Universe can be described by knowing the time dependence of $a(t)$. This can be achieved by considering Einstein's field equations and first deriving the Friedmann equation,
\begin{equation}
H^2(t)=\left(\frac{\dot{a}(t)}{a(t)}\right)^2=\frac{8\pi G \rho}{3}-\frac{K}{a^2(t)}+\frac{\Lambda}{3},
\label{eqn:Friedmann}
\end{equation}
where $H(t)$ is the Hubble parameter (as seen previously in equation (\ref{eqn:hubble})), $G$ is Newton's gravitational constant, $\rho$ is the energy density of the Universe, $\Lambda$ is known as the cosmological constant, and the dot represents a derivative with respect to time $t$. The cosmological constant is a free parameter (although constrained by observations) allowed by Einstein's equations, which is believed to be responsible for the observed late-time acceleration of the Universe.

The density $\rho$ of the Universe is generally taken to have two main components, radiation and matter, $\rho=\rho_{r}+\rho_{m}$. The radiation component consists primarily of photons and neutrinos (which travel at relativistic speeds if light enough). The matter component consists not only of ``ordinary'' baryonic matter (taken to consist in cosmology of protons, neutrons and electrons) but also a relatively unknown ``dark matter''. The presence of dark matter can be inferred from galaxy rotation curves, which indicate the presence of an unobserved diffuse matter in order to account for the apparent gravity, as well as being required to explain the speed at which galaxies formed in the early Universe. However, little is currently known about what dark matter is made of, and there are many models. The fact that we can't see it strongly suggests that it does not interact with the electromagnetic force, indicating that it is not baryonic. Secondly, dark matter must be ``cold'' (or at least not very warm) or the velocity dispersion of the dark matter particles would have disrupted galaxy formation. The simplest model of the Universe is referred to as $\Lambda CDM$ - so that in addition to the known components of matter and radiation, the energy density of the Universe has components of the cosmological constant $\Lambda$ and cold dark matter (CDM).

The Friedmann equation is often written in terms of the density parameter $\Omega$, given by $\Omega=\rho/\rho_c$, where $\rho_c$ is the critical density for which the Universe would be flat, $K=0$, given by 
\begin{equation}
\rho_c=\frac{3H^2}{8\pi G},
\end{equation}
where $\rho$ contains all contributions to the energy density including the cosmological constant. The cosmological constant and curvature terms can also be expressed in terms of the density parameter, and the Friedmann equation becomes
\begin{equation}
1=\Omega_{m}+\Omega_{r}+\Omega_\Lambda+\Omega_K,
\end{equation}
where $\rho_\Lambda=\frac{H^2 \Lambda}{8\pi G}$, and $\rho_K=\frac{-3H^2 K}{8\pi G}$. Current observational values from the \emph{Planck} satellite are $\Omega_m\approx0.31$, $\Omega_r\approx9\times10^{-5}$, $\Omega_\Lambda\approx0.69$, and $\Omega_K=(0\pm5)\times 10^{-3}$ \cite{Ade:2015xua}.

The fluid equation (also referred to as the continuity equation) can also be derived from Einstein's equations, and is needed to calculate the history of the Universe,
\begin{equation}
\dot{\rho}=-3H\left(\rho+p\right),
\label{eqn:fluid}
\end{equation}
where $p$ is the pressure. The pressure is often related to the density using the equation of state $\omega=p/\rho$. Equation (\ref{eqn:fluid}) can be used to relate the energy density to the scale factor as
\begin{equation}
\rho \propto a^{-3(1+\omega)}.
\end{equation}
Exact solutions are easily available when the universe is dominated either by radiation, matter, or the cosmological constant. When combined with the observed values today, this can then be used to determine the expansion history of the Universe.

\emph{Matter-domination} - on cosmological scales, matter is taken to have negligible pressure, and so $\omega=0$. The Universe has been matter dominated for most of its history, and it is only during this period that the formation of observed structure in the Universe is possible. The fluid equation, (\ref{eqn:fluid}) tells us 
\begin{equation}
\rho_{m}\propto a^{-3},
\end{equation}
As expected, the matter density scales inversely proportional to the volume of the Universe. The Friedmann equation, (\ref{eqn:Friedmann}), then gives the time dependence of the scale factor,
\begin{equation}
a(t)\propto t^{2/3}.
\end{equation}

\emph{Radiation-domination} - radiation pressure is related to the density by a factor $\omega=1/3$. Again, the fluid and Friedmann equations are solved to yield the following results:
\begin{equation}
\rho_{r} \propto a^{-4},
\end{equation}
\begin{equation}
a(t)\propto t^{1/2}.
\end{equation}
The Universe was initially radiation dominated, but quickly became dominated by the matter component because the radiation density decreases much faster than the matter density.

\emph{Cosmological constant domination} - also known as \emph{de Sitter space}, the cosmological constant is, unsurprisingly, constant and does not vary with time, corresponding to $\omega=-1$. Neglecting the density and curvature terms, the Friedmann equation predicts exponential expansion (or contraction for negative $\Lambda$)\footnote{Or the trivial solution, $H=\Lambda=0$},
\begin{equation}
a(t)\propto \exp(H t),
\end{equation}
where the Hubble parameter is, in this case, constant. Current observations indicate that the Universe is now entering a phase of domination by the cosmological constant - resulting in the expansion of the Universe accelerating. Evidence for this first came from the observation of type 1a supernovae \cite{Riess:1998cb}, where distant supernovae were found to be fainter than predicted - implying that the expansion is accelerating. At the time, this result was quite surprising as it had been expected that the expansion of the Universe would be slowing down. Since then, the fact that the Universe is accelerating has been corroborated by other evidence - and observations of the CMB, large-scale structure and gravitational lensing are all consistent with the existence of a cosmological constant. However, whilst it is known that the expansion of the Universe is now accelerating, the cosmological constant is only one possible explanation, though it is the simplest.

\begin{figure}[t]
\centering
	\includegraphics[width=\linewidth]{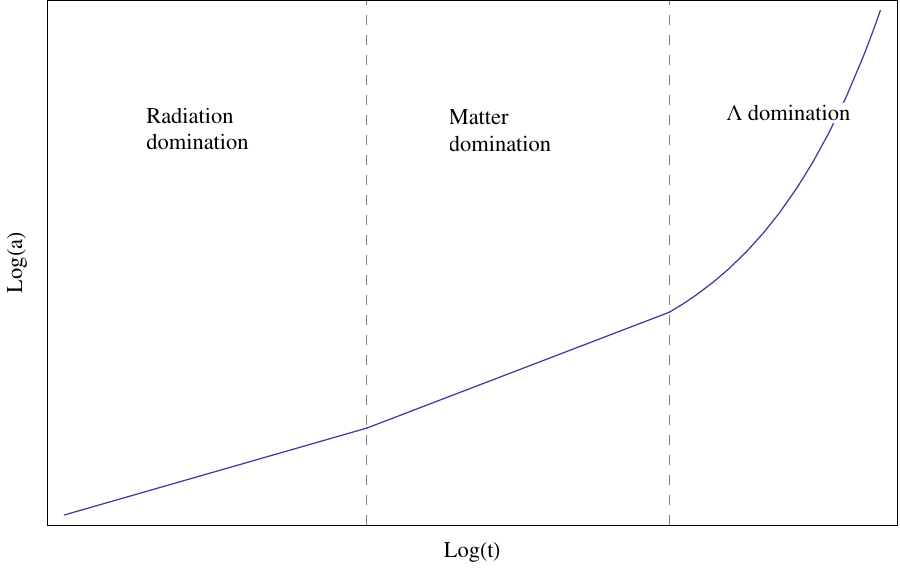}
 \caption[Expansion history of the Universe]{The expansion history of the Universe through the epochs of radiation-domination, $a\propto t^{1/2}$, matter-domination, $a\propto t^{2/3}$, and into the epoch of cosmological constant domination, $a\propto \exp(t)$.}
\label{fig:expansionHistory}
\end{figure}

Combining these results gives the final form of the Friedmann equation:
\begin{equation}
\frac{H^2(t)}{H_0^2}=\Omega_{r,0}a^{-4}(t)+\Omega_{m,0}a^{-3}(t)+\Omega_{k,0}a^{-2}(t)+\Omega_{\Lambda,0},
\end{equation}
where the subscript $0$ denotes today's value. This equation uses $\rho_m\propto a^{-3}$ even when the Universe is not matter dominated, as well as the other relations between the different densities and their dependence on the scale factor. Figure \ref{fig:expansionHistory} shows a schematic diagram of the size of the Universe and its expansion from shortly after the Big Bang till the present day.

We now have a (mostly) complete cosmological history of the Universe from shortly after the Big Bang up to the present - a brief (in cosmological terms) period of radiation domination during which the Universe underwent several phase transitions, followed by matter domination, and eventual cosmological constant domination. However, whilst the Big Bang theory successfully explains the evolution of the Universe, it fails to explain how it all began. As one looks further back into the past at higher temperatures and energy scales, our understanding of physics breaks down and we are not able to describe what happens at very early times including the initial singularity.




\subsection{Problems with the Big Bang theory}

In addition to the already mentioned inability to describe the initial singularity, there are several observations in apparent contradiction with predictions of the Big Bang theory, detailed below. It is, however, worth noting that these observations can be incorporated into the Big Bang theory by specifying extremely fine-tuned initial conditions - although this is not a particularly appealing or natural solution.

\subsubsection{The flatness problem}

The Universe is observed today to be flat, or at least very close to it. In the Friedmann equation, $K$ represents the spatial curvature, with a positive, negative or zero value representing a closed, open or flat universe respectively. In the absence of a cosmological constant, equation (\ref{eqn:Friedmann}) can be rewritten as
\begin{equation}
\rho a^2-\frac{3a^2 H^2}{8\pi G}=\frac{3K}{8\pi G},
\end{equation}
which can in turn be rewritten in terms of the density parameter $\Omega$
\begin{equation}
(\Omega^{-1}-1)\rho a^2=\frac{3K}{8\pi G}.
\label{eqn:flatnessProblem}
\end{equation}
Since the right-hand side of this equation contains only constants, the left-hand side must also be constant,
\begin{equation}
(\Omega^{-1}-1)\rho a^2=const.
\end{equation}
As the scale factor $a$ increases as the Universe expands, the density $\rho$ decreases - and whether the Universe is filled with radiation or matter, the density drops much faster than $a^2$ increases. The factor $\rho a^2$ therefore decreases rapidly as the Universe expands. In order for the left-hand side to remain constant, $(\Omega^{-1}-1)$ must remain exactly zero, or else grow rapidly. Therefore, for the Universe to be close to flat today, then $\Omega$ must have been extremely finely tuned to be close to unity initially - any small initial deviation from unity would be rapidly amplified resulting in a significant amount of curvature. To match the current observed flatness, $\Omega-1=0$ to 2 significant figures, the density of the Universe in the Planck era (when energies were close to the Planck scale) must have been equal to the critical density to over $60$ significant figures. The Big Bang theory offers no natural explanation as to why this should have been the case.

\subsubsection{The horizon problem}

\begin{figure}[t]
\centering
	\includegraphics[width=\linewidth]{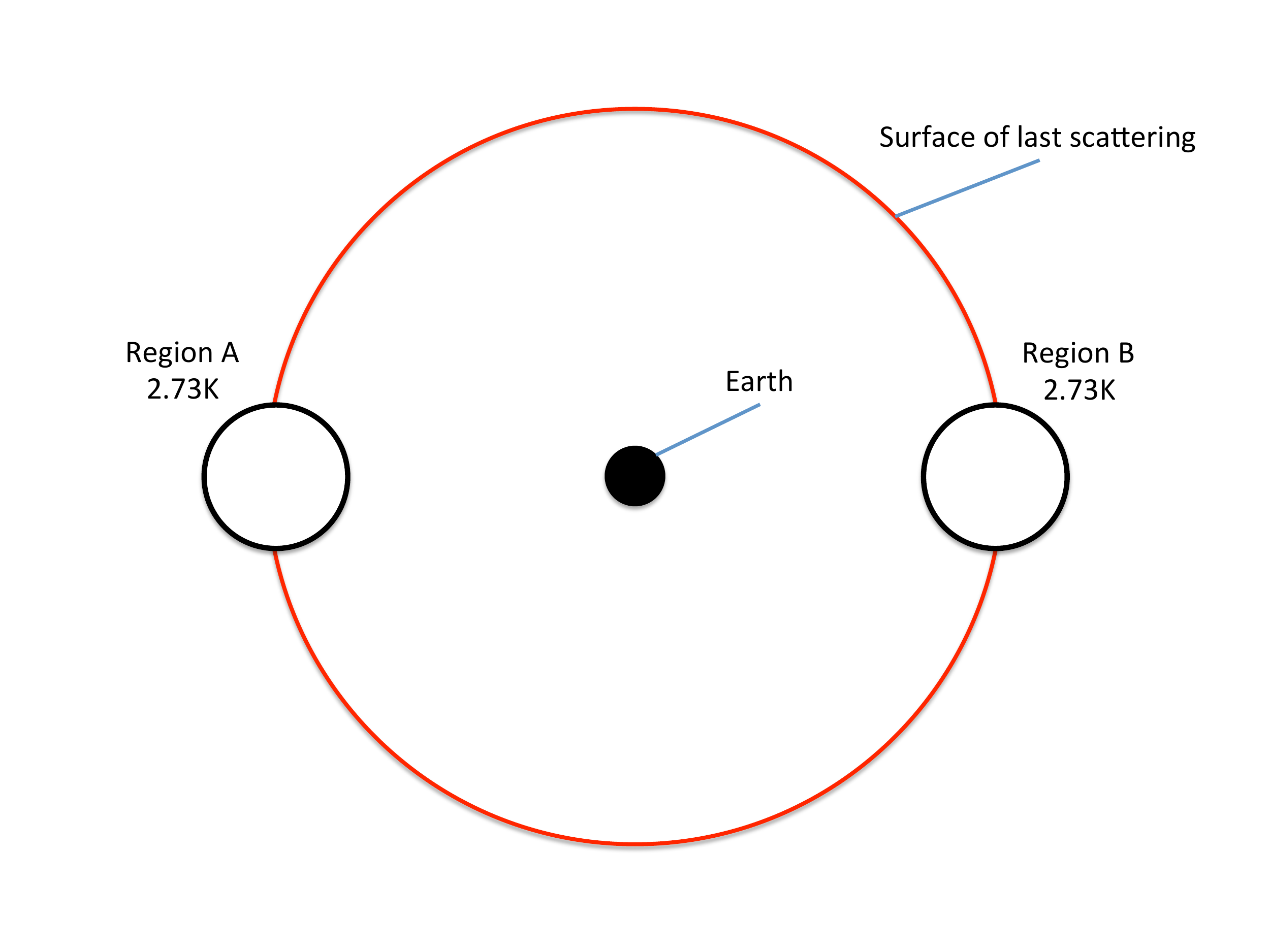}
 \caption[The horizon problem]{2 different regions on the surface of last scattering are observed, on Earth, to have the same temperature. The photons were emitted when the Universe was much younger (around $300~000$ years after the Big Bang). The smaller black circles show the particle horizon at that time (the maximum distance light could have travelled since the Big Bang). It is therefore impossible for the regions A and B to have been in causal contact, and there is no reason to expect them to be in thermal equilibrium. NB. Distances shown are not accurate.}
\label{fig:horizonProblem}
\end{figure}

When looking at astronomical objects, distances also correspond to times - the light from distant objects has been travelling for a longer time, and we are therefore looking at it at a more distant time in the past. If we now consider two galaxies in the night sky in opposite directions, each $10$ billion light years away from the Earth, then light from the first galaxy will not yet have had time to reach the second galaxy as the Universe is only $13.7$ billion years old. Each galaxy is therefore unaware of the existence of the other. The same thing applies to different regions in the CMB. The CMB is believed to have originated approximately $300~000$ years after the Big Bang, and light would only have been able to travel around $900~000$ light years in that time (this is greater than $2\times300~000$ light years due to the use of comoving units and the expansion of the Universe). It is therefore expected that two regions, A and B, on the surface of last scattering separated by a distance greater than this would not have been in causal contact at the time of decoupling (A would have been outside B's horizon - and was therefore unobservable) - and so there is no reason to expect these different regions to have reached thermal equilibrium. Figure \ref{fig:horizonProblem} shows a schematic diagram of the horizon problem. The CMB should not, then, be observed at the same temperature across the entire night sky. In contrast to this expectation, the CMB is observed to be almost perfectly isotropic, with a temperature of $2.723$K, and is uniform to $1$ part in $10^5$.

\subsubsection{The relic problem}
Also sometimes referred to as the \emph{magnetic monopole problem}, the relic problem stems from the fact that many Grand Unified Theories (GUTs) predict the existence of stable relic particles (the most famous of which is likely magnetic monopoles) and topological defects should have formed in the extremely hot and energetic early Universe. Such particles would have been produced in great abundances and would persist until today. Historically, magnetic monopoles in particular were predicted to have been produced in such numbers that they would be the dominant component of the Universe \cite{Zeldovich:1978wj,Preskill:1979zi} (more modern theories do not predict this abundance of magnetic monopoles, though other relics are still predicted depending on the model). Not only is this not the case, but also all searches for such relics have failed to detect any.

\section{Introduction to cosmological inflation}
Consisting of a (brief) epoch of accelerated expansion prior to the radiation- and matter-dominated epochs of the Universe, cosmological inflation was first proposed by \cite{Guth:1980zm} and \cite{Sato:1981ds} independently in 1981, and a revised model, dubbed ``new inflation,'' was soon proposed by \cite{Linde:1981mu} and \cite{Albrecht:1982wi}. Inflation elegantly resolves the observed problems with the Big Bang theory listed above - although still fails to explain the origin of the Universe, as the beginning of inflation is not understood. A universe dominated by some fluid with an equation of state $\omega$ will now be considered, and the values of $\omega$ required for accelerated expansion will be calculated. 

In absence of a cosmological constant, the acceleration equation (which can be calculated from the Friedmann equation (\ref{eqn:Friedmann}) and the fluid equation (\ref{eqn:fluid}), or directly from Einstein's equations) is given by 
\begin{equation}
\frac{\ddot{a}}{a}=-\frac{4\pi G}{3}\left(\rho+3p \right).
\end{equation}
In order for the acceleration of the Universe $\ddot{a}$ to be positive, the factor $(\rho+3p)$ must therefore be negative. Assuming positive energy density $\rho$, we require
\begin{equation}
w< -\frac{1}{3}.
\end{equation}
Assuming that the energy density $\rho$ is always positive, a fluid with negative pressure is required. Whilst we have already seen that the cosmological constant has an equation of state, $w=-1$, which satisfies this condition, it could not have been responsible for inflation as it is observed to be very small today and would have been strongly sub-dominant in the early universe. However, some type of fluid with $w=-1$ could have been responsible - and the energy density during inflation would have been constant for such a fluid. One of the most important consequences of inflation is the evolution of cosmological horizons, as discussed below.

\subsection{Cosmological horizons during inflation}
\label{subsec:horizons}

There are different definitions of the cosmological horizon:
\begin{itemize}
\item{The Hubble radius, $H^{-1}$, defines the boundary of causal processes as it defines the distance which light can travel during one Hubble time. The terms Hubble radius and horizon will be used interchangeably to refer to $H^{-1}$. Two points separated by less than one Hubble radius at a given time can be expected to come to thermal equilibrium.}

\item{The particle horizon is the (maximum) distance that a particle (travelling at the speed of light) can have travelled (since the start of the Universe), and represents the furthest distance at which we can retrieve information from the past. Events outside the particle horizon respective to an observer can therefore not be observed.
}

\item{The event horizon is the largest comoving distance that light emitted at a given time could \emph{ever} reach an observer in the future. If the expansion of the Universe is accelerating, then this will be a finite distance. The photons will never reach an observer further than the event horizon due to the observer's acceleration away from the source of the photons.
}
\end{itemize}

Whilst the particle horizon tells us the distance from which photons could have reached us by now, it is generally not possible to observe photons from such a distance for two reasons. The first is evident: we can only see photons from after the epoch of decoupling; the photons visible in the CMB are the oldest in the universe. The second reason is that photons from outside the Hubble radius are extremely red-shifted and have very low energies making them very quickly unobservable. 

Because it determines the scale of causal interactions, the region inside our current Hubble radius is often referred to as the ``observable'' Universe. The comoving Hubble radius is given by $(aH)^{-1}$. This region is only (a small) part of the whole Universe, which we are unable to communicate with at the moment. Comoving scales with a wavevector $k>aH$ are referred as \emph{sub-horizon}, and are in causal contact. Larger comoving scales, $k<aH$, are referred to as super-horizon, and are not in causal contact. The comoving Hubble radius during and after inflation will now be considered.

\begin{figure}[t]
\centering
	\includegraphics[width=\linewidth]{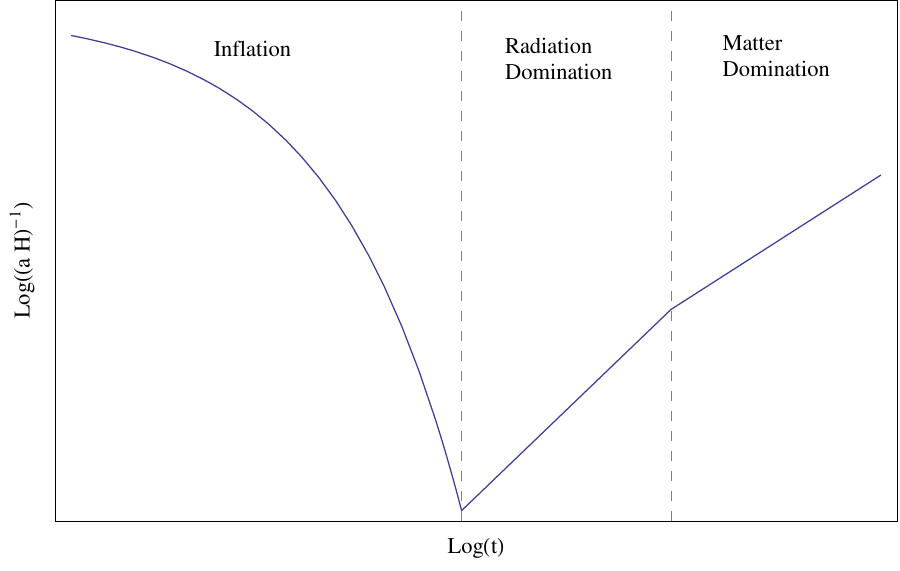}
 \caption[Comoving Hubble radius]{The evolution of the comoving Hubble radius $(aH)^{-1}$ over time. The comoving Hubble radius shrinks rapidly during inflation, before growing during the radiation- and matter-dominated epochs.}
\label{fig:horizons}
\end{figure}

During exponential inflation, $a(t)\propto \exp(H t)$, the Hubble parameter is constant, and the comoving Hubble radius shrinks as
\begin{equation}
(aH)^{-1}\propto\exp(-Ht).
\end{equation}
During radiation domination after inflation ends, the comoving Hubble radius grows as
\begin{equation}
(aH)^{-1}\propto (t-t_i)^{1/2},
\end{equation}
where $t_i$ is the time of the initial singularity assuming radiation domination for the entire history of the Universe. Likewise, during the matter dominated epoch, the horizon grows as
\begin{equation}
(aH)^{-1}\propto (t-t_i)^{1/3}.
\end{equation}
The horizon therefore shrinks during inflation, before growing during the following epochs of radiation and matter domination. Figure \ref{fig:horizons} shows the evolution of the comoving Hubble radius over time.

\subsection{How inflation solves the problems with the Big Bang}
 When considering the flatness problem, we now return to equation (\ref{eqn:flatnessProblem}). During radiation or matter domination, the factor $\rho a^2$ decreases rapidly as the Universe expands. However, during inflation, this factor now increases with time, and as the Universe expands we see
\begin{equation}
| \Omega^{-1}-1 | \rightarrow 0.
\end{equation}
Thus $\Omega=1$ is now an attractor solution, and for a large enough amount of inflation we obtain a Universe that is extremely close to flat no matter the initial curvature. Inflation therefore naturally provides an explanation for the observed flatness of today's Universe.

Inflation also solves the horizon problem as a small patch of the Universe that is initially in causal connection, and can therefore reach thermal equilibrium prior to or at an early stage of inflation, can expand to a size greater than that of the currently observable Universe. Figure \ref{fig:horizons} shows how the comoving horizon shrinks during inflation before growing during radiation- and matter-domination, meaning a comoving scale $R$ observed to be entering the horizon today was at some point in the distant past inside the horizon. All of the photons we see in the CMB were therefore once in causal contact, and were able to reach thermal equilibrium.

In addition, inflation can explain the lack of relic particles in the Universe. Because such particles are only predicted to form at extremely high energies, if inflation ends at a lower energy scale, then these particles would have been diluted by the rapid expansion of the Universe during inflation - and therefore be very difficult to detect today. However, the end of inflation can lead to the production of topological defects such as monopoles, cosmic strings and domain walls in some models. As these are not observed, viable inflationary models must not predict an over-abundance of such defects; this is discussed further in \cite{Sakellariadou:2007bv}.

In order to solve the problems the problems with Big Bang cosmology, it is believed that inflation must have lasted a minimum of $50-60$ $e$-folds \cite{Lyth:2009zz}, although there is no upper bound on the duration of inflation.

\subsection{Single scalar field inflation}

Current observations suggest the cosmological constant is very small, and so will have a negligible effect on dynamics during inflation, and the Universe rapidly approaches flatness after several $e$-folds of inflation, so for simplicity, we will therefore set $\Lambda=K=0$ when considering inflation. For inflation to occur, the dominant component of the energy density is required to have an equation of state $\omega<1/3$. Arguably the simplest suitable candidate is a single scalar field, referred to as the ``inflaton'' - though there are many models which predict the required behaviour, and which may or may not produce observably different characteristics. Some of these models will be discussed later in more detail in the context of primordial black hole formation. We will now consider an inflationary epoch dominated by a single, homogeneous scalar field $\phi$ minimally coupled to gravity. The equation of motion for the scalar field is then given by
\begin{equation}
\ddot{\phi}+3 H\dot{\phi}+\frac{\d V(\phi)}{\d\phi}=0,
\label{eqn:eqnsMotion}
 \end{equation}
where $H$ is the Hubble parameter, $V(\phi)$ is the potential of the scalar field, and a dot denotes a time derivative. The second term in the equation behaves as ``Hubble drag'' - a friction-like term that acts to slow down the evolution of the scalar field as a result of the expansion of the Universe. Assuming all other components of the Universe are negligible, the Friedmann equation can be written as
\begin{equation}
H^2=\frac{1}{3M_P^2}\left( \frac{1}{2}\dot{\phi}^2+V\right),
\label{eqn:inflationFriedmann}
\end{equation}
where $M_P$ is the \emph{Planck} mass. This can be differentiated to give,
\begin{equation}
\dot{H}=-\frac{\dot{\phi}^2}{2M_P^2}.
\end{equation}

Expressions for the density and pressure of the inflaton will now be calculated. The energy-momentum tensor $T_{\mu\nu}$ is in general given by,
\begin{equation}
T_{\mu\nu}=-\frac{\partial \mathcal{L}_M}{\partial g^{\mu\nu}}+g_{\mu\nu} \mathcal{L}_M,
\end{equation}
where $g_{\mu\nu}$ is the metric, and $\mathcal{L}_M$ is the Lagrangian density, $\mathcal{L}_M=(\nabla\phi)^2-2V(\phi)$. The energy-momentum tensor describing the inflaton is then
\begin{equation}
T_{\mu\nu}=\partial_\mu \phi \partial_\nu \phi-g_{\mu\nu}\left( \frac{1}{2}(\nabla\phi)^2 +V(\phi) \right).
\end{equation}
Comparing this to the energy-momentum tensor for a perfect fluid,
\begin{equation}
T_{\mu\nu}=(\rho+p)U_\mu U_\nu-p g_{\mu\nu,}
\end{equation}
the density $\rho$ and pressure $p$ of the inflaton are given by
\begin{equation}
\rho=\frac{1}{2}\dot{\phi}^2+V(\phi),
\end{equation}
\begin{equation}
p=\frac{1}{2}\dot{\phi}^2-V(\phi).
\end{equation}
It can be seen that the inflaton will have a negative pressure in the case that $V(\phi)>\frac{1}{2}\dot{\phi}^2$, and will have equation of state $\omega\approx -1$ if the potential energy dominates the kinetic energy
\begin{equation}
V(\phi) \gg \frac{1}{2}\dot{\phi}^2.
\end{equation}
In which case the inflaton would roll slowly down the potential, and undergo quasi-de Sitter expansion - though it is important that the kinetic term is not exactly equal to zero, as this would correspond to a cosmological constant (and pure de Sitter expansion) and inflation would not end.

In order to solve the problems with the Big Bang theory, a large amount of inflation is required - and the inflaton is required to accelerate sufficiently slowly so that inflation lasts long enough,
\begin{equation}
\bigg| \frac{\d V(\phi)}{\d \phi} \bigg| \gg |\ddot{\phi}|.
\end{equation}
The inflaton potential is therefore close to flat and relatively constant during inflation. With these assumptions, we can then simplify equations \ref{eqn:eqnsMotion} and \ref{eqn:inflationFriedmann}:
\begin{equation}
3 H\dot{\phi}+\frac{\d V(\phi)}{\d\phi}=0,
\end{equation}
\begin{equation}
H^2=\frac{1}{3M_P^2}V.
\end{equation}
The ``slow-roll'' parameters can then be defined as
\begin{equation}
\epsilon_H = -\frac{\dot{H}}{H^2},
\end{equation}
and
\begin{equation}
\eta_H = \frac{\dot{\epsilon}_H}{\epsilon_H H}.
\end{equation}
The slow-roll approximation corresponds to $\epsilon_H \ll 1$ and $|\eta_H|\ll 1$. These parameters can also be defined in terms of the potential $V$ rather than the Hubble parameter $H$
\begin{equation}
\epsilon_V = \frac{M_P^2}{2}\left( \frac{V'}{V}\right)^2,
\label{eqn:epsilon}
\end{equation}
\begin{equation}
\eta_V = M_P^2\frac{V''}{V},
\end{equation}
where the prime denotes a derivative with respect to $\phi$. The parameters are related as $\epsilon_H\approx\epsilon_V$ and $\eta_H=\eta_V-\epsilon_V$. This definition of the slow-roll parameters is useful because they provide information about the inflaton potential - the slope is described by $\epsilon_V$, and $\eta_V$ describes the curvature. Assuming that the inflaton is not initially located too near the minimum of its potential, inflation occurs as the inflaton slowly rolls down the potential, and will achieve the required amount of inflation if the potential is sufficiently flat for enough time. The simplest form of the potential is $V \propto \phi^n$, although there are many different models.

\begin{figure}[t]
\centering
	\includegraphics[width=\linewidth]{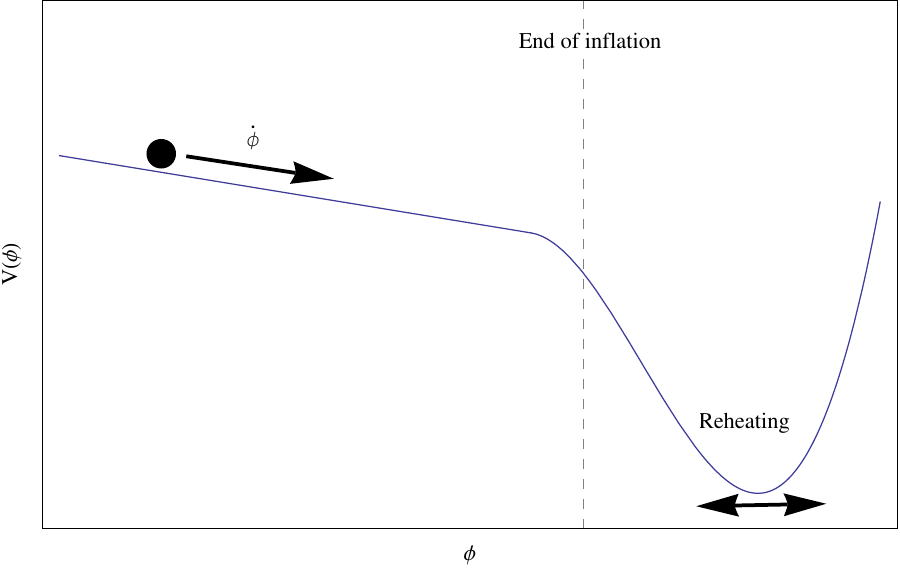}
 \caption[Potential of the inflaton]{An example of the inflaton potential. During inflation, the potential must be sufficiently flat to allow slow-roll and for enough inflation to happen. At the end of inflation, the inflaton reaches the minimum of its potential and oscillates about the minimum during reheating.}
\label{fig:potential}
\end{figure}

Inflation comes to an end when $\epsilon_H$ equals unity. The scalar field continues to roll down its potential until it reaches its minimum and then begins to oscillate about its equilibrium value. A process known as ``reheating'' now occurs. Little is known about reheating, and the precise mechanics depend upon the model being considered. In all cases the result is the recovery of the ``Hot Big Bang,'' where the Universe begins to evolve as described in the Big Bang theory - though now with a natural explanation for the initial conditions. The coupling of the inflaton to other particles causes the decay of the inflaton into standard model particles - releasing all its energy into reheating the Universe. Figure \ref{fig:potential} shows the behaviour of the potential during an epoch of slow-roll inflation followed by reheating.

\section{Cosmological perturbation theory}

Inflation as described so far provides a good description of how the Universe began, and solves the described problems with the Big Bang. However, such a universe would be perfectly homogeneous - there would be no stars, no galaxies, no planets, and ultimately, no life. However, inflation neatly resolves this issue by providing a mechanism for the generation of cosmological perturbations. The prediction of such perturbations is one of the great successes of cosmological inflation as it provides the seeds of the origin of structure in the Universe. 

What field is responsible for inflation and the production of cosmological perturbations is still an ongoing debate - the same field may have been responsible for both, or there may have been multiple fields for example. However, the majority of viable models describe scalar fields responsible for the origin of cosmological perturbations. In section \ref{sec:NG} the use of non-Gaussian statistics as a method for distinguishing between such models will be discussed, but in this section only the simplest model will be discussed. There are also alternatives to the paradigm of cosmological inflation (including, for example, bouncing cosmologies \cite{Battefeld:2014uga}), although these will not be discussed here.

\subsection{Quantum fluctuations in the vacuum}

Quantum field theory tells us that all fundamental particles are excitations of an associated field permeating all of space. Even in the absence of any particles the field still exists. If the uncertainty principle is considered, $\Delta E\Delta t > h/ 2\pi$, it can be seen that this allows an excitation of the field of energy $\Delta E$ lasting a very short time $\Delta t$ - and so the vacuum may be considered not to be empty - but contains extremely short-lived ``imaginary'' particles which disappear back into the vacuum. If the vacuum is considered over a macroscopic period of time, these particles are only measured as an effect on the zero-point energy of the vacuum. Such virtual particles typically have no physical effect (the Casimir effect being a notable exception) - although may be a possible explanation for the origin of the observed cosmological constant\footnote{However, the predicted value for such a cosmological constant is typically many orders of magnitude too large to account for the very small observed cosmological constant (i.e. \cite{Weinberg:1988cp})}. However, such quantum fluctuations can explain the origin of cosmological perturbations.

In previous sections, it was discussed that inflation ``smoothes out'' the Universe as required to solve the problems with the Big Bang theory - the horizon at the end of inflation lies inside a larger spatially flat region of the Universe which is in thermal equilibrium. However, on a quantum level, it also results in small quantum fluctuations. Quantum fluctuations $\delta\phi$ in the scalar field driving inflation, the inflaton $\phi$, are free to oscillate inside the comoving Hubble horizon - and quickly disappear back into the vacuum. However, as discussed in section \ref{subsec:horizons}, the comoving Hubble horizon shrinks rapidly during inflation - and so excited modes can quickly pass outside the causal horizon. The speed at which the wavelength of the fluctuation becomes larger than the causal horizon can be so fast that the fluctuation cannot propagate to disappear back into the vacuum. The fluctuation then ``freezes out'' with a non-zero amplitude and becomes a classical density perturbation. These modes then remain super-horizon for the remainder of inflation, and in the absence of other fields, are unable to evolve and remain constant. Once inflation ends, the comoving Hubble horizon begins to grow, and the now classical perturbations re-enter the horizon and can again begin to evolve. These perturbations are later observed in the form of anisotropies in the CMB, and eventually provide the seeds for the growth of galaxies and clusters in the Universe. The study of these perturbations is used to constrain the dynamics of inflation, and some of the earliest attempts include \cite{Guth:1982ec,Bardeen:1983qw,Sasaki:1986hm}.

\subsection{Perturbing the metric}

Perturbations to the metric will now be considered. During inflation, it will be assumed that the universe is entirely dominated by the inflaton scalar field $\phi(\mathbf{x},t)$, and decompose this into a background value $\phi_0(t)$, plus a small perturbation $\delta\phi(\mathbf{x},t)$:
\begin{equation}
\phi(\mathbf{x},t)=\phi_0(t)+\delta\phi(\mathbf{x},t).
\end{equation}
Given the very small amplitude of perturbations observed in the CMB, this is a reasonable approach. The perturbations $\delta\phi$ arise from quantum fluctuations during inflation and results in different regions of the universe having slightly different densities. Because inflation will end in a small patch of the Universe when the energy density reaches a specific value and the slow-roll parameters become large, this means that inflation ends at slightly different times in different regions of the universe, on a ``uniform-density'' hypersurface. Each small region can be considered independently in what is known as the ``separate universe approach'' \cite{Wands:2000dp} - with the perturbations to the scalar field taking effect simply as a small time shift. An observer in a small over-dense patch of the universe would observe exactly the same evolution as an observer in a small under-dense patch, merely observing it to occur slightly later. It is only when inflation ends and the two regions again come into causal contact that each observer would be able to identify the fact that they are in an over- or under-dense region. This will be an important consideration when considering the time at which primordial black holes form.

In the same manner as the scalar field, perturbations to the metric will now be considered. In this thesis, only a scalar perturbation to the spatial component of the metric will be considered - although both vector and tensor perturbations are also possible (see \cite{Lyth:2009zz}). The unperturbed component of the metric is the standard FLRW metric describing a flat universe:
\begin{equation}
\d s^2 = \d t^2+a^2(t)\mathrm{e}^{2\zeta}\d \mathbf{X}^2,
\end{equation}
where $\mathbf{X}$ represents the spatial coordinates, and $\zeta$ is referred to as the curvature perturbation in the uniform-density slicing. On super-horizon scales the curvature perturbation is often related to the gauge-invariant quantity, the comoving curvature perturbation $\mathcal{R}$ as (discussed in \cite{Lyth:2009zz})
\begin{equation}
\mathcal{R}=\zeta-H\frac{\delta\rho}{\dot{\rho}},
\end{equation}
where $H$ is the Hubble parameter, $\delta\rho$ is the perturbation in the energy density, and $\dot{\rho}$ is the time derivative of the background density. In the uniform-density slicing, there is no density contrast by definition, and $\mathcal{R}=\zeta$. In addition, in the super-horizon limit in other gauges the density perturbation rapidly shrinks to zero (this is an important consideration and is discussed further in chapter \ref{chap:paper1}) - and so in the super-horizon limit, the equality is again reached, $\mathcal{R}=\zeta$. For this reason, the symbols $\mathcal{R}$ and $\zeta$ are often used interchangeably within this thesis. Note that the sign conventions of $\zeta$ and $\mathcal{R}$ are not always consistent between different sources in the literature - throughout this thesis the sign conventions defined here are used.

\subsection{The power spectrum}

Because of the random nature of cosmological perturbations it is necessary to observe statistical measures of the perturbations. Due to (the assumptions of) statistical isotropy and homogeneity, the 2-point function is not a function of the absolute position or orientation of the spatial coordinates, but only a function of the magnitude of the separation between the two points, $\Delta x$. The power spectrum $\mathcal{P}_\zeta$ is the Fourier transform of the 2-point correlation function
\begin{equation}
\langle  \zeta(\mathbf{k}_1)\zeta(\mathbf{k}_2)\rangle=(2\pi)^3 \delta^{(3)}\left( \mathbf{k}_1+\mathbf{k}_2 \right) P_\zeta(k),
\end{equation}
where $\delta\left( \mathbf{k}_1+\mathbf{k}_2 \right)$ is a delta function. The power spectrum can be taken as a measure of the amplitude of perturbations in Fourier space. It is often useful to use the dimensionless power spectrum
\begin{equation}
\mathcal{P}_\zeta(k)\equiv\frac{k^3}{2\pi^2}P_\zeta(k),
\end{equation}
and the power spectrum is often parameterized as
\begin{equation}
\mathcal{P}_\zeta(k)=\mathcal{A}_s\left( \frac{k}{k_*} \right)^{n_s-1+\frac{1}{2}\alpha_s \mathrm{ln}\left( \frac{k}{k_*}\right)},
\end{equation}
where $\mathcal{A}_s$ is the amplitude of the power spectrum, $k_*$ is an arbitrary pivot scale, $n_s$ is referred to as the spectral index, and $\alpha_s$ is the running of the spectral index, given by $\alpha_s=\d n_s /\d\mathrm{ln} k$. Note that this equation is an expansion in $k$, and is therefore only expected to be valid for a limited range of scales. The power spectrum $\mathcal{P}_\zeta$ is said to be scale invariant if there is zero running and $n_s=1$, a blue spectrum refers to $n_s>1$ and a red spectrum refers to $n_s<1$.

For the simplest model and a Gaussian distribution, the power spectrum contains all of the statistical information - with higher-order correlation functions depending on the 2-point function. An important result here is that for a Gaussian distribution, because the 2-point correlation function is zero unless $\mathbf{k}_1=-\mathbf{k}_2$ , modes of different scales are therefore uncorrelated. All of the odd correlation functions $\langle \zeta^{n_{\textrm{odd}}}\rangle$ vanish and $\langle\zeta\rangle=0$ can be ensured as the homogeneous component of $\zeta$ can always be absorbed into the unperturbed background.

\subsection{Observational constraints}

The cleanest/simplest way to observe the power spectrum resulting from inflation is by the study of the oldest observable light in our Universe, the CMB. Over recent years there have been many observations of the CMB, the most notable of these have been the satellites COBE, WMAP and \emph{Planck} mentioned previously, although there have also been ground and balloon based observations. Successive missions have brought greater resolution of the CMB, as can be seen in figure \ref{fig:satellites}, and a corresponding increase in information available. Data from these satellites is often considered alongside data from other experiments and sources to provide the tightest constraints, such as baryonic acoustic oscillations (BAOs) and supernovae.

\begin{figure}[t]
\centering
	\includegraphics[width=\linewidth]{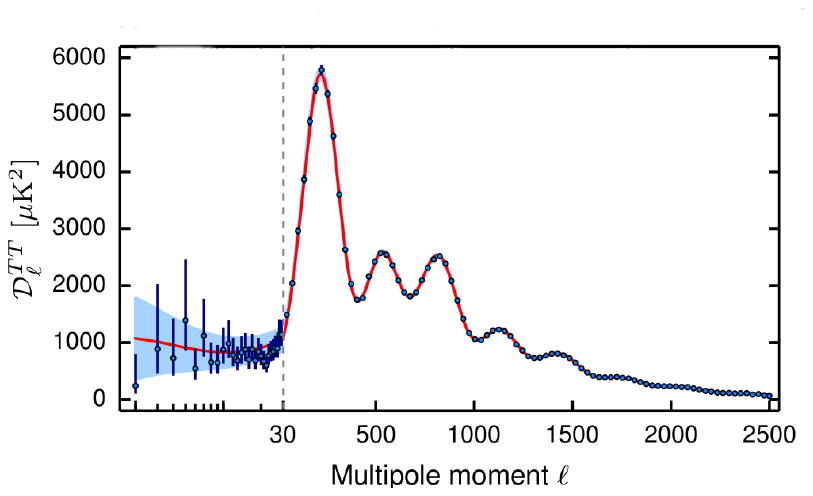}
 \caption[The power spectrum as observed by the \emph{Planck} satelite]{The power spectrum as observed by the \emph{Planck} satellite. Excellent agreement is seen with the best fit for a $\Lambda \mathrm{CDM}$ model as shown by the red line. Image credit: \cite{Ade:2015lrj}.}
\label{fig:planckPower}
\end{figure}

The cosmological perturbations sourced during inflation are seen as temperature anisotropies in the CMB, and the curvature perturbation $\zeta$ is often used for calculations during inflation because it translates well into the observed temperature anisotropies in the CMB. The 2-point correlation function are typically described by an angular power spectrum $C^{TT}_l$, where $l$ is the multipole number. As in \cite{Ade:2015lrj}, the angular power spectrum is related to the 3-dimensional power spectrum by the transfer function $\Delta_{l,T}(k)$:
\begin{equation}
C^{TT}_l=\int\limits_0^\infty \Delta_{l,T}(k)^2\mathcal{P}_\mathcal{R}(k).
\end{equation}
A larger multipole moment $l$ roughly corresponds to a larger $k$. Figure \ref{fig:planckPower} shows the constraints on the power spectrum from the \emph{Planck} satellite along with the best fit for a $\Lambda \mathrm{CDM}$ model, with numerous free parameters fitted to the data.


The data from the \emph{Planck} satellite provides a great deal of information about the Universe, although here only the parameters relevant to the primordial power spectrum will be considered. The results are presented relative to a pivot scale of $k_*=0.05\mathrm{Mpc}^{-1}$. Constraints on the power spectrum parameters given by \cite{Ade:2015lrj} are as follows: the amplitude of the power spectrum is
\begin{equation}
\mathcal{A}_s=(2.20\pm0.10)\times10^{-9},
\end{equation}
 and the spectral index
\begin{equation}
n_s=0.968\pm0.006,
\end{equation}
which represents a a $5.4\sigma$ deviation from scale invariance ($n_s=1$). The constraints on the running of the spectral index are 
\begin{equation}
\alpha_s=-0.003\pm0.007,
\end{equation}
which is consistent with zero. It is therefore a (possibly) surprising result that the entire spectrum of primordial perturbations on cosmological scales can be completely described by only 2 parameters - the amplitude and spectral index of the power spectrum.

\begin{figure}[t]
\centering
	\includegraphics[width=\linewidth]{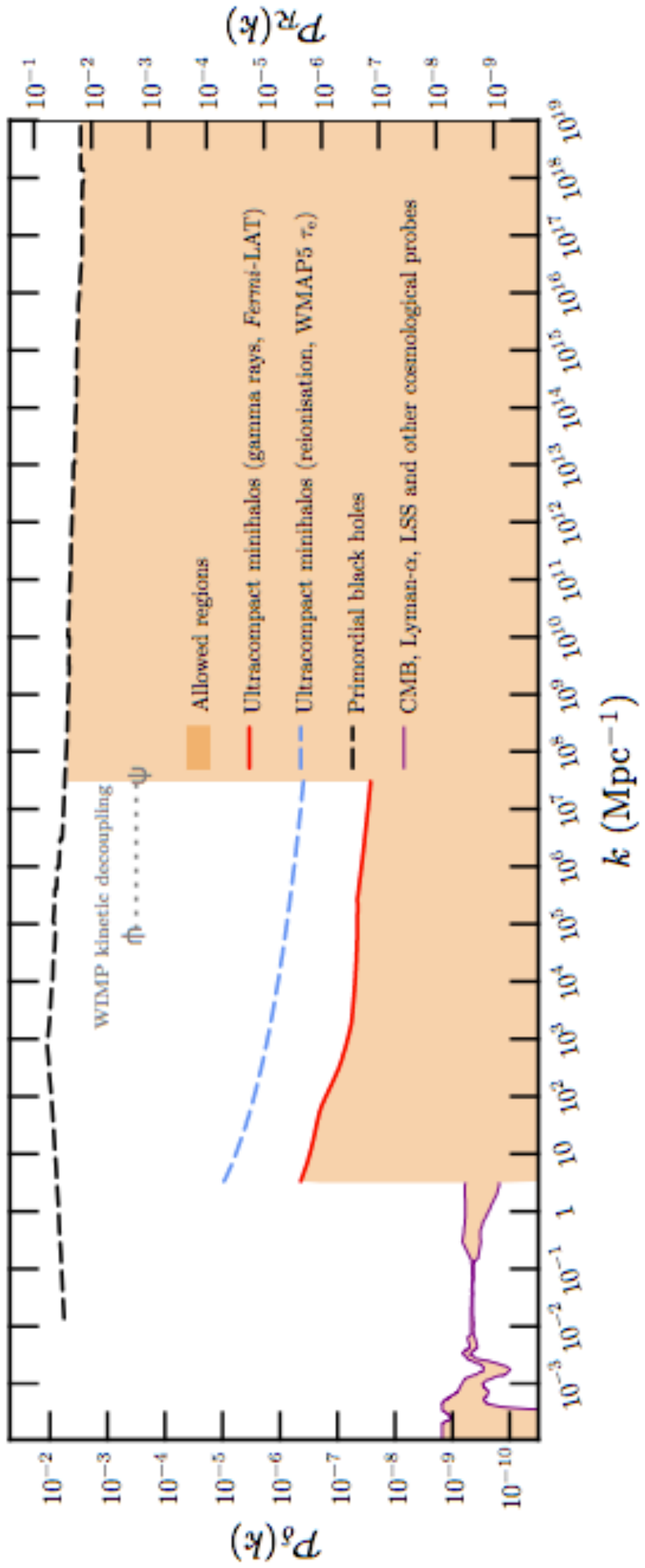}
 \caption[Constraints on the power spectrum at all scales]{Constraints on the power spectrum at all scales. Whilst constraints from the CMB (on the far left of the plot) are the strongest, they only constrain a relatively small range of scales. By contrast, constraints arising from PBHs are much weaker but span a much greater range of scales. Image credit: \cite{Bringmann:2011ut}.}
\label{fig:powerConstraints}
\end{figure}

However, constraints on the power spectrum and the primordial universe from the CMB and large scale structure only span a relatively small range of scales, probing only the largest 8-10 e-folds inside todays Hubble horizon. However, as seen previously, inflation needs to have lasted at least 50-60 e-folds in order to solve the problems with the big bang theory. Figure \ref{fig:powerConstraints} shows all of the constraints on the primordial power spectrum from all sources. Notably, constraints exist on the small-scale power spectrum from PBHs - though they are much weaker and provide only an upper bound.





\subsection{Comparison of data with single scalar field inflation}
To quickly recap, the simplest of model of inflation typically assumes:
\begin{itemize}
\item the universe is dominated by a single scalar field during inflation,
\item the field is slowly rolling, and the slow-roll parameters are therefore small,
\item the potential has a form similar to that shown in figure \ref{fig:potential},
\item minimal coupling to gravity,
\item perturbations sourced in the Bunch-Davies vacuum.
\end{itemize}

If these assumptions are true, then inflation predicts that a number of conclusions can be reached \cite{Lyth:2009zz}:
\begin{itemize}
\item the flatness problem, horizon problem and relic problem all have a natural explanation,
\item perturbations exist on scales greater than the Hubble horizon,
\item the power spectrum will be observed to be almost scale invariant,
\item very small running of the spectral index,
\item perturbations will be adiabatic, with the perturbations in the different fluids (matter, radiation) being explainable as simply a difference in the expansion of different regions of the universe,
\item the production of (possibly unobservably small) primordial gravitational waves in addition to the scalar density perturbations,
\item a Gaussian distribution of the density perturbations.
\end{itemize}

It is evident then, that current observations are consistent with this model. However, many more complicated models are also consistent with these constraints, some of which will be discussed in section \ref{subsec:PBHmodels}. Many of these models predict some amount of non-Gaussianity within the current constraints - and which may be very non-Gaussian on scales not observed in the CMB.


\section{Non-Gaussianity and higher order statistics}
\label{sec:NG}

The simplest model is not necessarily the correct one, and embedding inflation within a high energy theory, such as string theory, generally involves considering a more complicated model for inflation - and cosmological observations therefore provide an opportunity to study physics at much higher energies than is possible on Earth. Whilst there are many different models of inflation consistent with current observations, such models may also predict a small amount of non-Gaussianity, providing a tool to break the degeneracy between models of inflation. A significant detection of any non-Gaussianity would rule out the standard model of slow-roll, minimally-coupled, single-field inflation \cite{Komatsu:2009kd}.

For a Gaussian distribution, all of the statistical information is encoded in the power spectrum. However, for non-Gaussian distributions higher-order correlation functions can contain extra information. The bispectrum and trispectrum are, respectively, the Fourier transforms of the 3- and 4-point correlation functions. The bispectrum, for example, has the form
\begin{equation}
\langle\zeta(\mathbf{k}_1)\zeta(\mathbf{k}_2)\zeta(\mathbf{k}_3)\rangle=(2\pi)^3 \delta^{(3)}(\mathbf{k}_1+\mathbf{k}_2+\mathbf{k}_3)B(k_1,k_2,k_3),
\end{equation}
where the $\delta$ function ensures that the correlation function is zero unless the 3 vectors $k_n$ sum to zero - the triangle closure condition, as the momentum vectors make up the 3 sides of a closed triangle. The bispectrum is a function of 3 vectors, but different templates are often considered which peak in different triangular configurations of the momentum vectors. Common choices include the ``equilateral'' type, peaking when $k_1\approx k_2\approx k_3$, ``folded'' type, $k_1\approx2k_2\approx2k_3$, and ``squeezed'' type, $k_1\ll k_2\approx k_3$. For the majority of this thesis, only local-type non-Gaussianity, which peaks in the squeezed limit, will be considered - although other types are considered in chapter \ref{chap:paper5}. For local-type non-Gaussianity, the curvature perturbation $\zeta$ can be expressed as a Taylor series style expansion,
\begin{equation}
\zeta=\zeta_G+\frac{3}{5}\fNLloc\left(\zeta_G^2-\langle\zeta_G^2\rangle\right)+\cdots,
\label{eqn:local}
\end{equation}
where $\fNLloc$ (and higher order terms) are known as the non-Gaussianity parameter(s) and parameterize the relative importance of the higher order terms and correlation functions. The $\langle\zeta_G^2\rangle$ term is needed such that the expectation value of $\zeta$ is zero, $\langle\zeta\rangle$=0. In the local model, $\fNL$ is related to the bispectrum as
\begin{equation}
B(k_1,k_2,k_3)=\frac{6}{5}\fNLloc\left( P(k_1)P(k_2)+P(k_1)P(k_3)+P(k_2)P(k_3) \right).
\end{equation}
However, higher-order terms in the expansion (\ref{eqn:local}) can also be considered,
\begin{equation}
\zeta=\zeta_G+\frac{3}{5}\fNLloc\left(\zeta_G^2-\langle\zeta_G^2\rangle\right)+\frac{9}{25}g_{\mathrm NL}^{\mathrm local}\zeta_G^3+\frac{27}{125}h_{\mathrm NL}^{\mathrm local}\left(\zeta_G^4-\langle\zeta_G^4\rangle\right)+\frac{81}{625}i_{\mathrm NL}^{\mathrm local}\zeta_G^5+\cdots,
\end{equation}
although such higher order terms are often neglected due to the very small nature of cosmological perturbations. However, such terms can be relevant for PBHs, as discussed in chapter \ref{chap:paper2}.

The presence of non-Gaussianity affects the skewness and kurtosis of the probability density function for the primordial perturbations (see chapter \ref{chap:paper2}) - and can result in more or less structure/PBHs forming in the Universe even if the power spectrum remains unchanged. In addition, modal coupling between modes of different scales is also introduced when non-Gaussianity is considered - indeed, non-Gaussianity is often synonymous with modal coupling. Chapters \ref{chap:paper3} and \ref{chap:paper4} discuss further the implications of modal coupling in the context of PBH formation.



\subsection{Current observational bounds on non-Gaussianity}

By searching for a bispectrum or trispectrum in observations of the CMB, bounds can be placed on the non-Gaussianity parameters. The strongest current constraints are from the \emph{Planck} satellite \cite{Ade:2015ava} and for $\fNL$ the constraints at the $68\%$ confidence level are
\begin{equation}
\fNLloc=0.8\pm5.0,
\end{equation}
\begin{equation}
\fNLeq=-4\pm43,
\end{equation}
\begin{equation}
\fNLorth=-26\pm21,
\end{equation}
and in the local model, $g_{\mathrm NL}^{\mathrm local}$ is constrained as
\begin{equation}
g_{\mathrm NL}^{\mathrm local}=(-9.0\pm7.7)\times 10^4.
\end{equation}
Whilst the current bounds are consistent with zero non-Gaussianity, it is worth considering the following 2 facts: 
\begin{itemize}
\item Many inflationary models predict $\fNL$ of order unity, so are within the current bounds. It is theoretically possible to obtain much tighter constraints, $\Delta\fNL=\mathcal{O}(0.1)$, from future surveys such as SPHEREx and Euclid which will observe the distribution of matter and large scale structure. Such a measurement can be used to distinguish between competing models. 
\item The constraints on the non-Gaussianity parameters are only applicable to the scales observed in the CMB. There are no constraints on non-Gaussianity on smaller scales relevant to PBHs.
\end{itemize}

\section{Primordial black holes}

Primordial black holes (PBHs) are black holes that may have formed in the very early Universe. Whilst there are several different formation mechanisms, for example from cosmic strings (e.g. \cite{Hawking:1987bn}) and bubble collisions (e.g. \cite{Hawking:1982ga}), only PBHs which form from the collapse of large density perturbations will be considered in this thesis (see section \ref{subsec:PBH formation}). Such perturbations are generated during inflation, and quickly become super-horizon, see section \ref{chap:paper1}, and then once inflation ends, the horizon begins to grow and perturbations begin to reenter the horizon. If the density perturbation is large enough, it will then collapse almost immediately to form a PBH. Black holes formed from the collapse of stellar objects have a minimum mass, around 1.4 solar masses, as they need to be massive enough for gravity to overcome the neutron degeneracy pressure \cite{Chandrasekhar:1931ih}. PBHs on the other hand can form with almost any mass because at the time of formation the universe is much denser - PBHs typically form before the creation of neutrons and there is no need to overcome the neutron degeneracy pressure. However, PBHs do have a hypothetical minimum mass equal to the Planck mass, $m_{Pl}\approx10^{-5}\mathrm{g}$, for which the black hole radius is equal to the Planck length. They will typically form with roughly the horizon mass at the time of formation, with mass related to the time of formation as \cite{Green:2014faa}
\begin{equation}
m_{PBH}\approx m_H\approx10^{15}\left( \frac{t}{10^{-23}\mathrm{s}} \right)\mathrm{g}.
\end{equation}

Whilst PBHs have not been detected, there are upper bounds of the abundance of PBHs, and many different methods have been used to search for PBHs of different masses, as discussed in detail in section \ref{subsec:PBH search}. As realised by \cite{Hawking:1974rv}, black holes give out thermal radiation, known as Hawking radiation, and will eventually evaporate (possibly leaving behind Planck mass relics\footnote{At small masses, the Compton wavelength (which represents the smallest distance at which a mass can be localised) begins to exceed the Schwarzschild radius - and no black hole can be described. The minimum mass of a black hole is therefore approximately the Planck mass. \cite{Lee:2002bd} discusses a thermodynamic bound on the minimum mass of PBH which can form - again finding that the minimum mass PBH that can form is approximately equal to the Planck mass.}). The lighter the black hole is, the more radiation it emits and the faster the black hole evaporates. PBHs forming with a mass lighter than $m_{PBH}\approx10^{15}\mathrm{g}$ will be have evaporated by today, and their abundance can be constrained by looking for the effects of the radiation from their evaporation upon the Universe. More massive PBHs will persist until today, and their abundance is typically constrained by their gravitational effects on their surrounding.

PBHs have most often been used in cosmology because of their ability to constrain the small scales of the early universe, as can be seen in figure \ref{fig:powerConstraints}. Because the mass of a PBH depends upon the horizon scale at the time of formation, different mass PBHs can be related to specific scales in the early Universe. The abundance of PBHs depends on the primordial power spectrum, as detailed in \ref{chap:paper1}, and constraints on the abundance of PBHs can therefore be used to place constraints on the primordial power spectrum - and many calculations typically assume a Gaussian distribution (i.e. \cite{Josan:2009qn,Green:1997sz}). Whilst the constraints from PBHs span a much larger range of scales than constraints from the CMB, they are typically much weaker - as again seen in figure \ref{fig:powerConstraints}.

The possible existence of PBHs can also have other implications - they can provide a mechanism for reheating the Universe following the end of inflation which requires only gravitational interactions \cite{Hidalgo:2011fj}, and provide the seeds for the growth of super-massive black holes and galaxies \cite{Duechting:2004dk,Mack:2006gz,Khlopov:2008qy}. Perhaps most interestingly, they are also a viable candidate for dark matter (\cite{Hawking:1971ei} first proposed the idea of a large number of low mass gravitationally collapsed objects) - and could comprise the majority of the matter in the Universe, although there is only a narrow range of mass scales for which this could be the case, as seen in figure \ref{fig:abundanceConstraints}.

\subsection{Primordial black hole formation}
\label{subsec:PBH formation}

The possible formation of PBHs was first postulated by \cite{Carr:1974nx}. The authors considered a spherically symmetric perturbation, and by considering that in order to collapse the perturbation must be larger than its Jeans length, obtained a minimum amplitude for the density perturbation which could form PBHs. This is stated in terms of the density contrast $\delta$, the ratio of the density perturbation $\delta\rho$ to the background density $\bar{\rho}$,
\begin{equation}
\delta=\frac{\delta\rho}{\bar{\rho}}.
\end{equation}
It was found that the critical value for the density contrast, given at the time of horizon crossing, is equal to the equation of state, $\delta_c=\omega$ (and $\omega=\frac{1}{3}$ during radiation domination). If the density contrast is greater than this value, then the perturbation would collapse to form a PBH.

In order to calculate the abundance of PBHs forming in the early universe, it is important to know the minimum amplitude of perturbation necessary to form a PBH, referred to as the critical, or threshold, value - and as will be seen in chapter \ref{chap:paper1}, the abundance of PBHs depends exponentially on this quantity. Normally stated in terms of the density constrast, this is typically calculated in the linear regime (although it is necessary for calculations of PBH formation themselves to be highly non-linear), which greatly simplifies the calculation of their abundance. There has been extensive study of PBH formation - including simulations and analytic calculations. Most simulations of PBH formation have numerically evolved a perturbation in the density contrast to investigate PBH formation, although notably \cite{Shibata:1999zs} investigated a metric perturbation. In order to simplify the calculations, spherical symmetry is normally always assumed (although \cite{Kuhnel:2016exn} recently considered ellipsoidal collapse). A hoop conjecture is typically used to determine whether or not a PBH has formed (which is again greatly simplified by the assumption of spherical symmetry), by searching for a radius where the mass contained within a sphere of such radius is smaller than the Schwarzschild radius.

In the context of PBH formation, a density perturbation that is initially super-horizon is considered, and in the super-horizon limit is small and can be treated as a linear perturbation. \cite{Polnarev:2006aa} describes the quasi-homogeneous solution, which is used to define the initial conditions from a spatial curvature profile in their simulations. As the universe evolves and the horizon grows, the perturbation grows and quickly becomes non-linear. 

Once the perturbation reenters the horizon, it will typically either quickly collapse to form a PBH or dissipate - although an exception to this is the phenomena of critical gravitational collapse. In this situation, if the density contrast is very close to critical, PBH formation can be a drawn out process, lasting many Hubble times. As the object collapses, the outer layers are expelled, resulting in a self-similar solution that is always on the verge of forming a black hole - discussed in detail in \cite{Musco:2012au,Hawke:2002rf}. Many simulations have noted a powerful outgoing shock immediately following the formation of a PBH, for example \cite{Niemeyer:1999ak} and \cite{Musco:2004ak} - although this is not always seen, as in \cite{Musco:2008hv}. The presence of such a shock means that a new PBH is surrounded by an underdense region - and there is not significant accretion of matter onto the black hole \cite{Niemeyer:1999ak}. 

The possibility that a small-scale perturbation superposed on a large-scale perturbation could result in the double formation of PBHs was considered by \cite{Nakama:2014fra}, finding that the presence of one mode has little effect on the other if the separation in scales is large enough. Such a process would result in the smaller PBH being ``eaten'' by the larger PBH as it forms resulting in fewer PBHs, although this effect can be neglected if PBHs are sufficiently rare in the early universe and modal coupling can be neglected.

There is generally good agreement on the critical value, $\delta_c\approx0.45$ on comoving hypersurfaces, although this value depends on the initial profile of the perturbation. Most papers investigate several different shapes for the initial perturbation and calculate a critical value for each profile, although \cite{Nakama:2013ica} performed an extensive analysis of the effect of different initial profiles. Different spherically symmetric configurations, parameterized by 5 variables, were evolved numerically to determine if a PBH would form. It was found that the formation or non-formation of a PBH could be determined by two ``master'' parameters corresponding to the averaged overdensity in the center and the width of the transition region at the outer edge of the perturbation.

The formation of PBHs has also recently been investigated analytically. \cite{Harada:2013epa} considered a spherically symmetric compensated top-hat profile, such that the central overdensity is compensated for by a surrounding underdensity resulting in a flat universe. A critical value was found given by:
\begin{equation}
\delta_c=\frac{2}{3}\mathrm{sin}^2\left( \frac{\sqrt{3}\pi}{6} \right)\approx 0.4135,
\end{equation}
which is consistent with the values obtained from simulations. 

The mass of the PBH formed is typically of order the horizon mass, $m_H$, at the time of formation, although it has been noted in several papers (i.e. \cite{Hawke:2002rf, Niemeyer:1997mt, Musco:2012au}) that the mass of the resulting PBH, $m_{PBH}$, depends upon the initial amplitude of the density contrast, following a scaling law
\begin{equation}
\frac{M_{PBH}}{M_H}=K\left( \delta-\delta_c\right)^\gamma.
\end{equation}
The exact values for $K$ and $\gamma$ vary and depend on the shape of the initial profile, but are given approximately by $K=4$ and $\gamma=0.35$ \cite{Musco:2008hv}. In this thesis it will be assumed for simplicity that PBHs form with the horizon mass, as discussed in section \ref{sec:PBHmass}, although \cite{Kuhnel:2015vtw} discusses the implications of such a scaling law upon the calculated abundance of PBHs and constraints on the power spectrum.

The physical scale of the PBH formed will also be roughly equal to the horizon scale at the time of reentry. Whilst an exact calculation requires accounting for the non-linear evolution of a perturbation which forms a PBH, a relatively simple calculation can be used to estimate the physical size of the PBH. In order to collapse and form a PBH, a region must be overdense - but for simplicity a ``perturbation'' which has exactly the critical density, $\rho=\frac{3H^2}{8\pi G}$ (such that $\Omega=1$), will be considered. At horizon entry, the scale of the perturbation is equal to the Hubble radius, $r_H=H^{-1}$. The mass contained within the (spherical) horizon is therefore given by $m_H=\frac{4}{3}\pi r_H^3\rho=(2HG)^{-1}$. Assuming that exactly $100\%$ of the horizon mass falls into the PBH, the size of the black hole is then given by the Schwarzschild radius, $r_s=2GM=H^{-1}$ - and the physical scale of the resulting PBH is therefore equal to the horizon scale at the time of re-entry. Note that this is only an order of magnitude calculation - and has not accounted for the  overdensity necessary to form a PBH, non-linear evolution of the horizon due to the horizon, and the exact amount of matter falling into the PBH.



 

\subsection{The search for primordial black holes}
\label{subsec:PBH search}

Many methods have been used to search for direct or indirect evidence of PBHs. These methods fall into 2 broad categories - either searching for evidence of their evaporation, or searching for the effects of their gravity upon their surroundings. Whilst no evidence has yet been found of their existence, there are strong constraints on the abundance of different mass PBHs, which can be used to place constraints on the primordial power spectrum (i.e. \cite{Josan:2009qn}) as will be discussed in more detail in chapter \ref{chap:paper1}. The constraints on the abundance of PBHs are normally stated in terms of the fraction of the total energy density of the Universe making up PBHs at the time of formation
\begin{equation}
\beta(m_{PBH})=\frac{\rho_{PBH}}{\rho_{total}}.
\end{equation}
After forming, the energy density contained in PBHs evolves as matter, $\rho_{PBH}\propto a^{-3}$, where $a$ is the scale factor of the Universe. However, since PBHs typically form during the radiation dominated epoch of the Universe, the total energy density evolves as radiation, $\rho_{total}\propto a^{-4}$. This means that even if the fraction of the energy of the Universe contained in PBHs is initially very small it will become larger over time and can become significant at late times.

\begin{figure}[t]
\centering
	\includegraphics[width=\linewidth]{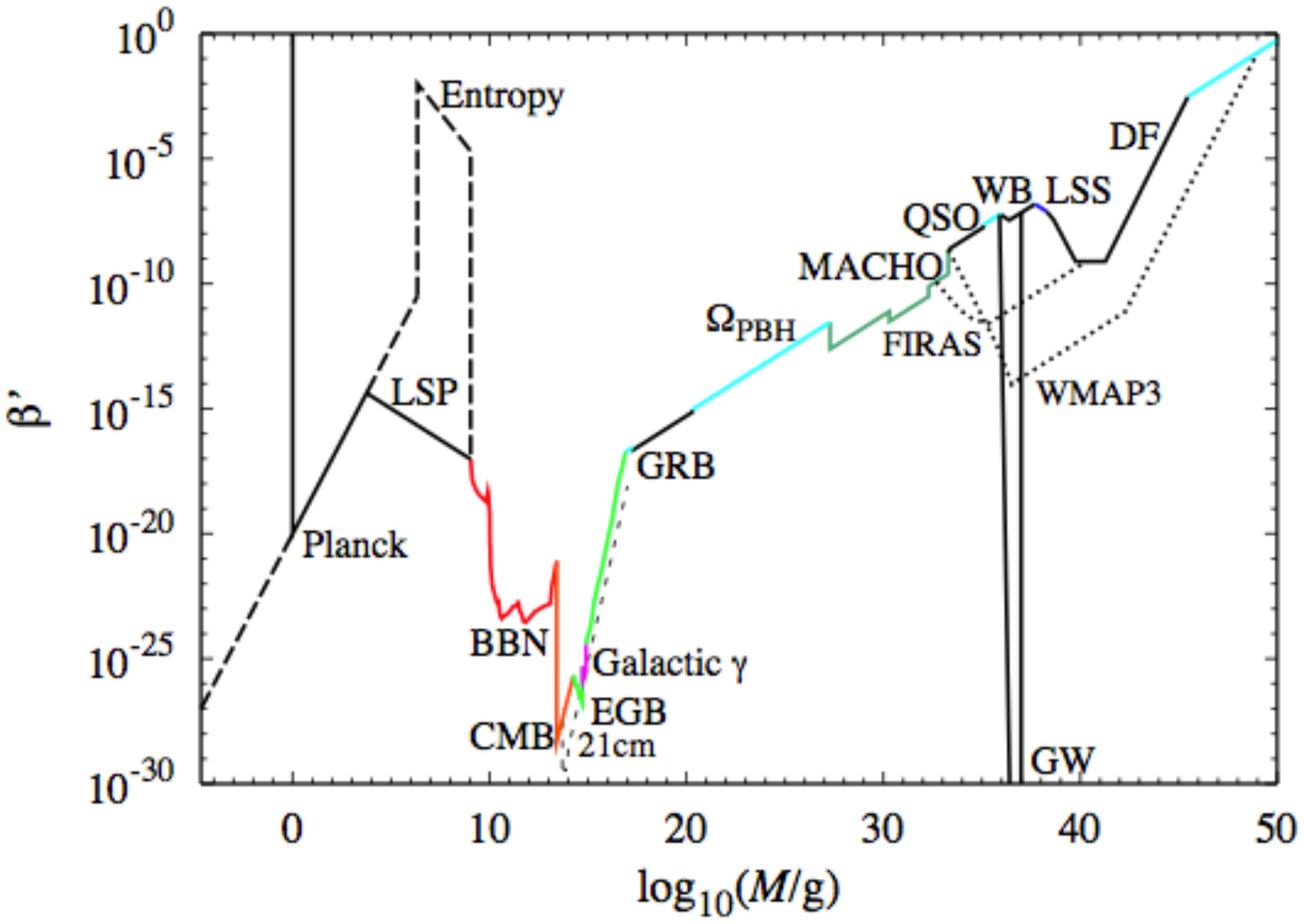}
 \caption[Constraints on the abundance of primordial black holes]{Constraints on the abundance of PBHs of different masses. The acronyms are lensing of gamma-ray bursts (GRB), stars (MACHO), quasars (QSO), and compact radio sources (RS), wide  binary  disruption  (WB),  globular  cluster  disruption  (GC),  dynamical  friction  (DF),  disk  heating (DH),  generation  of large-scale structure through Poisson fluctuations (LSS), accretion effects on the CMB (FIRAS, WMAP3), and gravitational waves (GW). Image credit: \cite{Carr:2009jm}.}
\label{fig:abundanceConstraints}
\end{figure}

Only a brief review of the constraints will be given here - a more detailed discussion of the constraints can be found in \cite{Carr:2009jm}, and figure \ref{fig:abundanceConstraints} shows the constraints on the abundance of PBHs as a function of their mass from that paper. \cite{Josan:2009qn} also discusses the constraints on PBH abundance and includes a calculation of the constraints on the primordial curvature perturbation power spectrum. The constraints typically apply to an integral of the PBH mass function over the range of masses that the constraint applies to, and the constraints given here will assume the PBH forms equal to the horizon mass.

\subsubsection{Evaporation constraints}

As shown by \cite{Hawking:1974rv}, black holes give out radiation (now known as Hawking radiation) and will eventually evaporate. The smaller the black hole, the more radiation is given out by the black hole and the faster the subsequent evaporation. The result is that whilst stellar mass black holes emit very little Hawking radiation and evaporate on a timescale many times greater than the current age of the Universe, PBHs may have formed with much smaller masses and have already evaporated by today. PBHs forming with a mass less than $m_{PBH}\approx10^{15}\mathrm{g}$ will have evaporated by today, with more massive PBHs still surviving (although such a statement assumes negligible accretion of mass since their formation, including accretion from the thermal background of the Universe). 

\begin{enumerate}
\item Diffuse $\gamma$-ray background: PBHs with masses in the range $2\times10^{13}\mathrm{g}<m_{PBH}<5\times 10^{14}\mathrm{g}$ will have evaporated between a redshift of $z\approx 700$ and the present day, and the products of their evaporation would contribute to the diffuse $\gamma$-ray background.

\item Cosmic rays: PBHs evaporating today would result in a large amount of cosmic rays, and their abundance can be constrained by the abundance of cosmic rays. The constraints obtained in this method are essentially equivalent to those obtained from the diffuse $\gamma$-ray background.

\item Hadron injection: PBHs forming with a mass $10^{8}\mathrm{g}<m_{PBH}<10^{10}\mathrm{g}$ would have evaporated before the end of nucleosynthesis and the products of their evaporation would have affected the abundance of light elements. Lighter PBHs would have evaporated too early and would not have affected the abundance of elements formed during nucleosynthesis.

\item Photodissociation of deuterium: PBHs forming with a mass in the range $10^{10}\mathrm{g}<m_{PBH}<10^{13}\mathrm{g}$ would have evaporated between the end of nucleosynthesis and recombination. Photons produced during their evaporation would have caused the dissociation of deuterium and changed the abundance observed today.

\item CMB spectral distortions: the CMB is observed to be very close to a black body spectrum, but PBHs with an initial mass $10^{11}\mathrm{g}<m_{PBH}<10^{13}\mathrm{g}$ would have evaporated close to the time of recombination and could have resulted in spectral distortions.

\item Stable massive particles: many extensions to the standard model of particle physics include the addition of stable or long-lived massive particles. PBHs with mass $m_{PBH}<10^{11}$g can emit these particles during their evaporation, and their abundance is constrained by the present day abundance (or lack thereof) of such massive particles. However, such constraints on the PBH abundance are dependent on the existence of such particles.

\item Present day relic density: black hole evaporation could leave a stable Planck mass relic - the expected theoretical lower bound for black hole mass. If this is the case then the present day density of such relics is limited to the density of cold dark matter. Again, this constraint is dependent upon the existence of Planck mass relics, but provides unique constraints on very light PBHs.
\end{enumerate}

\subsubsection{Gravitational constraints}
PBHs which initially formed with a mass greater than approximately $m_{PBH}>10^{15}$g would still exist today. Due to their small size relative to the distances involved, it is still very hard to directly observe such PBHs (although there is some discussion on the possibility that the recent detection of gravitational waves from merging black holes by LIGO \cite{Abbott:2016blz} may represent such an observation \cite{Bird:2016dcv}). PBHs remaining in the haloes of galaxies can be considered as massive compact halo objects (MACHOs), and constraints on their abundance can also be applied to PBHs.

\begin{enumerate}
\item Density of cold dark matter: the present day density of PBHs must not exceed the observed density of cold dark matter. Crucially, there is narrow window of masses, $10^{20}\mathrm{g}<m_{PBH}<10^{25}\mathrm{g}$, for which this is the only constraint and PBHs of such masses could make up the entirety of dark matter (although there has been debate about constraints on this window arising from neutron stars, see point 5).

\item Gravitational lensing: if there is a large density of PBHs (or other compact objects) in the Universe then this would result in the lensing of distant point sources. PBHs of different masses would result in a strong signal from different sources. Sources that may be lensed include $\gamma$-ray bursts, quasars, stars and radio sources.

\item Disruption of wide binaries: the interaction of wide binary stars (two stars orbiting each other at a large distance) with MACHOs can change the orbits of such systems. Observations of such stars provide constraints on PBHs in the range $10^{3}m_\odot<m_{PBH}<10^{8}m_\odot$ (where $m_\odot$ is one Solar mass).

\item Disk heating: MACHOs travelling across the galactic disk will heat the disk, increasing the velocity dispersion of stars within the disk. Such an effect leads to a constraint on the abundance of PBHs with mass $m_{PBH}>3\times 10^6m_\odot$.

\item Disruption of neutron stars: when a PBHs passes close to, or through, a neutron star it may get gravitationally captured by the neutron star \cite{Pani:2014rca}. If a PBH is captured within the neutron star, the neutron star is quickly destroyed by accretion onto the neutron star. The fact that neutron stars are observed in regions with a high dark matter density suggests that PBHs cannot make up a large fraction of dark matter in the mass range $10^{17}\mathrm{g}<m_{PBH}<10^{24}\mathrm{g}$. However, there has been significant debate about the validity of this constraint (e.g. \cite{Capela:2014qea}), and it is generally believed that the actual constraint should be much weaker. In order for a significant number of neutron stars to cature an orbiting PBH, they need to be able to absorb a lot of energy from the orbit of the PBH - and it is now thought that this process is not as efficient as first calculated, meaning that neutron stars are less likely to capture PBHs and thus be disrupted.

\item Gravitational waves: large density perturbations in the early universe generate second order tensor perturbations (gravitational waves). Such gravitational waves would result in a discrepancy in the timing of signals from pulsars. This results in a constraint on the amplitude of gravitational waves and density perturbations, resulting in a constraint on the number of PBHs formed within the mass range $10^{35}\mathrm{g}<m_{PBH}<10^{37}\mathrm{g}$.

\item X-rays and the CMB: after forming, PBHs can accrete matter, and X-rays given off during this accretion can produce an observable effect in the CMB. Such constraints apply to PBHs in the range $5\times10^{4}m_\odot<m_{PBH}<5\times10^{6}m_\odot$, although the constraints are dependent on the model for the accretion of matter.

\end{enumerate}

\begin{table}[]
\centering
\caption[Primordial black hole abundance constraints]{Constraints on the abundance of primordial black holes. To calculate these constraints it has been assumed that the power spectrum is scale invariant, $n_s=1$, over the range in question.}
\label{tab:PBHconstraints}
\begin{tabular}{|l|l|l|}
\hline
Description                     & Mass range                                                                                                                                 & Constraint on $\beta(m_{PBH})$                                                 \\ \hline
Hadron injection                & \begin{tabular}[c]{@{}l@{}}$10^8\mathrm{g}<m_{PBH}<10^{10}\mathrm{g}$\\ $10^{10}\mathrm{g}<m_{PBH}<3\times 10^{10}\mathrm{g}$\end{tabular} & \begin{tabular}[c]{@{}l@{}}$<10^{-20}$\\ $<10^{-22}$\end{tabular}              \\ \hline
Photodissociation of deuterium  & $10^{10}\mathrm{g}<m_{PBH}<10^{13}\mathrm{g}$                                                                                              & $<3\times 10^{-22}\left(\frac{m_{PBH}}{10^{10}\mathrm{g}}\right)^{1/2}$        \\ \hline
CMB spectral distortions        & $10^{11}\mathrm{g}<m_{PBH}<10^{13}\mathrm{g}$                                                                                              & $<10^{-21}$                                                                    \\ \hline
Stable massive particles        & $m_{PBH}<10^{11}\mathrm{g}$                                                                                                                & $<10^{-18}\left(\frac{m_{PBH}}{10^{10}\mathrm{g}}\right)^{-1/2}$               \\ \hline
Present day relic density       & $m_{PBH}<5\times 10^{14}\mathrm{g}$                                                                                                        & $<4\left(\frac{m_{PBH}}{5\times 10^{14}\mathrm{g}}\right)^{3/2}$               \\ \hline
Density of CDM                  & $m_{PBH}>5\times 10^{14}\mathrm{g}$                                                                                                        & $<2\times 10^{-19}\left(\frac{m_{PBH}}{5\times 10^{14}\mathrm{g}}\right)^{1/2}$ \\ \hline
Lensing of $\gamma$-ray bursts  & $10^{-16}m_{\odot}<m_{PBH}<10^{-13}m_{\odot}$                                                                                              & $<10^{-19}\left(\frac{m_{PBH}}{5\times 10^{14}\mathrm{g}}\right)^{1/2}$         \\ \hline
Lensing of quasars              & $0.001m_{\odot}<m_{PBH}<60m_{\odot}$                                                                                                       & $<10^{-19}\left(\frac{m_{PBH}}{5\times 10^{14}\mathrm{g}}\right)^{1/2}$         \\ \hline
Lensing of radio sources        & $10^6 m_{\odot}<m_{PBH}<10^{8}m_{\odot}$                                                                                                   & $<6\times 10^{-20}\left(\frac{m_{PBH}}{5\times 10^{14}\mathrm{g}}\right)^{1/2}$ \\ \hline
Disruption of wide binaries     & $10^3 m_{\odot}<m_{PBH}<10^8 m_{\odot}$                                                                                                    & $<3\times 10^{-20}\left(\frac{m_{PBH}}{5\times 10^{14}\mathrm{g}}\right)^{1/2}$ \\ \hline
Disk heating                    & $m_{PBH}<3\times 10^{6}m_{\odot}$                                                                       & $<2\times 10^{6}\left(\frac{m_{PBH}}{5\times 10^{14}\mathrm{g}}\right)^{-1/2}$  \\ \hline
Gravitational waves             & $5\times 10^{4}m_{\odot}<m_{PBH}<5\times 10^{6}m_{\odot}$                                                                                  & $<5\times 10^{-36}$                                                            \\ \hline
\end{tabular}
\end{table}

Table \ref{tab:PBHconstraints} shows a summary of all the constraints described above.

\subsection{Primordial black hole forming models}
\label{subsec:PBHmodels}

The abundance of PBHs forming at a given epoch is strongly dependent on the power spectrum at the horizon-scale at that time (as discussed in more detail in chapter \ref{chap:paper1}). In order for a significant number of PBHs to form, the power spectrum needs to be of order $10^{-2}$. However, single scalar field models typically predict a spectral index $n_{s}-1\approx-0.04$ (where we have assumed the spectral index is exactly constant, current observational constraints give $n_s-1=-0.032\pm0.006$) giving a red spectrum, meaning that the amplitude of the power spectrum decreases as smaller scales are considered. Since PBHs form on much smaller scales than those visible in the CMB, single field inflation models therefore generically predict a negligible amount of PBHs.

For a large number of PBHs to form, the power spectrum needs to be orders of magnitude larger than is observed in the CMB - meaning that it must become large on small scales. There exists a multitude of models for inflation that predicts such behaviour, whilst being consistent with current cosmological observations. Such models include the running-mass model \cite{Stewart:1996ey,Leach:2000ea,Kohri:2007qn,Drees:2011hb}, axion inflation \cite{Bugaev:2013fya,Freese:1990rb}, a waterfall transition during hybrid inflation \cite{Linde:1993cn,Bugaev:2011wy, Lyth:2011kj,Halpern:2014mca}, the curvaton model \cite{Lyth:2001nq,Kawasaki:2012wr,Bugaev:2012ai,Kohri:2012yw}, and PBHs can be formed from particle production during inflation \cite{Erfani:2015rqv} or passive density fluctuations \cite{Lin:2012gs}. 

 Here, several possible models will be discussed: features in the potential of single scalar field potential, the curvaton model, hybrid inflation, and the running-mass model.

\subsubsection{Features in the potential of single scalar field inflation}
Whilst it was earlier noted that single scalar field inflation does not predict significant formation of PBHs, this statement is dependent on the form of the inflaton potential. The amplitude of the power spectrum at a given scale is a function of the inflaton potential at the time the scale exits the horizon during inflation. The power spectrum can therefore become larger (or smaller) on small scales, depending on the form of the potential. Recall equation (\ref{eqn:epsilon}) giving the slow-roll parameter $\epsilon_V$ as a function of the potential,
\begin{equation}
\epsilon_V = \frac{M_P^2}{2}\left( \frac{V'}{V}\right)^2.
\end{equation}
The Hubble parameter at a scale $k$ is given by,
\begin{equation}
\mathrm{ln}(H(N))-\mathrm{ln}(H_0)=-\int\limits_{0}^{N}\epsilon_V(\tilde{N})\mathrm{d}\tilde{N},
\end{equation}
where $N$ is the number of $e$-folds between a pivot scale $k_0$ and $k$, and $H_0$ is the Hubble parameter at the pivot scale. The power spectrum is then given by
\begin{equation}
\mathcal{P}_\zeta(N)=A_s\frac{H^2(N)}{\epsilon_V(N)}.
\end{equation}
The power spectrum is therefore a function of the inflaton potential - and features in the potential, such as plateaus, causing $\epsilon_V$ to become very small can cause the power spectrum to become large on corresponding scales, forming a large amount of PBHs. However, it is worth noting that this is a simplistic treatment, and is intended only to show the power spectrum is dependant on the inflaton potential, a more detailed discussion can be found in \cite{Kinney:2005vj}. A small value of $\epsilon_V$ may not necessarily imply a large power spectrum.

\subsubsection{The curvaton model}

In the curvaton model, in addition to the inflaton there is a second field called the curvaton. During inflation, the inflaton field dominates the energy density of the universe, and the universe evolves in the same way as discussed in single scalar field inflation. At the end of inflation, the inflaton decays into radiation, whilst the curvaton field persists for some time and comes to dominate the energy density of the universe before decaying. As a result, the observed cosmological perturbations are sourced from quantum fluctuations in the curvaton field rather than the inflaton field. 

There have been several different versions of the curvaton model, for example \cite{Bugaev:2012ai,Kohri:2012yw}. Here we will briefly describe the model described by \cite{Kawasaki:2012wr}. In this model, the observed large-scale perturbations in the CMB are generated by the inflaton, whilst the small-scale perturbations are generated by the curvaton. The curvature perturbation power spectrum is given by the sum of the components from the inflaton and curvaton
\begin{equation}
\mathcal{P}_\zeta(k)=\mathcal{P}_{\zeta,inf}(k)+\mathcal{P}_{\zeta,curv}(k).
\end{equation}
If the spectral index of $\mathcal{P}_{\zeta,curv}(k)$ is $n_c>1$, the inflaton can be responsible for CMB scale perturbations (matching the prediction from single scalar field inflation, $n_s\approx0.96$) whilst the curvaton generates large-amplitude perturbations on small-scales, potentially resulting in a large number of PBHs forming.

\subsubsection{Hybrid inflation}
There is only a narrow window of masses (see figure \ref{fig:abundanceConstraints}) for which PBHs of a single mass can make up the entire observed density of dark matter. If this is the case then a narrow peak in the power spectrum is required - so that the power spectrum is large on the scale corresponding to the required mass, but small on other scales. Hybrid inflation can be considered an appealing model in this context as it naturally predicts a peak in the power spectrum at some small scale. \cite{Clesse:2015wea} also discusses the possibility of a broad peak in the power spectrum, forming a smaller amount of PBHs at each mass, but spread over a range of mass scales, thus evading some of the existing constraints.

In the standard picture of hybrid inflation, as described in \cite{Lyth:2010zq}, the inflating universe contains the inflation $\phi$ and a second scalar field $\chi$, dubbed the ``waterfall'' field. The potential is given by
\begin{equation}
V(\phi,\chi)=V_0+V(\phi)+\frac{1}{2}m^2(\phi)\chi^2+\frac{1}{4}\lambda\chi^4,
\end{equation}
where $\lambda$ is the self coupling of $\phi$, $V_0$ is a constant, $V(\phi)$ is the potential of $\phi$, and the effective mass $m(\phi)$ of the waterfall field $\chi$ is given by
\begin{equation}
m^2(\phi)=g^2\phi^2-m^2=g^2(\phi^2-\phi_c^2),
\end{equation}
with $0<\lambda\ll 1$ and $0<g\ll 1$. Until inflation nears its end, the waterfall field $\chi$, which has a non-zero minimum, is held fixed at the origin by its interactions with $\phi$. The displacement of $\chi$ from its minimum gives a constant contribution to $V$, dominating the total and resulting in single-field slow-roll inflation. However, when $\phi$ falls below $\phi_c$ the effective mass $m(\phi)$ becomes negative and eventually ends inflation. This process is called the waterfall, and ends when inflation ends. The power spectrum peaks on the scales exiting the horizon during the waterfall, and can result in the production of PBHs of the corresponding mass, as described in \cite{Lyth:2011kj}.

Hybrid inflation predicts a very small amount of non-Gaussianity, $\fNL\approx n_s-1\approx 0.04$ when on cosmological scales when the waterfall field is not active, which is too small to be observable in the CMB. However, as discussed in chapter \ref{chap:paper4}, such a value of $\fNL$ can be ruled out in the case that dark matter is made up of PBHs.

\subsubsection{The running mass model}

The running-mass model of inflation is a relatively simple model which emerges naturally in the context of super-symmetric extensions to the Standard Model of particle physics. The model is of the single-field but can have large variations in the inflaton potential and the slow-roll parameters - and can thus the power spectrum can become large on small-scales, forming a significant number of primordial black holes. The potential $V_\phi$ can be written generically as
\begin{equation}
V_\phi=V_0+\frac{1}{2}m^2_\phi(\phi)\phi^2,
\end{equation}
where $\phi$ is the scalar inflaton field. It is noted that this potential would lead to eternal inflation, and in the running-mass model, a hybrid-inflation type waterfall mechanism is invoked to end inflation. \cite{Drees:2011hb} works with the potential given by
\begin{equation}
V_\phi=V_)+\frac{1}{2}m^2_\phi(\phi)\phi^2+\frac{1}{2}c\phi^2\log\left(\frac{\phi}{\phi_*}\right)+\cdots,
\end{equation}
where $\phi_*$ is the field value at some local non-vanishing extremum of the potential, and $c$ is given by $c\equiv \frac{\mathrm{d}m^2_\phi}{\mathrm{d}\log\phi}\left|_{\phi=\phi_*}\right.$. Again it is worth noting that this equation is an expansion in $\log\left(\frac{\phi}{\phi_*}\right)$ - higher order terms can be included and the expression is likely only valid for a limited range of $\phi$. This form of the potential then allows for the calculation of the slow-roll parameters, as well as the spectral index and running of the spectral index. The power spectrum is then calculated (to a given order), and it is possible for the power spectrum to become large at small-scales, and produce PBHs. The bounds on the abundance of PBHs can then be used to place constraints on the parameter space for the running-mass model.

\section{Outline of papers}

The remainder of this thesis is composed of 5 published papers investigating the use of PBHs to constrain the early universe, separated into 5 chapters, and a conclusion. The format of the rest of this thesis is as follows:
\begin{itemize}

\item The first paper, chapter \ref{chap:paper1}, describes the calculation of the abundance of PBHs from the primordial curvature perturbation power spectrum and calculates constraints on the power spectrum parameters (amplitude, spectral index, running) arising from the constraints on the abundance of PBHs. 

\item Chapter \ref{chap:paper2} extends the calculation carried out in \cite{Byrnes:2012yx}, which considered the effect of the non-Gaussianity parameters $\fNL$ and $\gNL$ separately on the constraints on the power spectrum. In this chapter, local type non-Gaussianity up to $5^{th}$-order terms are considered simultaneously, and the effect of the resulting skewness and kurtosis of the distribution is calculated. It is shown that the constraints on the power spectrum are strongly dependent on the non-Gaussianity parameters to all orders, and generic relations between the non-Gaussianity parameters are also considered. In addition, the abundance of PBHs in the curvaton model is calculated, and it is shown that the power spectrum constraints are strongly dependent upon the density parameter of the curvaton $\Omega_\chi$ at the time it decays.

\item Chapter \ref{chap:paper3} further extends the calculation by accounting for the modal coupling to super-horizon modes that occurs in the presence of non-Gaussianity. It is found that such modal coupling typically increases the amount of PBHs forming, and tightens constraints on the power spectrum.

\item Chapter \ref{chap:paper4} further considers the implications of modal coupling in the context of PBH formation. By considering the effect of modes visible in the CMB upon the abundance of PBHs in different regions of the universe, which would be observed as cold dark matter isocurvature modes, strong constraints can be placed on the non-Gaussianity parameters if dark matter is composed of dark matter.

\item Chapter \ref{chap:paper5} considers the effect of different types of bispectrum upon the abundance of PBHs. Several different shapes of the bispectrum are considered using a numerical method to simulate density maps of the early universe and calculate the abundance of PBHs, and derive corresponding constraints on the power spectrum.

\item Chapter \ref{chap:conc} summarizes the main conclusions of the papers as well as describing possible directions for future research.

\end{itemize}

\newpage

\chapter{Calculating the mass fraction of primordial black holes}
\label{chap:paper1}
\begin{center}


Sam Young $^{1}$, Christian T. Byrnes $^{1}$, Misao Sasaki $^{2}$\\[0.5cm]
$^{1}$Department of Physics and Astronomy, Pevensey II Building, University of Sussex, BN1 9RH, UK\\
$^{2}$ Yukawa Institute for Theoretical Physics, Kyoto University, Kyoto 606-8502, Japan\\[0.5cm]


\end{center}

We re-inspect the calculation for the mass fraction of primordial black holes (PBHs) that are formed from primordial perturbations, finding that performing the calculation using the comoving curvature perturbation $\mathcal{R}_{c}$ in the standard way vastly overestimates the number of PBHs, by many orders of magnitude. This is because PBHs form shortly after horizon entry, meaning modes significantly larger than the PBH are unobservable and should not affect whether a PBH forms or not - this important effect is not taken into account by smoothing the distribution in the standard fashion. We discuss alternative methods and argue that the density contrast, $\Delta$, should be used instead as super-horizon modes are damped by a factor $k^{2}$. We make a comparison between using a Press-Schechter approach and peaks theory, finding that the two are in close agreement in the region of interest. We also investigate the effect of varying the spectral index, and the running of the spectral index, on the abundance of primordial black holes.

\newpage

\section{Introduction}

It is believed that primordial black holes (PBHs) could have formed in the early universe from the collapse of large density fluctuations, and if so, could have observational implications - either from their gravitational effects, or the effects of their Hawking radiation (see \cite{Carr:2009jm,Josan:2009qn} for recent lists of the constraints). They have not been observed, but this fact is enough that they can be used to constrain the early universe (i.e. \cite{Bugaev:2012ai,Young:2013oia,Green:1997sz,Byrnes:2012yx,Shandera:2012ke}) - and provide the only known tool for probing the primordial universe on extremely small scales (i.e. \cite{Scott:2012kx}). However, the constraints from PBHs on small scales are much weaker than those on cosmological scales, for example, the constraints from the cosmic microwave background from the \emph{Planck} satellite.

During inflation, the Hubble horizon shrinks on a comoving scale, and quantum fluctuations become classical density perturbations once they exit the horizon. Once inflation ends, the horizon begins to grow and perturbations begin to reenter the horizon. If a perturbation is large enough, it will collapse to form a PBH almost immediately after horizon reentry - and there has been extensive research into the nature of this collapse and how large a perturbation must be in order to collapse \cite{Shibata:1999zs,Hawke:2002rf,Niemeyer:1999ak,Musco:2004ak,Musco:2008hv,Musco:2012au,Niemeyer:1997mt}.

Calculations for the critical value of the density contrast, $\Delta$, or comoving curvature perturbation, $\mathcal{R}_{c}$, above which a region will collapse to form a PBH are typically of order 0.5 or 1 respectively - and so an insignificant number of PBHs will form unless the power spectrum on small scales is much larger than on large scales, by several orders of magnitude. This is possible in several models, such as the running mass model \cite{Drees:2011hb}, axion inflation \cite{Bugaev:2013fya}, a waterfall transition during hybrid inflation \cite{Bugaev:2011wy}, from passive density fluctuations \cite{Lin:2012gs}, or during inflation with small field excursions \cite{Hotchkiss:2011gz}. For a recent summary of PBH forming models see \cite{Green:2014faa}. Alternatively, the constraint on the formation criteria can be relaxed during a phase transition in the early universe, causing PBHs to form preferentially at that mass scale (i.e. \cite{Jedamzik:1999am}).

In this paper, we will review the calculation of the PBH abundance. The calculation typically computes the fraction of the universe that is above the critical value - in terms of $\Delta$ or $\mathcal{R}_{c}$. This is typically done using the theory of peaks, which calculates the number density of peaks above the critical value, or a Press-Schechter approach, which computes the volume of the universe above the critical value. In order to calculate the abundance of PBHs on different scales, the distribution is convolved with a smoothing function to smooth out modes smaller than the horizon, whilst leaving the horizon and super-horizon modes. When $\mathcal{R}_{c}$ is used to do the calculation in this manner, the super-horizon modes have a large impact on the calculation - we will argue that they should not affect the calculation and that using $\mathcal{R}_{c}$ can be misleading and give errors of many orders of magnitude compared to using $\Delta$. 

In Section 2, we will discuss the formation criteria for PBHs explaining these arguments, and in Section 3, we will briefly review the calculation of the mass of a PBH dependent on the scale it forms at. In Section 4 we discuss the different ways the abundance of PBHs and constraints on the early universe can be calculated for different models. We conclude our arguments in Section 5.

\section{Formation criteria}

The abundance of PBHs is normally stated in terms of $\beta$, the mass fraction of  the Universe contained within PBHs at the time of their formation. Typically, $\beta$ is given as a function of their mass (which, we will see later, is a function of the time at which they form) - so that $\beta$ can be used to describe the mass spectrum of PBHs. In order to determine whether a region of the early universe will collapse to form a PBH, then typically either the density or curvature of that region is compared to a threshold value, which is typically calculated from numerical simulations.

Traditionally, the density contrast $\Delta=\frac{\delta\rho-\rho}{\rho}$ had been used to calculate $\beta$. However, following the paper by \cite{Shibata:1999zs} which calculated the threshold value in terms of a metric perturbation $\psi$, and the paper by \cite{Green:2004wb}, it became more common to use the comoving curvature perturbation $\mathcal{R}_{c}$ (for example, \cite{Bugaev:2012ai,Shandera:2012ke})\footnote{The comoving curvature perturbation $\mathcal{R}_{c}$ is equal to the curvature perturbation on uniform density slices $\zeta$ on super-horizon scales, and because sub-horizon modes are smoothed out, it is common to use $\zeta$ instead of $\mathcal{R}_{c}$.}.

\begin{figure}[t]
\centering
	\includegraphics[width=\linewidth]{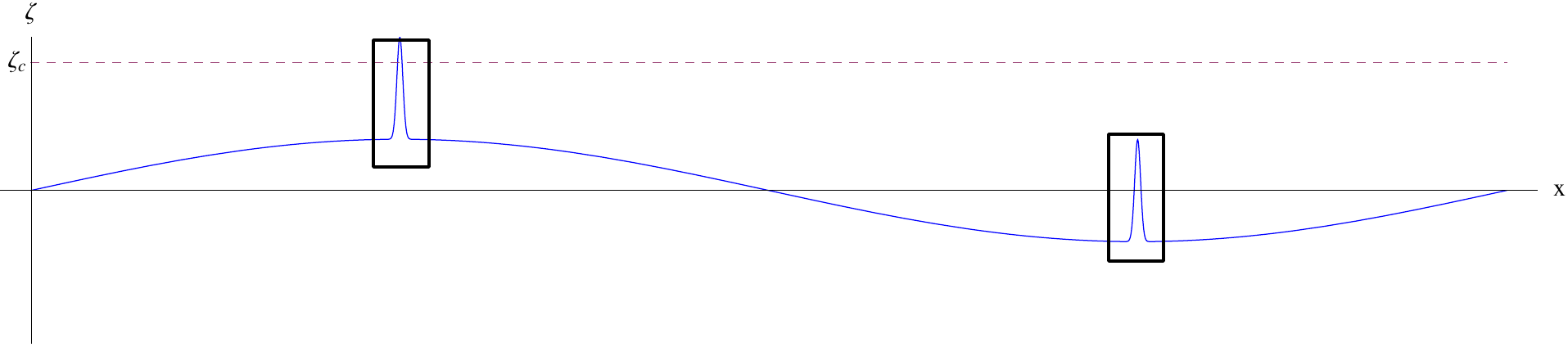}
 \caption[Superposition of small- and large-scale curvature perturbations]{Here, as an example, we show a universe with two sharp (Gaussian) peaks in $\mathcal{R}_{c}$ which sit on top of a long wavelength mode. The two thick black boxes represent the size of the visible universe to an observer at the centre of the peaks at the time of PBH formation, whilst the dotted red line represents the hypothetical threshold value for collapse. Both universes appear the same locally to each observer, and so the evolution of each patch should be identical (until the long wavelength becomes observable).}
\label{dontusezeta}
\end{figure}

In figure \ref{dontusezeta} we demonstrate the danger of using $\mathcal{R}_{c}$ to calculate $\beta$. By simply comparing the height of either peak, one would be drawn to the conclusion that the first (left hand) peak will collapse to form a PBH and the second (right hand) peak will not. However, because the long wavelength mode is well outside the horizon, it is unobservable at the expected time of collapse and invoking the separate universe approach (see \cite{Wands:2000dp}) means that it should not affect the local evolution of the universe. Therefore, the universe looks locally identical to observers at either peak - either both peaks should collapse to form a PBH or neither should\footnote{Note that we are assuming that a PBH will form shortly after entering the horizon, or not at all. It is possible for the PBH formation process to last several e-foldings after horizon entry \cite{Musco:2012au} in which case the long wavelength mode will become important, but only for values extremely close to the threshold value - although this is thought to be rare, see equation (\ref{close to critical}) (however, the effect of a perturbation sitting inside a much larger scale perturbation has not been well studied).}.

It should be noted that papers that have calculated a critical value in terms of $\mathcal{R}_{c}$ 
(i.e. \cite{Shibata:1999zs,Nakama:2013ica}) assume that $\mathcal{R}_{c}$ drops quickly to zero outside of the perturbation - so these values can be used if one assumes that there are no super-horizon perturbations affecting your calculation. Therefore it may be possible to use $\mathcal{R}_{c}$ to calculate $\beta$ if one takes care to exclude super-horizon modes from the calculation (one possibility is to simply subtract the long wavelength modes - although this is strongly dependent on what is considered to be a long wavelength.), and  in Section 4.5 we will consider an approximation where only the value of the power spectrum at horizon entry is used.

\begin{figure}[t]
\centering
	\includegraphics[width=\linewidth]{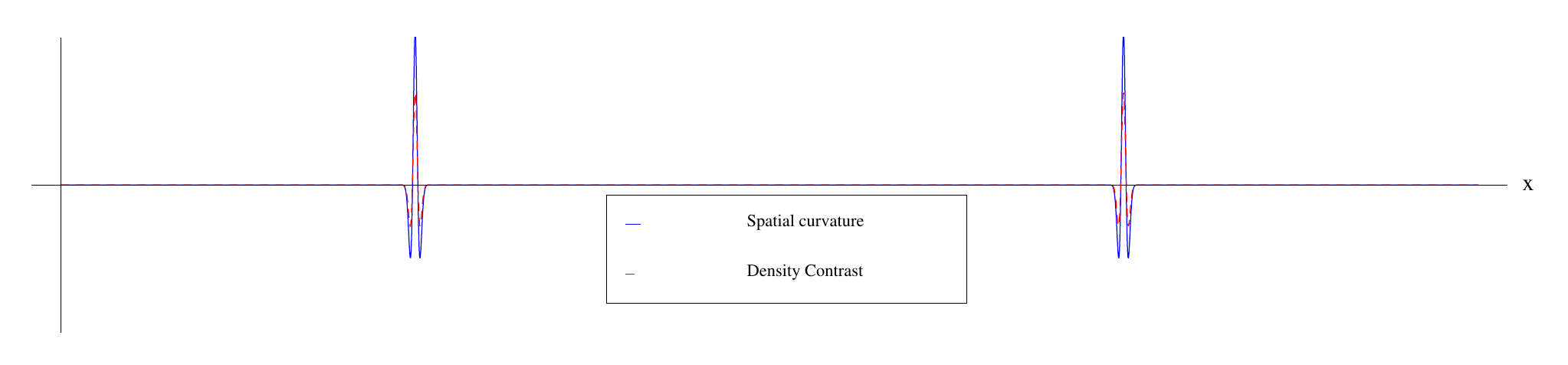}
 \caption[Superposition of small- and large-scale density and spatial curvature perturbations]{The same universe as shown in figure \ref{dontusezeta}, but this time showing the spatial curvature and the density contrast at the time the scale of the small peaks enter the horizon. We now see that both peaks look identical - and so should evolve in the same manner. We see that the peaks in the spatial curvature and density contrast are very similar, both having a Mexican hat profile (rather than the Gaussian shape in the comoving curvature perturbation) - note that the difference in the height of the peaks is due to the arbitrary scaling we have used in the figure.}
\label{curvature density}
\end{figure}

A more formal way to consider this to investigate the effect of super-horizon modes on local observables, such as the density contrast and the spatial 3-curvature. Figure \ref{curvature density} shows the same universe as figure \ref{dontusezeta} but in terms of the spatial curvature and density contrast.

\emph{Spatial curvature} - consider the perturbed, spatially flat FRW metric
\begin{equation}
\label{metric}
ds^{2}=-N^{2}dt^{2}+g_{ij}\left(dx^{i}+N^{i}dt\right)\left(dx^{j}+N^{j}dt\right); g_{ij}=e^{2\alpha}\delta_{ij},
\end{equation}
where we have chosen a comoving slicing, and
\begin{equation}
\alpha=\ln a(t) + \mathcal{R}_{c},
\end{equation}
 with $a(t)$ the scale factor of some flat background and $\mathcal{R}_{c}$ the comoving curvature perturbation. A constant value of $\mathcal{R}_{c}$ can be absorbed into the scale factor by defining
\begin{equation}
\bar{a}(t)=a(t)e^{\mathcal{R}_{c}},
\end{equation}
and so a constant $\mathcal{R}_{c}$ corresponds only to a rescaling of the spatial coordinates, as perhaps clear from the form of the metric (\ref{metric}). The spatial curvature is given by
\begin{equation}
R^{(3)}=-\frac{2}{e^{2\alpha}}\delta^{ij}\left(2\alpha,_{ij}+\alpha,_{i}\alpha,_{j}\right),
\end{equation}
and the spatial curvature of the metric is then
\begin{equation}
\label{spatial curvature}
R^{(3)}=-\frac{2}{e^{2\alpha}}\left(2\nabla^{2}\mathcal{R}_{c}+(\vec{\nabla}\mathcal{R}_{c})^{2}\right).
\end{equation}
If we consider a very long wavelength $\mathcal{R}_{c}$ mode, which appears constant on horizon scales, we see that the spatial curvature due to this mode is negligible due to the derivatives in Eq. (\ref{spatial curvature}).

\emph{Density contrast} - on comoving slices, there is a simple relation at linear order between the comoving curvature perturbation and the density contrast \cite{Green:2004wb}
\begin{equation}
\label{density contrast}
\Delta(t,k)=\frac{2(1+\omega)}{5+3\omega}\left(\frac{k}{aH}\right)^{2}\mathcal{R}_{c}(k),
\end{equation}
where $\omega$ is the equation of state $\omega=p/\rho$, which during radiation domination is $\frac{1}{3}$ ~\footnote{\cite{Josan:2009qn} derives an alternative formula valid on super- and sub-horizon scales during radiation domination,
\begin{equation}
\Delta(t,k)=-\frac{4}{\sqrt{3}}\left(\frac{k}{aH}\right)j_{1}\left(\frac{k}{\sqrt{3}aH}\right)\mathcal{R}_{c}(k),
\end{equation}
where $j_{1}$ is a spherical Bessel function. However, after smoothing, there is little difference between this and equation (\ref{density contrast}).}. In real space this is
\begin{equation}
\Delta(t,x)=\frac{2(1+\omega)}{5+3\omega}\left(\frac{1}{aH}\right)^{2}\nabla^{2}\mathcal{R}_{c}(x).
\end{equation}
Again, we see that this depends on the second derivative of $\mathcal{R}_{c}$ - and so the effect of super-horizon $\mathcal{R}_{c}$ modes is negligible. At linear order, the density contrast is therefore equivalent to the spatial curvature. However, there has been extensive research into the threshold value of $\Delta$ but not for $R^{(3)}$, we therefore advocate the use of the density contrast in order to calculate the mass fraction, $\beta$.

There has been extensive research on the threshold value for the density contrast above which a PBH will form. \cite{Carr:1975qj} was the first to derive a threshold value for the formation of PBHs, $\Delta_{c}\approx\omega$ where $\omega$ is the equation of state, by calculating the density necessary for gravity to overcome pressure forces. In recent years, numerical simulations of gravitational collapse have been used to investigate the collapse of different shapes of the initial density profile. \cite{Niemeyer:1999ak} studied initial shapes including Gaussian, Mexican hat, and polynomial, finding $\Delta_{c}\approx0.7$. \cite{Musco:2004ak,Musco:2008hv,Musco:2012au}\footnote{They note that the difference in value obtained by \cite{Niemeyer:1999ak} can be explained because they only considered a pure density perturbation imposed at the time of horizon crossing. Later work included only growing modes accounting for the effect of the perturbation in the velocity field.} later studied PBH formation, finding $\Delta_{c}\approx0.45$. More recently, \cite{Harada:2013epa} studied a top hat shape, finding an analytic formula $\Delta_{c}=\sin^{2}[\pi\sqrt{\omega}/(1+3\omega)]=0.41$ during radiation domination, and \cite{Nakama:2013ica} studied generalised shapes to determine the crucial parameters in the shape and size of an overdensity. See also \cite{Hawke:2002rf} \footnote{It was previously thought that there was an upper bound above which density perturbations would form a separate closed universe rather than a PBH, however, this has been shown not to be the case \cite{Kopp:2010sh}. This is relatively unimportant in practice, as the effect of an upper bound is negligible because higher peaks are exponentially suppressed.}.

\section{Primordial black hole mass}

\label{sec:PBHmass}

In order to calculate the mass spectrum, or mass function, of PBHs, it is necessary to relate the horizon scale at the time of formation to the mass of PBH formed. We will first review the calculation of the horizon mass carried out by \cite{Green:2004wb}. The horizon mass is
\begin{equation}
M_{H}=\frac{4\pi}{3}\rho(H^{-1})^{3}.
\end{equation}
In co-moving units, the horizon scale during radiation domination is $R=(aH)^{-1}\propto a$, and expansion at constant entropy gives $\rho\propto g_{*}^{-1/3}a^{-4}$ (where $g_{*}$ is the number of relativistic degrees of freedom, which is expected to be of order 100 in the early universe). This allows the horizon mass at a given reentry scale to be related to the horizon mass at matter radiation equality,
\begin{equation}
M_{H}=\frac{3}{2}M_{H,eq}(k_{eq}R)^{2}\left(\frac{g_{*,eq}}{g_{*}}\right)^{1/3},
\end{equation}
where we use $k_{eq}=0.07\Omega_{m}h^{2}$Mpc$^{-1}$, $g_{*,eq}\approx 3$ and $g_{*}\approx 100$. $M_{H,eq}$ is given by
\begin{equation}
M_{H,eq}=\frac{4\pi}{3}2\rho_{rad,eq}H_{eq}^{-3}=\frac{8\pi}{3}\frac{\rho_{rad,0}}{k_{eq}^{3}a_{eq}},
\end{equation}
where we take $a_{eq}^{-1}=24 000\Omega_{m}h^{2}$ and $\Omega_{rad,0}h^{2}=4.17\times10^{-5}$. Taking $\Omega_{m}h^{2}=0.14$ gives $M_{H,eq}=7\times10^{50}g$ (for this calculation, we have used the same numbers as \cite{Green:2004wb}).

Now that the horizon mass has been calculated, it remains to determine the fraction of the horizon mass which goes into the PBH, $f_{H}$. Several papers (for example, \cite{Hawke:2002rf,Niemeyer:1999ak}) have noted that, when the density is close to the critical value, the mass of PBH formed depends on the size of the over-density, obeying a simple power law,
\begin{equation}
f_{H}=C\left(\Delta-\Delta_{c}\right)^{\gamma},
\end{equation} 
where $C$ and $\gamma$ are constants - although the values calculated depend on the shape of the initial over-density. \cite{Chisholm:2006qc} summarises the different measurements, as well as discussing a minimum bound on the PBH mass from entropy constraints. Typical values for these parameters which we will consider here are $C=3$, $\Delta_{c}=0.5$, and $\gamma=0.3$. For these values, the mass of PBH formed is only significantly smaller than than the horizon mass, $M_{PBH}<0.1M_{H}$, for values of $\Delta$ in the range
\begin{equation}
0.5<\Delta<0.500012,
\label{close to critical}
\end{equation}
and so we will assume that PBHs form with a mass approximately equal to the horizon mass for the remainder of this paper. PBHs of significantly larger mass could form in regions where $\Delta$ is substantially larger than 0.5, but the abundance of these regions is exponentially suppressed, and are thus extremely rare.

\section{Primordial black hole abundance}

We will now discuss the calculation of the PBH mass fraction, $\beta$. The density contrast on a comoving slicing, $\Delta$, is smoothed on a given scale $R$, and the fraction of the universe with a density contrast above the critical value is calculated. The smoothed density contrast $\Delta(R,x)$ is calculated by convolving the density contrast with a window function $W(R,x)$:
\begin{equation}
\Delta(R,x)=\int_{-\infty}^{\infty}d^{3}x' W(R,x-x')\Delta(x').
\end{equation}
The variance of $\Delta(R,x)$ is given by
\begin{equation}
\label{density variance}
\langle\Delta^{2}\rangle=\int_{0}^{\infty}\frac{dk}{k}\tilde{W}^{2}(R,k)\mathcal{P}_{\Delta}(k),
\end{equation}
where $\tilde{W}(R,k)$ is the Fourier transform of the window function, and $\mathcal{P}_{\Delta}(k)$ is the density power spectrum. Using equation (\ref{density contrast}) this can be related to the comoving curvature perturbation power spectrum as,
\begin{equation}
\langle\Delta^{2}\rangle=\int_{0}^{\infty}\frac{dk}{k}\tilde{W}^{2}(R,k)\frac{4(1+\omega)^{2}}{(5+3\omega)^{2}}\left(kR\right)^{4}\mathcal{P}_{\mathcal{R}_{c}}(k).
\end{equation}

Throughout this paper, we will use a volume-normalised Gaussian window function, such that the Fourier transform is given by
\begin{equation}
\label{window function}
\tilde{W}(R,k)=\exp\left(-\frac{k^{2}R^{2}}{2}\right).
\end{equation}

In the remaining portion of this section, we discuss the difference between using a peaks theory or Press-Schechter approach, and the predicted mass spectra of PBHs for a scale invariant curvature spectrum, a power law spectrum and for a spectrum with a running of the spectral index.

\subsection{Peaks theory vs Press-Schechter}

The initial mass fraction of the Universe going into PBHs $\beta$ can be calculated either using a peaks theory approach, or a Press-Schechter approach. A comparison of these two methods was carried out by \cite{Green:2004wb} who compared the mass spectra calculated using the curvature perturbation, with peaks theory, and the density contrast, using a Press-Schechter approach. In their calculation it was necessary to assume a blue primordial power spectrum, $n_{s}>1$, and they found the two to be in close agreement\footnote{In the appendix, we correct their calculation, finding that calculating $\beta$ in the different methods disagree strongly.}. We will repeat the calculation here for the density contrast only - finding that using peaks theory or a Press-Schechter are not in as close agreement previously found in \cite{Green:2004wb} but still similar, to within a factor of order 10. 

To investigate the difference between the two methods, we will use a variable $\nu=\Delta/\sigma$\footnote{We note here that with peaks theory, the critical value is stated in terms of the peak value of a fluctuation, but in a Press-Schechter approach, it is the average value of the fluctuation. The relationship between the peak value and the average depends on the shape of the fluctuation - but typically, these are expected to differ only by a factor of order unity, with the peak value being higher. The difference in the critical value of the peak value and the average is therefore within the error of the predicted critical value from different sources. We also note the fact that looking for peaks above a certain value in a smoothed distribution is equivalent to looking for patches with an average density above that value - and so the distinction here is only a technical note.}, where $\sigma$ is the square root of the variance $\langle\Delta^{2}\rangle$ given by equation (\ref{density variance}) and is a function of the form of the power spectrum and the smoothing scale (the calculation of $\sigma$ is the same for either method).

In the theory of peaks, the number density of peaks above a height $\nu_{c}$ is given by \cite{Bardeen:1985tr}
\begin{equation}
n_{peaks}(\nu_{c},R)=\frac{1}{(2\pi)^{2}}\left(\frac{\langle k^{2}\rangle(R)}{3}\right)^{\frac{3}{2}}\left(\nu_{c}^{2}-1\right)\exp\left(-\frac{\nu_{c}^{2}}{2}\right),
\label{peaks}
\end{equation}
where $\langle k^{2}\rangle$ is the second moment of the smoothed density power spectrum
\begin{equation}
\langle k^{2}\rangle(R)=\frac{1}{\langle\Delta^{2}\rangle(R)}\int_{0}^{\infty}\frac{dk}{k}k^{2}\tilde{W}^{2}(k,R)\mathcal{P}_{\Delta}(k).
\end{equation}
If we assume a power law spectrum $\mathcal{P}_{\mathcal{R}_{c}}=A_{\mathcal{R}_{c}}(k/k_{0})^{n_{s}-1}$, and a Gaussian window function (equation (\ref{window function})), we obtain
\begin{equation}
\langle k^{2}\rangle(R)=\frac{n_{s}+3}{2R^{2}},
\end{equation}
assuming that $n_{s}>-3$. The number density of peaks above the threshold can be related to the density parameter $\Omega_{PBH,peaks}$ (which is equal to the mass fraction $\beta$ for a flat universe) by $\Omega_{PBH,peaks}(\nu_{c})=n_{peaks}(\nu_{c},R)M(R)/\rho$, where $M(R)$ is the mass of PBH associated with the horizon size $R$, $M(R)=(2\pi)^{3/2}\rho R^{3}$. Finally, we have
\begin{equation}
\beta_{peaks}(\nu_{c})=\Omega_{PBH,peaks}(\nu_{c})=\frac{(n_{s}+3)^{3/2}}{6^{3/2}(2\pi)^{1/2}}\nu_{c}^{2}\exp\left(-\frac{\nu_{c}^{2}}{2}\right).
\label{beta peaks}
\end{equation}

By contrast, the Press-Schechter calculation simply integrates the probability distribution function (PDF), 
\begin{equation}
P(\nu)=\frac{1}{\sqrt{2\pi}}\exp\left(-\frac{\nu^{2}}{2}\right), 
\end{equation}
over the range of values that form a PBH:
\begin{equation}
\beta_{PS}(\nu_{c})=2\int_{\nu_{c}}^{\infty}P(\nu)d\nu=2\int_{\nu_{c}}^{\infty}\frac{1}{\sqrt{2\pi}}\exp\left(-\frac{\nu^{2}}{2}\right)d\nu.
\end{equation}
This can be written in terms of the complimentary error function simply as
\begin{equation}
\beta_{PS}(\nu_{c})=\textrm{erfc}\left(\frac{\nu_{c}}{\sqrt{2}}\right),
\end{equation}
and using the asymptotic expansion of erfc$(\nu_{c})$ this can be written as
\begin{equation}
\beta_{PS}(\nu_{c})\approx\sqrt{\frac{2}{\pi}}\frac{1}{\nu_{c}}\exp\left(-\frac{\nu_{c}^{2}}{2}\right).
\end{equation}

\begin{figure}[t]
\centering
	\includegraphics[width=0.7\linewidth]{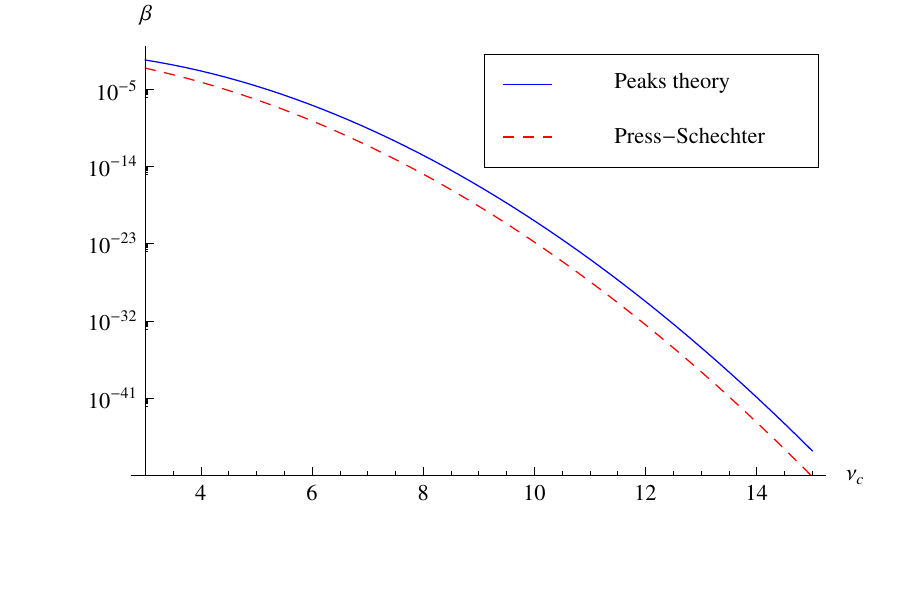}
 \caption[Comparison of peaks theory with a Press-Schechter calculation]{Here we compare the value of $\beta$ calculated using peaks theory or Press-Schechter against $\nu_{c}=\frac{\Delta_{c}}{\sigma}$.}
\label{peaks vs PS}
\end{figure}

Figure \ref{peaks vs PS} shows the difference in the predicted values of $\beta$ for either calculation - the two are in relatively close agreement (differing by a factor of order 10), whilst $\nu$ is not too large\footnote{However, this uncertainty in $\beta$ has little effect on the uncertainty of $\nu_{c}$ which would be calculated, as it depends only on $\log(\beta)$ (see \cite{Young:2013oia}).}. For larger values of $\nu_{c}$, $\beta_{peaks}$ is systematically higher than $\beta_{PS}$. However, the difference between these methods is small compared to the error due to uncertainties in the threshold value $\Delta_{c}$ (see figure \ref{spectral index} for an example).

\subsection{Scale invariant power spectrum}
In the case where the primordial curvature power spectrum is scale invariant, $\mathcal{P}(k)=A_{\mathcal{R}_{c}}$, where $A_{\mathcal{R}_{c}}$ is a constant, then the variance of the smoothed density field during radiation domination, $\omega=1/3$, is
\begin{equation}
\langle\Delta^{2}\rangle=\int_{0}^{\infty}\frac{dk}{k}\tilde{W}^{2}(k,R)\frac{4(1+\omega)^{2}}{(5+3\omega)^{2}}A_{\mathcal{R}_{c}}
=\frac{8}{81}A_{\mathcal{R}_{c}}.
\label{scale invariant density variance}
\end{equation}
Note that, as expected for a scale invariant spectrum, this is now independent of the smoothing scale $R$ 
 - and so predicts that $\beta$ is independent of the mass of the PBHs\footnote{It is also worth noting that for either a red or scale invariant power spectrum $\langle\mathcal{R}_{c}^{2}\rangle\rightarrow\infty$.}. Using peaks theory:
\begin{equation}
\beta=\frac{1}{2^{3/2}(2\pi)^{1/2}}\frac{81\Delta_{c}^{2}}{8A_{\mathcal{R}_{c}}}\exp\left(-\frac{81\Delta_{c}^{2}}{16A_{\mathcal{R}_{c}}}\right).
\label{scale invariant beta} 
\end{equation}

\subsubsection{Constraints on the power spectrum}

Using the relation between the (scale invariant) comoving curvature perturbation power spectrum and $\beta$, equation (\ref{scale invariant beta}), it is simple to calculate a constraint on the power spectrum from the constraint on $\beta$ at a given scale. We will here consider a constraint of size $\beta<10^{-20}$, with $\Delta_{c}=0.5$, and give the constraints one would calculated from peaks theory and Press-Schechter, seeing that the two are in very close agreement:
\begin{align}
\mathcal{P}_{\mathcal{R}_{c},peaks}<0.026, \nonumber \\
\mathcal{P}_{\mathcal{R}_{c},PS}<0.029.
\label{constraints}
\end{align}

\subsection{Power law power spectrum}

\begin{figure}[t]
\centering
	\includegraphics[width=\linewidth]{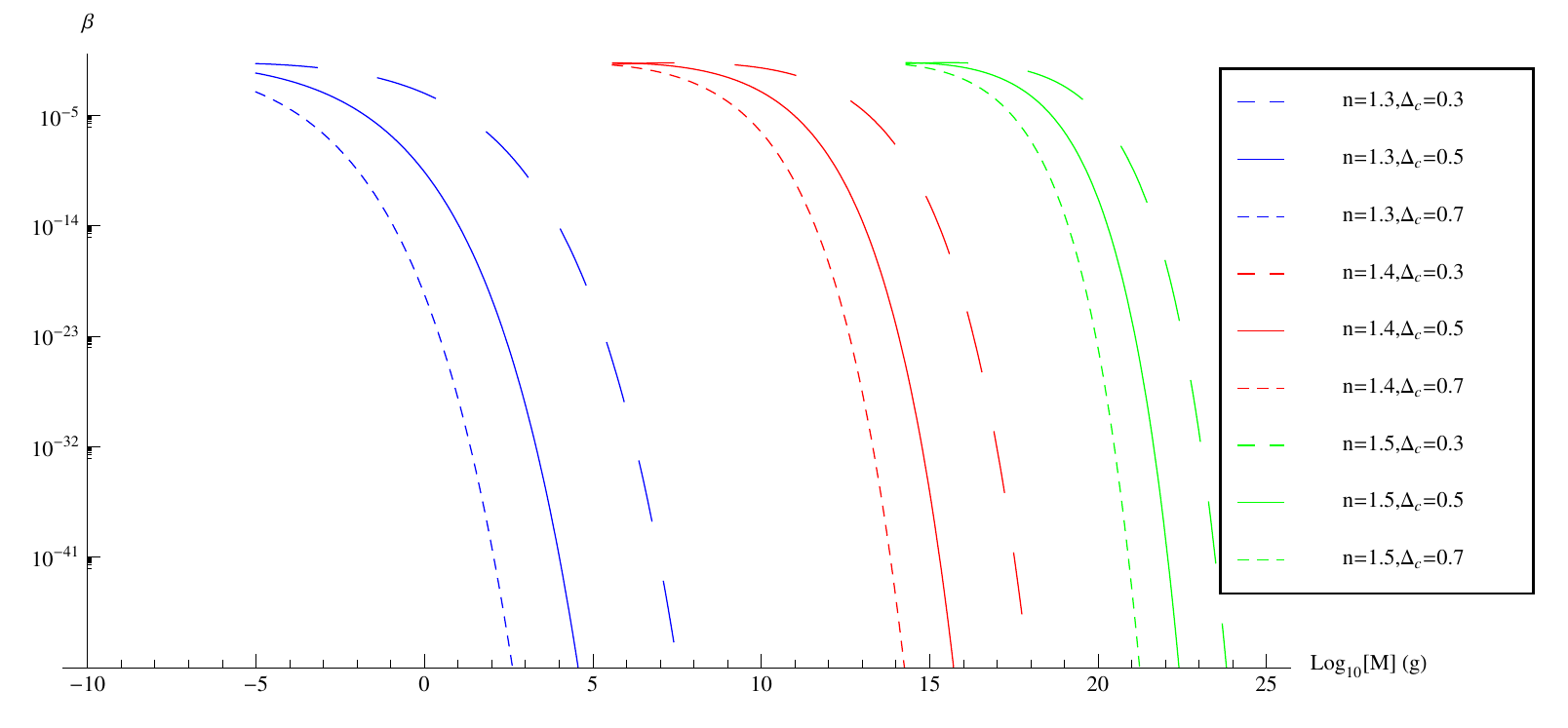}
 \caption[Primordial black hole mass spectra as a function of the spectral index and threshold value]{This figure shows the predicted PBH mass spectra for different values of $n_{s}$ and $\Delta_{c}$. A smaller spectral index produces PBHs of smaller masses. Note that the calculation has been artificially cut off when $\beta$ becomes large as it is only valid for rare peaks (where $\beta$ is small), as well as for PBHs smaller than the Planck mass ($M\approx 10^{-5}g$).}
\label{spectral index}
\end{figure}

In order to compare with \cite{Green:2004wb}, we will consider a power law spectrum (see also \cite{Drees:2011hb}). The form of the power spectrum is given by
\begin{equation}
\mathcal{P}_{\mathcal{R}_{c}}(k)=A_{0}\left(\frac{k}{k_{0}}\right)^{n_{s}-1},
\end{equation}
where $A_{0}$ is the amplitude of the power spectrum defined on some pivot scale $k_{0}$, and we will consider only blue spectra, $n_{s}>1$. In this case, the variance of the smoothed density field during radiation domination, given by equation (\ref{density variance}) is
\begin{equation}
\langle\Delta^{2}\rangle=\frac{8}{81}\frac{A_{0}}{(k_{0}R)^{n_{s}-1}}\Gamma\left(\frac{n_{s}+3}{2}\right),
\end{equation}
and $\beta$ is given by equation (\ref{beta peaks}). For the purposes of making a specific calculation we will take $A_{0}=2.2\times10^{-9}$ and $k_{0}=0.05$ Mpc$^{-1}$, loosely based on observations. Figure \ref{spectral index} shows the predicted mass spectra for a range of different spectral indexes $n_{s}$, and threshold values of the density contrast $\Delta_{c}$ - here, we only consider a blue spectrum (it is possible to consider a red spectrum on small scales in which case $\beta$ is larger for more massive PBHs, but a complicated model is needed to produce a significant number of PBHs and be consistent with observations). 

We can place a limit on the spectral index from the observational constraints on the abundance of PBHs - as has been done previously (for example, \cite{Green:1997sz}). Taking $\Delta_{c}=0.5$ and using the constraint $\beta<10^{-20}$ for PBHs in the mass range $10^{8}$g$<M_{PBH}<10^{10}$g \cite{Josan:2009qn}, the constraint on the spectral index is $n_{s}<1.34$. Because there is a minimum mass of PBHs, at the Planck mass, then we can also place a minimum value on $n_{s}$ which is required to form a significant number of PBHs. Approximately 70 $e$-foldings of inflation are required after todays horizon scale exited during inflation in order for the horizon to reach a sufficiently small scale corresponding to the Planck mass. Typical inflationary models predict that the current horizon scale exited the Hubble scale during inflation about 55 $e$-foldings  before the end of inflation \cite{Liddle:2003as}. In that case, the mass contained in the horizon scale at the end of inflation is approximately $e^{30}M_{\rm Planck}\sim 10^8$g. If we require that $\beta>10^{-20}$ for PBHs of mass $M_{PBH}=10^{-5}$g then the spectral index must be $n_{s}>1.26$. In order for a significant number of PBHs to form, then $n_{s}$ must lie in the range
\begin{equation}
1.26<n_{s}<1.34.
\end{equation}

\subsection{Running of the spectral index}

Over the large range of scales considered here, the spectral index is unlikely to be a constant. We will therefore consider a running of the spectral index, $\alpha$, defined as
\begin{equation}
\alpha=\frac{dn_{s}}{d\ln(k)},
\end{equation}
leading to an expression for the comoving curvature perturbation power spectrum given by
\begin{equation}
\mathcal{P}_{\mathcal{R}_{c}}(k)=A_{0}\left(\frac{k}{k_{0}}\right)^{n_{0}-1+\frac{1}{2}\alpha\ln(k/k_{0})},
\end{equation}
where $A_{0}$ and $n_{0}$ are the values of the power spectrum and spectral index respectively, defined at a pivot scale $k_{0}$. If values are given for parameters $k_{0}$, $A_{0}$, $n_{0}$ and $\alpha$ then the PBH mass spectra can be calculated as before, calculating the variance of the smoothed density contrast using equation (\ref{density variance}) and finding $\beta$ using equation (\ref{beta peaks}). 

\begin{figure}[t]
\centering
	\includegraphics[width=\linewidth]{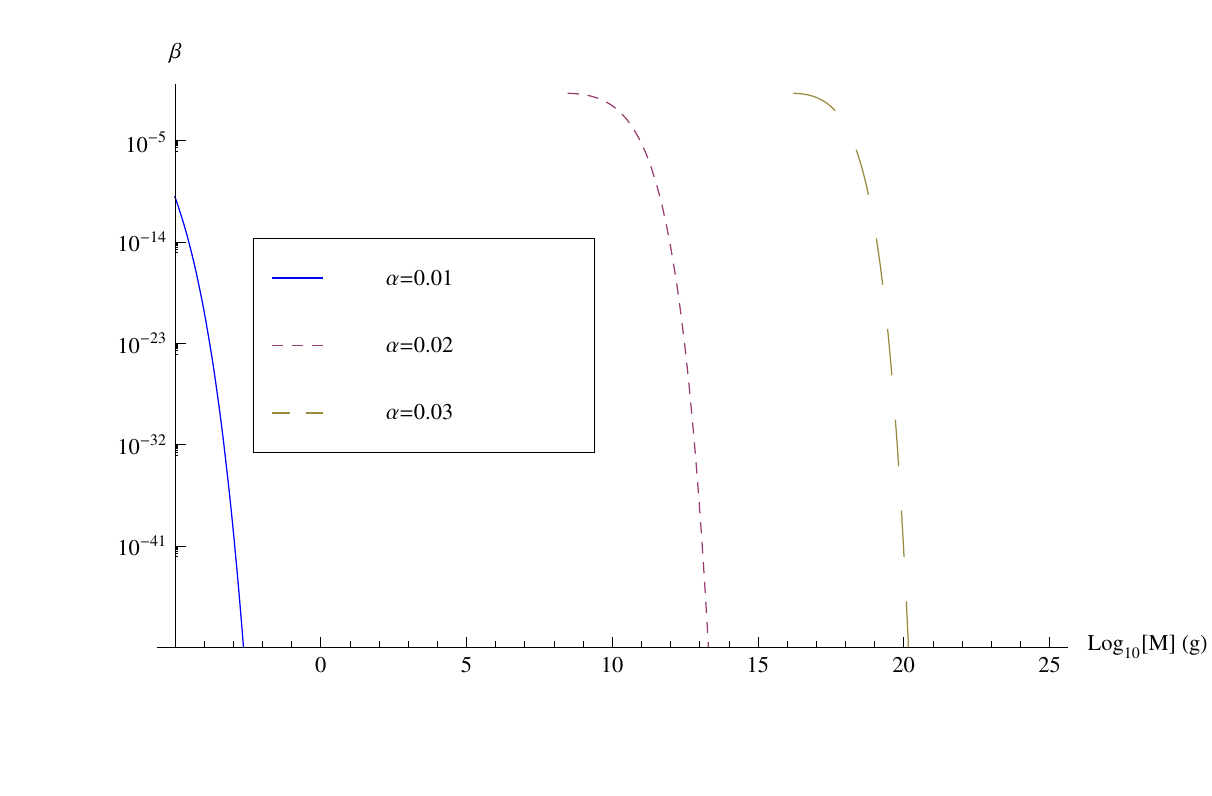}
\caption[Primordial black hole mass spectra as a function of the running of the spectral index]{This figure shows the predicted PBH mass spectra for different values of the running of the spectral index $\alpha$. Again, the calculation has been artificially cut off when $\beta$ becomes large.}
\label{running}
\end{figure}

The same as in the previous section, we will take $A_{0}=2.2\times10^{-9}$ and $k_{0}=0.05$ Mpc$^{-1}$. The \emph{Planck} collaboration \cite{Ade:2013uln} found a spectral index $n_{s}=0.9603\pm0.0073$, but no statistically significant running of the spectral index, $\alpha=-0.0134\pm0.0090$. We will therefore take $n_{0}=0.96$ and allow $\alpha$ to vary - see figure \ref{running}. A positive running is necessary to produce a significant number of PBHs, and the smallest value we will consider is $\alpha=0.01$.

PBHs of masses greater than $M_{PBH}\approx 10^{8}$g are well constrained by observations \cite{Josan:2009qn,Carr:2009jm}, and we see from figure \ref{running} that these values of the running produce too many PBHs, and would be ruled out by observational constraints. We therefore state an upper bound on the running of the spectral index, $\alpha<0.0162$ (again, using the constraint $\beta<10^{-20}$ for PBHs in the mass range $10^{8}$g$<M_{PBH}<10^{10}$g \cite{Josan:2009qn}). Although, again, we note that there is no reason to assume the running of the spectral index will be constant over a large range of scales.

We will not consider the running of the running in this paper, although it has been considered by \cite{Erfani:2013iea}, who places an upper limit on the running of the running by considering the non-production of (long lived) PBHs.

\subsection{Approximation using the comoving curvature perturbation power spectrum}
The power spectrum is, formally, the variance of the amplitude of the Fourier modes at a certain scale. Less formally, one can consider it to be the characteristic size of perturbations at that scale. We show in this section that one can quickly find an approximate value for the PBH mass fraction using the comoving curvature perturbation by only considering perturbations at the exact scale of horizon crossing, without using window functions - this is the approach used in previous papers \cite{Byrnes:2012yx, Young:2013oia}. At horizon crossing, the relation between the density contrast and the comoving curvature perturbation becomes even simpler, as the factor $(k/aH)=1$:
\begin{equation}
\Delta(t_{H},k)=\frac{2(1+\omega)}{5+3\omega}\mathcal{R}_{c}(k)=\frac{4}{9}\mathcal{R}_{c}(k),
\end{equation}
where $t_{H}$ is the time at horizon entry, and $\omega=1/3$ is the equation of state during radiation domination. As $\Delta$ is proportional to $\mathcal{R}_{c}$ at horizon entry, it is reasonable to assume that peaks in the smoothed density contrast correspond to peaks in the comoving curvature perturbation (ignoring other scales).

\begin{figure}[t]
\centering
	\includegraphics[width=\linewidth]{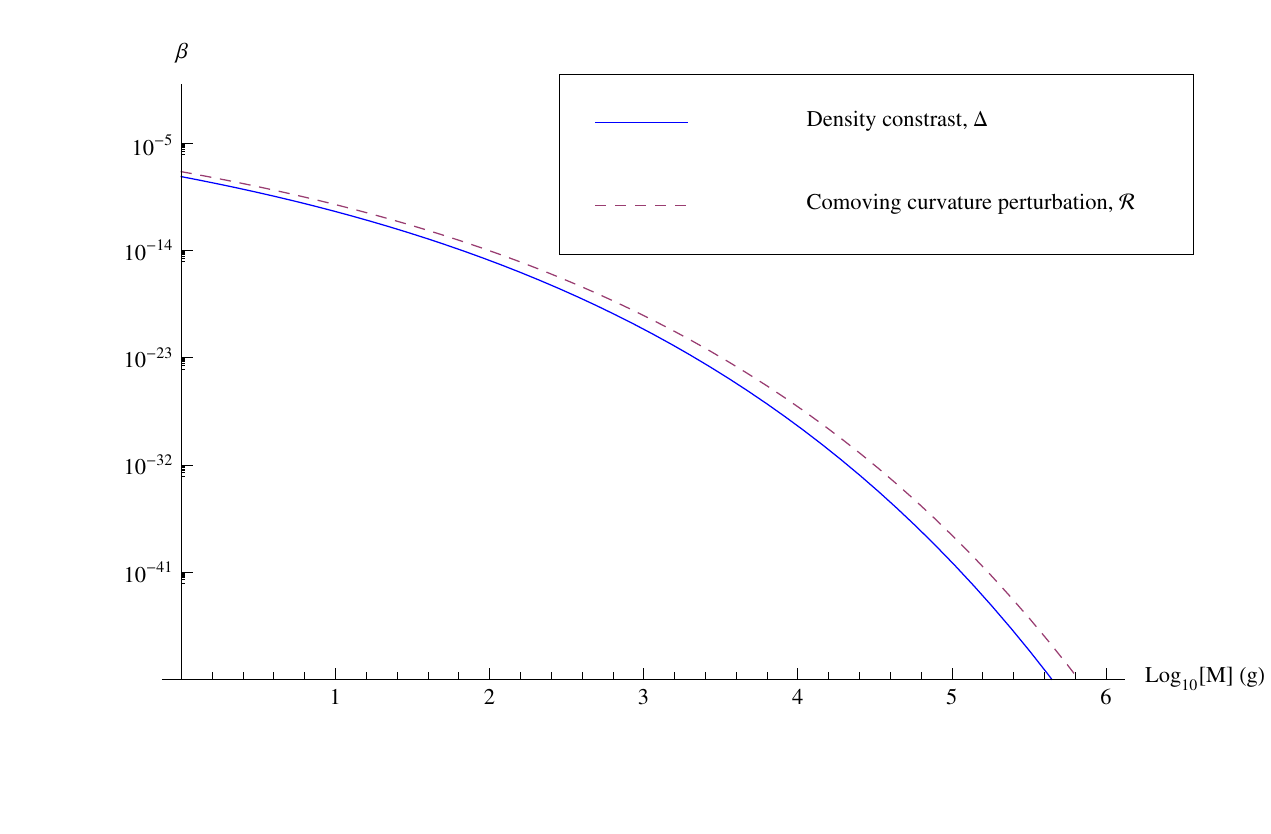}
\caption[Comparison of the primordial black hole mass spectra calculated from the density contrast and from the curvature perturbation]{We show the mass spectra of PBHs calculated, for a power law power spectrum $\mathcal{P}_{\mathcal{R}_{c}}(k)=A_{0}(k/k_{0})^{n_{s}-1}$, using the density contrast (method described in Section 4.3) and the comoving curvature perturbation (method described in Section 4.5). The values we have used in this figure are $A_{0}=2.2\times10^{-9}$, $k_{0}=0.05$Mpc$^{-1}$, $n_{s}=1.3$, $\Delta_{c}=0.4$ and $\mathcal{R}_{c,crit}=1.2$}
\label{zeta approx and density}
\end{figure}

We will assume that the power spectrum at a given scale gives the variance of the comoving curvature perturbation at that scale and use a Press-Schechter approach to calculate $\beta$:
\begin{equation}
\beta=2\int_{\mathcal{R}_{c,crit}}^{\infty}P(\mathcal{R}_{c})d\mathcal{R}_{c},
\end{equation}
where $P(\mathcal{R}_{c})$ is the (Gaussian) probability distribution function. Writing this in terms of the complimentary error function gives
\begin{equation}
\beta=\textrm{erfc}\left(\frac{\mathcal{R}_{c,crit}}{\sqrt{2 \mathcal{P}_{\mathcal{R}_{c}}}}\right).
\label{zeta approx}
\end{equation}
Compare this to the expression one would derive using the density contrast for a scale invariant power spectrum, where $\langle\mathcal{R}_{c}^{2}\rangle$ is given by equation (\ref{scale invariant density variance}),
\begin{equation}
\beta=\textrm{erfc}\left(\frac{9\Delta_{c}}{4\sqrt{\mathcal{P}_{\mathcal{R}_{c}}}}\right).
\label{beta spectral index}
\end{equation}
These two expressions will be exactly equal if $\Delta_{c}\approx\frac{2\sqrt{2}}{9}\mathcal{R}_{c,crit}$. However, these methods cannot be considered identical, which is evident if a power law spectrum is considered, $\mathcal{P}_{\mathcal{R}_{c}}(k)=A_{0}(k/k_{0})^{n_{s}-1}$. Equation (\ref{zeta approx}) is unchanged, but equation (\ref{beta spectral index}) becomes
\begin{equation}
\beta=\textrm{erfc}\left(\frac{9\Delta_{c}}{4\sqrt{\mathcal{P}_{\mathcal{R}_{c}}\Gamma{\left(\frac{3+n_{s}}{2}\right)}}}\right).
\end{equation}
However, provided that $\Gamma\left(\frac{3+n_{s}}{2}\right)\approx 1$ (which is satisfied if $n_{s}\approx 1$) and $\Delta_{c}=\frac{2\sqrt{2}}{9}\mathcal{R}_{c,crit}$, these two expressions will be approximately equal. Figure \ref{zeta approx and density} shows a specific example of these calculations, showing that they still agree closely.

We now compare the constraints on the power spectrum calculated in this method to the constraints calculated earlier (equation (\ref{constraints})). Using $\mathcal{R}_{c,crit}=1.2$ \cite{Shibata:1999zs,Green:2004wb}, and $\beta<10^{-20}$ gives the constraint
\begin{equation}
\mathcal{P}_{\mathcal{R}_{c}}<0.024,
\end{equation}
which is in close agreement with the previously calculated bound, equation (\ref{constraints}).

\section{Conclusions}

We have placed the calculation of the PBH abundance on a more solid grounding. Using the comoving curvature perturbation $\mathcal{R}_{c}$ can be misleading and care needs to be taken if one wishes to use $\mathcal{R}_{c}$ to perform this calculation, due to the effect of super-horizon modes. The problem with using $\mathcal{R}_{c}$ is most easily seen when one considers either a red or scale-invariant power spectrum, which causes the variance of $\mathcal{R}_{c}$ to diverge (it is possible to complete the calculation when a blue spectrum is considered but the results differ drastically from using $\Delta$, see Appendix). We therefore advocate the use of the density contrast to perform the calculation, which does not suffer from the same problem due to the $k^{2}$ dependence of super-horizon modes. In addition, calculations and simulations to calculate the critical threshold for collapse most often use $\Delta$. However, it is more convenient to calculate $\mathcal{R}_{c}$ when studying inflationary models, and finding the constraints on the small scale power spectrum from PBHs - an approximation for $\beta$ can be quickly calculated using $\mathcal{R}_{c}$ if the power spectrum, $\mathcal{P}_{\mathcal{R}_{c}}$, is used rather than using the variance, $\langle\mathcal{R}_{c}^{2}\rangle$ (although this can only ever be an approximation as modes of a similar scale can affect the production of PBHs - which this calculation ignores). It is therefore important that calculations using $\Delta$ or $\mathcal{R}_{c}$ give the same results, and we have provided a method for doing so.

We have considered both a Press-Schechter approach and a peaks theory approach, finding that there is a significant discrepancy between the two - however, this is dwarfed by the error due to uncertainty in the critical value of the density contrast above which PBHs are assumed to form, $\Delta_{c}$. In this paper, we use the peaks theory method, which has a better theoretical grounding. The implications of this paper will be explored further in future papers.

\section*{Acknowledgements}
SY is supported by an STFC studentship, and would like to thank Yukawa Institute for Theoretical Physics for its hospitality during a month long stay which was supported by the Bilateral International Exchange Program (BIEP). CB was supported by a Royal Society University Research Fellowship. The authors would like to thank Will Watson, Aurel Schneider, David Seery, Shaun Hotchkiss, Anne Green, Andrew Liddle, John Miller and Ilia Musco for useful discussions that brought about the production of this paper.

\section{Appendix}
For completeness, we include the calculation of the PBH mass fraction $\beta$ using the comoving curvature perturbation, and compare it to the calculation using the density contrast. This was initially done by \cite{Green:2004wb} who incorrectly calculated the density contrast power spectrum at the time of PBH formation - we will now correct the calculation. Assuming a blue power spectrum, $\mathcal{P}_{\mathcal{R}_{c}}=A_{0}\left(k/k_{0}\right)^{n_{s}-1}$ where $n_{s}>1$, the variance of the smoothed comoving curvature perturbation is
\begin{equation}
\langle\mathcal{R}_{c}^{2}\rangle(R)=\int^{\infty}_{0}\frac{dk}{k}\tilde{W}^{2}(k,R)\mathcal{P}_{\mathcal{R}_{c}}(k)=\frac{A_{0}}{2(k_{0}R)^{n_{s}-1}}\Gamma\left(\frac{n_{s}-1}{2}\right).
\end{equation}
The second moment of the power spectrum is given by
\begin{equation}
\langle k^{2}\rangle=\frac{1}{\langle\mathcal{R}_{c}^{2}\rangle(R)}\int^{\infty}_{0}\frac{dk}{k}k^{2}\tilde{W}^{2}(k,R)\mathcal{P}_{\mathcal{R}_{c}}(k)=\frac{n_{s}-1}{2R^{2}},
\end{equation}
leading us to the final expression for $\beta$ using equation (\ref{peaks}) for comoving curvature perturbation instead of density contrast:
\begin{equation}
\beta(R)=\frac{(n_{s}-1)^{3/2}}{6^{3/2}(2\pi)^{1/2}}\frac{\mathcal{R}_{c,crit}^{2}}{\langle\mathcal{R}_{c}^{2}\rangle(R)}\exp\left(\frac{\mathcal{R}_{c,crit}^{2}}{2\langle\mathcal{R}_{c}^{2}\rangle(R)}\right)
\end{equation}

\begin{figure}[t]
\centering
	\includegraphics[width=0.7\linewidth]{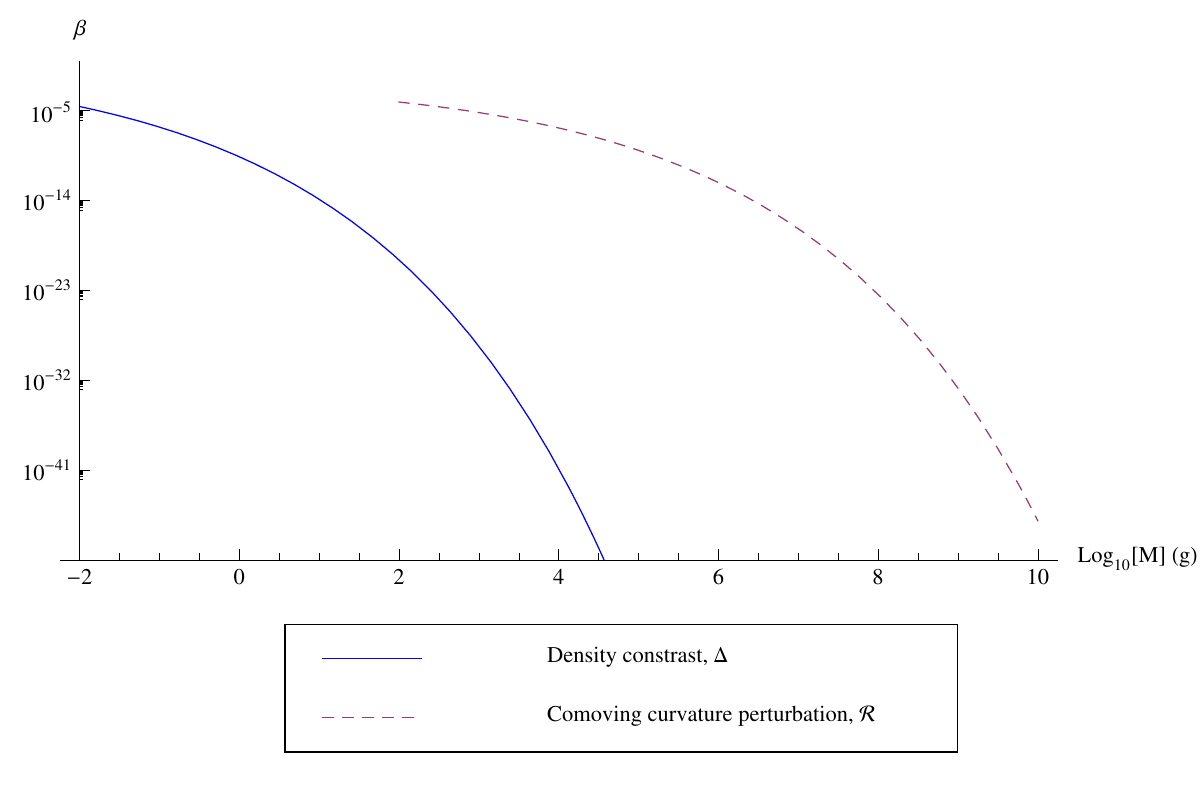}
\caption[Comparison of the current and previously used calculation of the primordial black hole abundance]{We have used $\Delta_{c}=0.5$, $\mathcal{R}_{c,crit}=1$, $n_{s}=1.3$, $A_{0}=2.2\times10^{-9}$ and $k_{0}=0.05$ Mpc$^{-1}$. Both curves represent the mass spectrum of PBHs formed from identical comoving curvature perturbation power spectra - but differ drastically due to the different methods used in the calculation.}
\label{density v curvature}
\end{figure}

The differences between this calculation and the calculation for the density contrast are shown in figure \ref{density v curvature} - we can see that they differ by many orders of magnitude.

\newpage

\chapter{Primordial black holes in non-Gaussian regimes}
\label{chap:paper2}

\begin{center}


Sam Young, Christian T.~Byrnes\\[0.5cm]
Department of Physics and Astronomy, Pevensey II Building, University of Sussex, BN1 9RH, UK\\[1cm]


\end{center}

Primordial black holes (PBHs) can form in the early Universe from the collapse of rare, large density fluctuations. They have never been observed, but this fact is enough to constrain the amplitude of fluctuations on very small scales that cannot be otherwise probed. Because PBHs form only in very rare large fluctuations, the number of PBHs formed is extremely sensitive to changes in the shape of the tail of the fluctuation distribution - which depends on the amount of non-Gaussianity present. We first study how local non-Gaussianity of arbitrary size up to fifth order affects the abundance and constraints from PBHs, finding that they depend strongly on even small amounts of non-Gaussianity and the upper bound on the allowed amplitude of the power spectrum can vary by several orders of magnitude. The sign of the non-linearity parameters ($f_{NL}$, $g_{NL}$, etc.) are particularly important. We also study the abundance and constraints from PBHs in the curvaton scenario, in which case the complete non-linear probability distribution is known, and find that truncating to any given order (i.e.~to order $f_{NL}$ or $g_{NL}$, etc.) does not give accurate results.

\newpage

\section{Introduction}

Primordial black holes (PBHs) have historically been used to study the small scales of the primordial universe. Whilst they have never been detected, this fact is enough to rule out or at least constrain many different cosmological models (see \cite{Carr:1994ar,Green:1997sz,Josan:2010cj,Peiris:2008be,PinaAvelino:2005rm}). Theoretical arguments suggest that PBHs can form from the collapse of large density perturbations during radiation domination \cite{Carr:1974nx}. If the density perturbation at horizon crossing exceeds a threshold value, then gravity will overcome pressure forces and that region collapses to form a PBH with mass of order the horizon mass.

There are tight observational constraints on the abundance of PBHs. These constraints come from their gravitational effects and results of the Hawking radiation from their evaporation. For recent updates and a compilation of the constraints see \cite{Josan:2009qn,Carr:2009jm,Harada:2005sc}. The various constraints place an upper limit on the mass fraction of the Universe contained within PBHs at the time of formation, $\beta$. The constraints vary from $\beta=10^{-27}$ to $\beta=10^{-5}$. These constraints can be used to constrain the primordial power spectrum on small scales, and hence models of inflation. Since PBHs form from the rare, large fluctuations in the extreme tail of the probability distribution function (PDF), any non-Gaussianity can significantly affect the number of PBHs formed. PBH formation can therefore be used to probe both the amplitude and non-Gaussianity of the primordial fluctuations on small scales.

In order for a significant number of PBHs to form, the power spectrum on small scales needs to be of order $10^{-2}$, orders of magnitude larger than on cosmic scales. Although a spectral index smaller than 1 has recently been observed by Planck, indicating a red spectrum, it is possible that the running of the spectral index turns up on smaller scales, and produces a lot of power at such scales. This is possible in models such as the running-mass model, the inflating curvaton and hybrid inflation \cite{Bugaev:2013vba,Bringmann:2001yp, Leach:2000ea, Kohri:2007qn, Alabidi:2009bk, Drees:2011hb, Kohri:2012yw,Linde:2012bt,Lin:2012gs}. Other possibilities include peaks in the power spectrum \cite{Bugaev:2010bb} or a phase transition after inflation \cite{Barrow:1992hq}.

The effects of non-Gaussianity on PBH formation were first studied by \cite{Bullock:1996at}, and \cite{Ivanov:1997ia} - reaching opposite conclusions on whether non-Gaussianity enhances or suppresses the number of PBHs formed. \cite{Lyth:2012yp} studied the constraints from PBH formation on the primordial curvature perturbation for cases where it has the form $\zeta=\pm\left(x^{2}-\langle x^{2}\rangle\right)$, where $x$ has a Gaussian distribution. The minus sign can be expected from the linear era of the hybrid inflation waterfall, where the positive sign might arise if $\zeta$ is generated after inflation by a curvaton-type mechanism. More recently, the effects of non-Gaussianity have been studied by \cite{Byrnes:2012yx}, who studied the effects of quadratic and cubic non-Gaussianity in the local model of non-Gaussianity, and \cite{Shandera:2012ke}, who considered small deviations from a Gaussian distribution, finding that whether PBH formation is enhanced or suppressed depends on the type of non-Gaussianity. The effects of non-Gaussianity in the curvaton model have also been studied recently by \cite{Bugaev:2013vba, Bugaev:2013vba}, who calculated constraints and PBH mass spectra for a chi-squared distribution. \cite{Seery:2006wk} showed how to obtain the probability distribution of the curvature perturbation working directly from the $n$-point correlation functions (which come from quantum field theory calculations) and discussed the possibility of using the constraints of PBHs to discriminate between models of inflation.

In this paper, we will go beyond earlier work and calculate the effects of arbitrarily large non-Gaussianity in the local model to 5th order, including terms of each type simultaneously. We also consider the curvaton model where a full non-linear solution for the curvature perturbation is available in the \emph{sudden decay approximation} \cite{Sasaki:2006kq}. It is found in this case that using a perturbative approach by deriving the non-Gaussianity parameters ($f_{NL}$, $g_{NL}$, etc.) and using the local model of non-Gaussianity disagrees strongly with the full solution - and so care needs to be taken when performing these calculations.

In Section 2, we review the calculation of the PBH abundance constraints in the standard Gaussian case. In Section 3, we review the work completed by \cite{Byrnes:2012yx} calculating the effects of quadratic and cubic non-Gaussianity in the local model, before extending this to higher orders. The expert reader may skip to Sec.~\ref{sec:higher-order}. In Section 4 we discuss the effects of a hierarchical scaling between the non-Gaussianity parameters ($g_{NL}\propto f_{NL}^{2}$, $h_{NL}\propto f_{NL}^{3}$, etc.), and in Section 5 we calculate the constraints on the primordial power spectrum in the curvaton model. We conclude with a summary in Section 6.

\section{PBHs in a Gaussian universe}

Whilst the condition required for collapse to form a PBH has traditionally been stated in terms of the smoothed density contrast at horizon crossing, $\delta_{hor}(R)$, we will follow \cite{Byrnes:2012yx} and work with the curvature perturbation, $\zeta$. PBHs form in regions where the curvature perturbation is greater than a critical value, $\zeta_c\simeq0.7-1.2$ \cite{Green:2004wb}. There is some uncertainty on the exact critical value, and it has a dependence upon the profile of the over density \cite{Shibata:1999zs,Ivanov:1997ia,Hidalgo:2008mv}. For simplicity, we will usually take $\zeta_c=1$, it would be straightforward to choose any other value if required.
It was initially thought that there was an upper limit on the amplitude of the fluctuation that would form a PBH, with larger fluctuations forming a separate universe, however, this has been shown not to be the case \cite{Kopp:2010sh}. Integrating over the fluctuations that would form PBHs, the initial PBH mass fraction of the Universe is:
\begin{equation}
\label{pressschecter}
\beta\equiv\frac{\rho_{PBH}}{\rho_{total}}{\Big|}_{formation}\simeq\int^{\infty}_{\zeta_{c}}P\left(\zeta\right)d\zeta,
\end{equation}
where $\zeta_{c}$ is the critical value for PBH production and $P(\zeta)$ is the probability distribution function. The above equation is not exact, for example due to the uncertainty in the fraction of mass within a horizon sized patch (whose average density is above the critical one) which will collapse to form a black hole. This is related to uncertainty of the overdensity profile and the critical value required for collapse, see e.g.~\cite{Niemeyer:1999ak,Yokoyama:1998xd,Hawke:2002rf,Musco:2012au} and references therein. Fortunately a numerical factor of order unity leads to only a small uncertainty in the constraints on $\sigma$ due to the logarithm, see  Eq.~(\ref{sigmaconstraint}). Order unity non-linearity parameters are much more important than a numerical coefficient multiplying the integral in (\ref{pressschecter}). For Gaussian fluctuations:
\begin{equation}
\label{gaussian}
P(\zeta)=\frac{1}{\sqrt{2\pi}\sigma}\exp\left(-\frac{1}{2}\frac{\zeta^{2}}{\sigma^{2}}\right),
\end{equation}
and so:
\begin{equation}
\label{gaussianintegral}
\beta\simeq\frac{1}{\sqrt{2\pi}\sigma}\int^{\infty}_{\zeta_{c}}\exp\left(-\frac{1}{2}\frac{\zeta^{2}}{\sigma^{2}}\right)d\zeta=\frac{1}{2} \textrm{erfc}\left(\frac{\zeta_{c}}{\sqrt{2}\sigma}\right).
\end{equation}
Because PBHs form in extremely rare large fluctuations in the tail of the probability distribution, one can use the large $x$ limit of $\textrm{erfc}(x)$ and show that \cite{Byrnes:2012yx}:
\begin{equation}
\frac{\sigma}{\zeta_{c}}\simeq\sigma={\cal P}_{\zeta}^{1/2}\simeq\sqrt{\frac{1}{2\ln\left(\frac{1}{\beta}\right)}}.
\label{sigmaconstraint}
\end{equation}
Note that $\sigma$ depends only logarithmically on $\beta$, this remains true once the effects of non-Gaussianity are taken into account. Taking $\zeta_{c}=1$, for $\beta=10^{-20}$ we obtain $\sigma=0.11$ and for $\beta=10^{-5}$ we obtain $\sigma=0.23$.

The variance of the probability distribution is related to the power spectrum of the curvature perturbation by $\sigma^{2}\approx {\cal P}_{\zeta}$. The constraints obtained in this manner differ by ${\cal O}(10\%)$ to those obtained from a full Press-Schechter calculation which includes a window function to smooth the curvature perturbation, as performed in \cite{Matarrese:2000iz,Bugaev:2012ai}. For $\beta=10^{-20}$ the full calculation gives ${\cal P}_{\zeta}^{1/2}=0.12$ \cite{Josan:2009qn}, as opposed to ${\cal P}_{\zeta}^{1/2}=0.11$ obtained with Eq.~(\ref{sigmaconstraint}). In the case of chi-squared non-Gaussianity, a calculation using the smoothed PDF has also been performed \cite{PinaAvelino:2005rm} and gives reasonable agreement with the approach we use here.

\section{PBHs and local non-Gaussianity}

We consider the effects of non-Gaussianity in the local model on the abundance of PBHs and the constraints we can place on the power spectrum. We will first review the work completed by \cite{Byrnes:2012yx} and discuss the effects of quadratic and cubic local non-Gaussianity, before moving onto the effects of higher order terms in Sec.~\ref{sec:higher-order}.

\subsection{Quadratic non-Gaussianity}

\begin{figure}[t]
\centering
\subfloat[\label{subfig-1:dummy}]{%
      \includegraphics[width=0.49\textwidth]{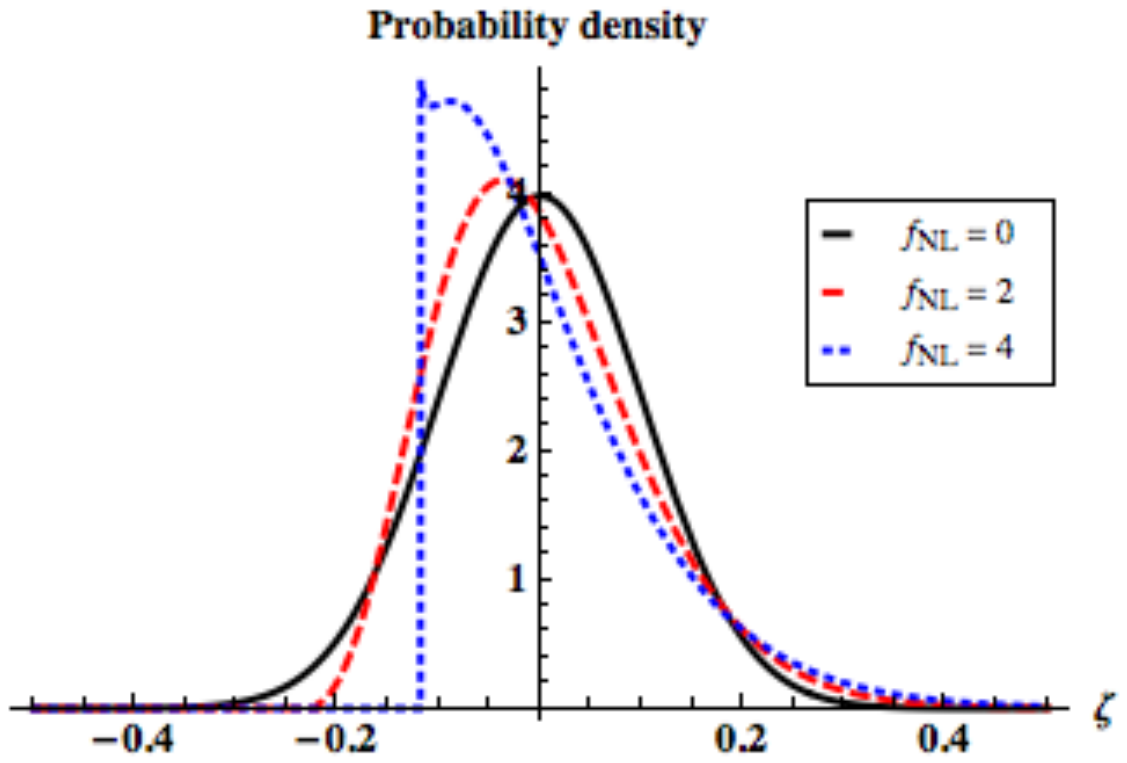}
    }
    \hfill
    \subfloat[\label{subfig-2:dummy}]{%
      \includegraphics[width=0.49\textwidth]{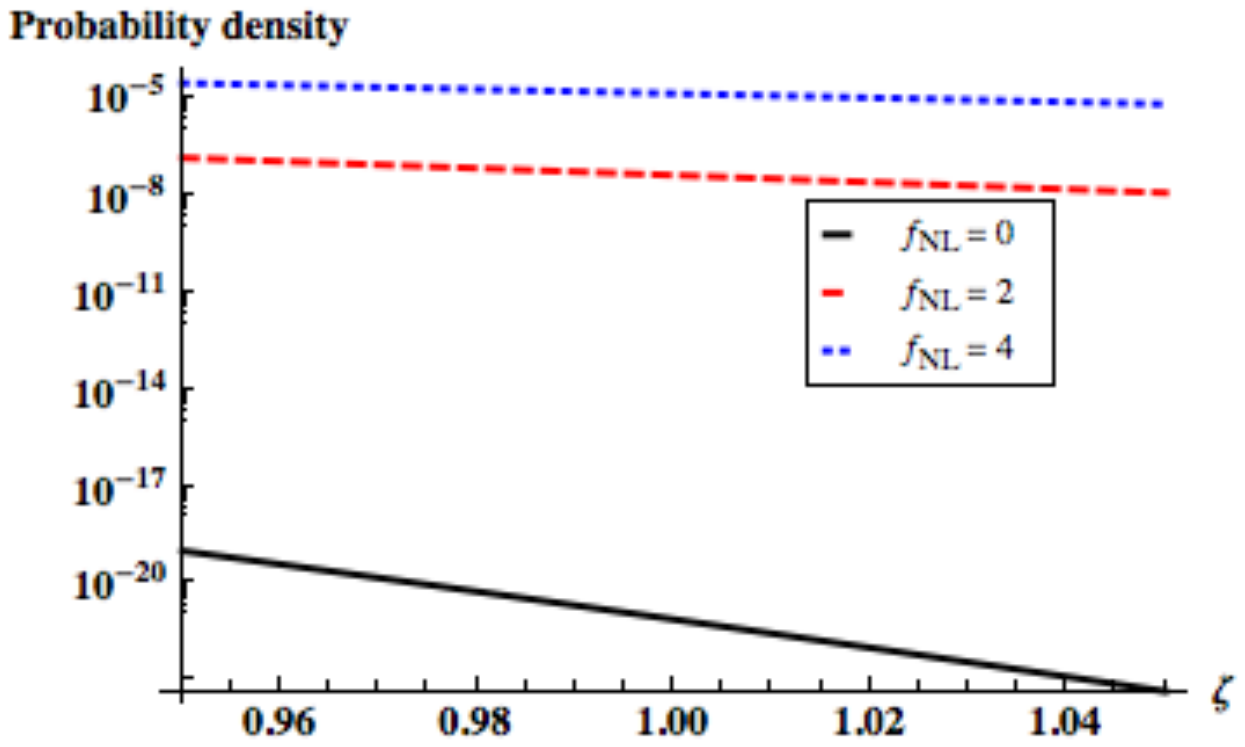}
    }
 \caption[The effect of $\fNL$ on probability distribution function]{The left plot shows the effect of positive $f_{NL}$ on the PDF. For negative $f_{NL}$ the PDFs are simply reflected in the y-axis. We see that the $f_{NL}$ term skews the distribution. The right plot shows the tail of the PDF where PBHs form - note that this is a logarithmic plot of the PDF. A relatively small change in $f_{NL}$ has a large effect on the number of PBHs produced - by many orders of magnitude. For these plots, we have taken $\sigma=0.1$.}
\label{fnlpdfs}
\end{figure}

We take the model of local non-Gaussianity to be
\begin{equation}
\label{quadraticNonGaussianity}
\zeta=\zeta_{g}+\frac{3}{5}f_{NL}\left(\zeta_{g}^{2}-\sigma^{2}\right).
\end{equation}
The $\sigma^{2}$ term is included to ensure that the expectation value for the curvature perturbation remains zero, $\langle\zeta\rangle=0$. Solving this equation to find $\zeta_{g}$ as a function of $\zeta$ gives two solutions
\begin{equation}
\zeta_{g\pm}(\zeta)=\frac{5}{6f_{NL}}\left[ -1 \pm \sqrt{1+\frac{12f_{NL}}{5} \left( \frac{3f_{NL}\sigma^{2}}{5}+\zeta\right)} \right].
\end{equation}
We can make a formal change of variable using
\begin{equation}
P_{NG}(\zeta)d\zeta=\sum_{i=1}^{n}\left|\frac{d\zeta_{g,i}(\zeta)}{d\zeta}\right|P_{G}\left(\zeta_{g,i}(\zeta)\right)d\zeta,
\label{changeofvariable}
\end{equation}
where $i$ is the sum over all solutions, to find the non-Gaussian probability distribution function (PDF). The non-Gaussian distribution is then given by:
\begin{equation}
\label{quadraticpdfeqn}
P_{NG}(\zeta)d\zeta=\frac{d\zeta}{\sqrt{2\pi}\sigma\sqrt{1+\frac{12f_{NL}}{5}\left(\frac{3f_{NL}\sigma^{2}}{5}+\zeta\right)}}\left(\epsilon_{+}+\epsilon_{-}\right),
\end{equation}
where
\begin{equation}
\epsilon_{\pm}=\exp\left(-\frac{\zeta_{g\pm}(\zeta)^{2}}{2\sigma^{2}}\right),
\end{equation}
and the initial PBH mass fraction is given by
\begin{equation}
\beta\simeq\int_{\zeta_{c}}^{\zeta_{max}}P_{NG}(\zeta)d\zeta.
\end{equation}
If $f_{NL}$ is positive (or zero) then $\zeta_{max}=\infty$, but if $f_{NL}$ is negative then $\zeta$ is bounded from above and $\zeta_{max}$ is given by
\begin{equation}
\zeta_{max}=-\frac{5}{12f_{NL}}\left(1+\frac{36f_{NL}^{2}\sigma^{2}}{25}\right).
\end{equation}
Figure \ref{fnlpdfs} shows the effect of $f_{NL}$ on the probability density function. The primary effect of $f_{NL}$ is to skew the distribution - for positive $f_{NL}$ we see a peak for negative values of $\zeta$, with a large tail for positive values (and vice versa for negative $f_{NL}$). The right panel shows a log plot of the effect of positive $f_{NL}$ on the tail of the PDF where PBH formation occurs. We see that, for positive $f_{NL}$, as $f_{NL}$ is increased the amplitude of the large tail increases dramatically. For negative values of $f_{NL}$, $\zeta$ is bounded from above, $\zeta<1$, and we would see no PBH formation for these values (by increasing $\sigma$ significantly, one can form PBHs for significantly negative $f_{NL}$, although we will see later that unless remarkable fine tuning occurs, this leads to an overproduction of PBHs).

We now use the observational constraints on $\beta$ to place constraints on the power spectrum. This is most easily calculated by making a transformation to a new variable $y$:
\begin{equation}
\label{y}
y=\frac{\zeta_{g\pm}(\zeta)}{\sigma},
\end{equation}
which has unit variance. For $f_{NL}>0$ we have
\begin{equation}
\beta\simeq\frac{1}{\sqrt{2\pi}}\left(\int^{\infty}_{y_{c+}}e^{-\frac{y^{2}}{2}}dy+\int^{y_{c-}}_{-\infty}e^{-\frac{y^{2}}{2}}dy\right),
\end{equation}
and for $f_{NL}<0$
\begin{equation}
\beta\simeq\frac{1}{\sqrt{2\pi}}\int^{y_{c+}}_{y_{c-}}e^{-\frac{y^{2}}{2}}dy,
\end{equation}
where $y_{c\pm}$ are the values of y corresponding to the threshold for PBH formation, $\zeta_{c}$:
\begin{equation}
y_{c\pm}=\frac{\zeta_{g\pm}(\zeta_{c})}{\sigma}.
\end{equation}
The expression for $\beta$ is then solved numerically using the tight and weak constraints, $\beta=10^{-20}$ and $10^{-5}$ respectively, to find a value for $\sigma$. The variance of $\zeta$ is then given by \cite{Boubekeur:2005fj,Byrnes:2007tm}
\begin{equation}
{\cal P}_{\zeta}=\sigma^{2}+4\left(\frac{3f_{NL}}{5}\right)^2\sigma^{4}\ln(kL),
\end{equation}
where the cut-off scale $L\approx1/H$ is of order the horizon scale, $k$ is the scale of interest and $\ln(kL)$ is typically ${\cal O}(1)$ (treating it as exactly 1 leads to percent level corrections, provided that $\sigma$ is small - we have numerically checked this).

\begin{figure}[t]
 \centerline{\includegraphics[scale=0.8]{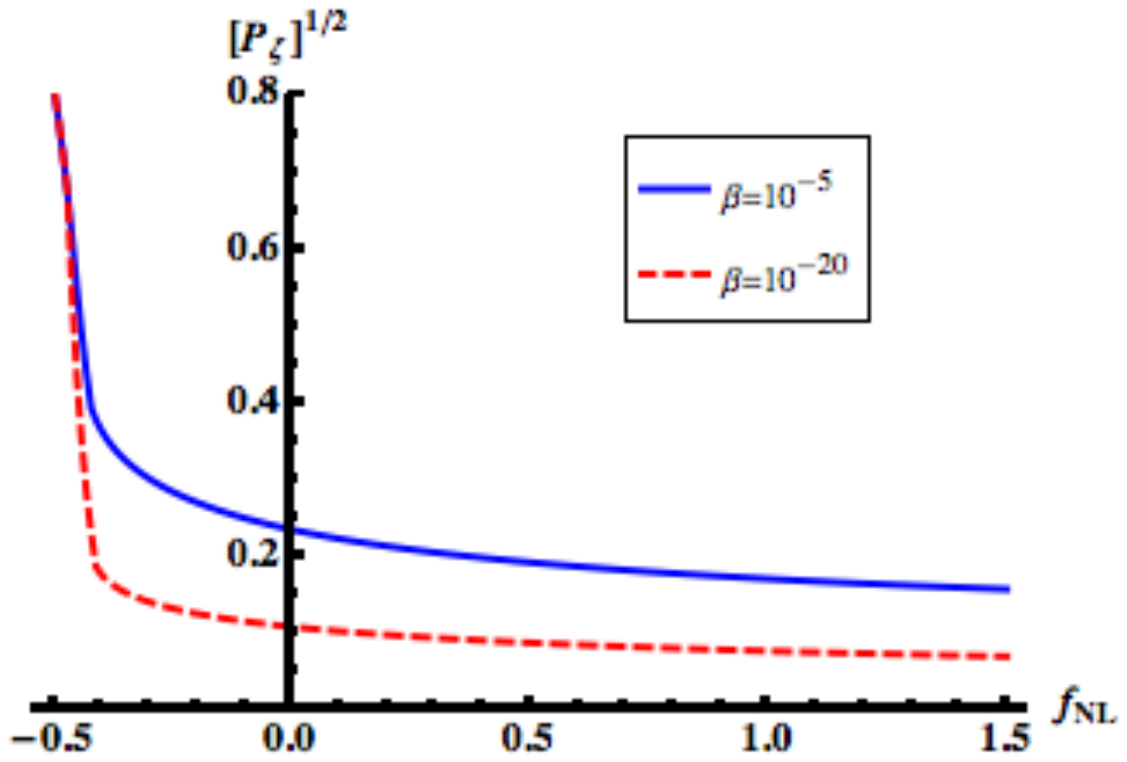}}
 \caption[Constraints on the power spectrum as a function of $\fNL$]{This plot shows how the constraints on the square root of the power spectrum due to PBHs depend on $f_{NL}$. The constraints for 2 values of $\beta$ are shown - note that, although $\beta$ changes by 15 orders of magnitude, the constraints only change by a factor of roughly 2.}
 \label{fnlconstraints}
\end{figure}

Figure \ref{fnlconstraints} shows how the constraints on the square root of the power spectrum change depending on the value of $f_{NL}$. For positive values of $f_{NL}$ we see that the constraints tighten (corresponding to an increase in the abundance of PBHs for a given value for the power spectrum, see Figure \ref{fnlpdfs}). For negative values, we see that the constraints weaken dramatically - this is because, unless $\sigma$ becomes large, no PBHs form at all. As $f_{NL}$ becomes significantly negative, we see that the constraints for $\beta=10^{-20}$ and $\beta=10^{-5}$ converge. Unless there is remarkable fine tuning in the size of the perturbations at small scales, there would either be far too many PBHs, or none. Using this method to calculate the constraints, as $f_{NL}$ becomes more negative the constraints on the power spectrum do flatten out at a value above 1 - however, the perturbative approach does not work when the perturbation amplitude is ${\cal O}(1)$ or higher, so these results cannot be trusted.

\subsection{Cubic non-Gaussianity}

The model of local non-Gaussianity is now taken to be
\begin{equation}
\label{eqn:cubicNG}
\zeta=\zeta_{g}+\frac{9}{25}g_{NL}\zeta_{g}^{3}.
\end{equation}
We follow the same process as before to calculate the PDFs and constraints on the power spectrum \cite{Byrnes:2012yx}. Care needs to be taken with the amount of solutions to Eq.~(\ref{eqn:cubicNG}). For $g_{NL}>0$, there is one solution for all $\zeta$. But for $g_{NL}<0$, there may be multiple solutions. For example, for $g_{NL}<0$, in the range
\begin{equation}
-\frac{2}{9}\sqrt{\frac{-5}{g_{NL}}}\leq\zeta\leq\frac{2}{9}\sqrt{\frac{-5}{g_{NL}}},
\end{equation}
there are 3 solutions to Eq.~\ref{eqn:cubicNG}. These solutions need to be taken into account when calculating PDFs or constraints on the power spectrum.

\begin{figure}[t]
\centering
\subfloat[Negative $g_{NL}$]{%
      \includegraphics[width=0.49\textwidth]{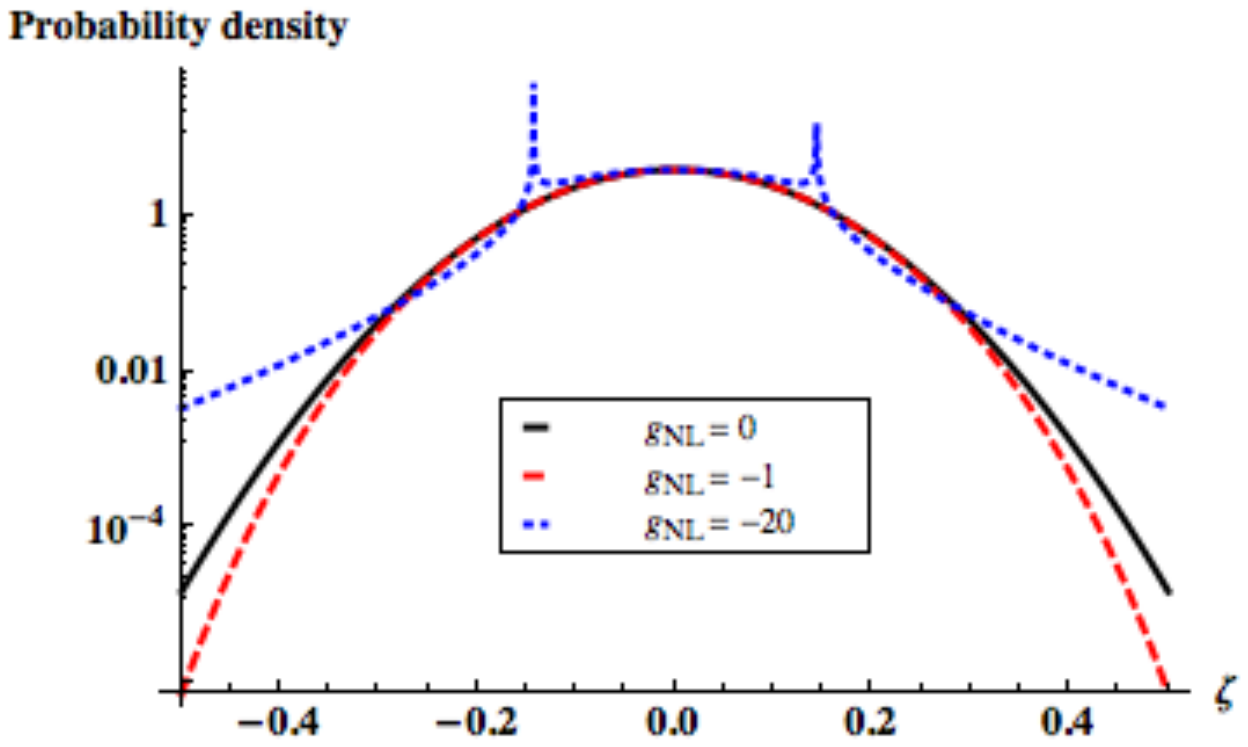}
    }
    \hfill
    \subfloat[Positive $g_{NL}$]{%
      \includegraphics[width=0.49\textwidth]{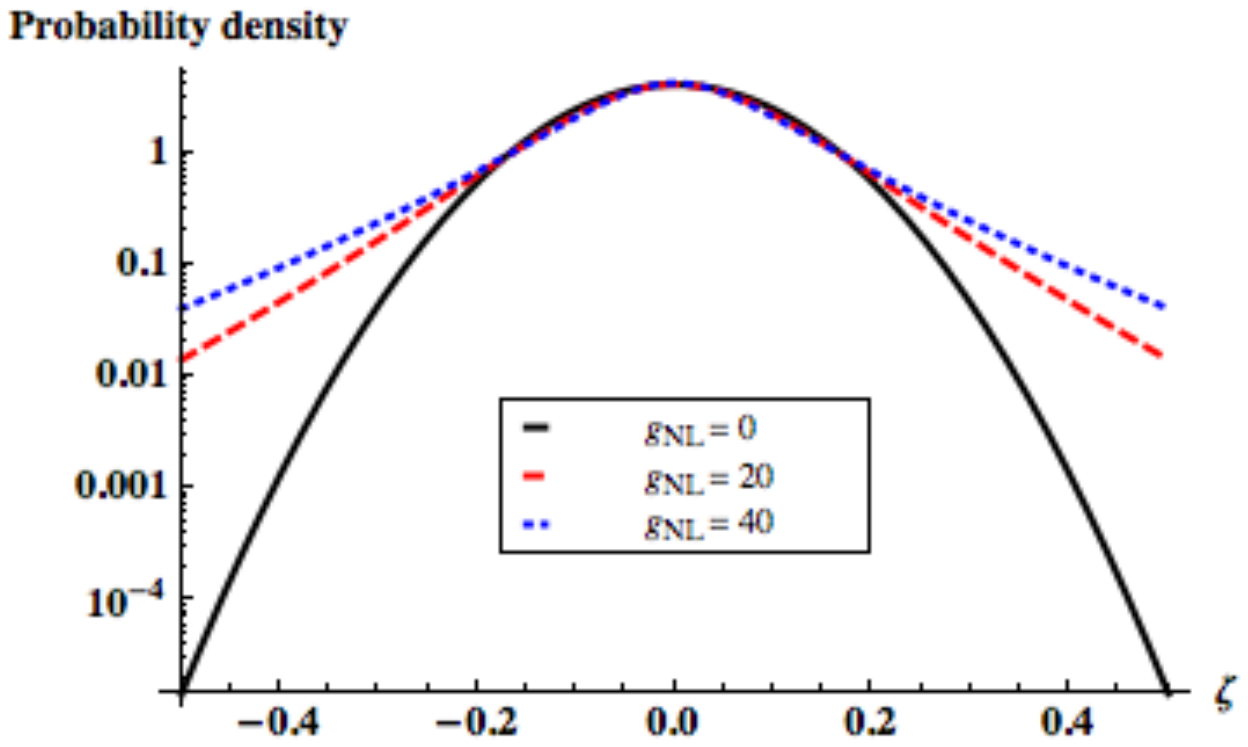}
    }
    \hfill
    \subfloat[Positive tail of the PDF]{%
      \includegraphics[width=0.49\textwidth]{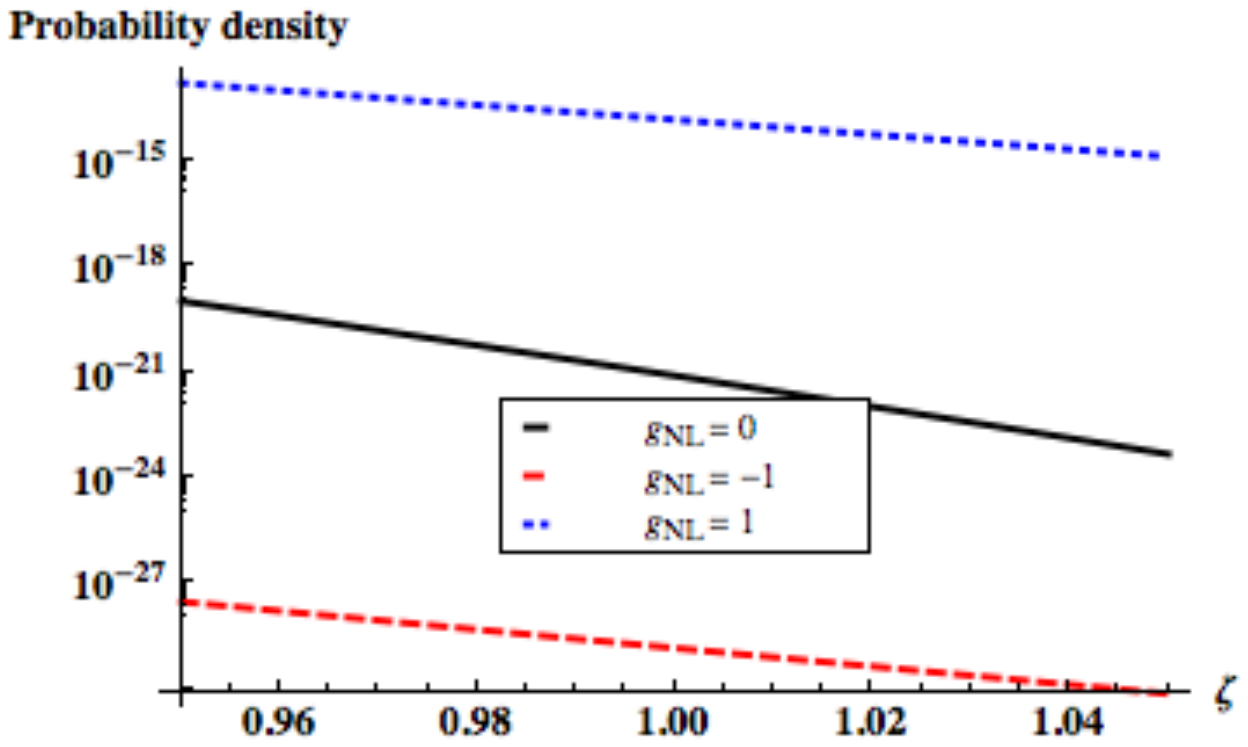}
    }
 \caption[The effect of $\gNL$ on probability density function]{The top left (right) plot shows the effect of negative (positive) $g_{NL}$ on the PDF. We see that $g_{NL}$ affects the kurtosis of the distribution. Positive $g_{NL}$ always gives a sharper peak with broader tails - enhancing PBH production. Large negative $g_{NL}$ has a similar effect - however, we see two sharp peaks in the distribution, due to the derivative in Eq.~(\ref{changeofvariable}) becoming infinite. For small negative $g_{NL}$ we see that the tails of the distribution are diminished. The bottom plot shows the tail of the PDF where PBHs form - again showing a very strong dependence on small amounts of non-Gaussianity, and again the sign of the non-Gaussianity is important. For these plots, we have again taken $\sigma=0.1$.}
 \label{gnlpdfs}
\end{figure}

Figure \ref{gnlpdfs} shows a log plot of the effects of $g_{NL}$ on the PDF. The upper left (right) panel shows the effect of positive (negative) $g_{NL}$. We see that $g_{NL}$ affects the kurtosis of the distribution - typically, serving to give a distribution which is more sharply peaked in the central region, but with larger tails. Positive $g_{NL}$ always serves to enhance the amplitude of the tails where PBHs form, as does large negative $g_{NL}$. However, for small negative $g_{NL}$ the tails of the PDF are diminished - leading to a lower PBH abundance (and consequently, weaker constraints). 

\begin{figure}[t]
\centerline{ \includegraphics[scale=0.8]{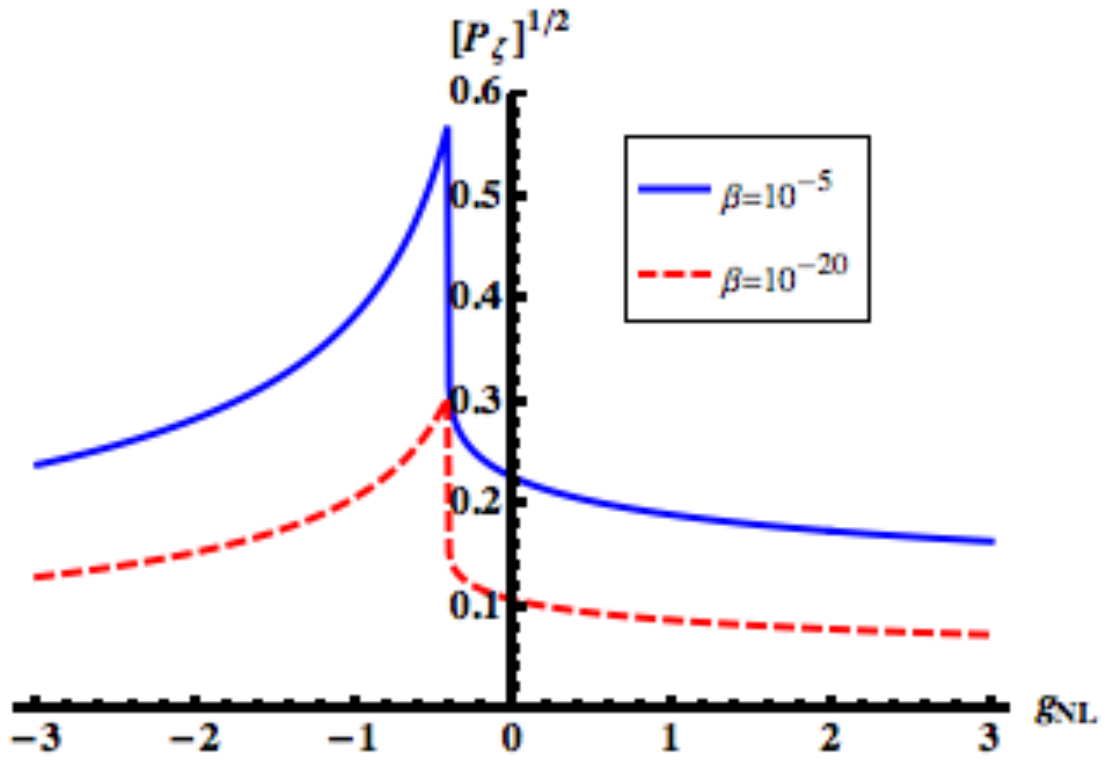}}
 \caption[Constraints on the power spectrum as a function of $\gNL$]{This plot shows how the constraints on the square root of the power spectrum due to PBHs depend on $g_{NL}$. }
\label{gnlconstraints}
\end{figure}

In order to calculate the constraints on the power spectrum, we again write an expression for $\beta$ to be solved. For positive $g_{NL}$ we have
\begin{equation}
\beta\simeq\frac{1}{\sqrt{2\pi}}\int_{y_{1}}^{\infty}e^{-\frac{y^2}{2}}dy.
\end{equation}
For $\frac{-20}{81}<g_{NL}<0$, there are 3 solutions to Eq.~(\ref{eqn:cubicNG}), and $\beta$ is given by
\begin{equation}
\label{betagnl}
\beta\simeq\frac{1}{\sqrt{2\pi}}\left(\int_{-\infty}^{y_{1}}e^{-\frac{y^2}{2}}dy+\int_{y_{2}}^{y_{3}}e^{-\frac{y^2}{2}}dy\right).
\end{equation}
Finally, for $g_{NL}<-\frac{20}{81}$, $\beta$ is given by
\begin{equation}
\beta\simeq\frac{1}{\sqrt{2\pi}}\int_{-\infty}^{y_{1}}e^{-\frac{y^2}{2}}dy.
\end{equation}
The limits on the integrals here ($y_{1}$, $y_{2}$, etc) are solutions for $y$ to Eq.~(\ref{eqn:cubicNG}). The variance in this model is given by \cite{Byrnes:2007tm}
\begin{equation}
{\cal P}_{\zeta}=\sigma^2\left(1+\frac{54}{25}g_{NL}\sigma^{2}\ln(kL)+27\left(\frac{9g_{NL}}{25}\right)^{2}\sigma^{4}\ln(kL)^{2}\right).
\end{equation}
Figure \ref{gnlconstraints} shows the constraints obtained for the cubic non-Gaussianity model. For small $g_{NL}$ we see that the constraints on the power spectrum are highly asymmetric between positive and negative $g_{NL}$. This is because for positive $g_{NL}$ an overdensity in the linear $\zeta$ regime is boosted by the cubic term - especially so in the tail of the PDF, and so the constraints tighten. However, for small negative $g_{NL}$ the opposite is the case and the two terms tend to cancel each other, and hence the constraints weaken dramatically. For very small negative $g_{NL}$, the 2nd term in the expression for $\beta$, Eq.~(\ref{betagnl}), dominates. As $g_{NL}\rightarrow-\frac{20}{81}$ from above, $y_{3}-y_{2}\rightarrow0$, and this term decreases rapidly so that the constraint on the power spectrum rapidly becomes weaker. As $g_{NL}$ becomes more negative, the first term in Eq.~(\ref{betagnl}) increases, and the constraints tighten again. As $g_{NL}$ becomes large, either positive or negative, then the cubic term in Eq.~(\ref{eqn:cubicNG}) dominates the expression, $\zeta\propto\pm\zeta_{g}^{3}$, and the constraints don't depend on the sign of $g_{NL}$. This is because the Gaussian PDF is invariant under a change of sign of $\zeta_{g}$, which is equivalent to changing the sign of $g_{NL}$ (in the case where the linear term is absent). For this reason, the constraints asymptote to the same value as $|g_{NL}|\rightarrow\infty$.

\subsection{Higher order terms}\label{sec:higher-order}

In this section, we consider the effects of higher order terms on the constraints that can be placed on the power spectrum. We take the model of local non-Gaussianity to be
\begin{equation}
\label{fullexpansion}
\zeta=\zeta_{g}+\frac{3}{5}f_{NL}\left(\zeta_{g}^{2}-\sigma^{2}\right)+\frac{9}{25}g_{NL}\zeta_{g}^3+\frac{27}{125}h_{NL}\left(\zeta_{g}^{4}-3\sigma^{4}\right)+\frac{81}{625}i_{NL}\zeta_{g}^5+\cdots .
\end{equation}
Higher order terms have a similar effect on the PDF as do the quadratic and cubic terms - even order terms introduce skew-like asymmetry to the PDF, whilst odd order terms affect kurtosis, and have similar effects on the tails of the PDFs.

\begin{figure}[t]
\centerline{ \includegraphics[scale=0.8]{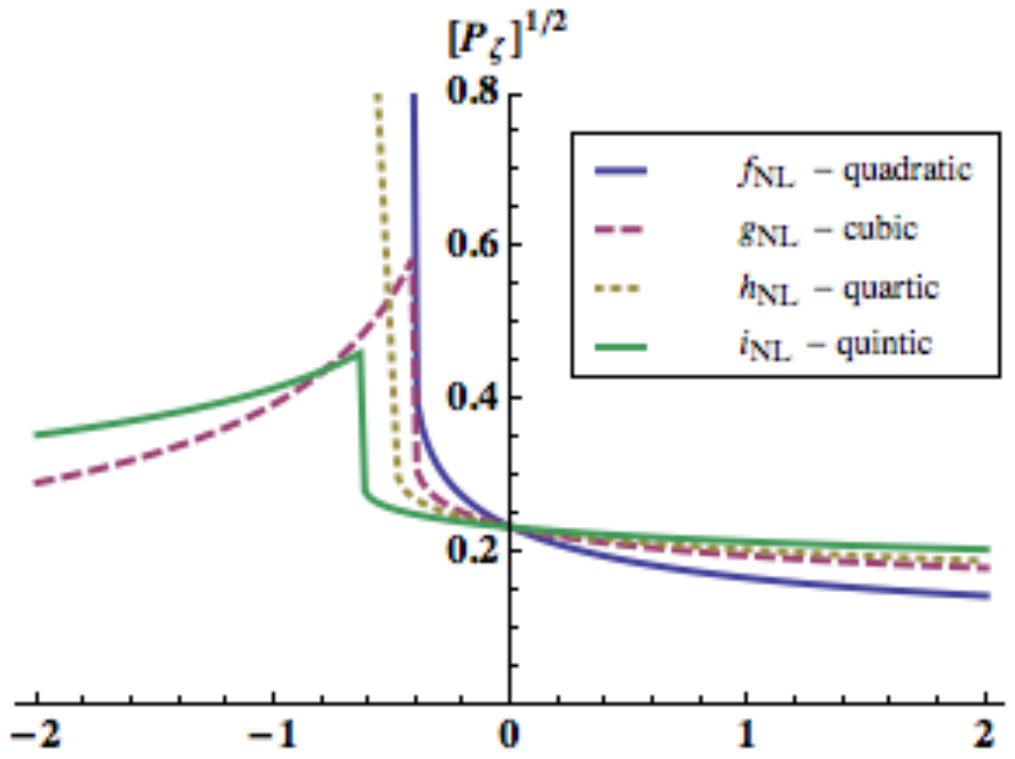}}
 \caption[Constraints on the power spectrum dependant on higher-order non-Gaussianity parameters]{Here we see how the constraints on the square root of the power spectrum depends on non-Gaussianity to $5^{th}$ order. We have considered the addition of each order term one at a time. Note that the even order terms display similar behaviour to each other, as do the odd order terms. The constraints here are shown for the case $\beta=10^{-5}$. Here, we have included only the linear term and one other term in Eq.~(\ref{fullexpansion}) for each order equation. The x-axis is either $f_{NL}$, $g_{NL}$, $h_{NL}$, or $i_{NL}$, depending on the order equation being used.}
\label{fghi}
\end{figure}

The number of solutions to $\zeta(\zeta_g)=1$ depends on the values of $f_{NL}$, $g_{NL}$, $h_{NL}$, etc. Because an analytic solution is not typically available for polynomial equations above 4th order, a numerical method was used to calculate the constraints on the power spectrum. Starting from the linear, purely Gaussian model, a value for $\sigma$ is calculated. The non-Gaussianity parameters are then varied slowly, and Eq.~(\ref{fullexpansion}) is solved using the previous value of $\sigma$ to find critical values of $\zeta_{g}$ required for PBH formation,
\begin{equation}
\zeta_{g}(\zeta_{c})=\zeta_{g1}, \zeta_{g2},\cdots .
\end{equation}
As before, a Gaussian variable $y$ with unit variance is used, Eq.~(\ref{y}), and an expression for $\beta$ is written. For example,
\begin{equation}
\beta\simeq\frac{1}{\sqrt{2\pi}}\left(\int_{y_{1}}^{y_{2}}e^{-\frac{y^2}{2}}dy+\int_{y_{3}}^{y_{4}}e^{-\frac{y^2}{2}}dy+...\right).
\end{equation}
This is then solved numerically to find a value for $\sigma$ and the variance is calculated. Provided that small enough steps are taken, and that $\sigma$ varies sufficiently slowly, the results obtained through this method are in excellent agreement to those obtained previously by an analytic method. Accounting for terms to $5^{th}$ order in $\zeta$ and including all orders in loops, using the techniques of \cite{Byrnes:2007tm} we find that the power spectrum is given by
%
%
\begin{eqnarray}
{\cal P}_{\zeta}&=&\sigma^{2}+\left(\frac{3}{5}\right)^2 \left( 4 f_{NL}^2+6 g_{NL}\right)\sigma^{4}\ln(kL)+ \left(\frac{3}{5}\right)^4 \left( 27 g_{NL}^2+ 48 f_{N}h_{NL}+30 i_{NL}\right) \sigma^{6}\ln(kL)^2 \nonumber \\
&&+\left(\frac{3}{5}\right)^6 \left( 240 h_{NL}^2+450 g_{NL} i_{NL}\right)\sigma^{8}\ln(kL)^3 +
\left(\frac{3}{5}\right)^{8}  2625 i_{NL}^2\sigma^{10}\ln(kL)^4.
\end{eqnarray}
Figure \ref{fghi} shows how the constraints on the power spectrum depend upon the non-Gaussianity parameters. Here, we consider the effects of each term in Eq.~(\ref{fullexpansion}) one at a time. Again, for higher order terms, we see similar behaviour to that seen for the quadratic and cubic non-Gaussianity. For even-order terms, the constraints become tighter for positive values, but weaken dramatically even for small negative values. For odd-order terms, the constraints become tighter for positive values, but for small negative values, the constraints initially weaken dramatically before tightening again. The constraints are most sensitive to small negative non-Gaussianity - where the positive tail of the PDF is strongly reduced, either due to a skew-like asymmetry in the PDF from even terms, or kurtosis type effects from the odd terms. 

\section{Hierarchical scaling}

In order to study the effects of the different types of local non-Gaussianity simultaneously, we introduce some hierarchical scaling relationship between the non-Gaussianity parameters. Here, we present the simple idea of a power law scaling between the terms:
\begin{equation}
\label{eqn:powerlaw}
g_{NL}\sim \alpha^{2} f_{NL}^{2},
h_{NL}\sim \alpha^{3} f_{NL}^{3},
i_{NL}\sim \alpha^{4} f_{NL}^{4}, \cdots,
\end{equation}
where $\alpha$ is a constant of order unity, and the model of local non-Gaussianity can be taken as
\begin{equation}
\zeta\sim\zeta_{g}+\frac{3}{5}f_{NL}\left(\zeta_{g}^{2}-\sigma^{2}\right)+\frac{9}{25}\alpha^{2}f_{NL}^{2}\zeta_{g}^3+\frac{27}{125}\alpha^{3}f_{NL}^{3}\left(\zeta_{g}^{4}-3\sigma^{4}\right)+\frac{81}{625}\alpha^{4}f_{NL}^{4}\zeta_{g}^5+\cdots.
\end{equation}
This type of relation can occur in several different models, including multi-brid inflation \cite{Lin:2010ua,Elliston:2012wm}, a similar scaling was used in \cite{Shandera:2012ke}.

\begin{figure}[t]
\centerline{ \includegraphics[scale=0.8]{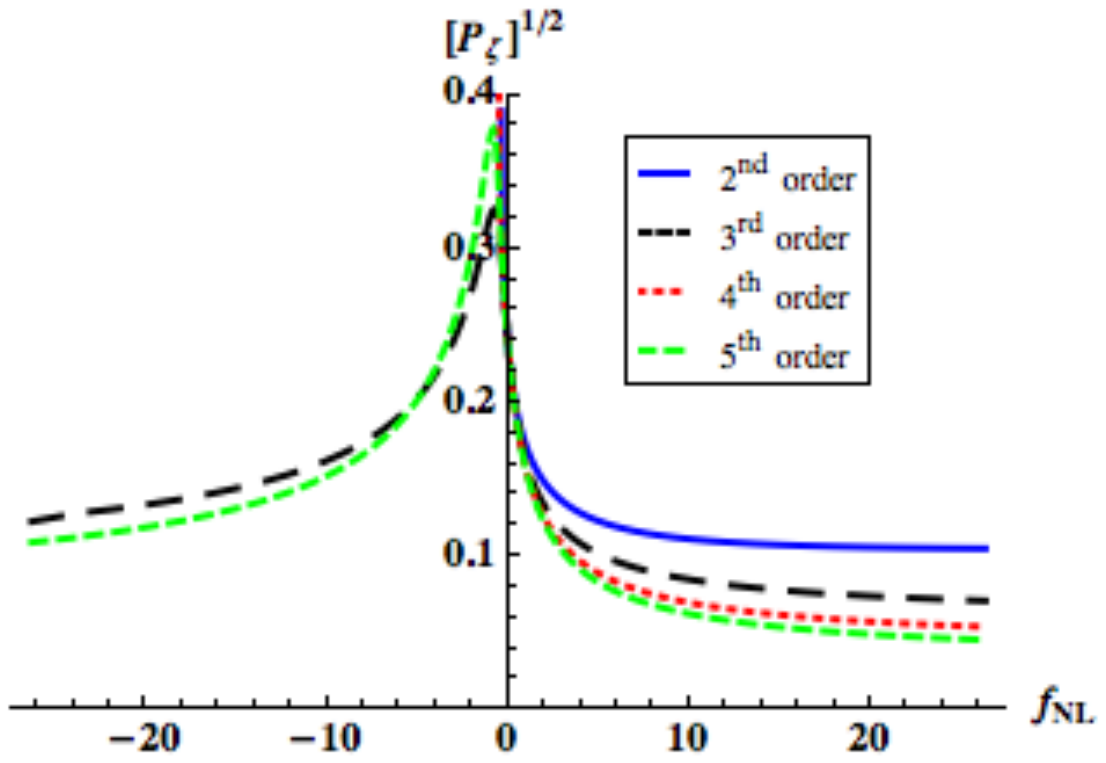}}
 \caption[Constraints on the power scaling and hierarchical scaling]{The constraints on the square root of the power spectrum in the case of a hierarchical power law between the non-Gaussianity parameters. The constraints here are shown for the case $\beta=10^{-5}$. We have used here the hierarchical power rule to different orders (up to $5^{th}$ order), and show the constraints obtained in each case change depending on $f_{NL}$. Note that we see two distinct behaviours - depending on whether the highest term on the expansion is odd or even - which give very different results for the case of negative $f_{NL}$.}
\label{fig:powerlaw}
\end{figure}

Figure \ref{fig:powerlaw} shows the effect of the hierarchical scaling to the constraints on the power spectrum to different orders, where we have taken $\alpha=1$ (modifying this term but keeping it of order unity does not significantly affect the results). When calculating to $n^{th}$ order, we have now included all terms up to and including the $n^{th}$ term (rather than just the single term in the previous section). Again, we see similar behaviour for the different order expansions - depending on whether the highest order term is even or odd.

For positive $f_{NL}$ the constraints tighten significantly as $f_{NL}$ increases, before converging to some constant as $f_{NL}\rightarrow\infty$. As $f_{NL}$ becomes large however, the highest-order term dominates Eq.~(\ref{fullexpansion}), and it is sufficient to take, for example, $\zeta\propto\zeta_{g}^{n}$. Note that the constraints found in this region depend on the order that Eq.~(\ref{fullexpansion}) is taken to - the constraints are slightly tighter for higher orders.

For negative $f_{NL}$, we see similar behaviour to that seen before when only a single term was considered. When the highest order term is even the constraints weaken dramatically as $f_{NL}$ becomes negative,  again requiring fine tuning to produce any PBHs without overproducing them. When the highest order terms are odd, we again see a peak where the constraints weaken for small negative values, before slowly tightening - however, the peak is now smoother. Again, as $|f_{NL}|\rightarrow\infty$ and for odd terms, the sign of the non-Gaussianity parameter does not matter, and the constraints approach the same value. Whilst this may not be obvious from figure \ref{fig:powerlaw}, if the axes were extended to large $f_{NL}$, of order $10^{4}$, we would see this to be the case.

\section{PBHs in the curvaton model}

Whilst the simplest inflationary models give rise to a nearly Gaussian distribution of the primordial curvature perturbation, multi-field models of inflation can lead to strong non-Gaussianity. One well motivated model is the curvaton model \cite{Sasaki:2006kq}. In this model, in addition to the field driving inflation, the inflaton $\phi$, there is a second light scalar field, the curvaton $\chi$, whose energy density is completely subdominant during inflation. At Hubble exit during inflation both fields acquire classical perturbations that freeze in. Here, the observed perturbations in the CMB and LSS, as well as perturbations on smaller scales, can result from the curvaton instead of the inflaton. At the end of inflation, the inflaton decays into relativistic particles (``radiation"). The curvaton energy density is still sub-dominant at this stage and carries an isocurvature perturbation - and at some later time, the curvaton also decays into radiation. Taking the curvaton to be non-relativistic before it decays, the energy density of the curvaton will decay slower than the energy density of the background radiation - and consequently the curvature perturbation due to the curvaton will become dominant. 

If the curvaton generates the perturbations on CMB scales, then in simple realisations of the curvaton scenario with a quadratic potential it cannot have a much larger amplitude of perturbations on smaller scales. However it is possible that a second stage of inflation has a dominant contribution to its perturbations from the curvaton model. Indeed if the curvaton mass, $m_{\chi}$, is reasonably heavy compared to the Hubble scale, then it will naturally have a blue spectrum giving the smallest scale perturbations the largest amplitude. The spectral index is given by $n_s-1=2 m^2_{\chi}/(3H^2)+2\dot{H}/H^2$, where all quantities should be evaluated at the horizon crossing time of the relevant scale \cite{Sasaki:2006kq}. Motivated by our discovery in the last section that truncating the PDF to any order in the non-linearity parameters can give a very bad approximation to the true result, a practical reason for studying the curvaton scenario is that this is a rare case in which the full non-linear PDF has been calculated. This allows us to check in a realistic and popular model whether the non-Gaussian corrections to the PDF are important, and whether just including the first few terms such as $f_{NL}$ or $g_{NL}$ would give an accurate result. We will see that the non-Gaussian corrections to all orders are always important when studying PBH formation.

\begin{figure}[t]
\centerline{ \includegraphics[scale=0.8]{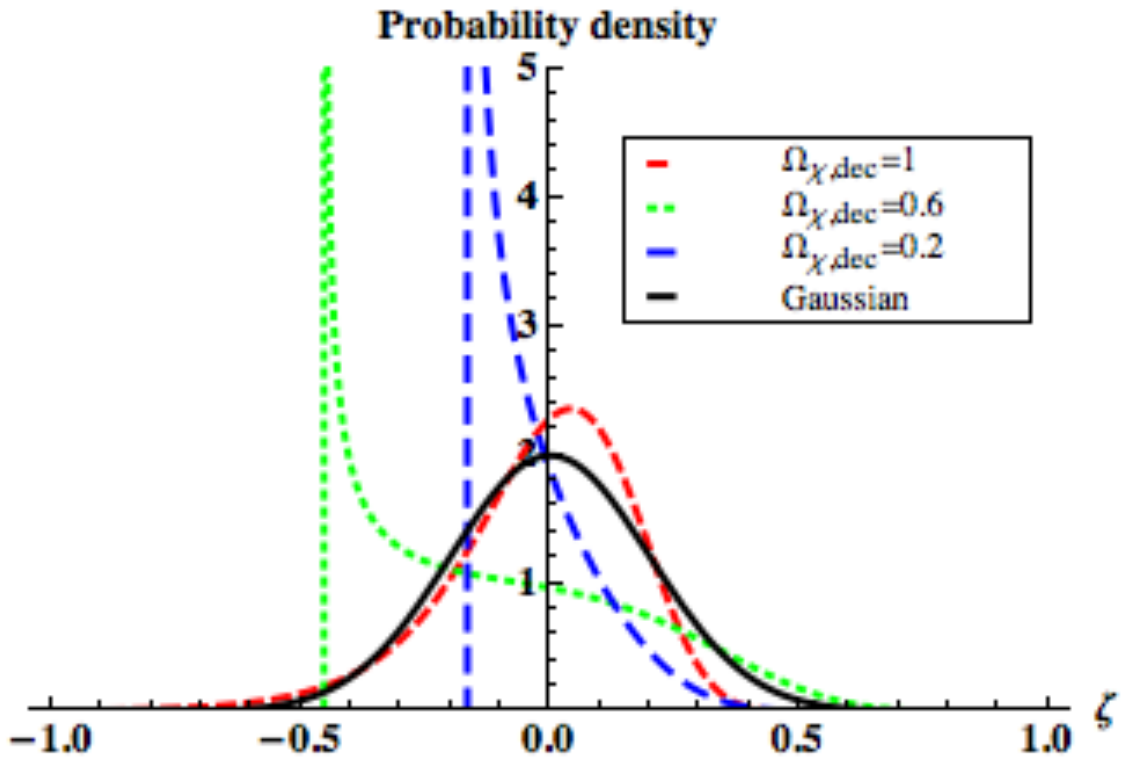}}
 \caption[Probability distributions in the curvaton model]{PDFs in the curvaton model. Here we see that, whilst $\Omega_{\chi,dec}\sim1$ the distribution is close to Gaussian. However, as $\Omega_{\chi,dec}$ the PDF becomes more non-Gaussian, enhancing the positive tail of the PDF. These have been calculated using a formal change of variable using \ref{curvatonequation} and \ref{nononlinear}. All the plots have a variance $\langle \zeta^{2}\rangle=0.04$.}
\label{curvatonpdfs}
\end{figure}

Here we use the result obtained by \cite{Sasaki:2006kq} in the \emph{sudden decay approximation}:
\begin{equation}
\label{curvrad}
\left(1-\Omega_{\chi,dec}\right)e^{4(\zeta_{r}-\zeta)}+\Omega_{\chi,dec}e^{3(\zeta_{\chi}-\zeta)}=1,
\end{equation}
where $\Omega_{\chi,dec}$ is the dimensionless curvaton density parameter for the curvaton at the decay time. Taking the curvature perturbation in the radiation fluid to be negligible, i.e. $\zeta_{r}=0$, Eq.~(\ref{curvrad}) reads
\begin{equation}
\label{curvatonequation}
e^{3\zeta_{\chi}}=\frac{1}{\Omega_{\chi,dec}}\left(e^{3\zeta}+(\Omega_{\chi,dec}-1)e^{-\zeta}\right).
\end{equation}
This gives the fully non-linear relation between the primordial curvature perturbation, $\zeta$, and the curvaton curvature perturbation, $\zeta_{\chi}$. Taking there to be no non-linear evolution between Hubble exit and the start of curvaton decay, the left hand side of Eq.~(\ref{curvatonequation}) is given by
\begin{equation}
\label{nononlinear}
e^{3\zeta_{\chi}}=\left(1+\frac{\delta_{1}\chi}{\bar{\chi}}\right)^{2},
\end{equation}
where $\delta_{1}\chi$ is the Gaussian perturbation in the curvaton field at Hubble exit, and $\bar{\chi}$ is the background value. Eq.~(\ref{curvatonequation}) is quartic in $e^{\zeta}$ and so this allows us to write an expression for the full curvature perturbation, $\zeta$, in terms of the Gaussian variable $\delta_{\chi}=\frac{\delta_{1}\chi}{\bar{\chi}}$, or equivalently write the Gaussian variable as a function of the curvature perturbation.
\begin{equation}
\delta_{\chi}=\delta_{\chi}(\zeta).
\end{equation}
Note that, for $\Omega_{\chi,dec}<1$, $\zeta$ is bounded from below, with the minimum value given by
\begin{equation}
\zeta_{min}=\frac{1}{4}\ln\left(1-\Omega_{\chi,dec}\right).
\end{equation}
Making a formal change of variable allows the PDF to be calculated. Figure \ref{curvatonpdfs} shows the PDFs obtained for different values of $\Omega_{\chi,dec}$. Whilst $\Omega_{\chi,dec}$ is close to unity, the PDF is close to Gaussian - however, the positive tail of the PDF is diminished, reducing PBH formation. As $\Omega_{\chi,dec}$ becomes smaller, the PDF becomes more strongly non-Gaussian, and the positive tail of the PDF is enhanced, increasing PBH formation.

Constraints on the power spectrum are obtained using the same method as before. Eq.~(\ref{curvatonequation}) is solved for $\zeta=\zeta_{c}$ to find the corresponding critical values of $\delta_{\chi}$, giving two solutions, $\delta_{c1}$ and $\delta_{c2}$, for all values of $\Omega_{\chi,dec}$. An expression for $\beta$ is written:
\begin{equation}
\beta\simeq\frac{1}{\sqrt{2\pi}\sigma}\left(\int^{\infty}_{\delta_{c1}}e^{-\frac{\delta_{\chi}^{2}}{2\sigma}}d\delta_{\chi}+\int^{\delta_{c2}}_{-\infty}e^{-\frac{\delta_{\chi}^{2}}{2\sigma}}d\delta_{\chi}\right).
\end{equation}
This expression is then solved numerically to find a value for $\sigma$ for a given value of $\beta$. Now that all of the necessary components have been found, the constraints on the power spectrum are calculated by finding the variance through numeric integration
\begin{equation}
{\cal P}_{\zeta}=\int_{\zeta_{min}}^{\infty}\zeta^{2}P_{NG}(\zeta)d\zeta=\int_{-\infty}^{\infty}\zeta(\chi_{g})^{2}P_{G}(\chi_{g})d\chi_{g},
\end{equation}
where $P_{NG}(\zeta)$ and $P_{G}(\chi_{g})$ are the non-Gaussian and Gaussian PDF's respectively. Care needs to be taken to ensure that the mean of $\zeta$ is zero during the calculation - if necessary defining a new variable with the mean subtracted, such that $\langle\zeta\rangle=0$.

\begin{figure}[t]
\centerline{ \includegraphics[scale=0.8]{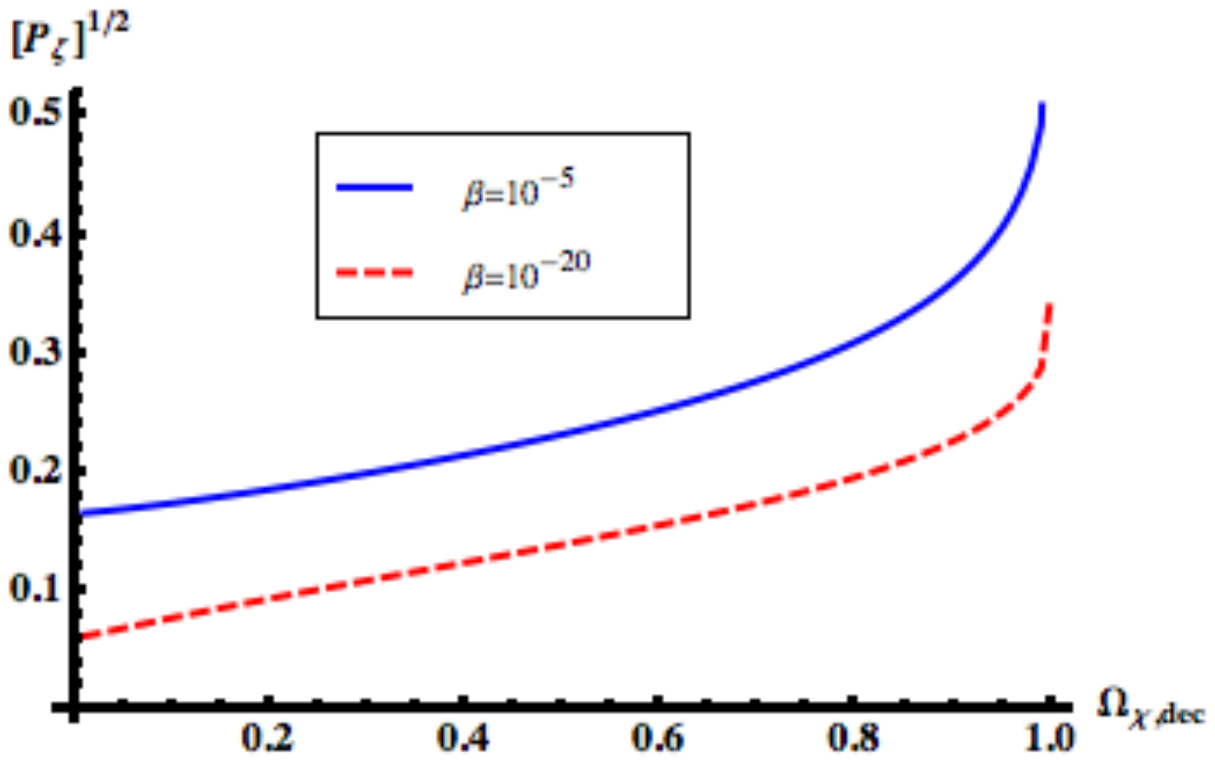}}
 \caption[Constraints on the power spectrum in the curvaton model]{Constraints on the square root of the power spectrum in the curvaton model. The constraints obtained for different constraints on $\beta$, the initial PBH mass fraction, as a function of $\Omega_{\chi,dec}$, the dimensionless curvaton density parameter at the time of decay.}
\label{curvatonconstraints}
\end{figure}

Figure \ref{curvatonconstraints} shows the constraints obtained for different values of $\beta$. When $\Omega_{\chi,dec}\sim1$, the constraints are weaker than in the Gaussian case even though the PDF is close to Gaussian - this is an example of even small amounts of non-Gaussianity having a large impact on the constraints. As $\Omega_{\chi,dec}\rightarrow0$, the constraints on the power spectrum become tighter, corresponding to an enhancement of the positive tail of the PDF.

\begin{figure}[t]
\centerline{ \includegraphics[scale=0.8]{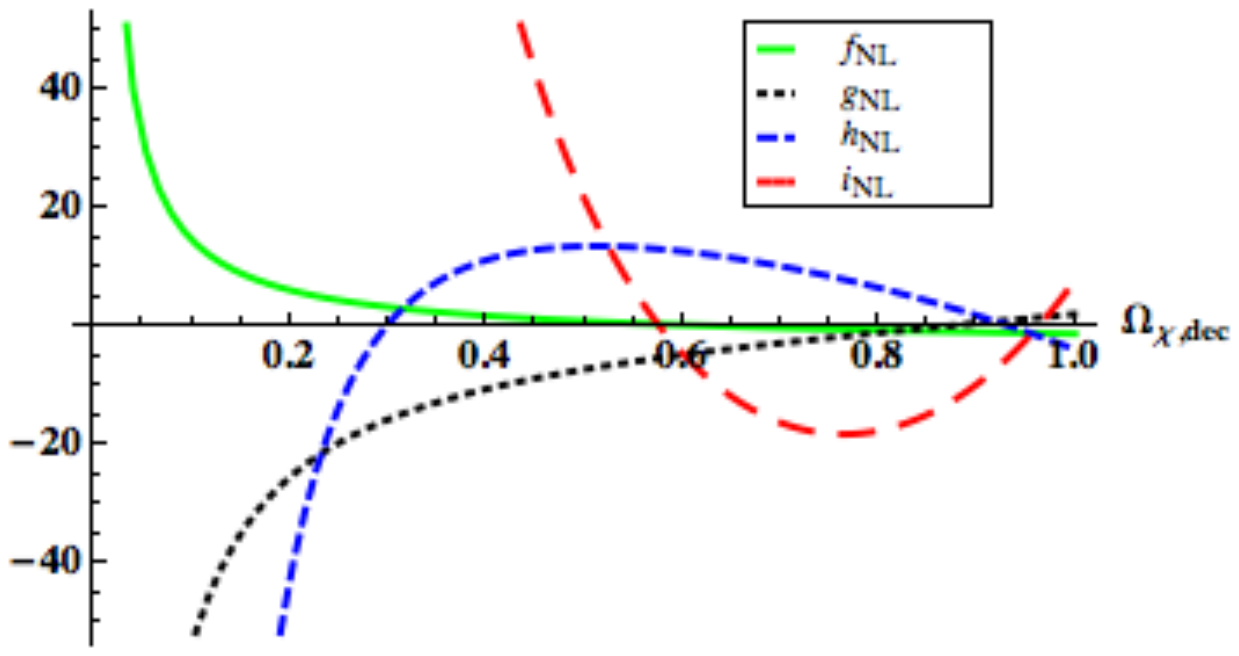}}
 \caption[Non-Gaussianity parameters in the curvaton model]{The non-Gaussianity parameters in the curvaton model.}
\label{curvatonparameters}
\end{figure}

It should be noted that, in this model, a full expansion for $\zeta$ can be obtained by performing a Taylor expansion of the solution to Eq.~(\ref{curvatonequation}) \cite{Sasaki:2006kq,Lin:2010ua}. Figure \ref{curvatonparameters} shows the non-Gaussianity parameters plotted as a function of $\Omega_{\chi,dec}$. Instead of using the full non-linear solution for $\zeta$, the calculation can be completed as in the previous section by using these solutions for the parameters. However, the results obtained in this manner typically do not match well with those obtained from an analytic solution - the contributions to the power spectrum from higher-order terms can become large and can be either positive or negative. This is due to the fact that, whatever order the expansion is carried out to, the Taylor expansion diverges from the analytic solution as $\zeta$ becomes large (of order unity or higher). For example, for $\Omega_{\chi,dec}=1$, $f_{NL}=-\frac{5}{4}$, and so a truncation at second order would not even come close to matching with the results obtained here. Comparing the constraints for $\beta=10^{-5}$ between Figs.~\ref{curvatonconstraints} and \ref{fnlconstraints}, notice that the Gaussian constraint of ${\cal P}_{\zeta}^{1/2}=0.23$ is reached for $\Omega_{\chi,dec}\simeq0.4$, but from figure \ref{curvatonparameters} we see that the non-linearity parameters are not typically small here, and so the matching is just coincidence. Hence we conclude that the non-Gaussianity of the curvaton model always has to be taken into account when calculating PBH constraints. 

\section{Summary}

The abundance of PBHs probes the tails of the PDF of primordial fluctuations, and is very sensitive to the effects of non-Gaussianity. We have calculated the effects of the local model of non-Gaussianity for terms up to 5th order, parameterized by $f_{NL}$, $g_{NL}$, $h_{NL}$, and $i_{NL}$. We have shown that any non-Gaussianity parameters of order unity can have a significant effect on the abundance of PBHs, and the constraints that can be placed on the power spectrum - due to the fact that the non-Gaussianity parameters have a large impact on the tails of the PDF.

The sign of the non-Gaussianity has a particularly strong effect. We see that positive terms of even order tighten the constraints significantly, but negative terms dramatically weaken the constraints, to the point where the curvature perturbation is order unity. Typically, when an even type of non-Gaussianity is considered, such as $f_{NL}$ or $h_{NL}$, if this term is negative and dominates the non-Gaussianity of the distribution, the amplitude of the primordial fluctuations will either be too small to form any PBHs, or so large that the Universe contains too many PBHs. Such a scenario would be incompatible with any future detection of PBHs. Odd-order terms, such as $g_{NL}$ or $i_{NL}$, tend to tighten the constraints regardless of their sign, but small negative terms can weaken the constraints dramatically over a small range of values. If PBHs were to be detected in the future, they could potentially rule out certain models and distributions. Care needs to be taken as truncations to set order in the model of non-Gaussianity used might not converge.

In the curvaton model, the PDF is relatively close to Gaussian if the Universe is dominated by the curvaton at the time of decay, $\Omega_{\chi,dec}\sim1$ - and in this case the constraints are weakened compared to a purely Gaussian distribution. As $\Omega_{\chi,dec}$ decreases, the distribution becomes more non-Gaussian, and the constraints on the power spectrum tighten significantly. Calculations obtained for the curvaton model by calculating the local non-Gaussianity parameters to e.g.~second or third order ($f_{NL}$ or $g_{NL}$) do not agree with those obtained using the full non-linear solution. Therefore, given a specific model, it may be necessary to calculate the full hierarchy (rather than truncating at a given order) before performing calculations, as we have done here for the curvaton model.

\section{Acknowledgements}
We would like to thank Anne Green and David Seery for useful discussions. SY thanks the STFC for financial support. CB is supported by a Royal Society University Research Fellowship.


\newpage

\chapter{The long-short wavelength mode coupling tightens primordial black hole constraints}
\label{chap:paper3}



\begin{center}


Sam Young, Christian T. Byrnes\\[0.5cm]
Department of Physics and Astronomy, Pevensey II Building, University of Sussex, BN1 9RH, UK\\[0.5cm]


\end{center}

The effects of non-Gaussianity on the constraints on the primordial curvature perturbation power spectrum from primordial black holes (PBHs) are considered. We extend previous analyses to include the effects of coupling between the modes of the horizon scale at the time the PBH forms and super-horizon modes. We consider terms of up to third order in the Gaussian perturbation. For the weakest constraints on the abundance of PBHs in the early universe (corresponding to a fractional energy density of PBHs  of $10^{-5}$ at the time of formation), in the case of Gaussian perturbations, constraints  on the power spectrum are $\mathcal{P}_{\zeta}<0.05$ but can significantly tighter when even a small amount of non-Gaussianity is considered, to $\mathcal{P}_{\zeta}<0.01$, and become approximately $\mathcal{P}_{\zeta}<0.003$ in more special cases. Surprisingly, even when there is negative skew (which naively would suggest fewer areas of high density, leading to weaker constraints), we find that the constraints on the power spectrum become tighter than the purely Gaussian case - in strong contrast with previous results. We find that the constraints are highly sensitive to both the non-Gaussianity parameters as well as the amplitude of super-horizon perturbations.

\newpage

\section{Introduction}
Theoretical arguments suggest that, if the right conditions are met, primordial black holes (PBHs) could have formed from the collapse of large density perturbations in the early universe. As a perturbation reenters the horizon, gravity can overcome the pressure forces and cause the perturbation to collapse to form a PBH with a mass of order the horizon mass. In order to collapse, then certain formation criteria need to be met, and this is normally stated in terms of the density contrast $\delta$ or the curvature perturbation $\zeta$. PBHs have traditionally been used to constrain the small scales of the early universe - and represent a unique window to constrain smallest scales. Whilst we have precision measurements from sources such as the cosmic microwave background (CMB) and large scale structure (LSS) (e.g.~the \emph{Planck} constraints on inflation \cite{Ade:2013uln}), these only place constraints on a handful e-folds of the largest scales inside the visible universe. PBHs can be used to place constraints on the power spectrum spanning around 50 e-folds, although the constraints from PBHs are typically much weaker \cite{Josan:2009qn}. Ultra compact mini-haloes (UCMHs) can also be used to probe small scales \cite{Bringmann:2011ut}, although these constraints depend on dark matter particles decaying into observable particles, and do not cover as large a range of scales as the constraints from PBHs.

PBHs have never been observed, either directly or indirectly, but there are tight observational constraints on the abundance of PBHs, and these are used to constrain the power spectrum, as will be described later. The constraints on the abundance of PBHs come from the effects of their evaporation on the early universe for small PBHs, or the effects of their gravity on the later universe for larger ones. The constraints are typically stated in terms of $\beta$, the mass fraction of the universe going into PBHs at the time of formation. The constraints on $\beta$ range from $\beta\lesssim 10^{-25}$ to $\beta<10^{-5}$, depending on the mass of PBH being considered. For recent updates and a compilation of the constraints see \cite{Josan:2009qn, Carr:2009jm}.

The constraints on the power spectrum coming from PBHs are typically of order $10^{-2}$, orders of magnitude larger than those observed on cosmic scales. Whilst a spectral index less than unity, $n_{s}\approx0.96$, has been observed (e.g.~\cite{Ade:2013uln}) on cosmic scales, suggesting the power spectrum should become smaller on small scales, it is nonetheless possible for it to become large on small scales and form a significant number of PBHs. This can be seen in numerous models, including the running mass model \cite{Drees:2011hb}, axion inflation \cite{Bugaev:2013fya}, a waterfall transition during hybrid inflation \cite{Bugaev:2011wy, Lyth:2012yp,Halpern:2014mca}, from passive density fluctuations \cite{Lin:2012gs}, or in inflationary models with small field excursions but which are tuned to produce a large tensor-to-scalar ratio on large scales \cite{Hotchkiss:2011gz}. See also \cite{Linde:2012bt,Torres-Lomas:2014bua,Suyama:2014vga}. For further reading and a summary of various models that can produce PBHs, see \cite{Green:2014faa}. Alternatively, the constraint on the formation criteria can be relaxed during a phase transition in the early universe, causing PBHs to form preferentially at that mass scale, e.g.~\cite{Jedamzik:1999am}.

The constraints from PBHs on the primordial power spectrum are highly sensitive to even small amounts of non-Gaussianity, and this has been studied extensively in the literature (e.g.~\cite{Bullock:1996at,Ivanov:1997ia,Seery:2006wk,Shandera:2012ke}), and in this paper we extend the calculation conducted by \cite{Byrnes:2012yx, Young:2013oia} to include the effects of large-scale inhomogeneities in the distribution caused by non-Gaussianity. 

In Section 2, we review how constraints on the abundance of PBHs can be used to constrain the power spectrum, and in Section 3 we review previous calculations of how local-type non-Gaussianity affects these constraints, as well as a more general discussion of the effects of non-Gaussianity. In Section 4, we describe how the presence of non-Gaussianity and large super-horizon modes can affect the abundance of PBHs which form on smaller scales, and apply this to the calculation of constraints in Sections 5 and 6, for quadratic and cubic type non-Gaussianity respectively. We finish with a discussion of key points in Section 7.

\section{Constraining the power spectrum}

Using the fact that PBHs have not been observed, one can place an upper limit on the primordial power spectrum on scales that could not otherwise be constrained. In this paper, this upper limit on the power spectrum, and its dependence upon non-Gaussianity, will be calculated. There are different constraints on the abundance of PBHs of different masses - and therefore different constraints on the primordial power spectrum \cite{Josan:2009qn}.

The abundance of PBHs is normally stated as the mass fraction of the universe contained within PBHs at the time of formation, $\beta$, and in a recent paper we showed how this can be calculated directly from the curvature perturbation power spectrum, $\mathcal{P}(\zeta)$, matching well with the traditional calculation (which calculates the abundance by using window functions to smooth the distribution). $\beta$ is given by
\begin{equation}
\beta=2\int_{\zeta_{c}}^{\infty}P(\zeta)d\zeta,
\end{equation}
where $\zeta_{c}$ is the threshold value for PBH formation, and $P(\zeta)$ is the probability density function (PDF) of $\zeta$. In the case of a Gaussian distribution of the curvature perturbation, this can be approximated as \cite{Byrnes:2012yx}
\begin{equation}
\beta=\mathrm{erfc}\left(\frac{\zeta_{c}}{\sqrt{2}\sigma}\right)\approx\mathrm{exp}\left(-\frac{\zeta_{c}^{2}}{2\sigma^{2}}\right).
\end{equation}
This can be rewritten to show how the constraints on $\beta$ give constraints on $\mathcal{P}_{\zeta}$,
\begin{equation}
\mathcal{P}_{\zeta}=\sigma^{2}=\sqrt{\frac{\zeta_{c}^{2}}{2\ln\left(1/\beta\right)}}.
\end{equation}
In this paper, we will take the threshold value for PBH formation to be $\zeta_{c}=1$ \cite{Shibata:1999zs,Green:2004wb}\footnote{In order to be consistent with calculations using the density contrast, it is preferable to use a larger value, $\zeta_{c}\approx1.2$ (the upper value found in \cite{Shibata:1999zs}), which matches better with the expected critical value of the density contrast, $\Delta_{c}\approx0.5$. However, whilst $\beta$ is extremely dependent on $\zeta_{c}$, the constraints on the power spectrum do not change significantly - and we use $\zeta_{c}=1$ in order to be consistent with previous papers.}. Significant uncertainty on  the critical value of collapse remains and the result depends on the density profile \cite{Musco:2004ak,Hidalgo:2008mv,Musco:2008hv,Nakama:2013ica,Harada:2013epa,Nakama:2014fra}. For $\beta<10^{-5}$ and $\beta<10^{-20}$, for a Gaussian distribution this gives the constraints $\mathcal{P}_{\zeta}<0.0513$ and $\mathcal{P}_{\zeta}<0.0115$ respectively.

\begin{figure}[t]
\centering
\subfloat[]{%
      \includegraphics[width=\textwidth]{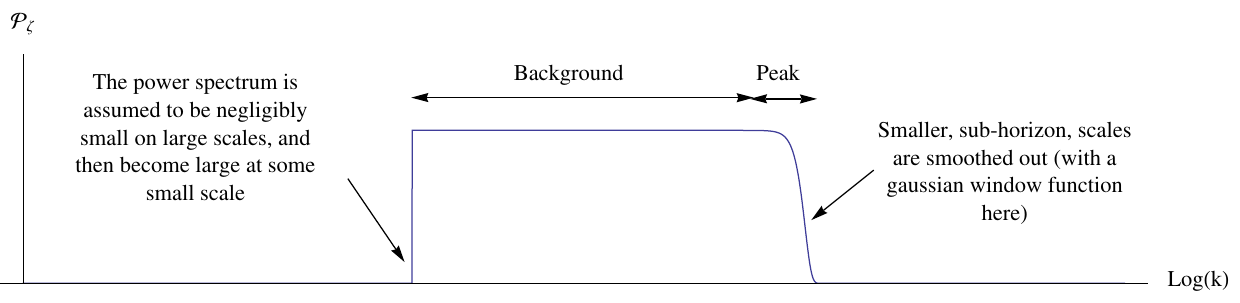}
    }
    \hfill
    \subfloat[]{%
      \includegraphics[width=\textwidth]{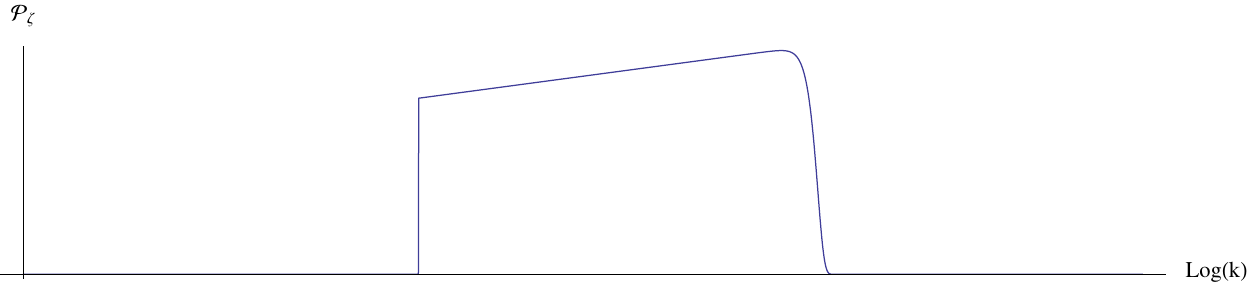}
    }
\caption[Model for the power spectrum]{The form of the power spectrum being used in this paper is shown. For simplicity, we assume that on large scales the power spectrum is negligibly small, $\mathcal{P}_{\zeta}\ll 1$, before quickly becoming large at some scale (in this case with a step function). The power spectrum is then assumed to be large down to arbitrarily small scales - although the effect of smoothing reduces the power on sub-horizon scales to be effectively zero. The ``peak'' scales correspond to the scale at which PBHs are forming at a given time (the horizon scale), where the 'background' scales are so large as to be unobservable. The top figure shows a flat spectrum, which is assumed to be the case for a Gaussian distribution. However, for a non-Gaussian distribution, the effect of coupling between modes will typically serve to increase the power on small scales, even when the amplitude of the Gaussian perturbations is scale invariant, as shown in the bottom figure.}
\label{powerSpectrum}
\end{figure}

In previous papers, we used PBH constraints to calculate how the constraints on the power spectrum depend on the amount of non-Gaussianity present (see section 3), in the local model of non-Gaussianity \cite{Byrnes:2012yx, Young:2013oia}. In this paper, we go beyond previous calculations, and account for large-scale inhomogeneities in the power spectrum caused by the non-Gaussian terms as documented in \cite{Byrnes:2011ri}. Whilst large super-horizon modes in the curvature perturbation do not affect the local evolution of the universe and therefore do not affect whether a region collapses to form a PBH or not \cite{Young:2014ana}, they can have an indirect effect due to their influence on smaller scale modes. In this paper, we will assume that the power spectrum becomes large below a certain scale (as demonstrated in Fig.~\ref{powerSpectrum}), and place constraints on the amplitude of this power spectrum from the constraints on the abundance of PBHs. The top power spectrum shown in Fig.~\ref{powerSpectrum} is scale invariant - which we assume to be the case for a Gaussian distribution. However, for a non-Gaussian distribution, the power spectrum increases as $k$ increases, which is due to the effects of modal coupling - so even though the Gaussian component of the perturbations is constant, overall the power spectrum increases. For a specific model, such a power spectrum is unlikely and a more suitable model for the power spectrum should be used.

\section{Review of non-Gaussian constraints}

It has previously been shown that the constraints which can be placed on the curvature perturbation power spectrum depend upon the distribution of perturbations present in the early universe (recent papers include \cite{Shandera:2012ke,Byrnes:2012yx,Young:2013oia,Bugaev:2012ai}), and that the mass fraction of the early universe going into PBHs, $\beta$, is strongly dependent on the amount of non-Gaussianity present. 

\begin{figure}[t]
\centering
\subfloat[$f_{NL}$]{%
      \includegraphics[width=0.49\textwidth]{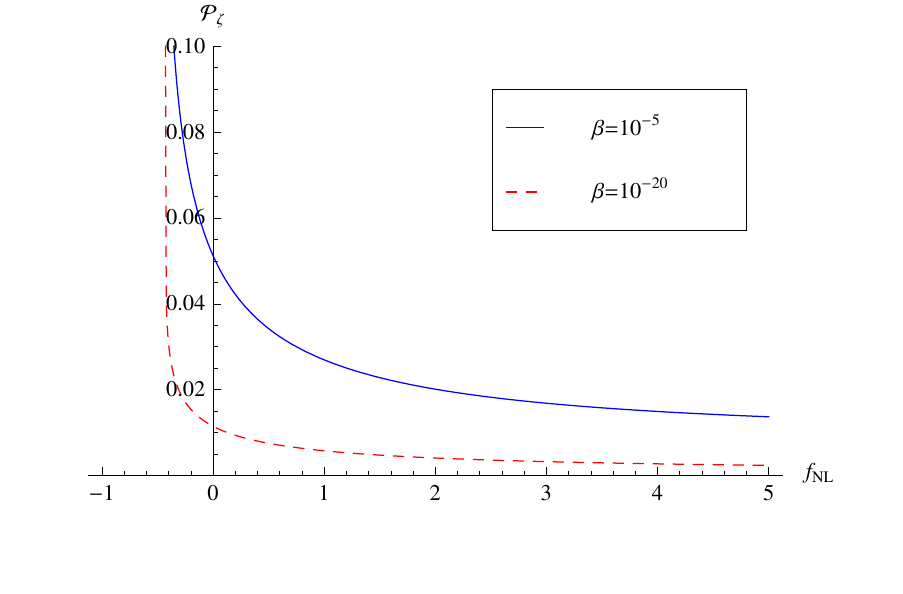}
    }
    \hfill
    \subfloat[$g_{NL}$]{%
      \includegraphics[width=0.49\textwidth]{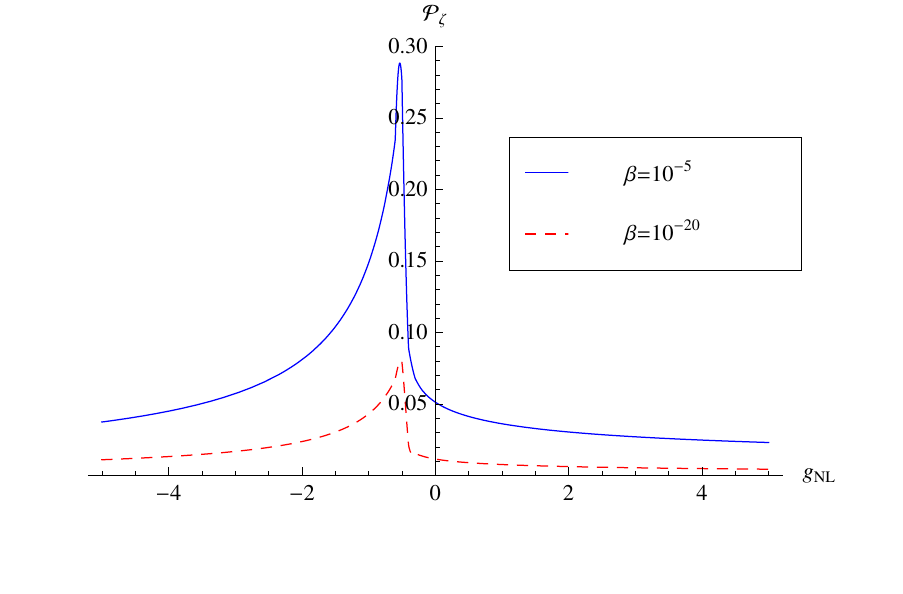}
    }
 \caption[Non-Gaussian dependence of constraints on the power spectrum from primordial black holes]{In the local model of non-Gaussianity, the constraints on the power spectrum, $\mathcal{P}_{\zeta}$, depend strongly upon the non-Gaussianity parameters. The left plot shows how constraints depend on $f_{NL}$ (assuming all higher order terms are zero). The constraints tighten significantly for positive $f_{NL}$ but weaken dramatically for negative $f_{NL}$. The right plot shows how constraints depend on $g_{NL}$ (assuming all higher order terms and the quadratic term are zero). For most values of $g_{NL}$ the constraints are tighter than the Gaussian case, but significantly weaker for small negative values of $g_{NL}$.}
\label{old constraints}
\end{figure}

In this paper, we will consider the local model of non-Gaussianity to third-order,
\begin{equation}
\zeta=\zeta_{G}+\frac{3}{5}f_{NL}\left(\zeta_{G}^{2}-\sigma^{2}\right)+\frac{9}{25}g_{NL}\zeta_{G}^{3}=h\left(\zeta_{G}\right),
\end{equation}
where $\sigma^{2}=\langle\zeta_{G}^{2}\rangle$. We define the solution to this equation as $\zeta_{G}=h^{-1}(\zeta)$, and $\beta$ can be expressed in terms of $h^{-1}(\zeta)$ \cite{Byrnes:2012yx}. Note that, whilst the meaning of $f_{NL}$ and $g_{NL}$ in this paper are the same as that used in observational cosmology of CMB and LSS, similar values of these parameters here have a much larger effect on the distribution than in the CMB or LSS. This is because the constraint on the amplitude of perturbations is much weaker - typically of order $10^{-1}$ rather than $10^{-5}$. Therefore, $f_{NL}\approx 1$ represents approximately a $10\%$ correction. We will here briefly review previous work by considering the case of positive $f_{NL}$ and zero $g_{NL}$, $h^{-1}(\zeta)$ has two solutions, given by
\begin{equation}
h_{\pm}^{-1}(\zeta)=\frac{-5\pm\sqrt{25+36f_{NL}^{2}\sigma^{2}+60f_{NL}\zeta}}{6f_{NL}}.
\end{equation}
$\beta$ can then be calculated by integrating over the PBH forming values of $\zeta_{G}$, giving \footnote{This is equivalent to integrating over the probability distribution function of $\zeta$: $\beta=2\int_{\zeta_{C}}^{\infty}P(\zeta)d\zeta$.}
\begin{equation}
\beta=\mathrm{erfc}(h_{+}^{-1}(\zeta_{c}))+\mathrm{erfc}(\lvert h_{-}^{-1}(\zeta_{c})\rvert).
\label{quadraticbeta}
\end{equation}
The full derivation can be seen in \cite{Byrnes:2012yx}. This expression can then be solved numerically for a given constraint on $\beta$, such as $\beta<10^{-5}$, to find a constraint on $\sigma$, and a constraint on the power spectrum can be calculated using \cite{Byrnes:2007tm}
\begin{equation}
\mathcal{P}_{\zeta}=\sigma^{2}+4\left(\frac{4f_{NL}}{5}\right)^{2}\sigma^{4}\mathrm{ln}(kL).
\end{equation}
Fig.~\ref{old constraints} shows how the constraints on the power spectrum depend upon the non-Gaussianity parameters $f_{NL}$ and $g_{NL}$ for $\beta=10^{-5}$ and $\beta=10^{-20}$. 
\begin{itemize}
\item{The $f_{NL}$ term affects the skew of the distribution - a positive $f_{NL}$ enhances the tail of the distribution, increasing PBH production, which means the constraints become tighter. For negative $f_{NL}$, the constraints weaken dramatically. There is a maximum value of $\zeta$ given by
\begin{equation}
\zeta_{max}=-\frac{5}{6f_{NL}}+\frac{3}{5}f_{NL}\left(\frac{25}{36 f_{NL}^{2}}-\sigma^{2}\right),
\end{equation}
which is a function of $\sigma$. In order for any PBHs to form, $\zeta_{max}$ must be greater than $\zeta_{c}$, and so for $f_{NL}<-\frac{5}{12}$, $\sigma$ must be above a certain value $\sigma_{c}$,
\begin{equation}
\sigma_{c}=\frac{\sqrt{-25-60f_{NL}}}{6f_{NL}}.
\end{equation}
If $\sigma$ (and so the power spectrum) is below this value, no PBHs are formed, but typically, if $\sigma$ is larger then too many PBHs form. This means that an extreme fine tuning of the power spectrum is required in order to generate a small but non-zero amount of PBHs.}
\item{The $g_{NL}$ term affects the kurtosis of the distribution. For positive $g_{NL}$. the tails of the probability density function are enhanced - meaning tighter constraints. For small negative values, the tails are diminished - meaning weaker constraints - but as $g_{NL}$ becomes more negative the tails become more enhanced - meaning constraints again become tighter.}
\end{itemize}

Similar behaviour is displayed for higher order terms - even terms have a similar effect as the quadratic term, and odd order terms have a similar effect to the cubic term. The effects of combining higher order terms was investigated \cite{Young:2013oia}, finding that for certain models displaying a simple relation between the non-Gaussianity parameters ($g_{NL}\propto f_{NL}^{2}$, $h_{NL}\propto f_{NL}^{3}$) the constraints calculated converge, but that care should be taken as this might not always be the case. 

\section{Large-scale inhomogeneities from non-Gaussianity}

In this section, we describe how the presence of local non-Gaussianity leads to a coupling between long and short wavelength modes, and thus how a mode which is greatly super-horizon at the time of PBH formation can have an effect on the distribution of PBHs on smaller scales. For a more detailed calculation and discussion of implications, the reader is directed to \cite{Byrnes:2011ri}.

\begin{figure}[t]
\centering
\subfloat[Superposition of short- and long-wavelength modes with modal coupling]{%
      \includegraphics[width=\textwidth]{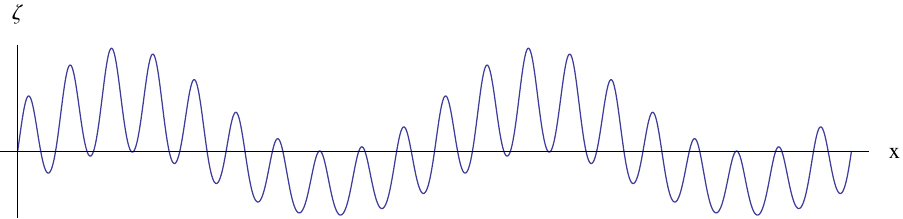}
    }
    \hfill
    \subfloat[Subtraction of the long-wavelength mode]{%
      \includegraphics[width=\textwidth]{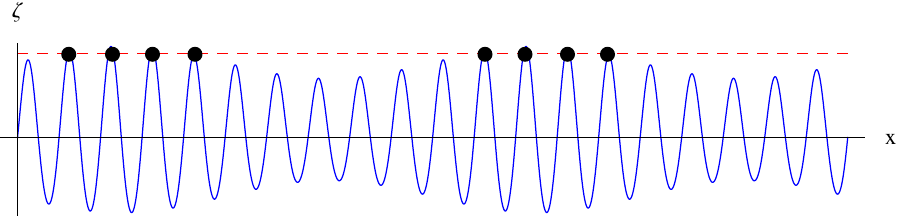}
    }
 \caption[Modal coupling of short- and long-wavelength modes]{The first (top) figure shows one arbitrary realisation of a universe containing exactly one long wavelength and one short wavelength Gaussian mode, and the corresponding non-Gaussian components where the universe contains quadratic non-Gaussianity. At the time when the short wavelength mode reenters the horizon after the end of inflation, the long wavelength mode is not yet visible - and will not affect the local evolution of the universe (i.e.~whether it forms a PBH or not). The second (bottom) plot shows the same universe with the long wavelength mode subtracted. $\zeta$ can now be used as a formation criterion for the formation of PBHs - if it is over a certain value, then that region will collapse to form a PBH. The dashed red line shows such a formation criterion, and the black circles represent areas which will collapse to form a PBH.}
\label{fig:large scale inhomos}
\end{figure}

We will consider a universe with a distribution in $\zeta$ described by the local model of non-Gaussianity (equation (\ref{eqn:quadratic local model})), but which contains exactly 2 Gaussian modes. We can therefore decompose the Gaussian component of $\zeta$ into its two components
\begin{equation}
\zeta_{G}=\zeta_{s}+\zeta_{l}.
\end{equation}
The first plot in Fig.~\ref{fig:large scale inhomos} shows one possible realisation of such a universe, with 2 Gaussian modes of arbitrary size. In this picture, the non-Gaussian components to not appear to be very important - they are small corrections to the existing Gaussian components.  However, as described in \cite{Young:2014ana}, super-horizon modes should not be considered when deciding if a region will collapse to form a PBH. We will study the time at which PBHs form on the scale of the shorter-scale mode (when that mode enters the horizon), and therefore neglect the components of $\zeta$ that depend only on the long wavelength mode. The second plot in figure\ref{fig:large scale inhomos} shows the relevant modes for formation of PBHs: the red dashed line represents a hypothetical formation criterion for PBHs and the black dots represent PBH forming regions. We note that in certain regions of the universe corresponding to peaks in the super-horizon mode, PBHs are produced in significant numbers, whilst in regions corresponding to troughs in the super-horizon mode, no PBHs would be produced.

The effect of different scale modes on the formation of primordial black holes has recently been investigated by \cite{Nakama:2014fra}, who investigated the case where a large perturbation which will collapse to form a PBH is itself superposed on a much larger perturbation which will also collapse to form a PBH upon reentry. The smaller PBH, which forms first, is swallowed by the second PBH as it forms, leading to a single large PBH. As expected, the first collapse is unaffected by the large-scale perturbation as it is outside the horizon at the time of collapse, and the second collapse is unaffected by the first due to the large scale difference between the two. Nakama also investigates the effect of sub-horizon modes on the possible collapse of a perturbation, finding that the presence of such modes lowers the threshold value for collapse - making the collapse of such a perturbation more likely. This a separate effect to the one which we are investigating in this paper - here, the effect of super-horizon modes on the distribution of horizon-scale perturbations is studied, whilst Nakama describes the effect of sub-horizon modes on the evolution of horizon-scale perturbations. The net result of the sub-horizon modes is to lower the formation threshold for PBHs, which would serve to further tighten the constraints derived in this paper.

\section{Inhomogeneous quadratic non-Gaussianity}
In the local model of non-Gaussianity, the curvature perturbation $\zeta$ is given to $2^{nd}$ order by
\begin{equation}
\label{eqn:quadratic local model}
\zeta=\zeta_{G}+\frac{3}{5}f_{NL}\left(\zeta_{G}^{2}-\langle\zeta_{G}^{2}\rangle\right),
\end{equation}
where $\zeta_{G}$ is a Gaussian variable, and it is necessary to subtract the $\langle\zeta_{G}^{2}\rangle$ term in the above expression so that the expectation value of $\zeta$ remains zero, $\langle\zeta\rangle=0$.

We will now use the peak-background split, separating the Gaussian component of the curvature perturbation $\zeta_{G}$ into a large scale  ``background'' perturbation $\zeta_{l}$ and a small-scale ``peak'' perturbation $\zeta_{s}$,
\begin{equation}
\zeta_{G}=\zeta_{l}+\zeta_{s}.
\end{equation}
The full expression for the curvature perturbation $\zeta$ then becomes
\begin{equation}
\zeta=\left(\zeta_{l}+\zeta_{s}\right)+\frac{3}{5}f_{NL}\left(\left(\zeta_{l}+\zeta_{s}\right)^{2}-\langle\left(\zeta_{l}+\zeta_{s}\right)^{2}\rangle\right).
\end{equation}
Terms which are independent of $\zeta_{s}$, and depend only on the large scale perturbation $\zeta_{l}$ can be neglected - as they are not visible at the time of PBH formation, leaving
\begin{equation}
\label{eqn:quadratic inhomogeneity}
\zeta=\left(1+\frac{6}{5}f_{NL}\zeta_{l}\right)\zeta_{s}+\frac{3}{5}\left(\zeta_{s}^{2}-\sigma_{s}^{2}\right).
\end{equation}
In a small patch of the universe, $\zeta_{l}$ will appear constant, and the above expression can be written in terms of new variables $\tilde{\zeta}_{G}$, $\tilde{\sigma}$ and $\tilde{f}_{NL}$, given by
\begin{equation}
\tilde{\zeta}_{G}=\left(1+\frac{6}{5}f_{NL}\zeta_{l}\right)\zeta_{s},
\end{equation}
\begin{equation}
\tilde{\sigma}=\left(1+\frac{6}{5}f_{NL}\zeta_{l}\right)\sigma_{s},
\end{equation}
\begin{equation}
\tilde{f}_{NL}=\left(1+\frac{6}{5}f_{NL}\zeta_{l}\right)^{-2}f_{NL}.
\end{equation}
This allows equation (\ref{eqn:quadratic inhomogeneity}) to be written in a form analogous to equation (\ref{eqn:quadratic local model}),
\begin{equation}
\zeta=\tilde{\zeta}_{G}+\frac{3}{5}\tilde{f}_{NL}\left(\tilde{\zeta}_{G}^{2}-\tilde{\sigma}^{2}\right)=\tilde{h}(\tilde{\zeta}_{G}).
\label{eqn:localQuadInhomo}
\end{equation}
Taking $\zeta_{l}$ to be constant in a given region of the universe, the mass fraction of the region going into PBHs $\tilde{\beta}$ can then be written in terms of the locally observable values $\tilde{f}_{NL}$, $\tilde{\zeta}_{G}$ and $\tilde{\sigma}$ in the same way as in equation (\ref{quadraticbeta}):
\begin{equation}
\tilde{\beta}=\mathrm{erfc}(\tilde{h}_{+}^{-1}(\zeta_{c}))+\mathrm{erfc}(\lvert \tilde{h}_{+}^{-1}(\zeta_{c})\rvert).
\end{equation}
However, this is still a function of $\zeta_{l}$, and to obtain the mass fraction of the entire universe going into PBHs, this should be integrated over $\zeta_{l}$
\begin{equation}
\label{quadratic beta}
\beta=\int_{-\infty}^{\infty}\tilde{\beta}(\zeta_{l})P(\zeta_{l})d\zeta_{l},
\end{equation}
where $P(\zeta_{l})$ is the (Gaussian) PDF of $\zeta_{l}$. Therefore, $\beta$ depends not only on the variance (power spectrum) of the small-scale perturbations (which is the scale PBH formation occurs at), but also on the variance of the large scale modes. In this paper, we assume the form of the power spectrum shown in Fig.~\ref{powerSpectrum}  - and therefore, the variance of the large-scale perturbations can be written as a function of the variance of the small-scale perturbations, depending on the number of e-folds one considers.

The variance of the large-scale perturbations is given by integrating the power spectrum multiplied by a smoothing function $W(k R)$, where $R$ is the smoothing scale, as follows
\begin{equation}
\langle\zeta_{l}^{2}\rangle=\int_{0}^{\infty}d\ln(k)W^{2}(k R)\mathcal{P}_{\zeta_{l}}(k).
\end{equation}
In practice, since we are assuming a scale invariant power spectrum (for the Gaussian components), which is zero below a certain value of $k$, then $\langle\zeta_{l}^{2}\rangle$ depends upon the number of e-folds $\mathcal{N}$ considered to be part of the background large-scale perturbation. We will approximate that
\begin{equation}
\label{efolds and variance}
\sigma_{l}=\sqrt{\langle\zeta_{l}^{2}\rangle}\approx \sqrt{\mathcal{N}} \sigma_{s},
\end{equation}
in order to derive constraints on the power spectrum from the constraints on the abundance of PBHs. Equation (\ref{efolds and variance}) can be substituted into equation (\ref{quadratic beta}), which can then be solved numerically to find a constraint on $\sigma_{s}$ from a constraint on $\beta$. The constraint on the power spectrum $\mathcal{P}_{\zeta}$ can then be calculated using \cite{Boubekeur:2005fj,Byrnes:2007tm}
\begin{equation}
\mathcal{P}_{\zeta}=\sigma_{s}^{2}+4\left(\frac{3}{5}f_{NL}\right)^{2}\sigma_{s}^{4}\ln\left(kL\right),
\end{equation}
where the cut-off scale $L\approx\frac{1}{H}$ is of order the horizon-scale, $k$ is the scale of interest. The factor $\ln\left(kL\right)$ can therefore become significant, as the power spectrum is taken to be large across a number of e-folds - and will be approximately equal to the number of e-folds being considered, $\mathcal{N}$ \cite{Suyama:2008nt,Kumar:2009ge}.

\begin{figure}[t]
\centering
\subfloat[$\beta=10^{-5}$]{%
      \includegraphics[width=0.49\textwidth]{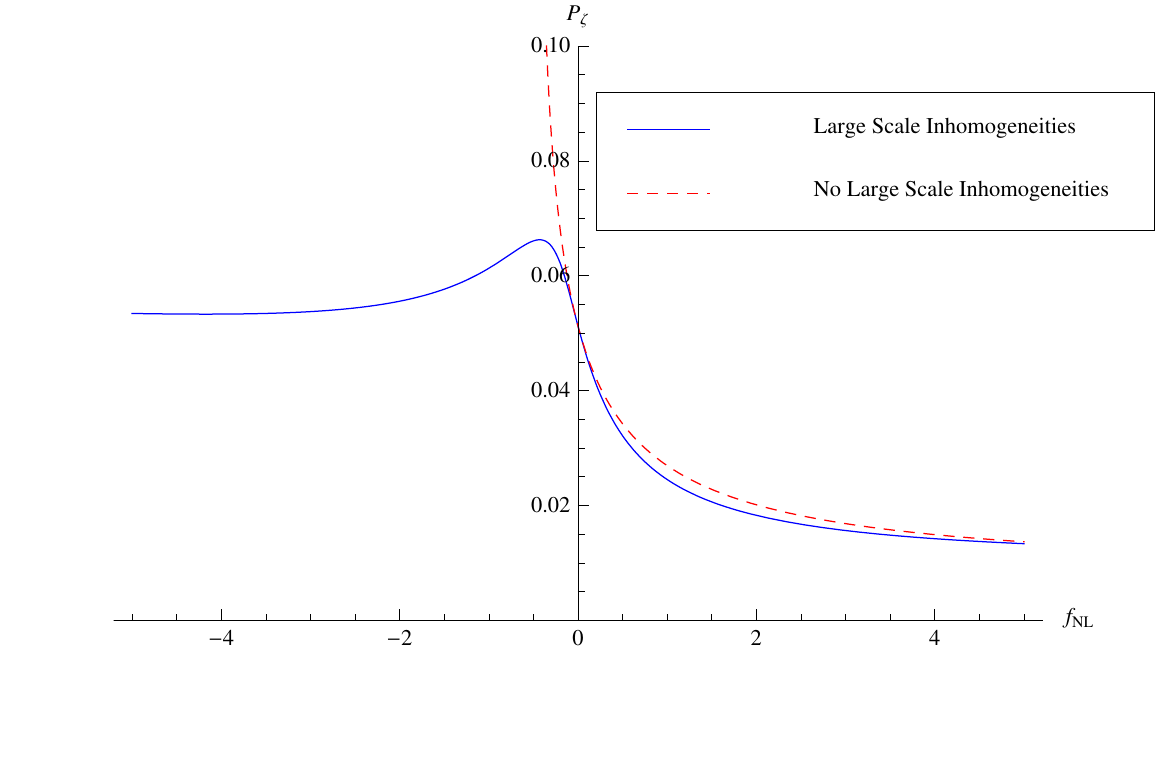}
    }
    \hfill
    \subfloat[$\beta=10^{-20}$]{%
      \includegraphics[width=0.49\textwidth]{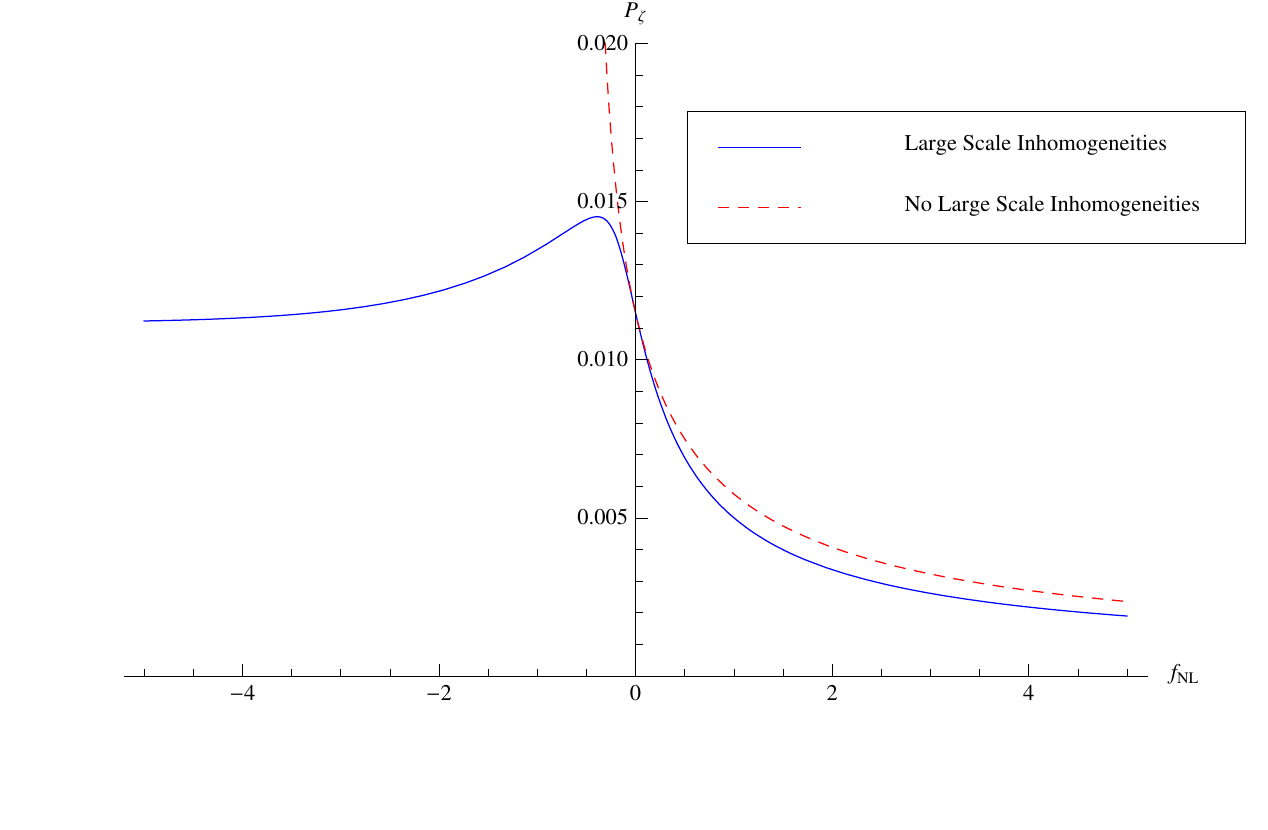}
    }
 \caption[Effect of modal coupling from $\fNL$ on the power spectrum constraints]{The constraints that can be placed upon the power spectrum are displayed - they depend significantly upon the value of the non-Gaussianity parameter, $f_{NL}$. The dotted red lines show the constraints calculated previously, where the effect of large-scale modes was not considered, and the solid blue lines show the constraints when they are included. The left plot (a) displays the constraints for $\beta<10^{-5}$ and the right plot (b) for $\beta<10^{-20}$. In this plot it is assumed that the variance of the Gaussian component of the large-scale perturbations is the same as that of the small-scale perturbations, $\langle\zeta_{l}^{2}\rangle=\langle\zeta_{s}^{2}\rangle=\sigma^{2}$. For positive $f_{NL}$ the constraints are tighter than the Gaussian case, and slightly stronger than in previous calculations ignoring modal coupling. For negative $f_{NL}$, the constraints are similar to the Gaussian case, and the dramatic weakening of the constraints as $f_{NL}$ becomes negative is no longer seen.}
\label{fnl constraints}
\end{figure}

Initially, we will consider a large-scale perturbation due to contributions from modes spanning only 1 e-fold - and so therefore, the variance of the large background perturbations is equal to that of the small-scale perturbations, $\sigma_{l}=\sigma_{s}$. The constraints are obtained by numerically solving equation (\ref{quadratic beta}) and allowing $f_{NL}$ to vary. The results are shown in Fig.~\ref{fnl constraints} for $\beta=10^{-5}$ and $\beta=10^{-20}$. We now note that, whilst the constraints still weaken slightly for small negative values of $f_{NL}$, the constraints become tighter again as $f_{NL}$ becomes more negative, quickly becoming similar to the Gaussian case - which was not seen in previous calculations \cite{Byrnes:2012yx,Young:2013oia} which neglected the long-short coupling (and hence are only valid if the power spectrum has a narrow peak). As $\lvert f_{NL}\rvert$ becomes large, the constraints asymptote to a constant value (which will be calculated in the next section).

Depending on the value of $\zeta_{l}$ in a given region of the universe, the production of PBHs can either be increased or decreased. However, the presence of large-scale perturbations always increases the total number of PBHs forming in the entire universe - meaning that the power spectrum can be constrained to a lower value so that PBHs are not overproduced. This can be demonstrated by considering what happens when $f_{NL}$ is negative - it was previously found that constraints become rapidly weaker when $f_{NL}$ is negative (where the large-scale background perturbations were not considered). This is due to the shape of the pdf of $\zeta$, which has a maximum value of $\zeta$ given by
\begin{equation}
\zeta_{max}=-\frac{5}{6f_{NL}}+\frac{3}{5}f_{NL}\left(\frac{25}{36f_{NL}^{2}}-\sigma^{2}\right).
\end{equation}
Unless there is fine tuning of the (local) power spectrum, this typically means that if $\sigma$ is small then no PBHs are formed, but above a critical value then so many PBHs form that the universe becomes dominated by them. However, in any given region, $\tilde{\sigma}$ and $\tilde{f}_{NL}$ are functions of $\zeta_{l}$. Therefore, depending on the value of $\zeta_{l}$, PBH production in a region can be either increased dramatically or reduced to zero. Overall, more PBHs would be produced in a universe containing such large-scale inhomogeneities - and so the power spectrum is more tightly constrained. A similar but less dramatic phenomenon occurs for positive $f_{NL}$ - meaning the power spectrum can be more tightly constrained for both positive and negative $f_{NL}$.

\begin{figure}[t]
\centering
	\includegraphics[width=0.7\linewidth]{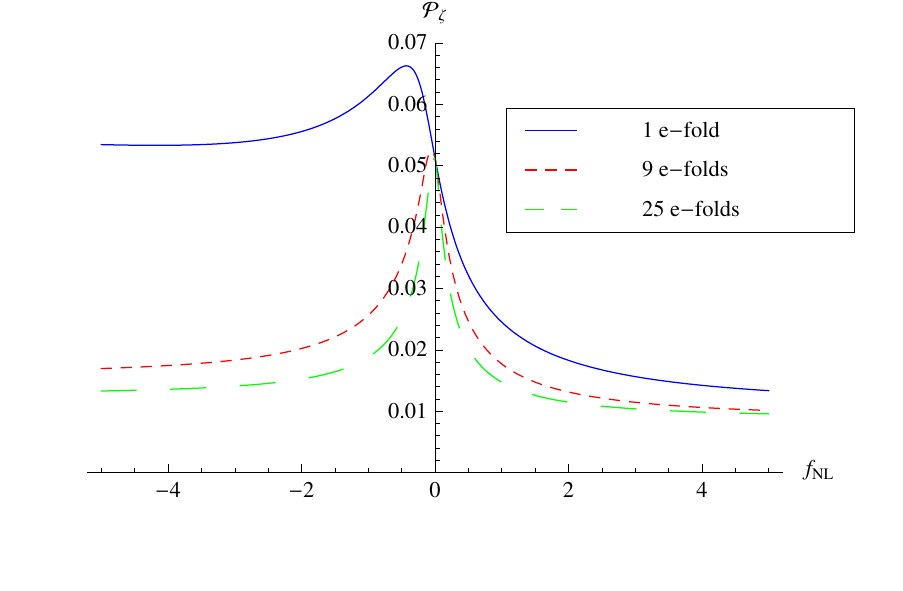}
 \caption[The effect of $\fNL$ and super-horizon modes on the power spectrum constraints]{The constraints which can be placed on the power spectrum from PBHs depend strongly on both the amount of non-Gaussianity and the amplitude of the background perturbations, given by $\langle\zeta_{l}^{2}\rangle=\mathcal{N}\mathcal{P}_{\zeta}$. This figure shows the constraint on $\mathcal{P}_{\zeta}$ from $\beta<10^{-5}$ as a function of $f_{NL}$ for $\mathcal{N}=1$, $9$ and $25$.}
 \label{many efolds quadratic}
\end{figure}

We will now consider what happens when a larger number of e-folds are considered to contribute to the background perturbation. In Fig.~\ref{many efolds quadratic} we show how the constraints change with the the variance of the background perturbations, considering the cases where the background is comprised from 9 e-folds, $\sigma_{l}=3\sigma_{s}$, and 25 e-folds, $\sigma_{l}=5\sigma_{s}$. If $f_{NL}$ is non-zero, the constraints on the small scales become much tighter as the variance on large scales increases. In order to explain this behaviour, it is useful to consider the case of large $f_{NL}$ where the linear term is dominated by the quadratic term in equation (\ref{eqn:quadratic local model}).

\subsection{Large $f_{NL}$}

If $f_{NL}$ becomes large enough such that the quadratic term dominates the linear term, we can simplify the expression for $\zeta$ to
\begin{equation}
\zeta=\pm\left(\zeta_{G}^{2}+\langle\zeta_{G}^{2}\rangle\right),
\label{quadraticNG}
\end{equation}
and performing the peak-background split as before, dropping the terms independent of $\zeta_{s}$, gives
\begin{equation}
\zeta=2\zeta_{l}\zeta_{s}\pm\left(\zeta_{s}^{2}+\sigma_{s}^{2}\right).
\end{equation}
Rewriting in terms of the variables one would observe locally
\begin{equation}
\tilde{\zeta}_{G}=2\zeta_{l}\zeta_{s},
\end{equation}
\begin{equation}
\tilde{\sigma}_{G}=2\zeta_{l}\sigma_{s},
\end{equation}
\begin{equation}
\tilde{f}_{NL}=\pm\frac{5}{12\zeta_{l}^{2}},
\end{equation}
which gives as before, see equation (\ref{eqn:localQuadInhomo}), 
\begin{equation}
\tilde{\zeta}=\tilde{\zeta}_{G}+\frac{3}{5}\tilde{f}_{NL}\left(\tilde{\zeta}_{G}^{2}-\tilde{\sigma}^{2}\right).
\end{equation}
However, we now note that, because the PDF of $\tilde{\zeta}_{G}$ is constant under a change of sign of $\tilde{\zeta}_{G}$, then the PDF of $\zeta$ is independent of the sign of $\zeta_{l}$. This can then be inserted as before into equation (\ref{quadratic beta}), which can then be solved numerically to find an upper limit on the power spectrum - this is the value that the constraints asymptote to in Fig.~\ref{fnl constraints} or \ref{many efolds quadratic}. Because the variance of the background depends on the number of e-folds it is comprised of, the constraints on the power spectrum depend on the number of e-folds between the horizon scale during PBH formation and the largest scale on which the power spectrum is enhanced, $\mathcal{N}$, see Fig.~\ref{powerSpectrum}.

\begin{figure}[t]
\centering
	\includegraphics[width=0.7\linewidth]{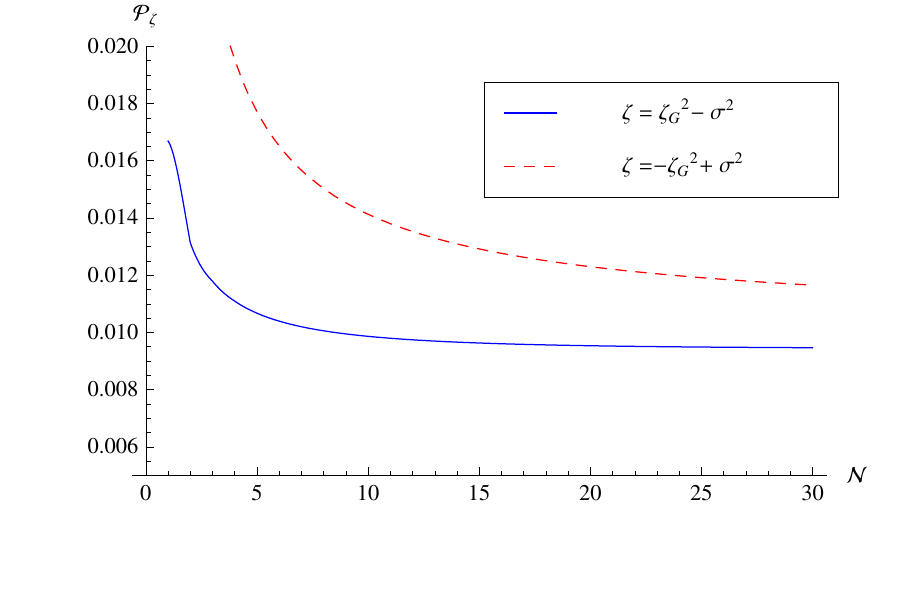}
 \caption[Constraints on the power spectrum for a $\pm\chi^2$ distribution]{The constraints on the power spectrum for the quadratic case, $\zeta\approx\zeta_{G}^{2}$, are shown for $\beta<10^{-5}$ as a function of the number of e-folds of fourier modes, $\mathcal{N}$, making up the background perturbation, with $\langle\zeta_{l}^{2}\rangle=\mathcal{N}\mathcal{P}_{\zeta}$. For small $\mathcal{N}$ the constraints are much weaker for the negative case than for the positive case, and both tighten significantly as $\mathcal{N}$ becomes large. As $\mathcal{N}$ becomes very large, both will eventually asymptote to the same constant value, $\mathcal{P}_{\zeta}<9.8 \times10^{-3}$.}
 \label{quadratic limit}
\end{figure}

Fig.~\ref{quadratic limit} shows how the constraints become tighter as more e-folds are considered. For a small number of e-folds, so that $\langle\zeta_{l}^{2}\rangle$ is not too large, the constraints are much weaker for the negative quadratic case. However, as more e-folds are considered, the constraints become much closer - this is because, in universes where $\langle\zeta_{l}^{2}\rangle$ is large, then $\tilde{f}_{NL}=\pm\frac{5}{12\zeta_{l}^{2}}$ is typically small. One can therefore approximate $\tilde{\zeta}$ as Gaussian\footnote{Surprisingly, starting from a completely non-Gaussian distribution with a large-scale non-Gaussian background, the small scales appear almost Gaussian (although even small amounts of non-Gaussianity have a very large effect on $\beta$). See \cite{Nelson:2012sb} for further reading.} - and the sign of the quadratic term in equation (\ref{quadraticNG}) is unimportant. Even for the weakest constraints on the abundance of PBHs, $\beta<10^{-5}$, the constraints on the power spectrum drop to $\mathcal{P}_{\zeta}<\mathcal{O}(10^{-2})$, around 5 times tighter than for the Gaussian case, and 2 orders of magnitude tighter for $f_{NL}<0$ compared to when modal-coupling is not considered.

Rather than being purely hypothetical, there are models that predict such a distribution. For example $\zeta=-(g^{2}-\langle g^{2}\rangle)$ (with $g$ a Gaussian variable) could be expected from the linear era of the hybrid inflation waterfall \cite{Lyth:2012yp}. The power spectrum in this model is expected to become large on some small scale before inflation ends, and peak at some value before decreasing again. In addition, $\zeta=g^{2}-\langle g^{2}\rangle$ could be predicted from a curvaton-type scenario (e.g.~\cite{Suyama:2008nt,Bugaev:2012ai,Peloso:2014oza}).

\section{Inhomogeneous cubic non-Gaussianity}

The local model of non-Gaussianity with a cubic term (assuming $f_{NL}=0$) is given by
\begin{equation}
\zeta=\zeta_{G}+\frac{9}{25}g_{NL}\zeta_{G}^{3}.
\label{eqn1:cubicNG}
\end{equation}
We again use the peak-background split, $\zeta=\zeta_{s}+\zeta_{l}$, such that
\begin{equation}
\zeta=\left(1+\frac{27}{25}g_{NL}\zeta_{l}^{2}\right)\zeta_{s}+\left(\frac{27}{25}g_{NL}\zeta_{l}\right)\zeta_{s}^{2}+\left(\frac{9}{25}g_{NL}\right)\zeta_{s}^{3}+\mathcal{O}(\zeta_{l}),
\end{equation}
where again, the terms dependent only on $\zeta_{l}$ are unimportant in the context of PBH formation, and are neglected. $\zeta_{l}$ appears constant in a small patch of the universe, and this can be rewritten in terms of $\tilde{\zeta}_{G}$, $\tilde{\sigma}$, $\tilde{f}_{NL}$ and $\tilde{g}_{NL}$.
\begin{equation}
\tilde{\zeta}_{G}=\left(1+\frac{27}{25}g_{NL}\zeta_{l}^{2}\right)\zeta_{s},
\end{equation}
\begin{equation}
\tilde{\sigma}=\left(1+\frac{27}{25}g_{NL}\zeta_{l}^{2}\right)\sigma_{s},
\end{equation}
\begin{equation}
\tilde{f}_{NL}=\left(\frac{9}{5}g_{NL}\zeta_{l}\right)\left(1+\frac{27}{25}g_{NL}\zeta_{l}^{2}\right)^{-2},
\end{equation}
\begin{equation}
\tilde{g}_{NL}=g_{NL}\left(1+\frac{27}{25}g_{NL}\zeta_{l}^{2}\right)^{-3}.
\label{eqn:bar gnl}
\end{equation}
Therefore, equation (\ref{eqn1:cubicNG}) can be rewritten as
\begin{equation}
\zeta=\tilde{\zeta}_{G}+\frac{3}{5}f_{NL}\left(\tilde{\zeta}_{G}^{2}-\tilde{\sigma}^{2}\right)+\frac{9}{25}\tilde{g}_{NL}\tilde{\zeta}_{G}^{3},
\end{equation}
where the $-\tilde{\sigma}^{2}$ term has been inserted manually to ensure $\langle\zeta\rangle=0$. An expression for the abundance of PBHs in a given region of the universe, $\tilde{\beta}$, can be derived in terms of $\tilde{\sigma}$, $\tilde{f}_{NL}$ and $\tilde{g}_{NL}$ - see \cite{Byrnes:2012yx} for details, we do not give the full calculation here. Again, in order to derive the complete expression for the abundance of PBHs in the entire universe, it is necessary to integrate over $\zeta_{l}$ as before,
\begin{equation}
\beta=\int_{-\infty}^{\infty}\tilde{\beta}(\zeta_{l})P(\zeta_{l})d\zeta_{l}.
\end{equation}
This expression can then be solved numerically to derive a constraint on $\sigma$ from a constraint on $\beta$. The constraint on the power spectrum, $\mathcal{P}_{\zeta}$ can then be calculated using \cite{Byrnes:2007tm}
\begin{equation}
\mathcal{P}_{\zeta}=\sigma^{2}+6\left(\frac{9g_{NL}}{25}\right)\sigma^{4}\ln(kL)+27\left(\frac{9g_{NL}}{25}\right)^{2}\sigma^{6}\ln(kL)^{2}.
\end{equation}

\begin{figure}[t]
\centering
\subfloat[$\beta=10^{-5}$]{%
      \includegraphics[width=0.49\textwidth]{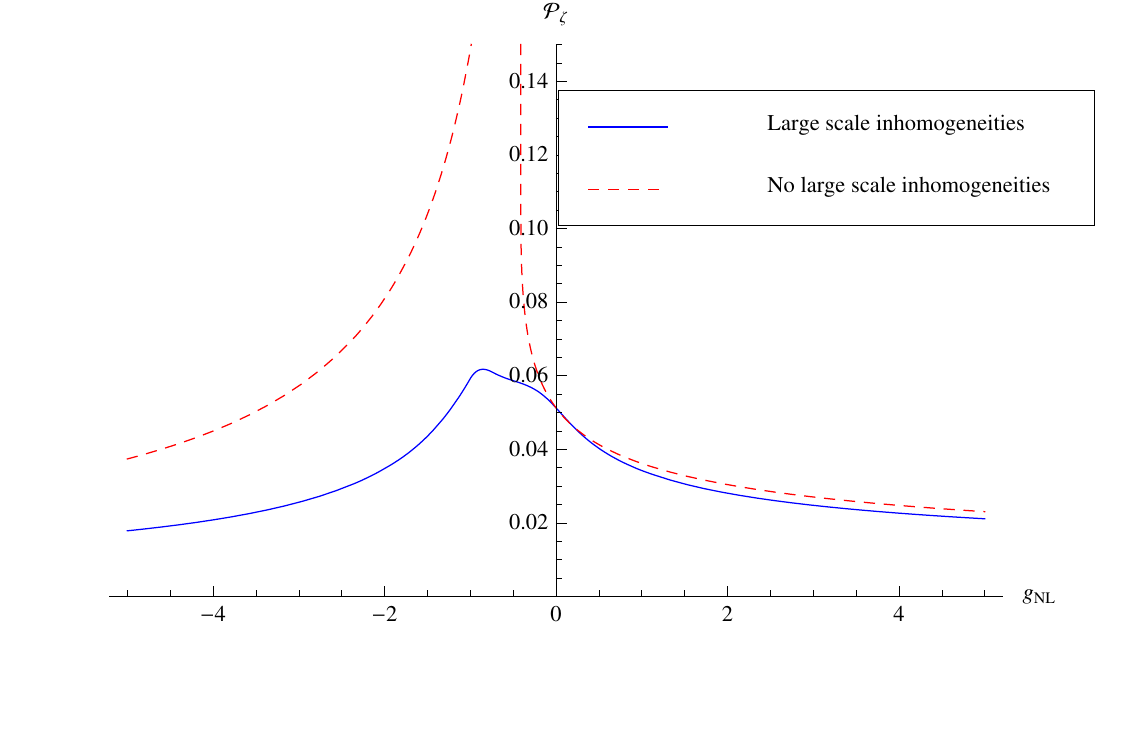}
    }
    \hfill
    \subfloat[$\beta=10^{-20}$]{%
      \includegraphics[width=0.49\textwidth]{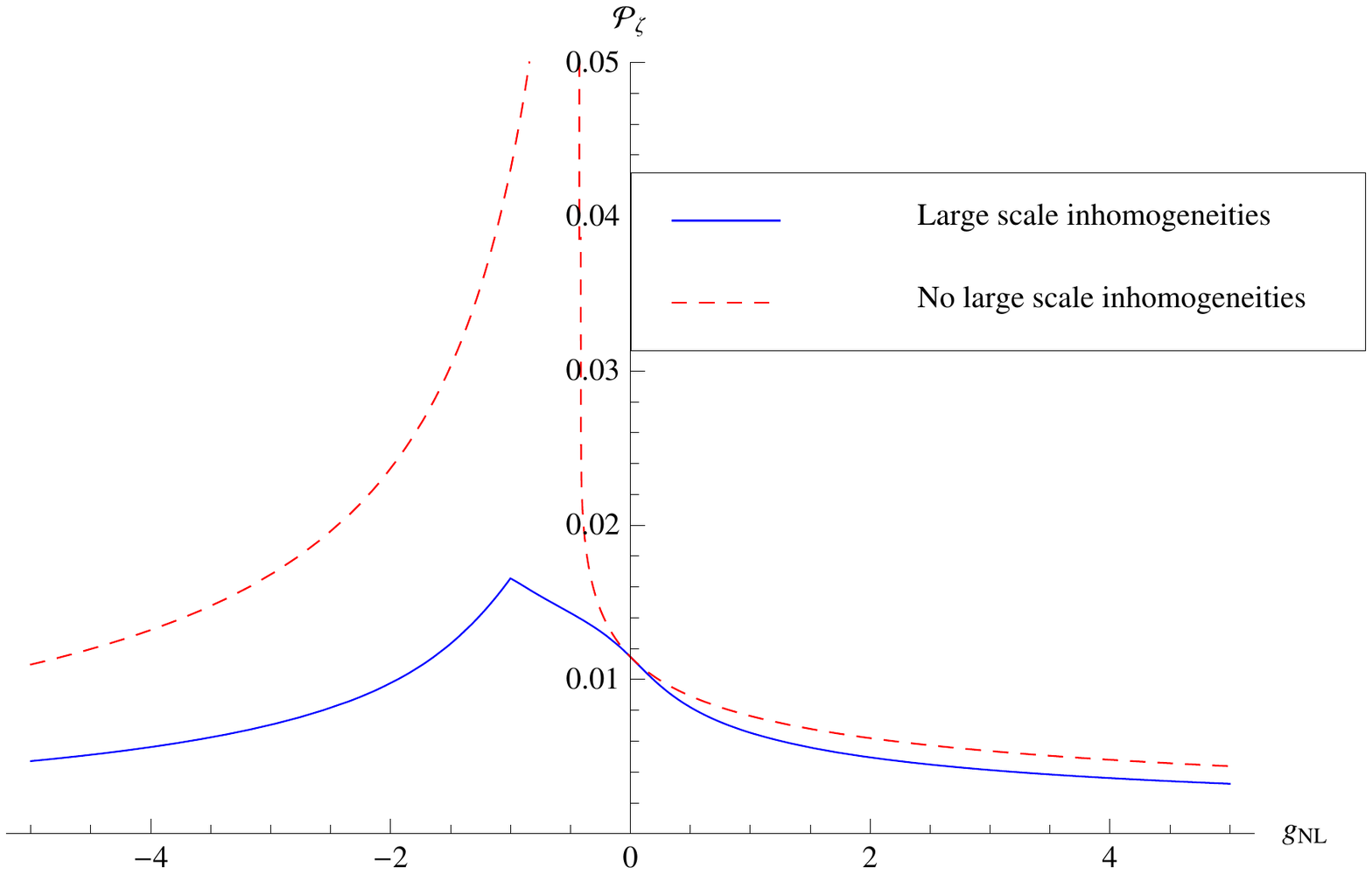}
    }
\caption[Effect of modal coupling from $\gNL$ on the power spectrum constraints]{The constraints that can be placed upon the power spectrum are displayed - they depend significantly upon the value of the non-Gaussianity parameter, $g_{NL}$. The dotted red lines show the constraints calculated previously, where the effect of large-scale modes was not considered, and the solid blue lines show the constraints when they are included. The left plot (a) displays the constraints for $\beta<10^{-5}$ and the right plot (b) for $\beta<10^{-20}$. In this plot it is assumed that the variance of the Gaussian component of the large-scale perturbations is the same as that of the small-scale perturbations, $\langle\zeta_{l}^{2}\rangle=\langle\zeta_{s}^{2}\rangle=\sigma^{2}$. Typically, the constraints tighten significantly when there is any non-Gaussianity present - with a slight weakening for small negative $g_{NL}$. The constraints are significantly tighter than previously calculated, and do not display as sharp a peak for small negative $g_{NL}$ where the constraints became rapidly weaker.}
\label{gnlCons}
\end{figure}

Fig.~\ref{gnlCons} shows how the constraints on the power spectrum depend on $g_{NL}$ for $\beta=10^{-5}$ and $\beta=10^{-20}$. Again, we see that constraints become tighter as the non-Gaussianity parameter $g_{NL}$ becomes large. However, the sharp peak seen in previous calculations is now smoothed out, and the constraints are significantly tighter - this is because only for a small range of values of $g_{NL}$ is the production of PBHs significantly reduced (seen by the region in which the constraints weaken in Fig.~\ref{old constraints}), but the background perturbations cause $g_{NL}$ to vary, see equation (\ref{eqn:bar gnl}). As seen in previous papers, as $\lvert g_{NL}\rvert$ becomes large, the constraints asymptote to the same value for negative or positive $g_{NL}$ - which is as expected (this will be explored in the next section).

\begin{figure}[t]
\centering
	\includegraphics[width=0.7\linewidth]{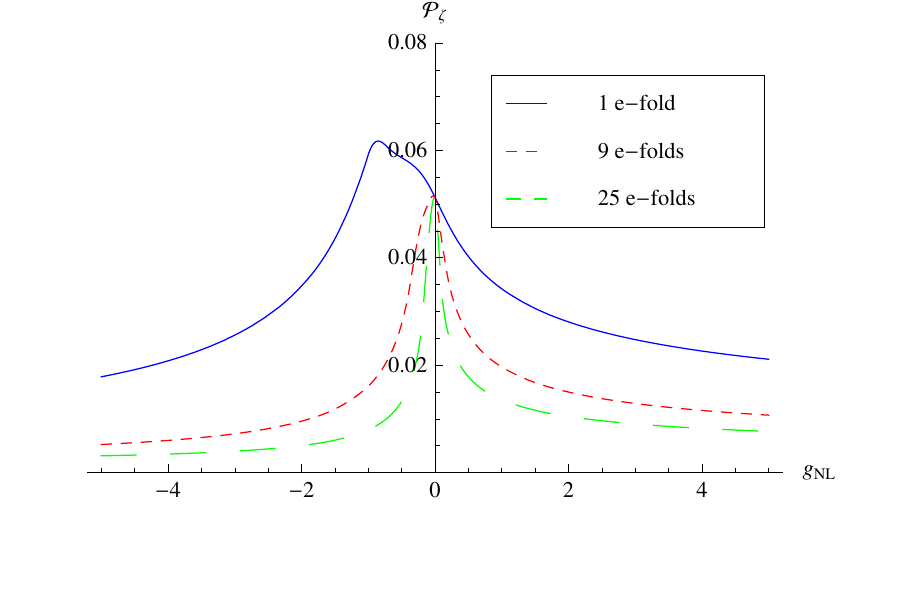}
\caption[The effect of $\fNL$ and super-horizon modes on the power spectrum constraints]{As in the quadratic case, the constraints which can be placed on the power spectrum from PBHs depend strongly on both the amount of non-Gaussianity and the amplitude of the background perturbations, given by $\langle\zeta_{l}^{2}=\mathcal{NP}_{\zeta}$. This figure shows the constraint on $\mathcal{P}_{\zeta}$ from $\beta < 10^{-5}$ as a function of $g_{NL}$ for $\mathcal{N} =1$, $9$ and $25$, becoming much tighter as more e-folds are considered.}
\label{efolds gnl}
\end{figure}

We will now again consider the constraints if the background perturbations consist of multiple e-folds of perturbations. Fig.~\ref{efolds gnl} shows the resultant constraints obtained if the background perturbations consist of 1, 9, or 25 e-folds, as before. When more e-folds are considered, the constraints become much tighter - only for small negative $g_{NL}$ do the constraints weaken slightly, but for all other values of $g_{NL}$ the constraints become significantly tighter, $\mathcal{P}_{\zeta}<\mathcal{O}(10^{-3})$ for even small values of $g_{NL}$.

\subsection{Large $g_{NL}$}

\begin{figure}[t]
\centering
	\includegraphics[width=0.7\linewidth]{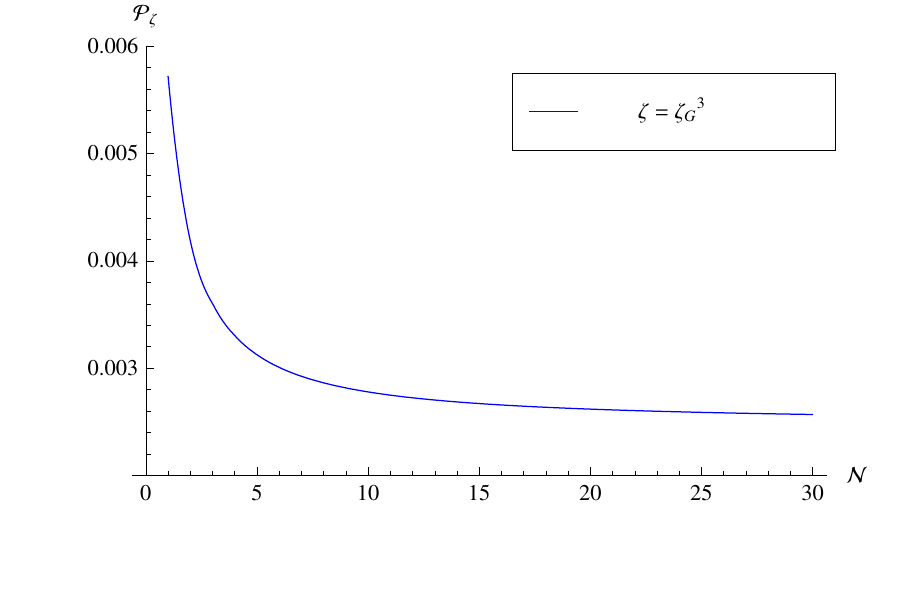}
 \caption[Constraints on the power spectrum for a $\pm\chi^3$ distribution]{The constraints on the power spectrum for the cubic case, $\zeta\approx\pm\zeta_{G}^{3}$, are shown for $\beta<10^{-5}$ as a function of the number of e-folds of fourier modes, $\mathcal{N}$, making up the background perturbation, with $\langle\zeta_{l}^{2}\rangle=\mathcal{NP}_{\zeta}$. Similar to the quadratic case, the constraints tighten significantly as the number of e-folds being considered increases, eventually reaching a constant for large $\mathcal{N}$ at $\mathcal{P}_{\zeta}<2.5\times10^{-3}$.}
\label{cubic limit}
\end{figure}

We will now consider the case where the cubic term dominates, and $\zeta$ can be expressed as
\begin{equation}
\zeta_{\pm}=\pm \zeta_{G}^{3}.
\end{equation}
In the cubic case, the sign does not matter - because a Gaussian distribution is symmetric, the PDF of $\zeta_{+}$ and $\zeta_{-}$ is the same, and we will therefore drop the dependence on the sign and discuss only the positive case. Completing the peak-background split and isolating the short scale gives
\begin{equation}
\zeta=3\zeta_{l}^{2}\zeta_{s}+3 \zeta_{l}\left(\zeta_{s}^{2}-\sigma_{s}^{2}\right)+\zeta_{s}^{3},
\end{equation}
where we have inserted the $\sigma_{s}^{2}$ term manually. Again, defining effective short-scale parameters:
\begin{equation}
\tilde{\zeta}_{G}=3\zeta_{l}^{2}\sigma_{s},
\end{equation}
\begin{equation}
\tilde{\sigma}=3\zeta_{l}^{2}\sigma_{s},
\end{equation}
\begin{equation}
\tilde{f}_{NL}=\frac{5}{3}\left(3\zeta_{l}^{2}\right)^{-2},
\end{equation}
\begin{equation}
\tilde{g}_{NL}=\frac{25}{9}\left(3\zeta_{l}^{2}\right)^{-3}.
\end{equation}
We note that as $\zeta_{l}$ becomes large, the small-scale observable universe will appear more Gaussian. The constraints on the power spectrum $\mathcal{P}_{\zeta}$ can then be computed numerically as before from constraints on the mass fraction of PBHs $\beta$, as a function of the number of e-folds considered in the background perturbation, $\mathcal{N}$ - the results can be seen in Fig.~\ref{cubic limit}. We see that, for a moderate number of e-folds considered, the constraints drop to $\mathcal{P}_{\zeta}<\mathcal{O}(10^{-3})$, eventually tightening to $\mathcal{P}_{\zeta}<2.4\times10^{-3}$.

\section{Conclusions}
We have extended the calculation for the abundance of PBHs, defined in terms of the mass fraction of the universe forming PBHs at the time of formation $\beta$, when there is non-Gaussianity present to include the effect of coupling between large-scale super-horizon modes and smaller horizon scale perturbations. We see that non-Gaussianity typically increase the overall amount of PBHs that would form - with some regions of the universe producing significantly more PBHs than other regions. A realisation of such a universe - containing significant non-Gaussianity and a broad peak in the power spectrum at scales significantly smaller than those visible in the CMB is possible in hybrid inflation, and in particular from the waterfall transition of $\mathcal{N}$-field hybrid inflation \cite{Halpern:2014mca}.

Observational constraints on $\beta$, which range from $\beta<10^{-5}$ to $\beta<10^{-20}$, can then be used to place an upper constraint on the primordial curvature perturbation power spectrum, $\mathcal{P}_{\zeta}$. We have investigated the constraints which can be placed on the power spectrum dependent on the amount of non-Gaussianity present and the coupling between modes, for a simple model of the power spectrum. Because non-Gaussianity typically increases PBH formation, the constraints on $\mathcal{P}_{\zeta}$ are typically much tighter - and we show that the constraints from PBHs may be significantly tighter than calculated in previous work. The presence of non-Gaussianity and large super-horizon modes have a large impact on the constraints - and when there is significant non-Gaussianity the constraints can become tighter by several orders of magnitude. The effect of simultaneously having a non-zero $f_{NL}$ and $g_{NL}$ has also been considered, although the analysis has not been explicitly included in this paper. It is again found that small negative values of $f_{NL}$ or $g_{NL}$ weaken the constraints slightly, but typically the constraints become stronger.

In this paper, we have considered local-type (squeezed) non-Gaussianity, which includes a significant coupling between the modes \cite{Komatsu:2009kd}. We would expect results to be similar for flattened-type non-Gaussianity as there is still a significant coupling between modes of different lengths (albeit weaker than in the local model). However, for equilateral type non-Gaussianity (which is peaked in the limit of all three modes having the same wavelength) we would not expect significant coupling between large and short scales, so the results would be expected to more closely reflect previous analyses in which large amplitude perturbations on only one scale were considered. However there have not been any detailed studies made of how non-Gaussianity of non-local shapes effects the bounds on PBHs.

The main source of error in the calculation arises from the uncertainty in the formation criterion, which lies in the range $0.7<\zeta_{c}<1.2$ - and this has a very large effect on the calculated value for $\beta$, which can easily vary by several orders of magnitude \cite{Young:2014ana}. However, the effects on the constraints calculated are much less drastic, and the error due to the uncertainty in $\zeta_{c}$ is expected to be of order $10\%$. There is also uncertainty of how intermediate modes should be handled, which are currently excluded from the calculation - how long does a mode have to be before it is considered to be part of the background? The size of this cut-off scale can have a non-negligible effect on the constraints calculated - although how important the effect is depends on the specific form of the power spectrum being considered. In this paper, we have avoided this uncertainty by considering the background perturbations to result from a given number of e-folds of modes.

We also note that the Taylor-type expansion of $\zeta$ in terms of $f_{NL}$ and $g_{NL}$, which we have used here, may not give an accurate result for the constraints. It was shown in a previous paper \cite{Young:2013oia} that higher orders terms can have a significant effect, and care should therefore be taken to ensure that results are valid when calculating constraints for a specific model.

\section{Acknowledgements}
The authors would like to thank Misao Sasaki for useful discussion during the production of this paper. 
SY is supported by an STFC studentship, and would like to thank Yukawa Institute for Theoretical Physics for its hospitality during a month long stay which was supported by the Bilateral International Exchange Program (BIEP). CB is supported by a Royal Society University Research Fellowship.


\newpage

\chapter{Signatures of non-gaussianity in the isocurvature modes of primordial black hole dark matter}
\label{chap:paper4}



\begin{center}


Sam Young, Christian T. Byrnes\\[0.5cm]
Department of Physics and Astronomy, Pevensey II Building, University of Sussex, BN1 9RH, UK\\[0.5cm]


\end{center}

Primordial black holes (PBHs) are black holes that may have formed very early on during the radiation dominated era in the early universe. 
We present here a method by which the large-scale perturbations in the density of primordial black holes may be used to place tight constraints on non-Gaussianity if PBHs account for dark matter (DM). The presence of local-type non-Gaussianity is known to have a significant effect on the abundance of primordial black holes, and modal coupling from the observed CMB scale modes can significantly alter the number density of PBHs that form within different regions of the universe, which appear as DM isocurvature modes. Using the recent \emph{Planck} constraints on isocurvature perturbations, we show that PBHs are excluded as DM candidates for even very small local-type non-Gaussianity, $|f_{NL}|\approx0.001$ and remarkably the constraint on $g_{NL}$ is almost as strong. Even small non-Gaussianity is excluded if DM is composed of PBHs. If local non-Gaussianity is ever detected on CMB scales, the constraints on the fraction of the universe collapsing into PBHs (which are massive enough to have not yet evaporated) will become much tighter.

\newpage

\section{Introduction}
Primordial black holes (PBHs) are black holes that theoretical arguments suggest might have formed from the direct gravitational collapse of large density perturbations very shortly after the end of inflation. PBHs may theoretically form with any mass, although their abundance is typically well constrained by observations. Whilst PBHs with mass lower than $10^{15}$g would have evaporated by today (with the possible exception of Planck mass relics), more massive PBHs would still survive, and represent a viable dark matter (DM) candidate.

Many efforts have been made to observe PBHs, and whilst they have not yet been seen, this has led to many corresponding constraints on their abundance in different mass ranges \cite{Carr:2009jm}. The constraints typically assume that PBHs form at a single mass scale and are stated in terms of the mass fraction of the universe going into PBHs at the time of formation, $\beta$. There exists only a narrow window in which PBHs of a single mass could make up the entirety of DM, with other scales being excluded by observations. It is noted that there has been a recent claim that the tidal capture of PBHs by neutron stars could be used to exclude the remaining window (apart from Planck mass remnants) \cite{Pani:2014rca}, but this has been refuted in \cite{Capela:2014qea,Defillon:2014wla}. The results presented here can also be applied if DM is composed of smaller PBHs which have all but evaporated by today leaving Planck mass remnants which may make up DM \cite{Carr:1994ar}. Whilst this mass range is not explicitly considered, it is certainly not ruled out by observations, and the results presented here are almost independent of the PBH mass.

In order for a significant number of PBHs to form, the power spectrum on small scales needs to be significantly larger than observed in the CMB - of order $10^{-2}$ in the case of Gaussian perturbations. This is possible in many models of inflation, including the running mass model \cite{Drees:2011hb}, axion inflation \cite{Bugaev:2013fya}, a waterfall transition during hybrid inflation \cite{Bugaev:2011wy, Lyth:2012yp,Halpern:2014mca}, from passive density fluctuations \cite{Lin:2012gs}, or in inflationary models with small field excursions but which are tuned to produce a large tensor-to-scalar ratio on large scales \cite{Hotchkiss:2011gz}. See also \cite{Linde:2012bt,Torres-Lomas:2014bua,Suyama:2014vga}, and a summary of various models which can produce PBHs is presented in \cite{Green:2014faa}. Alternatively, the constraint on the formation criteria can be relaxed during a phase transition in the early universe, causing PBHs to form preferentially at that mass scale \cite{Jedamzik:1999am} - although such an effect will not be considered here.

PBHs have traditionally been used to investigate the early universe by placing a constraint on the small-scale power spectrum from the corresponding constraint on their abundance \cite{Green:1997sz,Josan:2009qn,Shandera:2012ke}. In this paper, large-scale fluctuations in the PBH density caused by local-type non-Gaussianity are considered. If DM is composed entirely, or partially, of PBHs, these perturbations will be seen as isocurvature modes in cold dark matter (CDM) - upon which there are tight constraints from the recent \emph{Planck} data release \cite{Ade:2015lrj}.

The isocurvature perturbations are formed in a highly non-linear manner in this model. PBHs form shortly after horizon reentry during radiation domination, with an energy density exponentially sensitive to the amplitude of the power spectrum. Observational constraints imply that at most one region in a million collapsed into a PBH so the large-scale radiation density is almost unaffected, but if PBHs form DM then the amplitude of the DM perturbation is extremely sensitive to the modal coupling. Using this mechanism, CDM (with zero pressure) is formed in a universe which could have previously have been made up entirely of radiation and hence had no isocurvature perturbation prior to PBH formation. Such an effect is impossible within linear perturbation theory \cite{Wands:2002bn}.

In a previous paper,  the peak-background split was used to investigate the effect of modal coupling on the constraints that can be placed on the small-scale power spectrum \cite{Young:2014oea}. In this paper we use the same mechanism to investigate the extent to which modal coupling produces CDM isocurvature modes and discuss the implications of such an effect. Even if the initial conditions are adiabatic, which has been shown to be the case in single-field inflation, if there is modal coupling then the conversion of radiation into CDM (by collapse into PBHs) can have different efficiencies in different regions of the universe, which introduces isocurvature modes in the CMD after inflation has ended.

Even single-field inflation generates a small value of $f_{NL}$ with magnitude comparable to the spectral index \cite{Maldacena:2002vr} - which apparently could therefore rule out single-field inflation as a mechanism to create PBH DM. However, it has been argued that this is a result of gauge choice \cite{Pajer:2013ana,Tanaka:2011aj}, and that for our purposes the effective $f_{NL}=0$ in single-field inflation. It is therefore assumed in this paper that $f_{NL}$ can be arbitrarily close to zero.

Throughout, we will assume $f_{NL}$ to be scale invariant whilst the power spectrum becomes several orders of magnitude larger at small-scales - which is likely to be unrealistic given a specific model. However, this is a conservative approach, because if $|f_{NL}|$ were to become larger at some small scale, it would not weaken the constraints derived here, but would be likely to strengthen them. Even if the bispectrum was exactly zero when all three modes have sub CMB scales, the modal coupling between the CMB and PBH scales would still effect the amplitude of the power spectrum on PBH scales and the constraints which we derive would not be significantly weakened. In such a case, the perturbations within a region smaller than we can probe on the CMB would be Gaussian, but the variance would vary between different patches, in a way completely correlated to the long wavelength perturbation.   

Shortly prior to the release of this paper, \cite{Tada:2015noa} released a paper discussing a similar effect and the use of PBHs as biased tracers. We confirm their results and extend the calculation to account for the non-Gaussianity parameter $g_{NL}$ as well as $f_{NL}$, the effect of intermediate modes (between the CMB- and PBH-scales), and make use of the more recent results from the \emph{Planck} 2015 data release. Because all surviving PBHs necessarily behave as at least a subdominant DM component today, we also show how the allowed fraction of PBHs can be constrained more tightly than previously realised, under the presence of even small non-Gaussianity.

The layout of this paper is as follows: in section 2, the calculation of the PBH abundance, in both the Gaussian and non-Gaussian case, is reviewed. In section 3, modal coupling and how the peak-background split may be used to investigate its effects on PBH abundance is discussed. In section 4, the calculation is applied to the formation of CDM isocurvature modes and place constraints on the non-Gaussianity parameters in the case of PBH DM, and the calculation is extended to include simultaneous $f_{NL}$ and $g_{NL}$, intermediate modes, and the case where PBHs only make up a portion of the DM. We conclude with a summary of our arguments in section 6.

\section{Calculating the abundance of primordial black holes}

The abundance of PBHs is normally stated in terms of $\beta$: the energy fraction of the universe going into PBHs at the time of formation. The standard calculation used in the literature uses a Press-Schechter approach, although it has been shown that, for a Gaussian distribution, this matches well when the theory of peaks is used. It has been argued that the density contrast, rather than the curvature perturbation, should be used - although an approximation using the curvature perturbation works very well if care is taken to exclude super-horizon modes from the calculation, and this simplifies the calculation greatly. In this section, we will briefly review the calculation, as well as the main sources of error, for both Gaussian and non-Gaussian cases.

When a perturbation reenters the horizon, if its amplitude exceeds a certain threshold, or critical, value, then gravitational forces will overcome pressure forces and the region will collapse to form a primordial black hole. There has been extensive research to calculate the threshold value \cite{Niemeyer:1999ak,Hawke:2002rf,Musco:2004ak,Musco:2008hv,Harada:2013epa,Nakama:2013ica}, which is typically stated in terms of the density contrast. The critical value of the density perturbation is believed to be $\delta_{c}\approx0.45$. However, in this paper the curvature perturbation is used, and the corresponding critical value is $\zeta_{c}\approx1$ - within the range found by \cite{Shibata:1999zs}, and is consistent with using the density contrast \cite{Young:2014ana}.

The main source of uncertainty in the critical value is due to the unknown shape of primordial perturbations - and this is the largest source of error in the calculation of the abundance. However, whilst the effect on the calculated value of the abundance is large, the effect of this uncertainty on derived parameters is relatively small. For example, an error of $\mathcal{O}(10\%)$ in the threshold value results in an error of several orders of magnitude in the calculated $\beta$ but only an error of $\mathcal{O}(10\%)$ in the constraint on the power spectrum \cite{Byrnes:2012yx,Young:2013oia}. In this paper, because our results depend only on the relative abundance of PBHs in different regions of the universe, the conclusions are not sensitive to small changes in the threshold value.

Using a Press-Schechter approach, the mass fraction of the universe going into PBHs at the time of formation is given by integrating over the probability density function (PDF),
\begin{equation}
\beta=\int\limits_{\zeta_c}^{\infty}P(\zeta)d\zeta.
\end{equation}
In the case of a Gaussian distribution, the probability density function is
\begin{equation}
P(\zeta)=\frac{1}{\sqrt{2 \pi \sigma^{2}}}\exp\left(-\frac{\zeta}{2\sigma^{2}}\right),
\end{equation}
Where $\sigma^{2}$ is the variance of perturbation amplitude at the PBH forming scale. $\beta$ can therefore be written in terms of the complimentary error,
\begin{equation}
\beta=\mathrm{erfc}\left(\frac{\zeta_{c}}{\sqrt{2\sigma^{2}}}\right).
\end{equation}
Expanding using the large-x limit of $\mathrm{erfc}(x)$, gives
\begin{equation}
\beta\approx\sqrt{\frac{2\sigma^{2}}{\pi\zeta_{c}^{2}}}\exp\left(-\frac{\zeta_{c}^{2}}{2\sigma^{2}}\right).
\end{equation}
This is valid only if the distribution is Gaussian, and because PBHs form in the extreme positive tail of the PDF, their abundance is very sensitive to any non-Gaussianity, which we discuss below.

\subsection{Calculating the abundance of PBHs in the presence of non-Gaussianity}
In the local model of non-Gaussianity, the curvature perturbation is given by
\begin{equation}
\zeta=\zeta_{G}+\frac{3}{5}f_{NL}\left(\zeta_{G}^{2}-\sigma^{2}\right)+\frac{9}{25}g_{NL}\zeta_{G}^{3}+...=h\left(\zeta_{G}\right),
\label{local NG}
\end{equation}
where $\sigma^{2}$ is the variance of the Gaussian variable $\zeta_{G}$, and is subtracted to ensure the expectation value of $\zeta$ is zero.

The calculation of the abundance of PBHs is most easily performed by calculating the values of $\zeta_{G}$ which correspond the critical value, $\zeta_{c}$, and integrating over the corresponding regions of the Gaussian PDF of $\zeta_{G}$ - the reader is directed to \cite{Byrnes:2012yx,Young:2013oia} for a full derivation. For example, let us consider the case where $g_{NL}$ and higher-order terms are zero:
\begin{equation}
\zeta=\zeta_{G}+\frac{3}{5}f_{NL}\left(\zeta_{G}^{2}-\sigma^{2}\right)=h\left(\zeta_{G}\right).
\end{equation}
$h^{-1}(\zeta_{c})$ therefore has two solutions, given by
\begin{equation}
h_{c\pm}^{-1}=h_{\pm}^{-1}(\zeta_{c})=\frac{-5\pm\sqrt{25+60\zeta_{c}f_{NL}+36\zeta_{c}^{2}f_{NL}^{2}\sigma^{2}}}{6f_{NL}}.
\label{hfnl}
\end{equation}
For positive $f_{NL}$
\begin{equation}
\label{pos fnl beta}
\beta=\sqrt{\frac{2}{\pi\sigma^{2}}}\left(\int\limits_{h_{c+}^{-1}}^{\infty}\exp\left(-\frac{\zeta_{G}^{2}}{2\sigma^{2}}\right)d\zeta_{G}+\int\limits_{-\infty}^{h_{c-}^{-1}}\exp\left(-\frac{\zeta_{G}^{2}}{2\sigma^{2}}\right)d\zeta_{G} \right),
\end{equation}
and for negative $f_{NL}$
\begin{equation}
\label{neg fnl beta}
\begin{split}
\beta & =\sqrt{\frac{2}{\pi\sigma^{2}}}\int\limits_{h_{c+}^{-1}}^{h_{c-}^{-1}}\exp\left(-\frac{\zeta_{G}^{2}}{2\sigma^{2}}\right)d\zeta_{G}\\
& = \sqrt{\frac{2}{\pi\sigma^{2}}}\left(\int\limits_{h_{c+}^{-1}}^{\infty}\exp\left(-\frac{\zeta_{G}^{2}}{2\sigma^{2}}\right)d\zeta_{G} - \int\limits_{h_{c-}^{-1}}^{\infty}\exp\left(-\frac{\zeta_{G}^{2}}{2\sigma^{2}}\right)d\zeta_{G} \right).
\end{split}
\end{equation}
Furthermore, if we make the assumption that $f_{NL}$ is small, $f_{NL}\ll1$, which we will show is justified in the case that DM is composed of PBHs (and is further verified by the findings of \cite{Tada:2015noa}), the above expressions can  be simplified further. In the expression of $\beta$ for positive and negative $f_{NL}$, the first term inside the brackets dominates, and $\beta$ can be written in terms of one complimentary error function,
\begin{equation}
\label{beta approx}
\begin{split}
\beta & =\sqrt{\frac{2}{\pi\sigma^{2}}}\int\limits_{h_{c+}^{-1}}^{\infty}\exp\left(-\frac{\zeta_{G}^{2}}{2\sigma^{2}}\right)d\zeta_{G}\\
& =\mathrm{erfc}\left(\frac{h_{c+}^{-1}}{\sqrt{2}\sigma}\right)\\
& \approx\sqrt{\frac{2\sigma^{2}}{\pi(h^{-1}_{c+})^{2}}}\exp\left(-\frac{(h^{-1}_{c+})^{2}}{2\sigma^{2}}\right).
\end{split}
\end{equation}
Deriving an analytic expression as shown here is not a necessary step, but it is a useful approximation, and we will later use this result to derive an analytic expression for bias factor and amplitude of isocurvature modes in the PBH density.

Although it is not shown here, the same calculation can be performed for the local model of non-Gaussianity containing $g_{NL}$ - the interested reader is again directed to \cite{Byrnes:2012yx,Young:2013oia} for a full discussion of the calculation. In the case where only a cubic and linear term are considered
\begin{equation}
\zeta=\zeta_{G}+\frac{9}{25}g_{NL}\zeta_{G}^{3}=h(\zeta_{G}),
\end{equation}
then $h^{-1}(\zeta_c)$ has up to three possible solutions, depending on the value of $g_{NL}$ and $\zeta_{c}$. However, assuming that $g_{NL}$ is small, $g_{NL}\ll1$, which again will be shown later, the expression is dominated by one $\mathrm{erfc}$ function as in equation (\ref{beta approx}), with a different expression for $h^{-1}(\zeta_c)$. To first order in $g_{NL}$
\begin{equation}
h^{-1}_{c}=\zeta_{c}-\frac{9\zeta_{c}^{3}g_{NL}}{25}.
\end{equation}


\section{Modal coupling and the peak-background split}
\begin{figure}[t]
\centering
\subfloat[Superposition of short- and long-wavelength modes with modal coupling]{%
      \includegraphics[width=\textwidth]{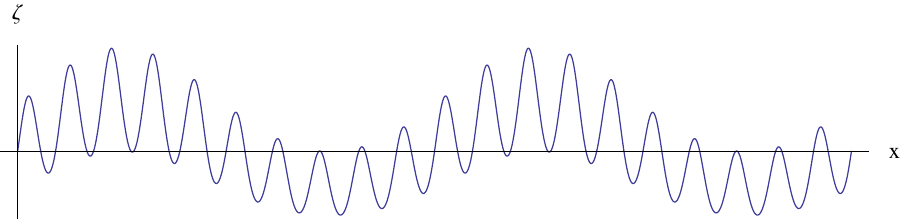}
    }
    \hfill
    \subfloat[Subtraction of the long-wavelength mode]{%
      \includegraphics[width=\textwidth]{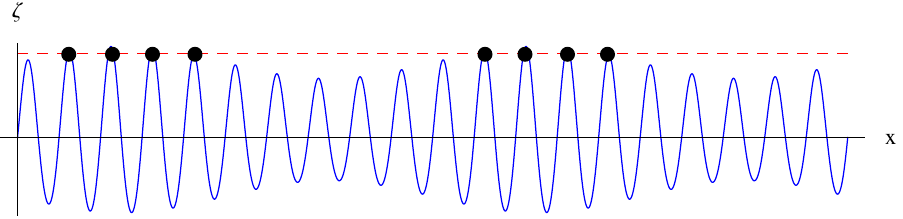}
    }
 \caption[Modal coupling of short- and long-wavelength modes]{The top plot shows an example of a universe containing only two modes. As an example of modal coupling, the amplitude of the short wavelength mode is a function of the long wavelength mode - the amplitude of the short-wavelngth mode is larger at the peak of the long-wavelength mode. At the time when short-wavelength mode enters the horizon, and PBHs at that scale form, the long-wavelength mode is not yet visible and will not affect whether a PBH forms or not. The bottom plot shows the same universe, but with the long wavelength mode subtracted, enabling $\zeta$ to be used a formation criterion for PBHs. The dashed red line shows the formation threshold for PBHs - regions where the curvature perturbation is greater than the formation threshold will collapse to form a PBH. The black circles represent areas which will collapse to form a PBH. It can be seen that a relatively small change in the amplitude of the small-scale mode can have a large impact on the number of PBHs forming in a region.}
\label{large scale inhomos}
\end{figure}

It has previously been shown \cite{Young:2014ana} that curvature perturbation modes which are a long way outside the horizon at the time of PBH formation have little effect on whether a PBH forms. This is due to the suppression of large-scale density modes by a factor $k^{2}$ relative to the curvature perturbation. In radiation domination:
\begin{equation}
\delta(t,k)=\frac{2(1+\omega)}{5+3\omega}\left(\frac{k}{aH}\right)^{2}\zeta(k)=\frac{4}{9}\left(\frac{k}{aH}\right)^{2}\zeta(k),
\end{equation}
where $\omega=1/3$ is the equation of state, and $(aH)^{-1}$ is the horizon scale at the time of PBH formation. However, long wavelength modes can have an indirect effect on the abundance of PBHs, $\beta$, due to modal coupling from non-Gaussianity. A long wavelength mode can affect both the amplitude and distribution of the small-scale perturbations which may form PBHs. In figure \ref{large scale inhomos}, we show how the coupling of long- and short-wavelength modes can affect the number of PBHs forming in different regions of the universe. At the peak of the long wavelength mode, the amplitude of the small-scale mode is increased, forming more PBHs, whilst the opposite occurs at the trough. 

How modal coupling can affect the constraints on the power spectrum at small scales from PBHs has been investigated \cite{Young:2014oea}, although it was assumed that all the modes involved were sub-CMB and potentially had a large amplitude. In this paper, we will go beyond previous work and study the case where the large-scale modes are observable in the CMB and hence very small. Despite the their small amplitude, we show that these perturbations have a remarkably large effect on observations. In this section, we will briefly review the calculation using the peak-background split to investigate modal coupling due to the local non-Gaussianity parameters $f_{NL}$ and $g_{NL}$, and in the following section, apply this to the abundance of PBHs and the creation of isocurvature modes.

\subsection{Quadratic non-Gaussianity, $f_{NL}$}
We will take the model of local non-Gaussianity, in terms of the curvature perturbation $\zeta$, to be described by
\begin{equation}
\label{quadratic local model}
\zeta=\zeta_{G}+\frac{3}{5}f_{NL}\left(\zeta_{G}^{2}-\sigma^{2}\right)=h(\zeta_{G}),
\end{equation}
where $\zeta_{G}$ is a Gaussian variable. It is necessary to subtract $\sigma^{2}=\langle\zeta_{G}^{2}\rangle$ so that the background (average) value of $\zeta$ remains zero. We will now employ the peak-background split, and write the Gaussian component as the sum of a long-(background) and short-(peak) wavelength component,
\begin{equation}
\zeta_{G}=\zeta_{l}+\zeta_{s}.
\end{equation}
Equation (\ref{quadratic local model}) then becomes:
\begin{equation}
\zeta=\left(\zeta_{l}+\zeta_{s}\right)+\frac{3}{5}f_{NL}\left(\left(\zeta_{l}+\zeta_{s}\right)^{2}-\langle\left(\zeta_{l}+\zeta_{s}\right)^{2}\rangle\right).
\end{equation}
However, terms which depend only on the long-wavelength mode do not affect PBH formation, and should not be considered when determining the abundance of PBHs. We therefore subtract those terms, leaving:
\begin{equation}
\label{quadratic inhomogeneity}
\zeta=\left(1+\frac{6}{5}f_{NL}\zeta_{l}\right)\zeta_{s}+\frac{3}{5}\left(\zeta_{s}^{2}-\sigma_{s}^{2}\right).
\end{equation}
We can now rewrite the expression in terms of new variables, $\tilde{\zeta}_{G}$, $\tilde{\sigma}$ and $\tilde{f}_{NL}$, and calculate the abundance of PBHs $\beta$ as described in section 2, as a function of the long wavelength mode, $\zeta_{l}$.
\begin{equation}
\label{local fnl variables}
\begin{split}
&\tilde{\zeta}_{G}=\left(1+\frac{6}{5}f_{NL}\zeta_{l}\right)\zeta_{s},\\
&\tilde{\sigma}=\left(1+\frac{6}{5}f_{NL}\zeta_{l}\right)\sigma_{s},\\
&\tilde{f}_{NL}=\left(1+\frac{6}{5}f_{NL}\zeta_{l}\right)^{-2}f_{NL}.\\
\end{split}
\end{equation}
Equation (\ref{quadratic inhomogeneity}) can then be written in a form analogous to equation (\ref{quadratic local model}),
\begin{equation}
\zeta=\tilde{\zeta}_{G}+\frac{3}{5}\tilde{f}_{NL}\left(\tilde{\zeta}_{G}^{2}-\tilde{\sigma}^{2}\right)=\tilde{h}(\tilde{\zeta}_{G}).
\label{localQuadInhomo}
\end{equation}
Therefore, both the amplitude and distribution of the small-scale perturbations are affected. In order to calculate the abundance of PBHs, the variables in equation (\ref{local fnl variables}) can then be inserted into equation (\ref{beta approx}).

\subsection{Cubic non-Gaussianity, $g_{NL}$}
Here, we will follow the same steps as for $f_{NL}$, to show how the presence of a cubic term causes modal coupling. For this section, we will assume $f_{NL}=0$, and $\zeta$ to be given by
\begin{equation}
\zeta=\zeta_{G}+\frac{9}{25}g_{NL}\zeta_{G}^{3}.
\label{cubicNG}
\end{equation}
Again, using the peak-background split, one obtains:
\begin{equation}
\zeta=\left(1+\frac{27}{25}g_{NL}\zeta_{l}^{2}\right)\zeta_{s}+\left(\frac{27}{25}g_{NL}\zeta_{l}\right)\zeta_{s}^{2}+\left(\frac{9}{25}g_{NL}\right)\zeta_{s}^{3}+\mathcal{O}(\zeta_{l}),
\end{equation}
where again, the terms dependent only on $\zeta_{l}$ are neglected because they don't have a significant effect on PBH formation. The above expression can then be rewritten in terms of new variables $\tilde{\zeta}_{G}$, $\tilde{\sigma}$, $\tilde{f}_{NL}$ and $\tilde{g}_{NL}$, given by
\begin{equation}
\begin{split}
&\tilde{\zeta}_{G}=\left(1+\frac{27}{25}g_{NL}\zeta_{l}^{2}\right)\zeta_{s},\\
&\tilde{\sigma}=\left(1+\frac{27}{25}g_{NL}\zeta_{l}^{2}\right)\sigma_{s},\\
&\tilde{f}_{NL}=\left(\frac{9}{5}g_{NL}\zeta_{l}\right)\left(1+\frac{27}{25}g_{NL}\zeta_{l}^{2}\right)^{-2},\\
&\tilde{g}_{NL}=g_{NL}\left(1+\frac{27}{25}g_{NL}\zeta_{l}^{2}\right)^{-3}.\\
\end{split}
\label{bar gnl}
\end{equation}
Equation (\ref{cubicNG}) can then be rewritten as
\begin{equation}
\zeta=\tilde{\zeta}_{G}+\frac{3}{5}f_{NL}\left(\tilde{\zeta}_{G}^{2}-\tilde{\sigma}^{2}\right)+\frac{9}{25}\tilde{g}_{NL}\tilde{\zeta}_{G}^{3}.
\end{equation}
An expression for the abundance of PBHs in a given region of the universe, $\tilde{\beta}$, can then be derived as shown in section 2. 

In this section, it has been shown that long wavelength modes can affect the amplitude of local small-scale perturbations and the non-Gaussianity parameters, and in the next section the effect of this on the abundance of PBHs within a given region will be discussed.

\section{The isocurvature modes of primordial black hole dark matter on CMB scales}

\begin{figure}[t]
\centering
\includegraphics[width=0.8\linewidth]{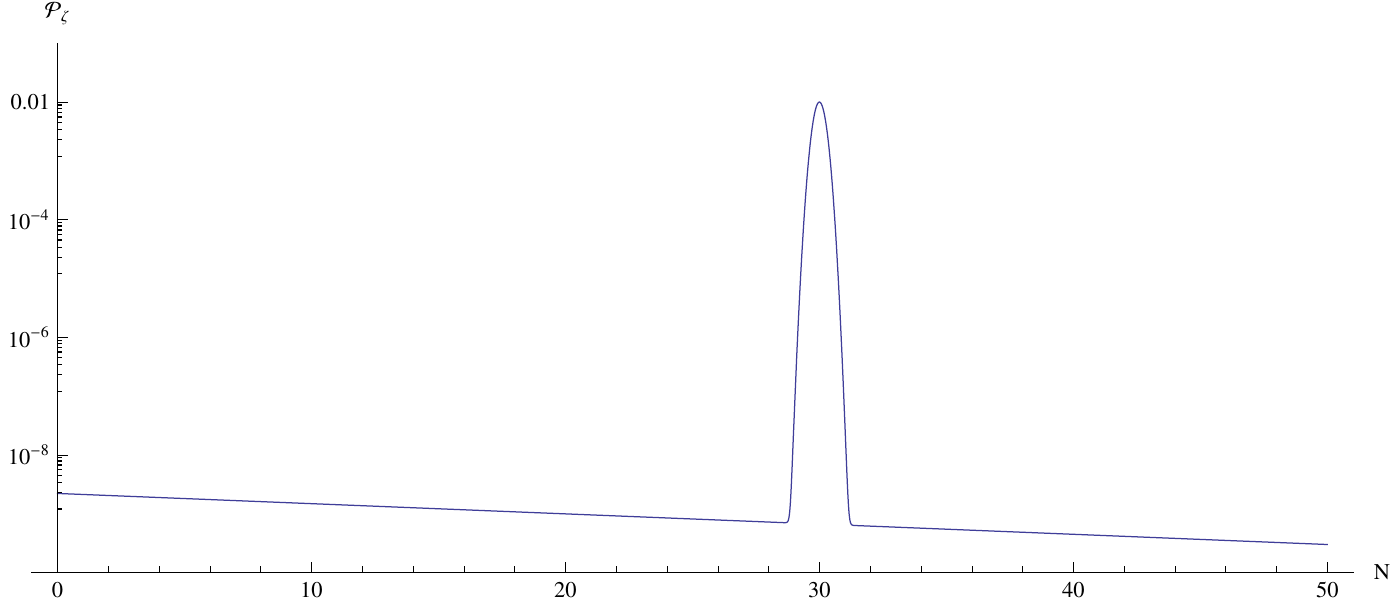}
\caption[A narrow peak in the power spectrum]{An example of a power spectrum containing a narrow peak. $N$ represents number of e-folds, with smaller scales represented by larger $N$. The power spectrum is small on most scales with a spectral index of $n_{S}=0.96$, compatible with observations of the cosmic microwave background (CMB) and large scale structure (LSS). The narrow peak in the power spectrum corresponds to the scale at which will PBHs form.}
\label{narrow peak}
\end{figure}

The abundance of PBHs in a region of the universe can be affected significantly by large-scale curvature perturbation modes in different regions of the universe. If PBHs make up DM, then these differences in the abundance of PBHs will appear as fluctuations in the density of DM. In the presence of local-type non-Gaussianity, the fluctuations in the DM can be significantly greater than the curvature perturbations responsible for producing them - and tight constraints can therefore be placed on the non-Gaussianity parameters if this is the case from the isocurvature constraints from \emph{Planck}. 

We will define the difference in the abundance of PBHs at the time of formation, $\delta_{\beta}$, as
\begin{equation}
\delta_{\beta}=\frac{\beta-\bar{\beta}}{\bar{\beta}},
\label{deltabeta}
\end{equation}
where $\beta$ and $\bar{\beta}$ are the perturbed and background values of the PBH abundance at the time of formation respectively. If the large-scale curvature perturbation $\zeta$ is small, it can be related to $\delta_{\beta}$ by a constant factor $b$ (referred to the scale dependent bias in \cite{Tada:2015noa}),
\begin{equation}
\delta_{\beta}=b \zeta_{l},
\end{equation}
where $b$ is a function of the non-Gaussianity parameters, the variance of the small-scale perturbations and the critical value for PBH formation $\zeta_{c}$. The factor $b$ therefore parameterizes the bias of PBHs to form in the presence of large-scale curvature perturbations.

In this section, we will consider the case where the power spectrum is very small on all scales, except for a narrow region where there is a sharp spike -  which is responsible for the production of PBHs of a mass corresponding to this scale\footnote{The mass of a PBH is roughly equal to the horizon mass at the time of formation. See \cite{Young:2014ana} for further discussion.}. An example of such a power spectrum is given in figure \ref{narrow peak}. We therefore ignore in this section the presence of perturbations of intermediate scales, but extend the calculation in the following section to account for when there is a broad peak in the power spectrum.

The abundance of PBHs at a later time on a comoving slicing will be affected by difference in their density at the time of formation, as well as by the difference in expansion since the time of formation - in denser regions of the universe, inflation ends and PBHs form slightly later, so even if the PBH density is constant at the time of formation, the density will not be constant. To first order in $\zeta$, the density of PBHs can be expressed as
\begin{equation}
\Omega_{PBH}=\left(1+b\zeta+3\zeta\right)\bar{\Omega}_{PBH},
\end{equation}
where the $3\zeta$ term is simply the adiabatic mode expected from the expansion of the universe, and $\bar{\Omega}_{PBH}$ is the background density of PBHs. The $b\zeta$ term therefore represents a deviation from the expected amplitude of the mode if it was purely adiabatic - it is an isocurvature mode, which will either be fully correlated, or fully anti-correlated depending on the sign of $f_{NL}$. If PBHs make up a significant fraction of the DM content of the universe, the constraints on isocurvature modes from \emph{Planck} can then be used to constrain $b$ - and therefore constrain the non-Gaussianity parameters\footnote{Note that the reverse is also true - for a given value of the non-Gaussianity parameters, an upper limit can be placed on the amount of DM which is made of PBHs}. For simplicity in this paper, except section 5.2, we will assume that DM is entirely composed of PBHs, and calculate corresponding constraints on the non-Gaussianity parameters $f_{NL}$ and $g_{NL}$.
On CMB scales, the constraints from \emph{Planck} on isocurvature modes can be used \cite{Ade:2015lrj}
\begin{equation}
100\beta_{iso}=\begin{cases} 0.13 &\mbox{, fully correlated} \\
0.08 &\mbox{, fully anti-correlated},
\end{cases}
\end{equation}
where
\begin{equation}
\beta_{iso}=\frac{\mathcal{P}_{iso}}{\mathcal{P}_{iso}+\mathcal{P}_{\zeta}}.
\end{equation}
The fully correlated modes correspond to positive $b$, whilst fully anti-correlated corresponds to negative $b$ (and positive/negative $f_{NL}$ and $g_{NL}$ respectively). The isocurvature power spectrum is related to the curvature perturbation power spectrum as
\begin{equation}
\mathcal{P}_{iso}=b^{2}\mathcal{P}_{\zeta},
\end{equation}
and we therefore obtain constraints on $b$ as
\begin{equation}
-0.028<b<0.036.
\end{equation}
This result will now be used to derive a result on the non-Gaussianity parameters.

\subsection{Isocurvature modes from $f_{NL}$}
In section 2, an expression for the abundance of PBHs at the time of formation $\beta$, was derived in terms of the non-Gaussianity parameter $f_{NL}$, the variance of the Gaussian component\footnote{$\sigma$ is related to the power spectrum as follows \cite{Byrnes:2007tm} 
\begin{equation} 
\mathcal{P}_{\zeta}=\sigma^{2}+\left(\frac{3}{5}\right)^2 \left( 4 f_{NL}^2+6 g_{NL}\right)\sigma^{4}\ln(kL)+ \left(\frac{3}{5}\right)^4 \left( 27 g_{NL}^2\right) \sigma^{6}\ln(kL)^{2}, 
\end{equation} 
where the higher-order terms from $g_{NL}$ have also been included, and $\ln(kL)$ is a factor of around unity. Note that, since the non-Gaussianity parameters are found to be very small, the higher-order terms will not have a significant impact, and to a good approximation $\mathcal{P}_{\zeta}=\sigma^{2}$.} $\sigma^{2}$, and the critical value for collapse $\zeta_{c}$ - equation (\ref{beta approx}), with $h^{-1}$ given by equation (\ref{hfnl}). However, this calculation assumes there is no coupling to large-scale modes (and is equivalent to the background value, $\bar{\beta}$, if large-scale perturbations are small - as is the case here). In section 3 it was shown how to account for the presence of a large-scale modes - namely, by using the transformed variables $\tilde{f}_{NL}$ and $\tilde{\sigma}$ instead, given by equation (\ref{local fnl variables}) - which calculates the perturbed  abundance $\beta$.

By combing equations (\ref{hfnl}), (\ref{pos fnl beta}), (\ref{neg fnl beta}), (\ref{local fnl variables}) and (\ref{deltabeta}), it is possible to derive an expression for $\delta_{\beta}$ in terms of $f_{NL}$, $\sigma_{s}$ (where the $s$ subscript has been adopted to denote the small PBH scale), and the critical value $\zeta_{c}$. Expanding the expression to first order in $\zeta$ gives the result
\begin{equation}
\delta_{\beta}=\frac{25+30\zeta_{c}f_{NL}+36f_{NL}^{2}\sigma_{st}^{2}-5\sqrt{25+60\zeta_{c}f_{NL}+36f_{NL}^{2}\sigma_{s}^{2}}}{3f_{NL}\sigma_{s}^{2}\sqrt{25+60\zeta_{c}f_{NL}+36f_{NL}^{2}\sigma_{s}^{2}}}\zeta,
\end{equation}
and therefore $b$ is given by
\begin{equation}
b=\frac{25+30\zeta_{c}f_{NL}+36f_{NL}^{2}\sigma_{st}^{2}-5\sqrt{25+60\zeta_{c}f_{NL}+36f_{NL}^{2}\sigma_{s}^{2}}}{3f_{NL}\sigma_{s}^{2}\sqrt{25+60\zeta_{c}f_{NL}+36f_{NL}^{2}\sigma_{s}^{2}}},
\label{bfnl}
\end{equation}
or to first order in $f_{NL}$
\begin{equation}
b=\frac{6}{5}\left(1+\frac{\zeta_{c}^{2}}{\sigma_{s}^{2}}\right)f_{NL}.
\label{simple bfnl}
\end{equation}
As expected, a positive $f_{NL}$, which boosts the power spectrum on small scales in areas of higher density, produces a positive bias, and fully correlated isocurvature modes in PBH DM\footnote{The second expression for $b$ corresponds to equation (14) in \cite{Tada:2015noa}. The more complicated expression, equation (\ref{bfnl}), is because a Gaussian distribution on small scales has not been assumed. The differences between the 2 calculations are discussed in Appendix B.}. Negative $f_{NL}$ has the opposite effect, and produces fully anti-correlated isocurvature modes.

\begin{figure}[t]
\centering
\subfloat[Negative $\fNL$]{%
      \includegraphics[width=0.49\textwidth]{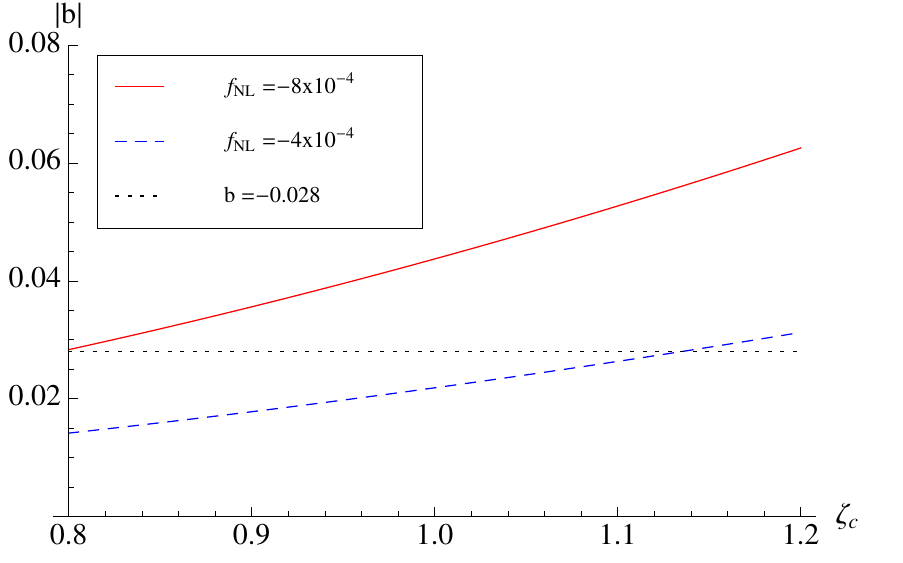}
    }
    \hfill
    \subfloat[Positive $\fNL$]{%
      \includegraphics[width=0.49\textwidth]{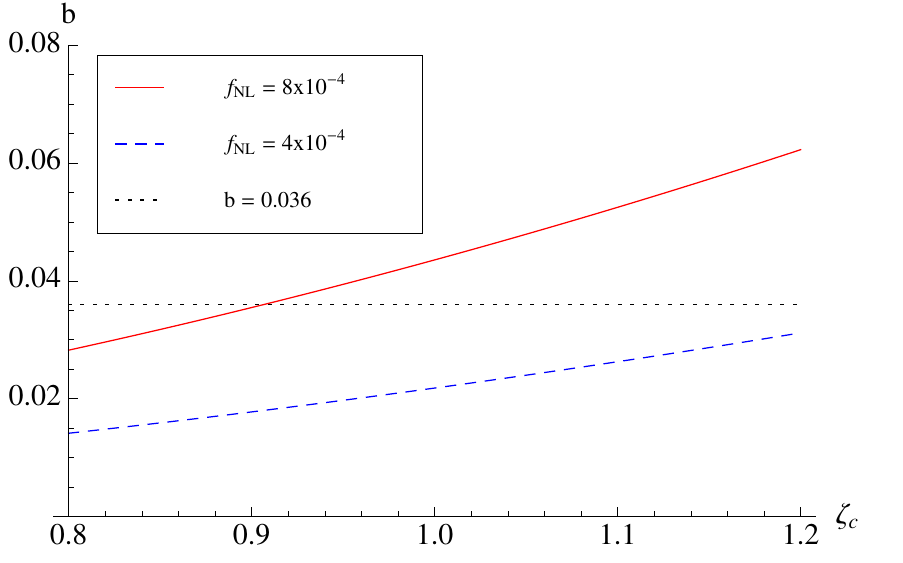}
    }
\caption[Primordial black hole bias from $\fNL$ and the threshold value]{The plots above show the effects of a different threshold $\zeta_{c}$ on the PBH bias $b$ arising from an $f_{NL}$ term. A larger value of $\zeta_{c}$ suggests a larger bias factor. The left plot shows the effect for negative $f_{NL}$ and the right plot for positive $f_{NL}$. The dotted black lines represent the constrains on $b$ from the constraints on isocurvature modes from \emph{Planck}. $|f_{NL}|=8\times10^{-4}$ is typically excluded whilst $|f_{NL}|=4\times10^{-4}$ is typically allowed. To generate these plots the value $\sigma=0.15$ has been used.}
\label{zetac vs b fnl}
\end{figure}

\begin{figure}[t]
\centering
\subfloat[Negative $\fNL$]{%
      \includegraphics[width=0.49\textwidth]{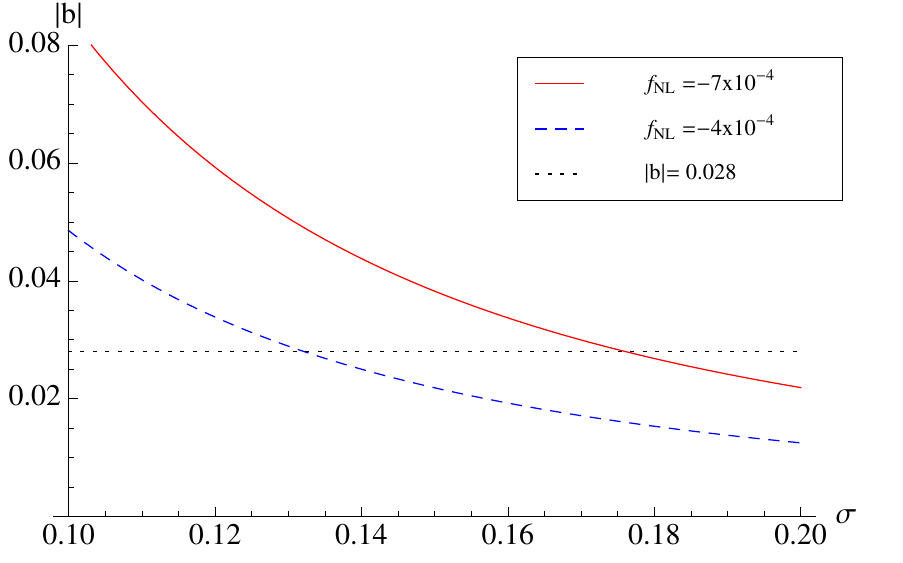}
    }
    \hfill
    \subfloat[Positive $\fNL$]{%
      \includegraphics[width=0.49\textwidth]{Paper4/sigmaVsBfnlNeg}
    }
\caption[Primordial black hole bias from $\fNL$ and the small-scale variance]{The effects of a different $\sigma$ on the PBH bias $b$ arising from an $f_{NL}$ term are investigated. A larger value of $\sigma$ suggests a smaller bias factor. The left plot shows the effect for negative $f_{NL}$ and the right plot for positive $f_{NL}$. The dotted black lines represent the constrains on $b$ from the constraints on isocurvature modes from \emph{Planck}. $|f_{NL}|=7\times10^{-4}$ is typically excluded whilst $|f_{NL}|=4\times10^{-4}$ is typically allowed. To generate these plots the value $\zeta_{c}=1$ has been used.}
\label{sigma vs b fnl}
\end{figure}

In order to investigate the constraints on the non-Gaussianity parameters, it is necessary to estimate values for the other parameters involved, and how these would affect the constraints. The variance of the small-scale perturbations and the critical value. 
\begin{itemize}
\item{First, $\zeta_{c}$ is considered: there is significant error in the exact value of the threshold value, due to uncertainty in the shape of the primordial perturbation which collapse to form PBHs. Most recent simulations have calculated the critical value in terms of the density contrast, finding $\delta_{c}\approx0.4$. This is consistent with the calculation here if the critical value of the curvature perturbation is related by a factor $\frac{4}{9}$, meaning $\zeta_{c}\approx 1$, which is consistent with the range of values found in \cite{Shibata:1999zs}. Figure \ref{zetac vs b fnl} shows how the factor $b$ depends on the critical value for different values of $f_{NL}$.}
\item{To calculate $\sigma_s$, it is necessary to first calculate the value of $\beta$ for which PBHs are otherwise unconstrained by observations and could be DM. The range of mass scales in which PBHs can form a significant fraction of DM is roughly $10^{17}\mathrm{g}<M_{PBH}<10^{24}\mathrm{g}$ \cite{Carr:2009jm}. The constraint on $\beta$ from the abundance of DM in this range are given by \cite{Josan:2009qn}
\begin{equation}
\beta<2\times10^{-19}\left(\frac{M_{PBH}}{f_{M}5\times10^{14}\mathrm{g}}\right)^{1/2},
\label{beta PBH mass}
\end{equation}
where $f_{M}$ is the fraction of the horizon mass which ends up inside the PBH\footnote{$f_{M}$ is a factor of order unity, which is neglected as it has very little effect on the calculated value of $\sigma_s$.}, and $M_{PBH}$ is the mass of the PBH. Assuming DM to be made up entirely of PBHs of a single mass scale within this range, $\beta$ can therefore range from $\beta<10^{-16}$ to $\beta<10^{-11}$. Assuming the most optimistic and pessimistic values for $\beta$ and $\zeta_{c}$, $\sigma_s$ is calculated to lie in the range $0.1<\sigma_s<0.2$ for close to Gaussian perturbations \cite{Byrnes:2012yx}. Figure \ref{sigma vs b fnl} displays how $b$ changes with $\sigma_s$.}
\end{itemize}

Smaller values of the variance of the small-scale perturbations, $\sigma_s^{2}$, would lead to tighter constraints on $f_{NL}$, whilst a smaller critical value $\zeta_{c}$ leads to tighter constraints on $f_{NL}$. Because a larger value of $\zeta_{c}$ implies a larger value of $\sigma_{s}$, these effects virtually cancel out - and the results presented below are therefore not sensitive to uncertainty in $\zeta_{c}$. 

Assuming PBH form at a single mass scale, the weakest constraint on $f_{NL}$ comes from considering the mass of the largest PBHs which could make up DM, which is taken to be $M_{PBH}=10^{25}$g, for which $\beta\approx10^{-14}$. If DM is made entirely of PBHs, the constraints on $f_{NL}$ are therefore
\begin{equation}
-4\times10^{-4}<f_{NL}<5\times10^{-4}.
\end{equation}
The results are not significantly different for PBHs of different mass. For example, for $M_{PBH}=10^{20}$g the constraints on $f_{NL}$ are
\begin{equation}
-3\times10^{-4}<f_{NL}<4\times10^{-4}.
\label{fnl-constraint}\end{equation}

\subsection{{Isocurvature modes from $g_{NL}$}}
In addition to $f_{NL}$, it is interesting to consider isocurvature modes arising from $g_{NL}$ and place constraints, or whether the effects of modal coupling from $g_{NL}$ could cancel the effects from $f_{NL}$. The effect of higher-order terms are beyond the scope of this paper.

\begin{figure}[t]
\centering
	\includegraphics[width=0.6\linewidth]{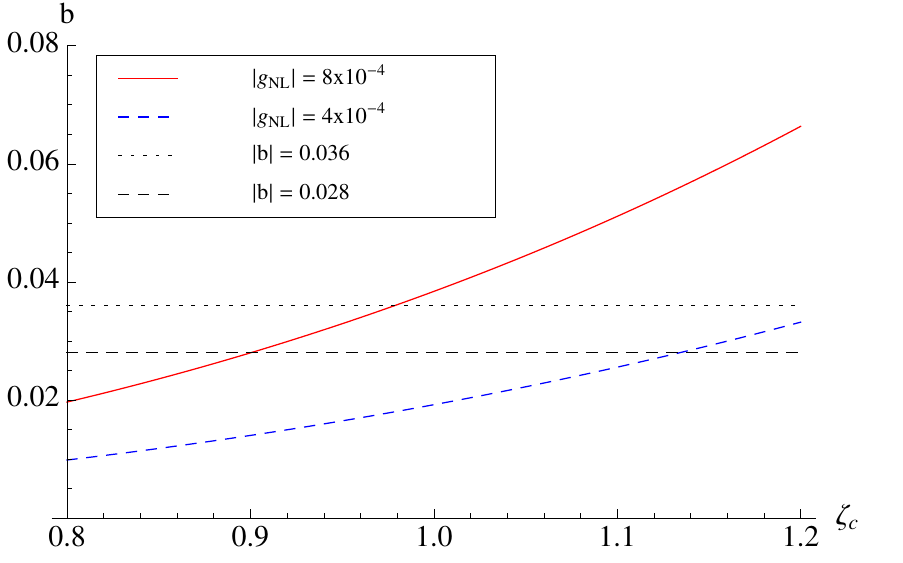}
\caption[Primordial black hole bias from $\gNL$ and the threshold value]{The plot above show the effects of a different threshold $\zeta_{c}$ on the PBH bias $b$ arising from a $g_{NL}$ term. A larger value of $\zeta_{c}$ suggests a larger bias factor. As the expression for $b$, equation (\ref{bgnl}), is anti-symmetric under a change of sign of $g_{NL}$, the results for negative and positive $g_{NL}$ are shown on one plot - but with different constraints on the amplitude of $|b|$, represented by the dotted black lines. $|g_{NL}|=8\times10^{-4}$ is typically excluded whilst $|g_{NL}|=4\times10^{-4}$ is typically allowed. To generate these plots the value $\sigma=0.15$ has been used.}
\label{zetac vs b gnl}
\end{figure}

\begin{figure}[t]
\centering
	\includegraphics[width=0.6\linewidth]{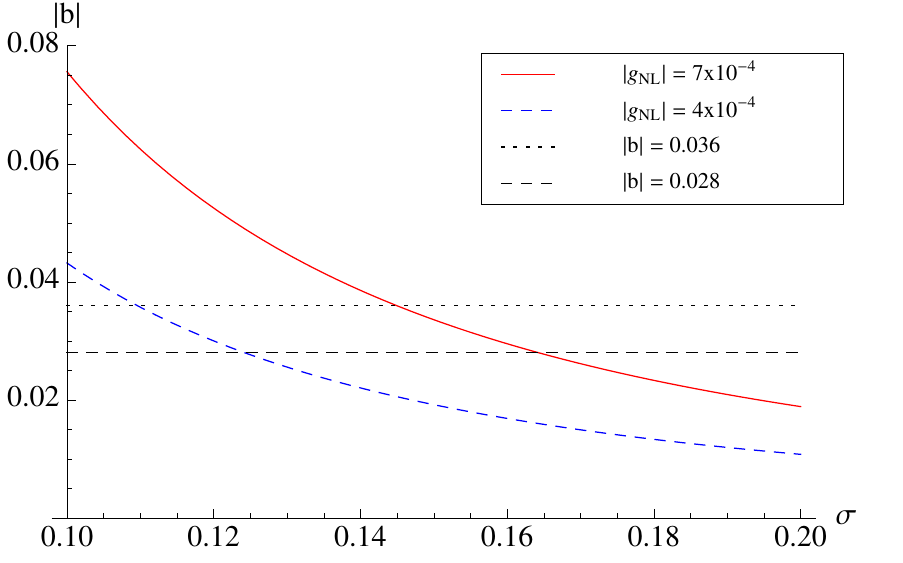}
\caption[Primordial black hole bias from $\gNL$ and the small-scale variance]{This plot shows the effects of a different $\sigma$ on the PBH bias $b$ arising from a $g_{NL}$ term. A smaller value of $\sigma$ suggests a larger bias factor. As the expression for $b$, equation (\ref{bgnl}), is anti-symmetric under a change of sign of $g_{NL}$, the results for negative and positive $g_{NL}$ are shown on one plot - but with different constraints on the amplitude of $|b|$, represented by the dotted black lines. $|g_{NL}|=7\times10^{-4}$ is typically excluded whilst $|g_{NL}|=4\times10^{-4}$ is typically allowed. To generate these plots the value  $\zeta_{c}=1$ has been used. This range of $\sigma$ is used because it is approximately the range of values required to generate the correct number of PBHs to form DM (assuming that perturbations are close to Gaussian).}
\label{sigmavsbgnl}
\end{figure}

The same derivation can be followed as that for $f_{NL}$, leading to an expression for $b$ to first order in $g_{NL}$
\begin{equation}
b=-\frac{27\left(\sigma_{s}^{2}-\zeta_{c}^{2}\right)\left(\sigma_{s}^{2}+\zeta_{c}^{2}\right)}{25\sigma_{s}^{2}\zeta_{c}}g_{NL}.
\label{bgnl}
\end{equation}
Again, as expected, positive $g_{NL}$ corresponds to fully correlated isocurvature modes, and negative $g_{NL}$ corresponds to fully anti-correlated isocurvature modes. The PBH bias factor $b$ is again a function of the non-Gaussianity parameter $g_{NL}$, the variance of the small-scale perturbations $\sigma_{s}^{2}$, and the formation threshold $\zeta_{c}$. The dependence of $b$ on $\zeta_{c}$ and $\sigma_{s}$ is shown in figures \ref{zetac vs b gnl} and \ref{sigmavsbgnl} respectively.

We see again that smaller values of $\sigma_s$ would lead to tighter constraints on $g_{NL}$, whilst a smaller $\zeta_{c}$ leads to tighter constraints on $g_{NL}$. However, unlike the case with $f_{NL}$, the constraint which can be placed on $g_{NL}$ depends on the value of $\zeta_{c}$, although only by a factor of $\mathcal{O}(10\%)$. The results presented below are the weakest constraints, corresponding to a low formation threshold, for PBHs of mass $10^{25}$g
\begin{equation}
-6\times10^{-4}<g_{NL}<7\times10^{-4}.
\end{equation}
Notice that these constraints are very comparable to those on $f_{NL}$, see (\ref{fnl-constraint}). The $f_{NL}$ term has an effect of $\mathcal{O}(10^{-5})$ on the small-scale power spectrum, whilst the $g_{NL}$ term only has an effect of $\mathcal{O}(10^{-10})$, and therefore, naively, the constraints on $g_{NL}$ would be expected to be roughly 5 orders of magnitude weaker than $f_{NL}$. However, a $g_{NL}$ term also has an effect on the small-scale $\tilde{f}_{NL}$, as seen in equation (\ref{bar gnl}), of $\mathcal{O}(10^{-5})$, and because the abundance of PBHs is extremely sensitive to non-Gaussianity, this causes significant isocurvature modes in the PBH DM. In the case where $\zeta_{l}=10^{-5}$ and $g_{NL}=10^{-3}$, then $\tilde{f}_{NL}\approx10^{-8}$. Such a small $\tilde{f}_{NL}$ nonetheless creates a perturbation in the PBH density of $\mathcal{O}(10^{-6})$, which represents an isocurvature mode of around $10\%$ of $\zeta$ - which is excluded by \emph{Planck}. Because the abundance of PBHs $\beta$ is sensitive to higher-order non-Gaussianity parameters \cite{Young:2013oia}, isocurvature modes are expected to rule out significant non-Gaussianity at higher orders as well - although a quantitative calculation is beyond the scope of this paper. Higher-order non-Gaussianity parameters are considered briefly in section 5.5.

\section{Further consideration of constraints from isocurvature modes}
In section 5.4, constraints were placed separately on $f_{NL}$ and $g_{NL}$ separately, assuming that DM was entirely composed of primordial black holes. In this section, the calculation is extended to account for more general models.

\subsection{Isocurvature modes from $f_{NL}$ and $g_{NL}$}
The presence of non-zero non-Gaussianity parameters has been shown to create significant isocurvature modes, which has led to very tight constraints on these parameters under the assumption that DM is composed entirely of PBHs. The calculation is now extended to account for non-zero $f_{NL}$ and $g_{NL}$ simultaneously - for example, it is possible that the effect of a large positive $f_{NL}$ and large negative $g_{NL}$ can cancel out, leaving a very small isocurvature mode.

Because the non-Gaussianity parameters may now become quite large, the full numeric calculation for the PBH abundance is used to derive a value for the PBH bias $b$, for example by using equations (\ref{pos fnl beta}) or (\ref{neg fnl beta}) rather than the much simpler equation (\ref{beta approx}).

\begin{figure}[t]
\centering
\subfloat[]{%
      \includegraphics[width=0.49\textwidth]{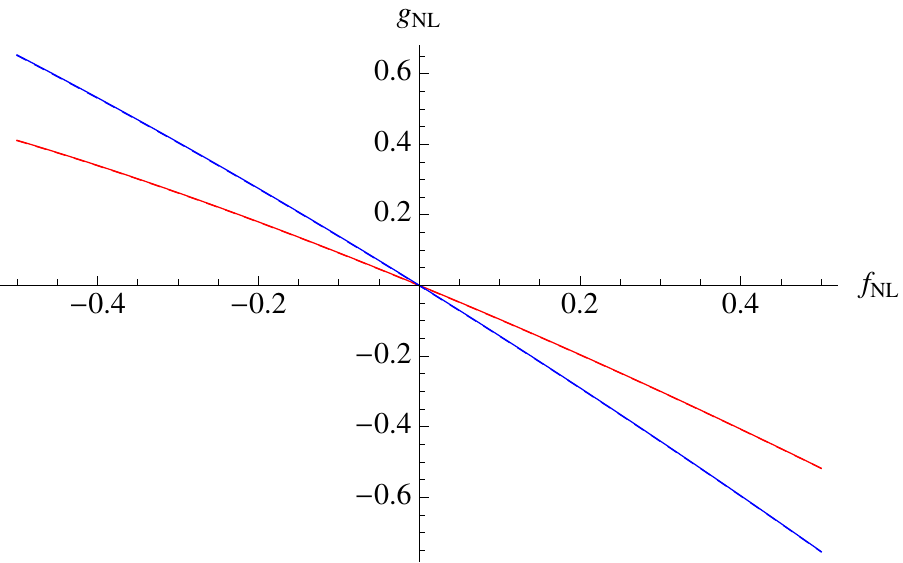}
    }
    \hfill
    \subfloat[]{%
      \includegraphics[width=0.49\textwidth]{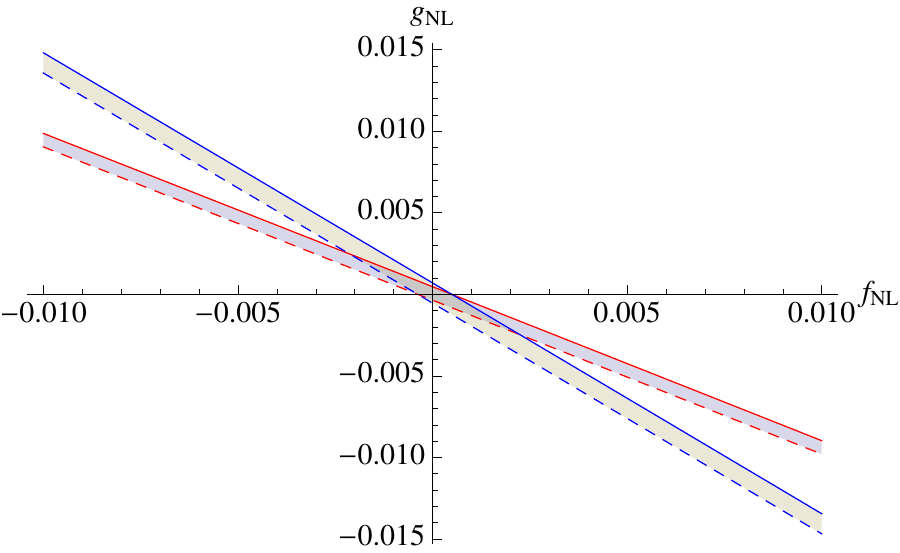}
    }
\caption[Allowed values of $\fNL$ and $\gNL$]{The constraints on simultaneous $f_{NL}$ and $g_{NL}$ are displayed. The right plot simply displays the central region of the plot on the left. The solid lines represent an upper limit from fully correlated isocurvature modes, whilst the dotted lines represent a lower limit from fully anti-correlated isocurvature modes. There is some uncertainty in the value of $g_{NL}$ given a value of $f_{NL}$ due to uncertainty in the critical value $\zeta_{c}$ - the blue lines are obtained using $\zeta_{c}=0.8$, and the red lines are obtained using $\zeta_{c}=1.2$. It can nonetheless be seen that large $f_{NL}$ or $g_{NL}$ are excluded unless very finely tuned. The shaded regions between the lines can be considered as $2\sigma$ contour plots from the \emph{Planck} constraints.}
\label{fnl vs gnl}
\end{figure}

Figure \ref{fnl vs gnl} shows the values of $g_{NL}$ that are permitted for different values of $f_{NL}$ for PBHs of mass $M_{PBH}=10^{25}$g. Whilst large values of $f_{NL}$ and $g_{NL}$ are allowed, there needs to be significant fine tuning to ensure that the resultant isocurvature modes are not excluded by the \emph{Planck} results - $g_{NL}$ needs to have the correct value to $\mathcal{O}(0.1\%)$. We note that there is some uncertainty in the value of $g_{NL}$ required for a given $f_{NL}$ due to the uncertainty in the formation threshold $\zeta_{c}$ - although this does not affect the conclusion that large non-Gaussianity parameters are not allowed unless very very finely tuned. This conclusion is expected to remain true for higher-order terms \cite{Young:2013oia}.

\subsection{Fractional primordial black hole dark matter}
So far, it has been assumed that DM is made entirely of PBHs. The calculation is now extended to account for the fact that PBHs may only make up a small fraction of DM, and this is parameterized by $r_{PBH}$, the ratio of PBH density to DM density.
\begin{equation}
r_{PBH}=\frac{\Omega_{PBH}}{\Omega_{DM}}.
\end{equation}
In this case, the density of DM is described by
\begin{equation}
\Omega_{DM}=\left(1+r_{PBH}b\zeta+3\zeta\right)\bar{\Omega}_{DM},
\end{equation}
and the relative amplitude of the isocurvature modes is now given by $r_{PBH}b$. Therefore, from the \emph{Planck} constraints on isocurvature modes instead give constraints on the factor $r_{PBH}b$,
\begin{equation}
-0.028<r_{PBH}b<0.036.
\end{equation}

\begin{figure}[t]
\centering
\subfloat[$\fNL$]{%
      \includegraphics[width=0.49\textwidth]{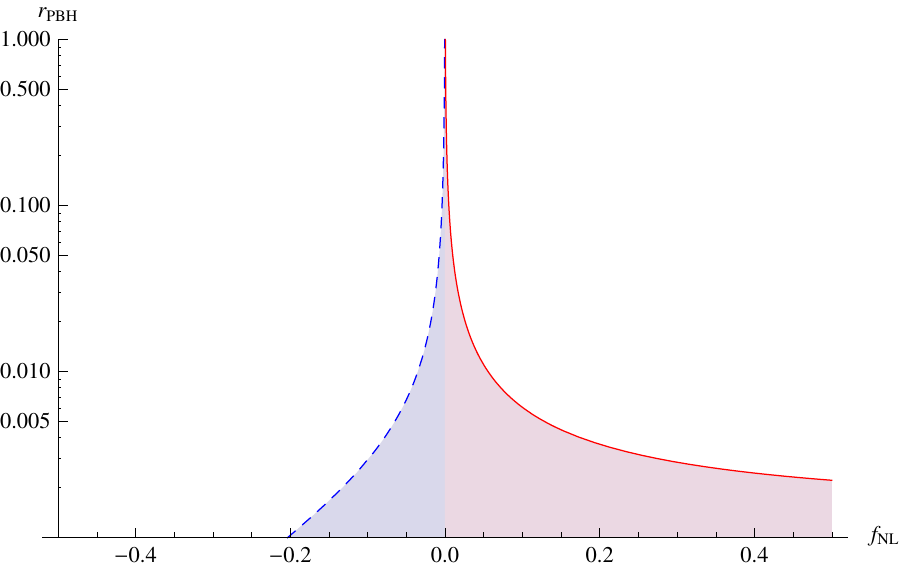}
    }
    \hfill
    \subfloat[$\gNL$]{%
      \includegraphics[width=0.49\textwidth]{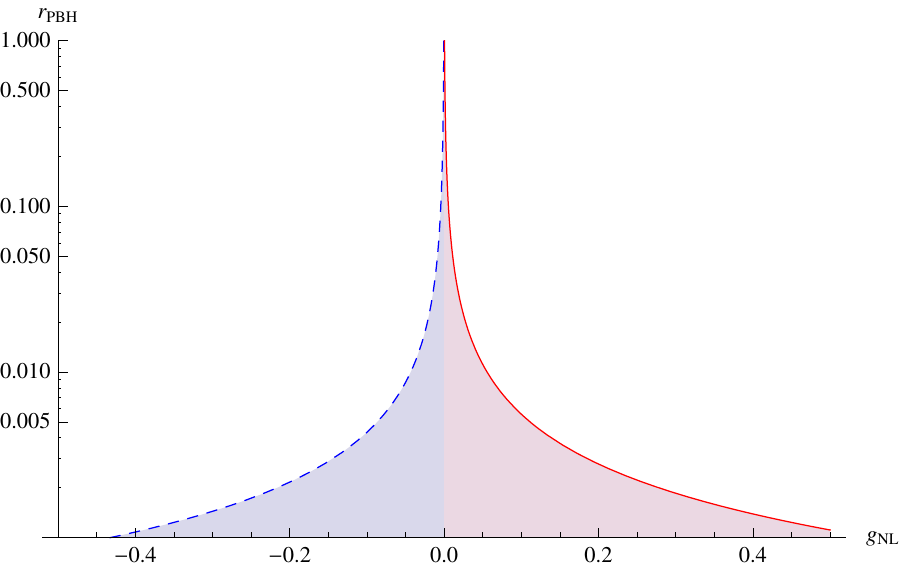}
    }
\caption[Constraints on the non-Gaussianity parameters for dark matter partially composed of primordial black holes]{In the case where PBHs only make up a small fraction of the DM content of the universe, parameterized by $r_{PBH}$, the constraints on $f_{NL}$ and $g_{NL}$ can become significantly weaker. This is due to the fact a large isocurvature mode in the PBH density would only translate into a small isocurvature mode in the DM density. The plots above show the allowed values of $f_{NL}$ and $g_{NL}$ for different values of $r_{PBH}$. Whilst the plots show the constraints for PBHs of mass $M_{PBH}=10^{25}$g, the constraints are not very sensitive to the PBH mass.}
\label{fnlgnl vs rpbh}
\end{figure}

The constraints which can be placed on the non-Gaussianity parameters therefore depend upon the PBH DM fraction, $r_{PBH}$. Figure \ref{fnlgnl vs rpbh} shows the allowed values of $f_{NL}$, $g_{NL}$ and $r_{PBH}$ if the PBH mass is $M_{PBH}=10^{25}$g. 
\begin{itemize}
\item{Large $r_{PBH}$: if PBHs make up a large fraction of DM then very tight constraints can be placed on the non-Gaussianity parameters, $f_{NL},~g_{NL}<\mathcal{O}(10^{-2})$.}
\item{Small $r_{PBH}$: if PBHs make up a small fraction of DM, $r_{PBH}<0.1$, then the constraints on $f_{NL}$ and $g_{NL}$ weaken significantly. However, the non-Gaussianity parameters only become larger than $1$ if $r_{PBH}<\mathcal{O}(10^{-3})$. In the case where $r_{PBH}$ is very small, the non-Gaussianity parameters can become large and it is crucial to account for the effect of a non-Gaussian distribution on the PBH forming scale, as done in this paper - as seen by the strong asymmetry for positive and negative $f_{NL}$.}
\end{itemize}

As $r_{PBH}$ becomes very small, $f_{NL}$ can become large and positive, but is still strongly restricted to not be large and negative. This is partly due to the fine tuning of the small-scale power spectrum necessary to produce a small but not too large number of PBHs when $f_{NL}$ is negative - even a very small amount of modal coupling can mean that this fine tuning is disrupted in different regions of the universe, causing large amounts of variation in the number density of PBHs forming. This effect is not seen unless the non-Gaussian distribution on small scales is accounted for. For $g_{NL}$, the constraints do not depend much on the sign of $g_{NL}$, and the small difference is due almost entirely to the difference in constraints from \emph{Planck} on fully, or fully anti-, correlated modes.

\subsection{Intermediate modes}

\begin{figure}[t]
\centering
\includegraphics[width=0.8\linewidth]{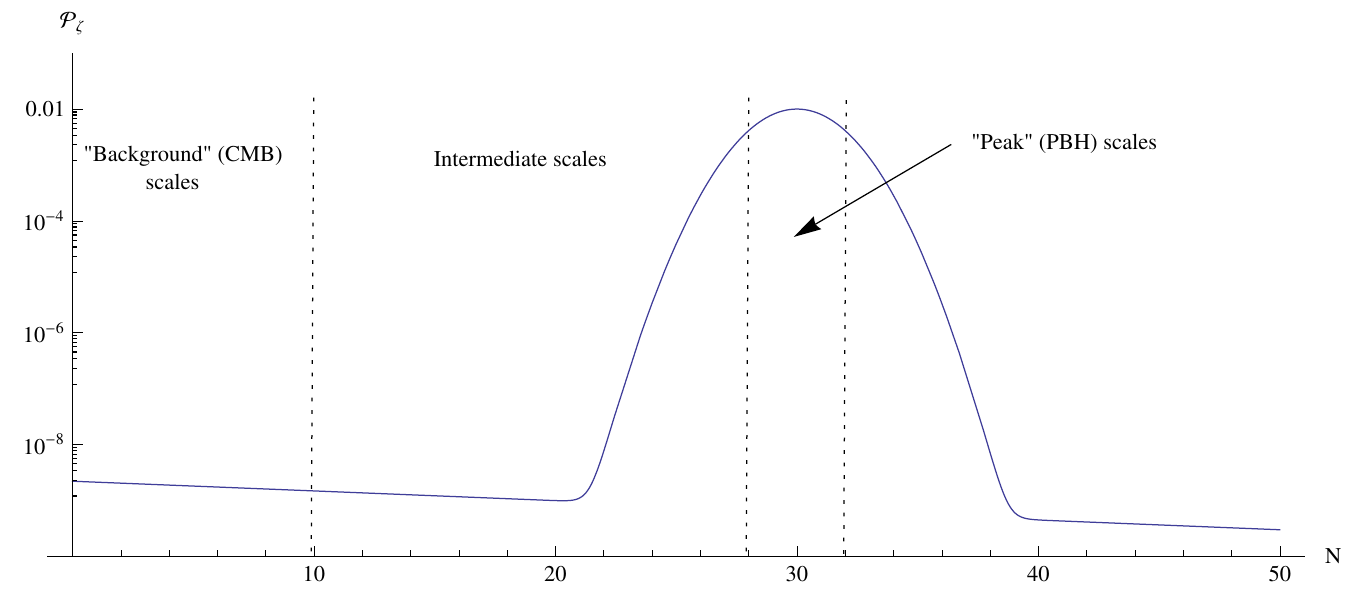}
\caption[A broad peak in the power spectrum]{An example of a power spectrum containing a broad peak. In this paper, there are 3 difference scales: the large "background" scales visible in the CMB, the small "peak" PBH forming scale (the exact scale of which depends on the mass PBH being considered), and the intermediate scales between the background and the peak. In such a case, the intermediate modes can have an effect on the PBH bias.}
\label{broad peak}
\end{figure}

The intermediate scales in between the large scales visible in the CMB and the small scale at which PBHs form have so far been ignored. This is a valid approximation if the power spectrum is small at all scales except for a narrow peak at the PBH forming scale, as in figure \ref{narrow peak}. However, this may not be the case if, for example, the power spectrum has a broad peak, as seen in figure \ref{broad peak}, or becomes blue at small scales. In this case, the abundance of PBHs, as well as the amplitude of isocurvature modes, can be significantly affected by the presence of perturbations on these intermediate modes.

If the power spectrum of the intermediate modes is not small, they will have a significant effect on the number of PBHs that form, as well as the isocurvature modes visible in PBH DM. This will be investigated in a similar to the peak-background split, and the curvature perturbation is split into short, intermediate, and long components:
\begin{equation}
\zeta_{G}=\zeta_{s}+\zeta_{i}+\zeta_{l}.
\end{equation}
The mass fraction of a given region of the universe going into PBHs is then calculated as before, as a function of $\zeta_{i}$ and $\zeta_{l}$, in addition to $f_{NL}$, $g_{NL}$, $\sigma_{s}$ and $\zeta_{c}$,
\begin{equation}
\beta=\beta\left(\zeta_{i},\zeta_{l}\right).
\end{equation}
However, the intermediate modes are too small scale to be observed in the CMB, and should therefore be averaged over:
\begin{equation}
\beta\left(\zeta_{l}\right)=\int\limits_{-\infty}^{\infty}\tilde{\beta}\left(\zeta_{i},\zeta_{l}\right)P(\zeta_{i})d\zeta_{i},
\end{equation}
where $\tilde{\beta}$ is the value of $\beta$ in different (intermediate-scale) regions of the universe, and $P(\zeta_{i})$ is the probability density function of $\zeta_{i}$, and is given by:
\begin{equation}
P(\zeta_{i})=\frac{1}{\sqrt{2\pi\langle\zeta_{i}^{2}\rangle}}\exp\left({-\frac{\zeta_{i}^{2}}{2\langle\zeta_{i}^{2}\rangle}}\right).
\end{equation}

In principle, $\langle\zeta_{i}^{2}\rangle$, can be obtained by integrating the power spectrum over the relevant range of scales. However, since this is unknown and model dependent, it is parameterized here by $r_{int}$, the ratio of the variance of intermediate modes $\langle\zeta_{i}^{2}\rangle$ to the variance of the short modes $\sigma_{s}^{2}$
\begin{equation}
r_{int}=\frac{\langle\zeta_{i}^{2}\rangle}{\sigma_{s}^{2}}.
\end{equation}

The value of $\langle\zeta_{i}^{2}\rangle$ can become larger than $\sigma_{s}^{2}$ due to the fact that many scales can contribute to $\zeta_{i}$, but only one scale contributes to $\zeta_{s}$. $\langle\zeta_{i}^{2}\rangle$ is calculated by integrating the power spectrum over the range of scales considered to be intermediate
\begin{equation}
\langle\zeta_{i}^{2}\rangle=\int\limits_{k_{min}}^{k_{max}}\frac{dk}{k}\mathcal{P}_{\zeta}(k),
\end{equation}
and can become large if the power spectrum is large over a significant range of this integration. In contrast, the PBH scale perturbations $\zeta_{s}$ are only composed of perturbations from one scale\footnote{Formally, $\sigma_{s}^{2}$ is given by integrating the power spectrum multiplied by a window function. However, provided that the spectral index is close to 1, or alternatively there is a peak spanning approximately 1 e-fold at the PBH scale, $\sigma_{s}^{2}$ is approximately equal to the power spectrum at that scale.}. Therefore, $\langle\zeta_{i}^{2}\rangle$ can become significantly larger than $\sigma_{s}^{2}$ even though the power spectrum has its largest value at the PBH scale. However, it is likely that in such a scenario, PBHs of multiple mass scales would be produced, which is discussed later.

\begin{figure}[t]
\centering
\subfloat[$\fNL$]{%
      \includegraphics[width=0.49\textwidth]{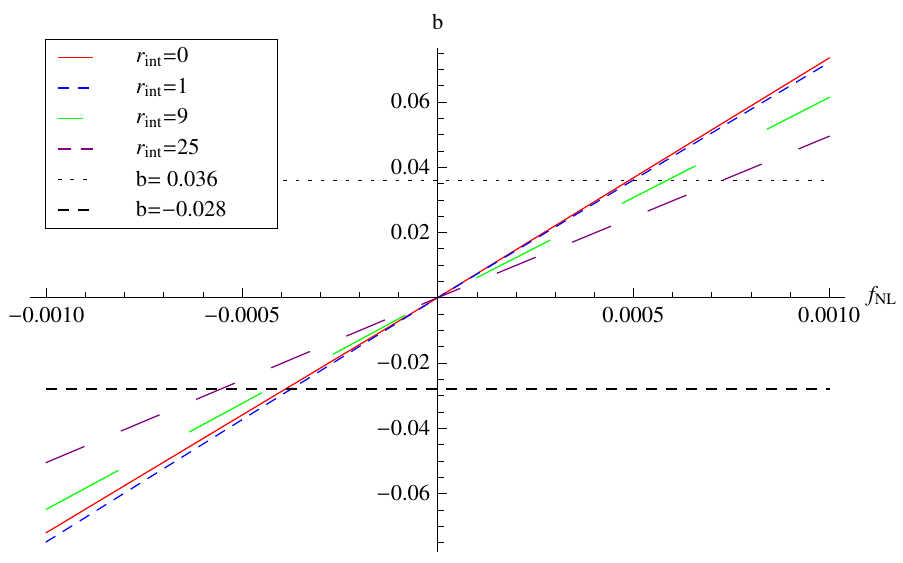}
    }
    \hfill
    \subfloat[$\gNL$]{%
      \includegraphics[width=0.49\textwidth]{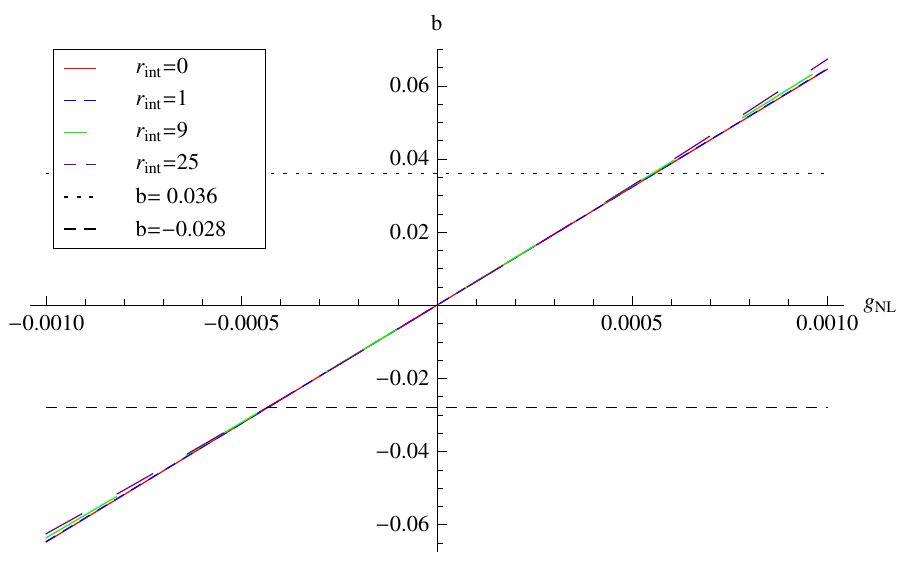}
    }
\caption[Effect of intermediate modes on the primordial black hole bias]{The effect of intermediate modes on the PBH bias $b$ is displayed for $M_{PBH}=10^{25}$g. The variance of the intermediate modes is parameterized by $r_{int}$, the ratio of $\langle\zeta_{int}^{2}\rangle$ to $\sigma_{s}^{2}$. The effect is negligible unless $r_{int}$ becomes large, in which case the PBH bias arising from an $f_{NL}$ term becomes significantly smaller, although has little effect for $g_{NL}$.}
\label{fnlgnl vs bint}
\end{figure}

The amplitude of the isocurvature modes therefore depends on the non-Gaussianity parameters, the small-scale power spectrum $\sigma_{s}^{2}$, the formation threshold $\zeta_{c}$, and $r_{int}$. A value for the PBH bias $b$ is then calculated numerically, figure \ref{fnlgnl vs bint} displays $b$ dependent on these variables. The effect of intermediate modes on the amplitude of isocurvature modes is relatively small for small $f_{NL}$ or $g_{NL}$ unless the variance of the intermediate scales is very large. The constraints on $f_{NL}$ can be weakened by a factor $\mathcal{O}(1)$, although the constraints on $g_{NL}$ are not significantly affected.

Note that a model where the power spectrum is large over a broad range of scales would likely also produce PBHs with a large range of masses, and vice versa. This fact does not affect the conclusions presented here, as the production of PBHs at all mass scales would be affected by bias in a similar way. We have shown that intermediate modes can significantly affect the PBH bias, although which modes are considered to be intermediate depends on the scale at which PBHs are forming, and therefore on the mass of PBHs forming. The exact constraints depend on the form of the power spectrum, and must therefore be calculated on a model by model basis, which goes beyond the scope of this paper - although the constraints will not be weaker than $f_{NL},~g_{NL}\lesssim10^{-3}$.

\subsection{Higher-order terms}
Whilst only the constraints on $f_{NL}$ and $g_{NL}$ have been calculated here, very tight constraints on higher-order non-Gaussianity parameters are also expected. In the same way that a $g_{NL}$ term has a small but significant effect on $f_{NL}$, equation (\ref{local NG}), higher-order terms affect the previous term. Because the mass fraction of the universe forming PBHs is extremely sensitive to non-Gaussianity parameters at higher orders \cite{Young:2013oia}, even very small changes to higher-order non-Gaussianity parameters due to modal coupling creates significant creates significant perturbations in the PBH density at large scales. As an example, we will consider a $5^{th}$ order term in local-type non-Gaussianity:
\begin{equation}
\zeta=\zeta_{G}+\frac{81}{625}i_{NL}\zeta_{G}^{5}.
\end{equation}
Utilising the peak-background split gives a $4^{th}$ order term at small scales, $\tilde{h}_{NL}$, given by
\begin{equation}
\tilde{h}_{NL}=3i_{NL}\zeta_{l}.
\end{equation}
Inserting $\zeta_{l}\approx10^{-5}$ and $i_{NL}=10^{-3}$ gives $\tilde{h}_{NL}\approx10^{-8}$. The modulation of the $\tilde{h}_{NL}$ by the long wavelength mode $\zeta_l$ then generates a perturbation in the density of PBHs forming, $\delta_{\beta}\approx10^{-6}$. In the picture of PBH DM, this results in a fully-correlated isocurvature mode, with a bias factor of $b\approx0.1$ - which is excluded by \emph{Planck}. Because it can be shown that high order terms have an effect on the preceding term which is linear in $\zeta$, tight constraints are expected on such non-Gaussianity parameters, only weakening slightly as higher-order terms are considered.

\section{Summary}
The effect of modal coupling under the presence of non-Gaussianity of the local type produces significant isocurvature modes in the density of PBHs in the early universe. If PBHs make up a significant fraction of DM, the constraints on isocurvature modes in cold DM from \emph{Planck} can be used to constrain the non-Gaussianity parameters - in this paper we have considered $f_{NL}$ and $g_{NL}$
Using the constraints from \emph{Planck} on isocurvature modes enables tight constraints to be placed on $f_{NL}$ and $g_{NL}$,
\begin{equation}
|f_{NL}|,|g_{NL}|<\mathcal{O}(10^{-3}),
\end{equation}
unless $f_{NL}$ and $g_{NL}$ have opposite signs and have been extremely finely tuned so that the effect from each term cancels. Cases where the constraints could become weaker have also been considered: if the power spectrum is large on scales between those visible in the CMB and the PBH forming scale, or if DM is only partially composed of PBHs, finding that under these conditions the constraints weaken very slightly (unless PBHs make up a very tiny fraction of DM). Therefore, the detection of significant numbers of PBHs would rule out significant local non-Gaussianity, and vice versa. Our constraints are almost independent of the PBH mass, and can also be applied to Planck mass relics that may be left behind from the evaporation of small PBHs.


The production of isocurvature modes can therefore be used to constrain PBH forming models that may otherwise be permitted. For example, we will consider here two models that may be ruled out as mechanisms to produce PBH DM:
\begin{itemize}
\item{Hybrid inflation: hybrid inflation typically predicts a non-zero $f_{NL}$, but there is some freedom in the exact value. \cite{Clesse:2013jra} predicts $f_{NL}\approx -1/N_{*}$, where $N_{*}$ is the number of e-folds between horizon exit of some pivot scale and the end of horizon. Inflation is believed to have lasted at least $50-60$ e-folds, which would give $f_{NL}=\mathcal{O}(10^{-2})$ - several orders of magnitude higher than allowed by the constraints presented here. \cite{Mulryne:2011ni} predicts that $f_{NL}$ can span a range of values from $10^{-2}$ to $10^5$ - the entire range of which would be ruled out as a method of producing PBH DM.}
\item{The curvaton: the amount of non-Gaussianity in the curvaton model depends on the density parameter, $\Omega_{\chi}$, of the curvaton, $\chi$, at the time it decays into radiation: $f_{NL}=-5/4$ if $\Omega_{\chi}=1$ \cite{Sasaki:2006kq}. Although higher-order local non-Gaussianity terms are generated, it is unlikely that these will generate small isocurvature perturbations to evade the constraints.}
\end{itemize}

There are, however, limitations to the calculations carried out in this paper. Notably, we have only considered local-type non-Gaussianity, and throughout it has been assumed that $f_{NL}$ and $g_{NL}$ are scale invariant. We have also only calculated the dependence of isocurvature modes on $f_{NL}$ and $g_{NL}$, and shown them to a roughly equivalent effect - with $g_{NL}$ having only a marginally smaller effect. Higher-order terms are therefore also likely to have a similar effect on isocurvature modes. We also note that it has recently been observed that sub-horizon perturbations at the time of PBH formation have an effect on whether a perturbation will collapse to form a PBH or not \cite{Nakama:2014fra}. The expected amplitude of these sub-horizon modes would be affected by modal coupling - and therefore affect the amount of PBHs forming, affecting the isocurvature modes. However, this effect is expected to be negligible whilst the non-Gaussianity parameters are very small.

\section{Acknowledgements}
SY is supported by an STFC studentship, and CB is supported by a Royal Society University Research Fellowship. We thank David Seery, John Miller, Yuichiro Tada and Shuichiro Yokoyama for useful discussions.

\section{Appendix A: Full expression for $\delta_{\beta}$ from a $g_{NL}$ term}
For completeness, the full expression for $\delta_{\beta}$  arising from a $g_{NL}$ term is included - though this expression is still only valid for small $g_{NL}$. This expression would replace the simpler equation (\ref{bgnl}).

\begin{equation}
\begin{split}
\delta_{\beta}=& \left(-50 3^{1/3} 10^{2/3}+\left(-27 \sqrt{g_{NL}} \text{$\zeta_{c}$}+\sqrt{300+729g_{NL} \text{$\zeta_{c}$}^2}\right)^{1/3} \left(45\ 3^{2/3} 10^{1/3} \sqrt{g_{NL}} \text{$\zeta_{c}$} \right.\right.\nonumber \\
& \left.\left.\left.-5\ 3^{1/6} 10^{1/3} \sqrt{100+243g_{NL} \text{$\zeta_{c}$}^2}+100 \left(-27 \sqrt{g_{NL}} \text{$\zeta_{c}$}+\sqrt{300+729g_{NL} \text{$\zeta_{c}$}^2}\right)^{1/3}\right.\right.\right.\nonumber \\
&  \left.\left.\left. -54g_{NL}\sigma^2 \left(-27 \sqrt{g_{NL}} \text{$\zeta_{c}$}+\sqrt{300+729g_{NL} \text{$\zeta_{c}$}^2}\right)^{1/3}\right)\right) \left(-25 3^{1/3} 10^{2/3} \sqrt{g^3 \left(100+243 g_{NL} \text{$\zeta_{c}$}^2\right)} \right. \right. \nonumber \\
& \left.\left. \left(-27 \sqrt{g_{NL}} \text{$\zeta_{c}$}+\sqrt{300+729 g_{NL} \text{$\zeta_{c}$}^2}\right)^{2/3}-250 \sqrt{3} g_{NL}^{3/2} \left(3^{1/6} 10^{1/3} \sqrt{100+243 g_{NL} \text{$\zeta_{c}$}^2}\right. \right. \right. \nonumber \\
& \left. \left. \left. \left. -20 \left(-27 \sqrt{g_{NL}} \text{$\zeta_{c}$} +\sqrt{300+729 g_{NL} \text{$\zeta_{c}$}^2}\right)^{1/3}\right)+225 g_{NL}^2 \text{$\zeta_{c}$} \left(30\ 3^{1/6} 10^{1/3}-6 \sqrt{100+243 g_{NL} \text{$\zeta_{c}$}^2} \right. \right. \right. \right. \nonumber \\
& \left. \left. \left. \left. \left(-27 \sqrt{g_{NL}} \text{$\zeta_{c}$} \right.+\sqrt{300+729 g_{NL} \text{$\zeta_{c}$}^2}\right)^{1/3}+3^{5/6} 10^{2/3} \left(-27 \sqrt{g_{NL}} \text{$\zeta_{c}$}+\sqrt{300+729 g_{NL} \text{$\zeta_{c}$}^2}\right)^{2/3}\right) \right. \right. \nonumber \\
& \left. \left. +243\ 3^{1/6} 10^{1/3} g_{NL}^3\sigma^2 \text{$\zeta_{c}$} \left(30+10^{1/3} \left(-81 \sqrt{g_{NL}} \text{$\zeta_{c}$}+3 \sqrt{300+729 g_{NL} \text{$\zeta_{c}$}^2}\right)^{2/3}\right) \right. \right. \nonumber \\
& \left. \left. -27\ 3^{1/3} g_{NL}^{5/2} \left(-450 3^{1/6} \text{$\zeta_{c}$}^2 \left(-27 \sqrt{g_{NL}} \text{$\zeta_{c}$}+\sqrt{300+729 g_{NL} \text{$\zeta_{c}$}^2}\right)^{1/3} \right. \right. \right. \nonumber \\
& \left. \left. +10^{1/3}\sigma^2 \sqrt{100+243 g_{NL} \text{$\zeta_{c}$}^2} \left(10\ 3^{1/3}+10^{1/3} \left(-27 \sqrt{g_{NL}} \text{$\zeta_{c}$}+\sqrt{300+729 g_{NL} \text{$\zeta_{c}$}^2}\right)^{2/3}\right)\right)\right) \nonumber \\
& \left(150\ 30^{1/3}\sigma^2 \sqrt{\frac{100}{g_{NL}}+243 \text{$\zeta_{c}$}^2} \left(-27 g_{NL}^2 \text{$\zeta_{c}$}+\sqrt{3} \sqrt{g^3 \left(100+243 g_{NL} \text{$\zeta_{c}$}^2\right)}\right)^{5/3} \left(-10 3^{1/3} \right. \right. \nonumber \\
& \left. \left. +10^{1/3} \left(-27 \sqrt{g_{NL}} \text{$\zeta_{c}$}+\sqrt{300+729 g_{NL} \text{$\zeta_{c}$}^2}\right)^{2/3}\right)\right)^{-1} \text{$\zeta$}.
\end{split}
\end{equation}

\section{Appendix B: Comparison with {\it ``Primordial black holes as biased tracers''}}
\begin{figure}[t]
\centering
\includegraphics[width=0.8\linewidth]{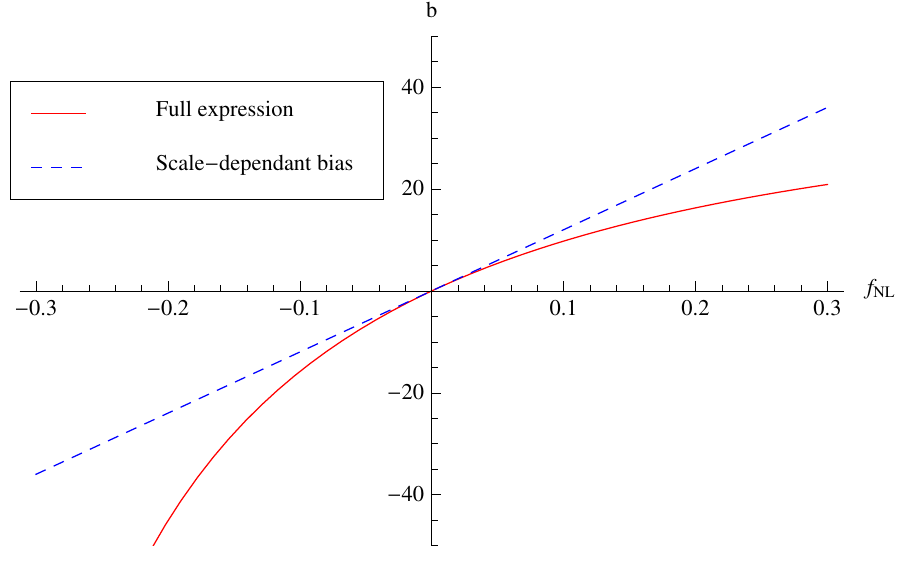}
\caption[Comparison of results with \cite{Tada:2015noa}]{A comparison of the results derived in this paper with those derived in \cite{Tada:2015noa}. The solid red line denotes the full expression for the PBH bias given by equation (\ref{bfnl}), and the dashed blue line represent the {\it scale-dependent bias} given by equation (14) in \cite{Tada:2015noa}. To make these plots, the values $\sigma_{s}=0.1$ and $\zeta_{c}=1$ have been used.}
\label{comparison}
\end{figure}

In their paper, {\it "Primordial black holes as biased tracers"}, \cite{Tada:2015noa} derive an expression for the {\it scale-dependent bias} given by
\begin{equation}
\Delta b(k)=2 f_{NL} \mathcal{M}_{l}^{-1}(k)\frac{\delta_{c}^{2}}{\sigma_{s}^{2}}.
\end{equation}
This is equivalent to equation (\ref{simple bfnl}) in this paper. The factor of $3/5$ difference is due to a different definition of $f_{NL}$, and the factor $\mathcal{M}_{l}^{-1}(k)$ is a result of their use of the density contrast rather than the curvature perturbation. The $+1$ in the brackets of equation (\ref{simple bfnl}) is a small correction and can be neglected. Therefore, the results for very small $f_{NL}$ in the 2 papers are equivalent. In figure \ref{comparison} the two expressions are compared. For $|f_{NL}|<\mathcal{O}(10^{-2})$ the two calculations match well, but diverge rapidly for larger $|f_{NL}|$.

It is therefore necessary to use the full calculation derived in this paper in situations where $f_{NL}$ could become larger than $10^{-2}$. Whilst such a large value of $f_{NL}$ is generally excluded by the constraints on isocurvature modes in the PBH DM scenario, it is relevant where higher-order terms are considered, or that PBHs form a sub-dominant component of DM.

\newpage

\chapter{Influence of large local and non-local bispectra on primordial black hole abundance}
\label{chap:paper5}




\begin{center}


Sam Young$^{1}$, Donough Regan$^{2}$, Christian T. Byrnes$^{3}$\\[0.5cm]
Department of Physics and Astronomy, Pevensey II Building, University of Sussex, BN1 9RH, UK\\[0.5cm]
$^{1}$S.M.Young@sussex.ac.uk, $^{2}$D.Regan@sussex.ac.uk, $^{3}$C.Byrnes@sussex.ac.uk \\[1cm]


\end{center}

Primordial black holes represent a unique probe to constrain the early universe on small scales - providing the only constraints on the primordial power spectrum on the majority of scales. However, these constraints are strongly dependent on even small amounts of non-Gaussianity, which is unconstrained on scales significantly smaller than those visible in the CMB. This paper goes beyond previous considerations to consider the effects of a bispectrum of the equilateral, orthogonal and local shapes with arbitrary magnitude upon the abundance of primordial black holes. Non-Gaussian density maps of the early universe are generated from a given bispectrum and used to place constraints on the small-scale power spectrum. When small, we show that the skewness provides an accurate estimate for how the constraint depends on non-Gaussianity, independently of the shape of the bispectrum. We show that the orthogonal template of non-Gaussianity has an order of magnitude weaker effect on the constraints than the local and equilateral templates.

\newpage

\section{Introduction}

Primordial black holes (PBHs) are black holes that may have formed very early on in the history of the universe from the collapse of density perturbations generated during inflation. During inflation, quantum fluctuations are stretched out by the rapid expansion of the universe, and quickly become larger than the Hubble horizon, becoming classical density perturbations. Once inflation ends, the perturbations begin to reenter the horizon, and if large enough, can collapse to form a PBH. Because such perturbations can reenter the horizon before baryogenesis, there is no need for such black holes to have a large enough mass to overcome degeneracy pressures - and the formation of PBHs with very small masses is possible.

Because PBHs form on small scales, they have often been used to constrain the smallest scales in the early universe. Precision measurements and constraints upon the primordial Universe are available from the cosmic microwave background (CMB) and large scale structure (for example, the constraints on inflation from \emph{Planck} \cite{Ade:2013uln}), but these only provide constraints on the largest 6-8 e-folds inside the visible universe - while inflation is expected to have lasted 50-60 e-folds. PBHs, on the other hand, provide constraints on a much greater range of scales, spanning around 50 e-folds, although the constraints are much weaker.

Many attempts have been made to detect PBHs, yet they remain undetected. However, a tight upper limit can be placed on the abundance of PBHs, which is typically stated in terms of the mass fraction of the Universe contained within PBHs at the time of formation, $\beta$. Constraints on $\beta$ vary greatly for PBHs of different mass, ranging from $\beta<10^{-5}$ to $\beta<10^{-25}$. For a summary of the constraints see \cite{Carr:2009jm}. Because the number of PBHs forming depends on the primordial power spectrum, constraints on the abundance of PBHs can be used to place bounds on the power spectrum \cite{Josan:2009qn}. These constraints on the power spectrum are typically of order $10^{-2}$, significantly weaker than constraints from the CMB.

In order for a significant number of PBHs to form, the power spectrum needs to be orders of magnitude larger than is observed in the CMB - meaning that it must become large on small scales. There are a range of models for inflation which do predict such behaviour, whilst being consistent with current cosmological observations. Such models include the running mass model \cite{Kohri:2007qn,Drees:2011hb}, axion inflation \cite{Bugaev:2013fya}, a waterfall transition during hybrid inflation \cite{Bugaev:2011wy, Lyth:2012yp,Halpern:2014mca}, from particle production during inflation \cite{Erfani:2015rqv}, inflationary models with small field excursions but which are tuned to produce a large tensor-to-scalar ratio on large scales \cite{Hotchkiss:2011gz}, and can be formed from passive density fluctuations \cite{Lin:2012gs}. See also \cite{Linde:2012bt,Torres-Lomas:2014bua,Suyama:2014vga}. For further reading and a summary of various models which can produce PBHs, see \cite{Green:2014faa}. Such models typically predict at least a small amplitude of non-Gaussianity - and it has been shown that constraints on the small-scale power spectrum are strongly dependent on non-Gaussianity \cite{Byrnes:2012yx} - and can vary by over an order of magnitude. 

Previous papers have used an analytic method to investigate the effects of non-Gaussianity - and were limited to either investigating local-type non-Gaussianity for which analytical results are available \cite{Bullock:1996at,Ivanov:1997ia,PinaAvelino:2005rm,Seery:2006wk,Shandera:2012ke,Lyth:2012yp,Byrnes:2012yx,Young:2013oia} or, in the case of \cite{Shandera:2012ke}, also to a small amplitude of equilateral non-Gaussianity. This paper goes beyond previous work to investigate the effects of three different bispectrum shapes of arbitrary size on the abundance of PBHs, and on the resulting constraints, by making use of non-Gaussian density maps. We make the first study of orthogonal non-Gaussianity, and show that, for a given value of $\fNL$, it has a much smaller effect on the constraints than the equilateral and local non-Gaussian templates. We explain this observation by calculating the skewness parameter as a function of all three bisepectral templates.


The paper is organised as follows: in section \ref{sec:sim_proc} the generation of the density maps and calculation of the PBH abundance is detailed. In section \ref{sec:constraints}, the constraints on the power spectrum as a function of the bispectrum are calculated. Section \ref{sec:conclusions} concludes with a summary of the results.

\section{Simulation Procedure}\label{sec:sim_proc}
\subsection{Generation of non-Gaussian density maps}

Methods for the simulation of a map incorporating an arbitrary bispectrum were developed by Regan et al. in a series of papers \cite{Fergusson:2010ia,Regan:2011zq,Schmittfull:2012hq} (see also \cite{Wagner:2010me}). Representing the primordial curvature in Fourier space as $\zeta(\bk)$, one may simulate the curvature of a Gaussian distribution using a random number generator with variance per scale, $k$, given by the power spectrum $P_\zeta(k)$ (and zero mean). For clarity we will denote the Gaussian map as $\zeta_G(\bk)$. The bispectrum $B_\zeta(k_1,k_2,k_3)$, given by the expectation value of the three point function
\begin{equation}
\langle \zeta(\bk_1)\zeta(\bk_2)\zeta(\bk_3)\rangle =(2\pi)^3 \delta_D(\bk_1+\bk_2+\bk_3) B_\zeta(k_1,k_2,k_3)\,,
\label{eq:bispdef}\end{equation}
may be simulated using the Gaussian maps by calculating $\zeta(\bk)=\zeta_G(\bk)+\fNL \zeta_B(\bk)$  where
\begin{equation}
\zeta_B(\bk)=\int \frac{d^3\bk_1}{(2\pi)^3} \frac{d^3\bk_2}{(2\pi)^3}\delta_D(\bk-\bk_1-\bk_2)\zeta_G(\bk_1)\zeta_G(\bk_2)\frac{B_\zeta^{\fNL=1}(k,k_1,k_2)}{2\left(P_\zeta(k)P_\zeta(k_1)+P_\zeta(k)P_\zeta(k_2)+P_\zeta(k_1)P_\zeta(k_2)\right)}\,.
\end{equation}
Here we define the quantity $\fNL\equiv 5 B_\zeta(k,k,k)/(18 P_\zeta(k)^2)$ such that $B_\zeta \equiv\fNL B_\zeta^{\fNL=1}$.
Direct implementation of this convolution is numerically prohibitive unless the bispectrum can be written in a separable form, i.e.~in the form $f(k_1)g(k_2)h(k_3)$ for arbitrary one dimensional functions $f,g,h$. This is possible for sufficiently smooth generic bispectra using techniques developed in \cite{Fergusson:2009nv,Regan:2010cn,Fergusson:2010gn}. In particular, a partial wave decomposition may be employed to write the bispectrum in the form
\begin{equation}
\frac{B_\zeta^{\fNL=1}(k_1,k_2,k_3)}{2\left(P_\zeta(k_1)P_\zeta(k_2)+P_\zeta(k_1)P_\zeta(k_3)+P_\zeta(k_2)P_\zeta(k_3)\right)} = \sum_{rst} \alpha_{rst}^Q q_{\{ r}(k_1) q_s(k_2) q_{t\}}(k_3)\,,
\end{equation} 
where the notation $\{ r s t\}$ refers to all symmetrised combinations of the labels $r,s,t$ - necessary due to symmetry of the bispectrum. The triple label indices may be partially ordered such that a single index $n\equiv \{ r s t\}$ may be used to enumerate the coefficients of the expansion in the form $\alpha_n^Q$. Calculation of these coefficients only requires an inner product on the space of bispectra - restricted due to the triangle condition imposed by the Dirac delta condition in~\eqref{eq:bispdef}. Interested readers are referred to  \cite{Regan:2013wwa} for further details of the decomposition procedure. Given this decomposition, calculation of the bispectrum map reduces to calculation of fast Fourier transforms with
\begin{align}
\zeta_B(\bk)&=\frac{1}{2}\sum_n \alpha_n^Q \int d^3 \bx e^{i\bk\cdot\bx} q_{\{ r}(k) M_s(\bx) M_{t\}}(\bx)\,,\\
{\rm where} \quad M_s(\bx)&=\int \frac{d^3\bk}{(2\pi)^3}e^{i\bk\cdot\bx} q_s(k) \zeta_G(\bk)\,.
\end{align}

In this paper we focus on the three standard bispectrum templates (local, equilateral, orthogonal) for which the respective bispectra are of the form
\begin{align}
B_\zeta^{\rm local}(k_1,k_2,k_3)&= \frac{6}{5}\fNL \left(P_\zeta(k_1)P_\zeta(k_2)+ P_\zeta(k_1)P_\zeta(k_3)+P_\zeta(k_2)P_\zeta(k_3)\right)\,,\\
B_\zeta^{\rm eq}(k_1,k_2,k_3)&= \frac{18}{5}\fNL \Big(-\left[P_\zeta(k_1)P_\zeta(k_2)+ 2\, {\rm perms}\right]- 2\left[ P_\zeta(k_1)P_\zeta(k_2)P_\zeta(k_3)\right]^{2/3} \nonumber\\
&\qquad\qquad\,\, +\left[P_\zeta^{1/3}(k_1)P_\zeta^{2/3}(k_2)P_\zeta(k_3)+ 5\, {\rm perms}\right]\Big)\,,\\
B_\zeta^{\rm orth}(k_1,k_2,k_3)&= \fNL \Big(3 B_\zeta^{\rm eq,\fNL=1}(k_1,k_2,k_3) -\frac{36}{5}\left[ P_\zeta(k_1)P_\zeta(k_2)P_\zeta(k_3)\right]^{2/3} \Big) .
\end{align}
For clarity we will, where necessary, distinguish $\fNL$ for the various shapes by writing $\fNL^{\rm local}, \fNL^{\rm eq}$ and $\fNL^{\rm orth}$. We note that for the local model, the map making procedure reduces to the simple form
\begin{equation}
\zeta(\bk)=\zeta_G(\bk) + \frac{3}{5}\fNL (\zeta_G\star\zeta_G)(\bk)\, ,
\label{localtype}
\end{equation}
where the symbol $\star$ indicates a convolution.

Our simulations are carried out on a grid of $128^3$ points, and employ a scale invariant power spectrum of the form $P_\zeta(k) = A_\zeta/k^3$. The amplitude $A_\zeta$ is given by the \emph{Planck} value $A_\zeta=4.75\times 10^{-8}$ but is boosted for either one $e$-fold of points (between $10$ and $27.2$ grid points in each dimension in Fourier space) or 2.5 $e$-folds (between grid points $10$ and $128$) to a much larger amplitude - typically of order $10^{-2}$ - required to form a significant number of PBHs; the boosted region of the power spectrum will be referred to as the peak in the power spectrum later in the paper. The amplitude of this boost is then tuned such that the required amount of PBHs would form. Calculation of the PBH abundance is discussed in the following section.

We restrict our analysis to the bispectrum, but note that generating a non-zero bispectrum inevitably results in non-zero higher n-point functions. For the local model, this corresponds to generating the minimum possible trispectrum with $\tau_{NL}=(6 f_{\rm NL}/5)^2$ and $g_{\rm NL}=0$ \cite{Byrnes:2006vq}, which \cite{Shandera:2012ke} calls the hierarchical scaling. Our simulations automatically take this into account. However, care should be taken in interpreting the large $f_{\rm NL}$ regime for the equilateral and orthogonal models for which the trispectrum may be of a different form.

\subsection{Calculation of PBH abundance}
As described in \cite{Young:2014ana} the abundance of primordial black holes should be computed using the density contrast rather than the primordial curvature perturbation, due to the damping of super-horizon modes by a factor $k^2$. In addition, it is necessary to account for the window function $W(R,x)$ with which the density contrast is smoothed on a given scale $R$.  

Assuming radiation domination, the relationship between the smoothed density fluctuation, $\Delta_R$, and the curvature perturbation, $\zeta$, is given by
 \begin{equation}
 \Delta_R(\bk) = \int \frac{d^3 \bk}{(2\pi)^3} \tilde{W}(R,k) \frac{4}{9} ( k R)^2 \zeta(\bk)\,,
 \end{equation}
where $\tilde{W}(R,k)$ denotes the Fourier transform of the window function. In this work we employ a volume-normalised Gaussian window function, such that\footnote{We shall drop the tilde in what follows and assume the window function is in Fourier space unless otherwise specified.}
\begin{equation}
\tilde{W}(R,k) = \exp\left(-\frac{k^2 R^2}{2}\right)\,.
\end{equation}

In order to compute the abundance, $\beta$, of PBHs we count the number of grid points for which the smoothed density exceeds the threshold, $\Delta_c$ at which PBHs form, i.e. such that $ \Delta_R(\bx) > \Delta_c$\,. Our computation of the variance is performed by Fourier transforming the smoothed density contrast to real space to obtain $\Delta_R(\bx)$, and then calculating
\begin{equation}
\mathcal{P}_{\Delta_R}=\langle \Delta_R(\bx)^2 \rangle\,,
\end{equation} 
where $\langle \dots \rangle$ represents the averaging over all grid points in real space. For ease of comparison to the literature, which do not employ a smoothing function, we note that for the rescaled density contrast, $\tilde{\Delta}_R=\exp(1/2)\Delta_R$, we obtain the approximate result $\mathcal{P}_{\tilde{\Delta}_R}\approx (4/9)^2\mathcal{P}_{\zeta}$ due to the function $(kR)^2 {W}(R,k)$ peaking with value $\exp(-1/2)$ in the boosted region. We will make use of this (accurate) approximation in the remainder of this paper. The threshold at which
PBHs form at any grid point $\bx$ is taken to be $\tilde{\Delta}_c\equiv \exp(1/2){\Delta}_c= 4/9$. This corresponds to a threshold $\Delta_c\simeq 1/3$, as used in previous theoretical predictions - though is slightly below the accepted value $0.45$ calculated from simulations \cite{Shibata:1999zs,Musco:2008hv,Nakama:2013ica}.

The variance of the Gaussian density map - denoted $\sigma$ for clarity of notation - may be evaluated as
\begin{equation}
\sigma^2 = \int \frac{dk}{k}\frac{A_\zeta(k)}{2\pi^2} (k R)^4\frac{16}{81}{W}(R,k)^2\,.
\end{equation}
In addition the skewness, $\skewR$, is calculated by employing the following expression
\begin{equation}\label{eq:skewness}
\skewR = \frac{\langle \Delta_R(\bx)^3 \rangle}{\langle \Delta_R(\bx)^2\rangle^{3/2}}\,.
\end{equation}
We shall, unless otherwise indicated, use $R=\sqrt{2}/k_{\rm peak}$, where $k_{\rm peak}$ represents the wavenumber approximately half an $e$-fold from the smallest scale on which the Gaussian amplitude is boosted (i.e. corresponding to $20$ grid points in Fourier space).

\section{Constraints on the small-scale power spectrum}\label{sec:constraints}
Bounds on the abundance of PBHs, $\beta$, can be used to constrain the curvature perturbation power spectrum. Previous constraints have been obtained using an analytic method \cite{Shandera:2012ke,Byrnes:2012yx,Young:2013oia,Young:2014oea}, and it has been shown that the constraints can depend strongly on non-Gaussianity. It is normally assumed that PBHs form with approximately the horizon mass, although it is well known that the mass of the PBH that forms depends on the amplitude of the overdensity - and the mass has been found to follow a scaling law. The effect of this was recently considered \cite{Kuhnel:2015vtw} and leads to a shift and broadening of the PBH masses, and an overall decrease of the mass contained in primordial black holes. However, the PBHs formed still have approximately the horizon mass (the peak in the mass formed is typically half the horizon mass), and has a very small effect on the derived constraints - and so the effect is neglected here.

The effects of local-type non-Gaussianity have previously been studied, and it was found that the constraints on the power spectrum, $\mathcal{P}_\zeta$, can vary by up to an order of magnitude when $\fNLloc$ changes from $-0.5$ to $0.5$. Initially, a power spectrum which peaks over a small range of scales was considered \cite{Byrnes:2012yx,Young:2013oia}. Because $\fNLloc$ has a strong effect on the tails of the distribution function where PBHs form, small changes in $\fNLloc$ have a very large effect on the abundance of PBHs. Positive $\fNLloc$ increases the amount of PBHs which form such that the constraints become gradually tighter as $\fNLloc$ increases. For negative $\fNLloc$ the constraints loosen, but become weaker very quickly as $\fNLloc$ decreases, with no PBHs formed unless the power spectrum becomes much larger.

Later, the case where the power spectrum spans a larger range of scales was considered, allowing for the effect of super-horizon modes \cite{Young:2014oea}. Super-horizon modes normally do not directly affect PBH formation as far as is known \cite{Nakama:2014fra}, but can have an indirect effect due to modal coupling to horizon scale modes. Overall, the effect of modal coupling increases PBH formation - tightening constraints on the power spectrum. Notably, for negative $\fNLloc$, whilst constraints still weaken for small negative values, they become stronger as $\fNLloc$ becomes larger. A full discussion can be seen in \cite{Young:2014oea}.

The method detailed in section \ref{sec:sim_proc} is used to calculate the abundance of PBHs, $\beta$, as a function of the power spectrum and bispectrum - and this can be used to place an upper limit on the power spectrum for a given upper limit on $\beta$. Due to the amount of resources required to generate large maps, we restrict ourselves to a relatively weak constraint, $\beta<10^{-4}$. Whilst this constraint is weaker than any of the existing constraints on PBH abundance, it allows for an easier investigation of the effects of non-Gaussianity. It has been shown that the effect of non-Gaussianity upon the power spectrum constraint is relatively large compared to the effect of the constraint on $\beta$ \cite{Young:2014oea}. In any case, we expect the qualitative lessons drawn from our results to hold for any smaller value of $\beta$, although a simulation with a larger grid would have to be made to calculate the precise constraints.



\begin{figure}
 \begin{minipage}{1\textwidth}
\centering
\subfloat[]{%
      \includegraphics[width=0.49\textwidth]{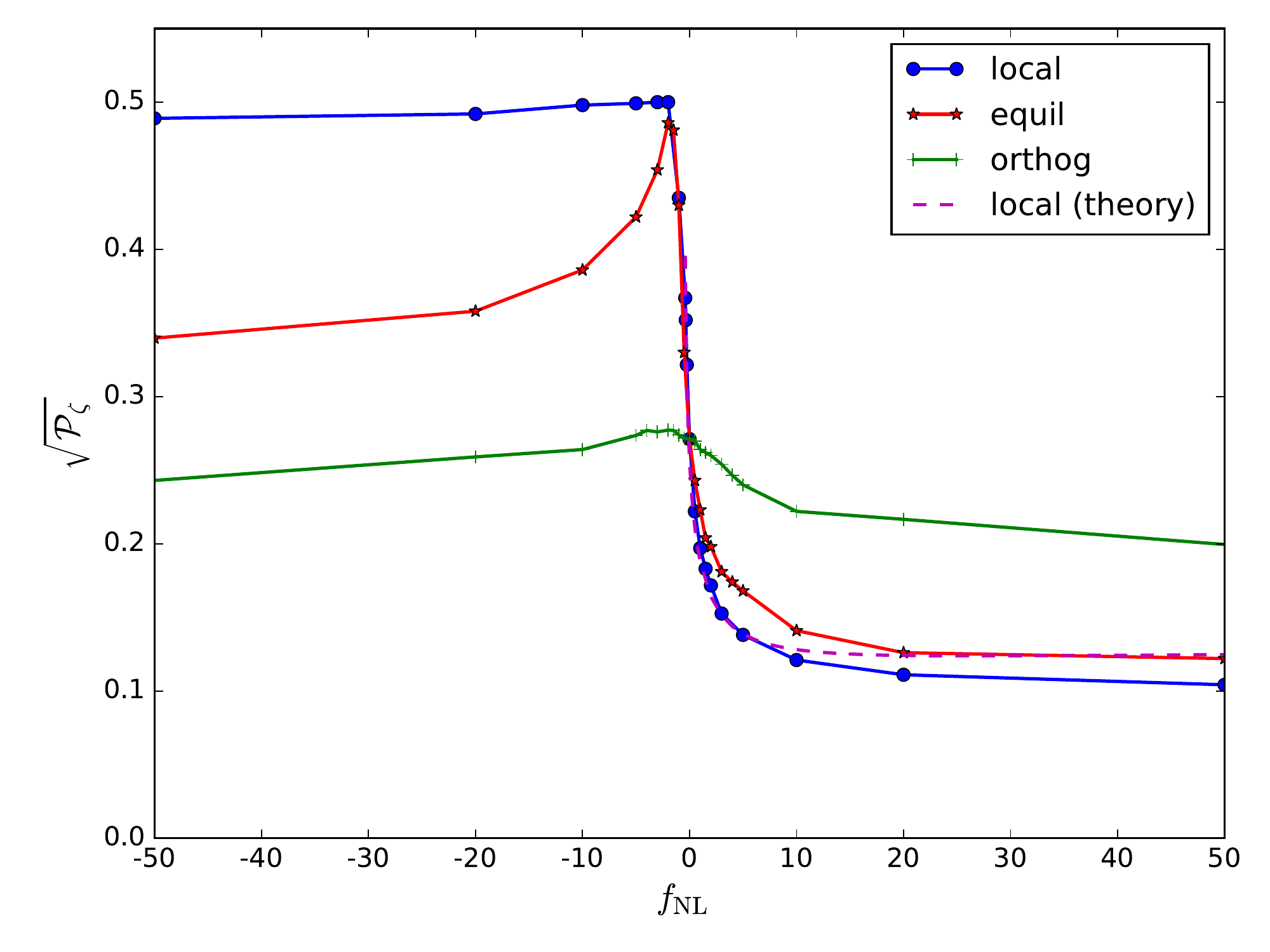}
    }
    \hfill
    \subfloat[]{%
      \includegraphics[width=0.49\textwidth]{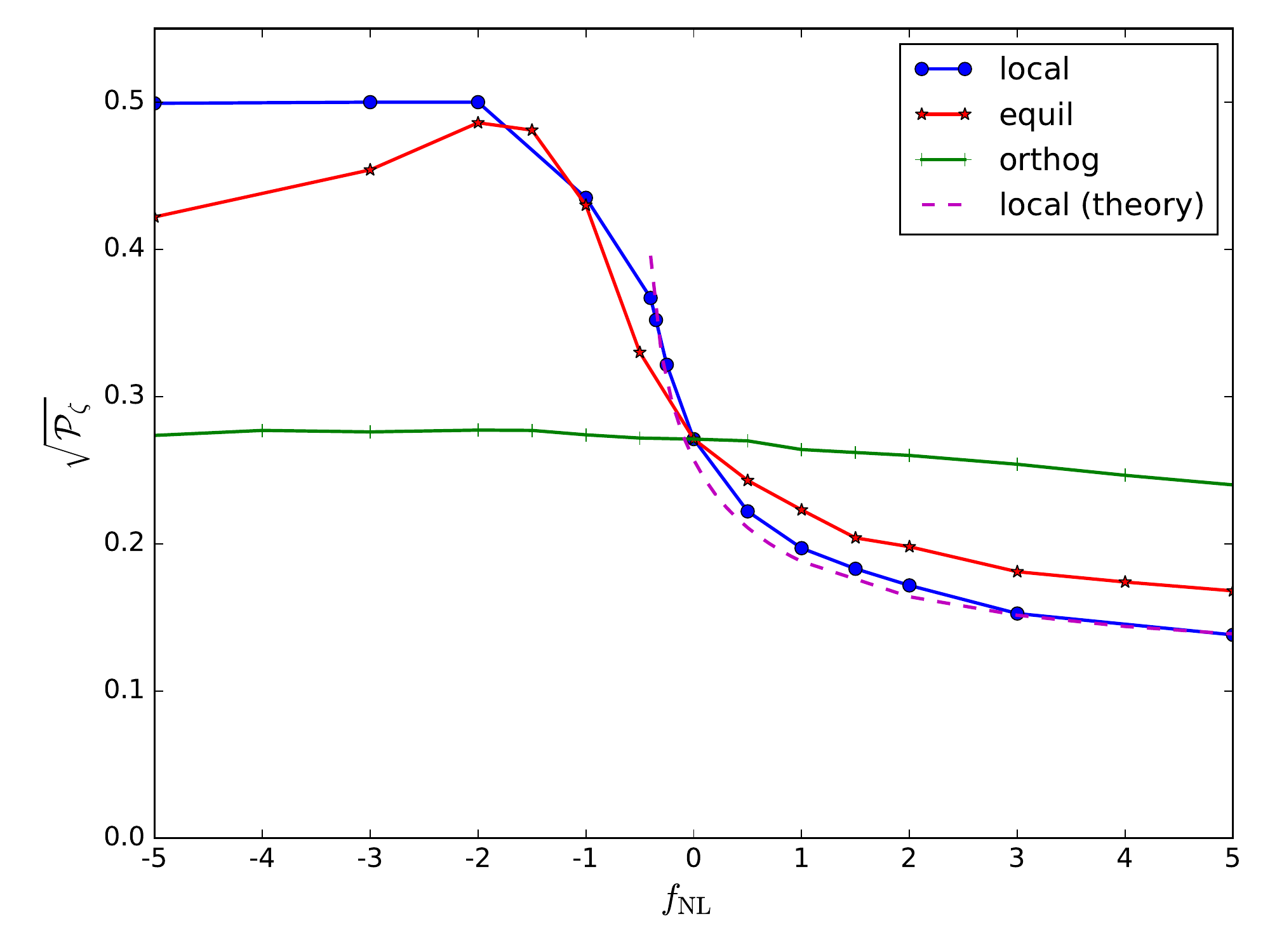}
    }
\caption[Power spectrum constraints for different bispectrum shapes]{The constraints on the power spectrum as a function of $\fNL$ for the different bispectrum shapes is plotted. The plots show the upper limit on the power spectrum peak, spanning 1 e-fold, for a constraint on the abundance of PBHs $\beta<10^{-4}$. The right plot simply shows the central region of the left plot. Constraints become quickly tighter for positive $\fNL$ in the local and equilateral configurations, and weaker for negative $\fNL$. For the orthogonal configuration however, constraints are only weakly dependent on the value of $\fNLorth$. The dotted line represents the theoretical prediction for the constraint on the power spectrum for the local model originally derived in \cite{Byrnes:2012yx}. There is strong agreement for small values of $\fNLloc$, but the results disagree for larger values - although the same qualitative behaviour is seen.}
\label{1_efold_plots}  
 \end{minipage}
 \begin{minipage}{1\textwidth}
\centering
\subfloat[]{%
      \includegraphics[width=0.49\textwidth]{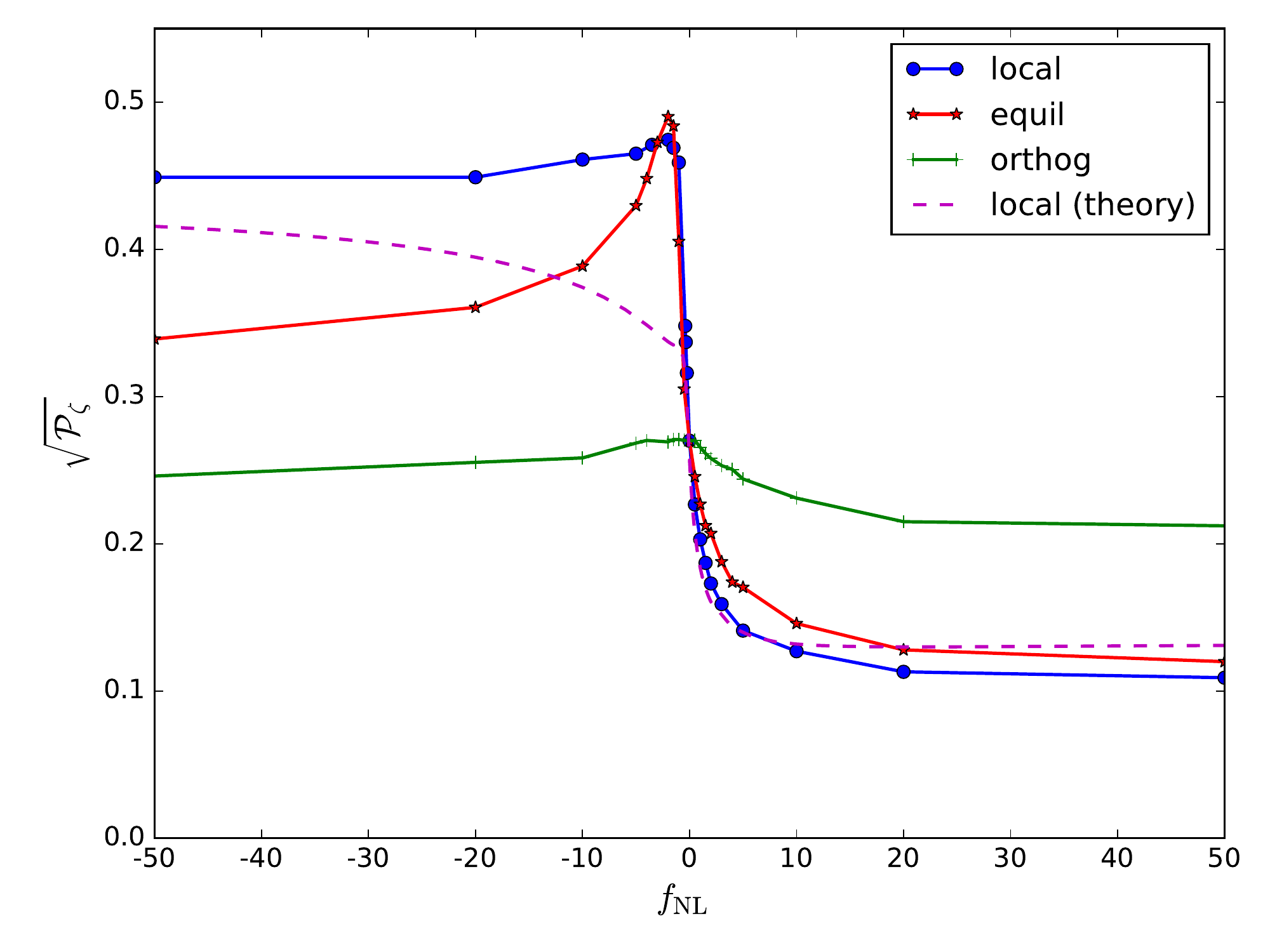}
    }
    \hfill
    \subfloat[]{%
      \includegraphics[width=0.49\textwidth]{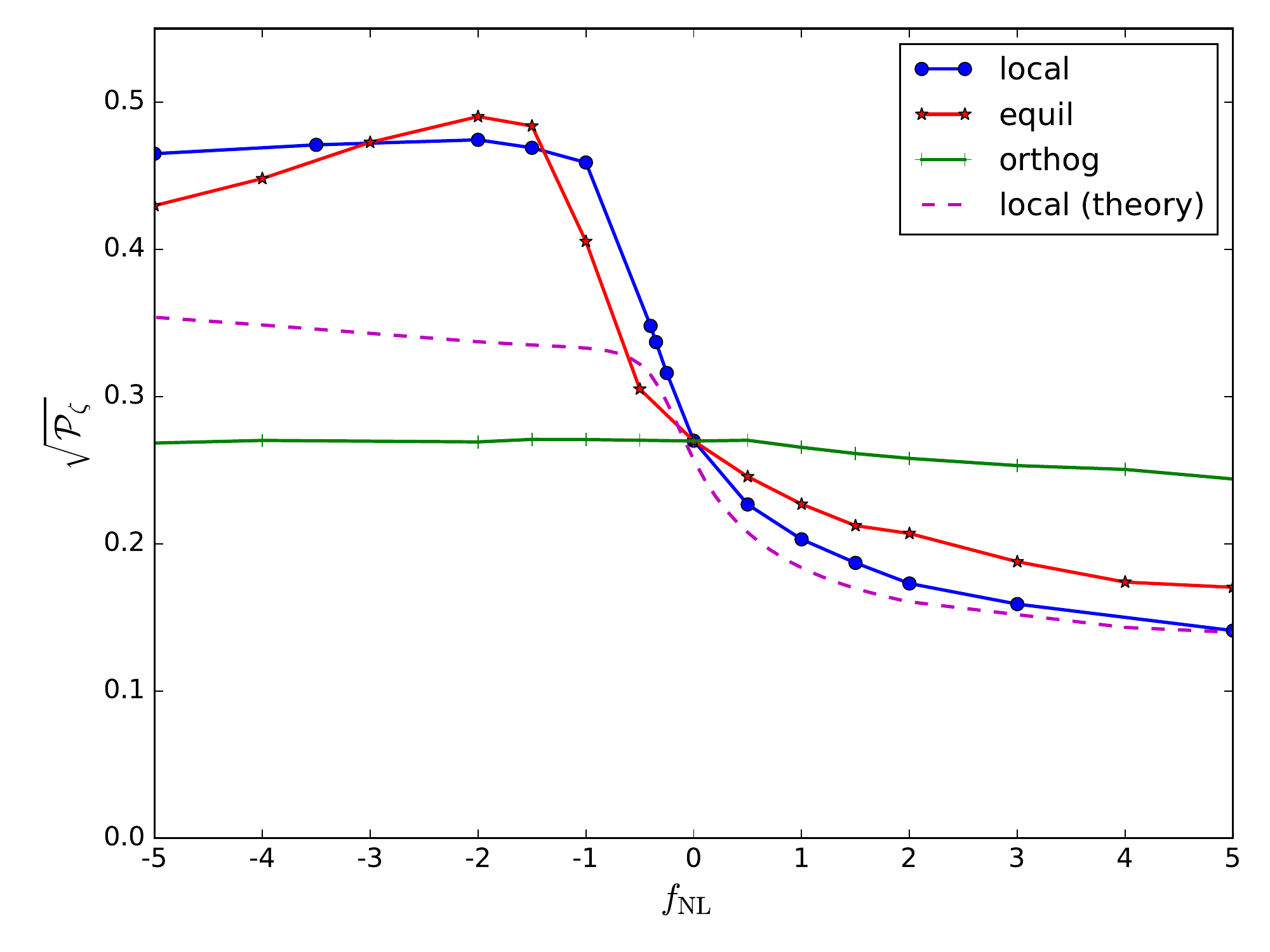}
    }
\caption[Effect of modal coupling to super-horizon modes]{The constraints on the power spectrum as a function of $\fNL$ for the different bispectrum shapes is plotted. The plots show the upper limit on the power spectrum peak, spanning 2.5 e-folds, for a constraint on the abundance of PBHs $\beta < 10^{-4}$. The constraints display the same behaviour as seen in figure \ref{1_efold_plots}, with the exception that the constraints in the local model are slightly tighter due to stronger modal coupling now that different scale modes are being considered - especially for negative values of $\fNLloc$. As expected, the theoretical line for the local model does not match well for negative values - this is because the peak-background split has been used which assumes a large separation in scales between the ``peak" modes and the ``background" modes, with intermediate modes neglected.}
\label{25_efold_plots} \end{minipage}
\end{figure}

Figure \ref{1_efold_plots} shows the constraints on the peak value of the power spectrum, spanning 1 e-fold, obtained for different values of $\fNL$ for the local, equilateral and orthogonal bispectrum shapes, as well as the theoretical predictions for the local-type (as calculated in \cite{Young:2014oea}, with no super-horizon modes present).  The lines show the maximum allowed amplitude of the power spectrum given a constraint on the abundance of PBHs, $\beta<10^{-4}$. There is good agreement in the local model with the theoretical prediction for small values of $\fNLloc$, but mild disagreement for larger values - the theoretical model slightly overestimates the constraints for large positive $\fNLloc$. This is due to the fact that the calculation of the power spectrum assumes a dominant Gaussian component (there is much stronger agreement for the Gaussian component of the power spectrum). The bounds typically become stronger for positive values of $\fNL$ but significantly weaker for negative values. As $\fNL$ becomes large and negative, constraints quickly reach a maximum value before becoming slightly tighter - due to the effect of modal coupling \cite{Young:2014oea}. This is not seen for the theoretical prediction, which does not account for the modal coupling - meaning the predicted constraints are much weaker. 

The exception is the orthogonal shape, with results showing that constraints on the power spectrum are relatively insensitive to orthogonal-type non-Gaussianity. This is due to the $\fNLorth$ having only a small effect on the skewness of the distribution, which will be discussed in more detail later in the paper. 

Another important note is that the theoretical calculation predicts the constraints on the power spectrum rapidly become weaker, and greater than unity, for negative $\fNL$ - and whilst the rapid weakening of constraints is still seen in the numerically generated constraints, they quickly reach some maximum value. In the case of equilateral- and orthogonal-type non-Gaussianity, the constraints then become stronger as more negative values of $\fNL$ are considered. This is believed to be due to the  strong signal in the bispectrum shapes when 3 modes of the same scale are considered - and so the effect of modal coupling tightens constraints, as discussed in more detail in \cite{Young:2014oea}. By contrast, the local-type peaks in the squeezed limit - when the modes considered are of significantly different scales - and so the effect of modal coupling is less important.

Figure \ref{25_efold_plots} shows the power constraints obtained for a peak in the power spectrum spanning 2.5 e-folds. Due to the computing resources required for a larger peak, we do not consider broader peaks than this in the power spectrum. The plot for the theoretical calculation for the local model now includes the effect of modal coupling to super-horizon modes with a large power spectrum spanning $\frac{1}{2}$ an e-fold \cite{Young:2014oea}. The constraints obtained are similar to the case where a narrower peak in the power spectrum is considered - positive $\fNL$ increases PBH abundance and tightens constraints, whilst negative $\fNL$ has the opposite effect. The most significant difference can be seen in the local model. Constraints are now slightly tighter than previously, and notably stronger for negative values of $\fNLloc$ (though we note the theoretical prediction is unchanged here - and still does not account for modal coupling). This is due to the fact that the local shape peaks in the squeezed limit when the modes are of different scales. The peak in the power spectrum is now broad enough that small-scale modes and large-scale modes are considered - allowing the effect of significant modal coupling. It has previously been noted that modal coupling typically increases PBH production, and tightens constraints \cite{Young:2014oea}. Thus, when a broader peak in the power spectrum is considered, constraints become tighter for the local shape - but remain largely unchanged for the equilateral and orthogonal shapes.

\subsection{Skewness}

We will now consider the skewness of the different bispectrum shapes, and show that when the non-Gaussianity and skewness parameters, $\fNL$ and $\skewR$, respectively, are small that the skewness alone can be considered to produce constraints on the power spectrum. However, as $\fNL$ and the skewness become large, the effects of the different bispectrum shapes must be considered. The skewness, given by equation~\eqref{eq:skewness}, may be computed for a given bispectrum using, 
\begin{align}
\langle \Delta_R(\bx)^2 \rangle&=\int d\ln k {W}(R,k)^2 \frac{k^3 P_{\zeta}(k)}{2\pi^2} \,,\nonumber\\
\langle \Delta_R(\bx)^3 \rangle&=\frac{2}{(2\pi)^4}\int\limits_0^\infty d\mathrm{ln}(k) k^3 {W}(R,k) \int\limits_0^\infty d\mathrm{ln}(q) q^3 {W}(R,q) \int\limits_{-1}^1 d\mu\, {W}(R,k_\mu){B_\zeta}(k,q,k_\mu)\,,
\end{align}
where $\mu=\mathrm{cos}(\theta)$, with $\theta$ representing the angle between ${\bf k}$ and ${\bf q}$. Calculating the skewness for the three bispectrum shapes being considered using this formula gives:
\begin{equation}
\skewR^{\rm local}=2.6\fNLloc\sqrt{\mathcal{P}_\zeta}\, ,
\end{equation}
\begin{equation}
\skewR^{\rm equil}=1.1\fNLeq\sqrt{\mathcal{P}_\zeta}\, ,
\end{equation}
\begin{equation}
\skewR^{\rm orthog}=0.07\fNLorth\sqrt{\mathcal{P}_\zeta}\, .
\end{equation}
We note that the numbers obtained here are slightly different than the values obtained by \cite{Shandera:2012ke} for the local and equilateral model, due to the choice of window functions, transfer functions and the form of the power spectrum. To our knowledge, this is the first time that the skewness has been calculated for the orthogonal model.

We see that the skewness is relatively large for the local and equilateral shapes but small for the orthogonal shape - which is why the constraints are less dependent on $\fNLorth$ than on $\fNLloc$ and $\fNLeq$. Note that the above analytic formulae are only correct whilst the skewness is small. 

\begin{figure}[t]
\centering
\includegraphics[width=0.7\linewidth]{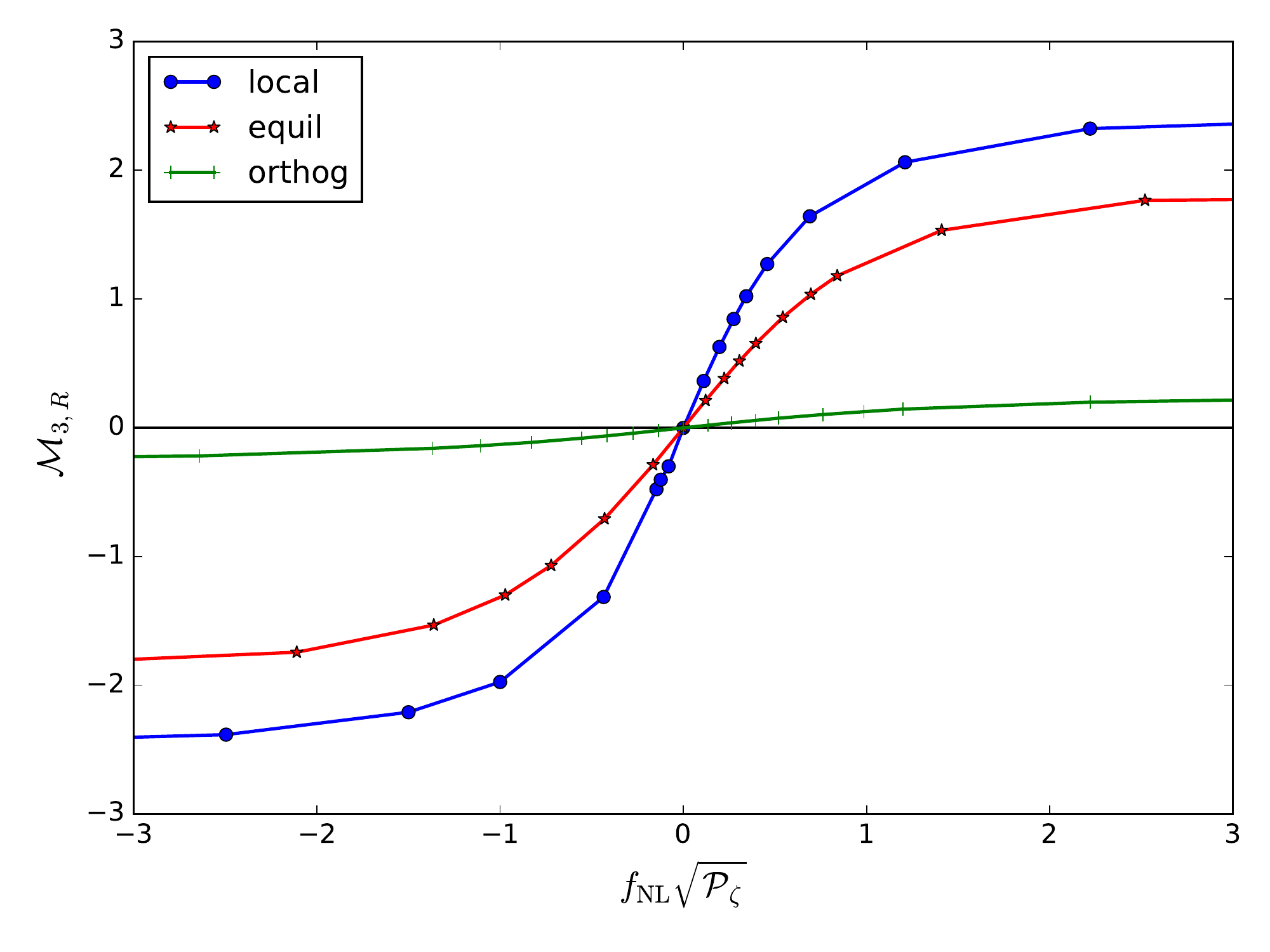}
\caption[Skewness of the distribution]{The skewness, $\skewR$, is plotted against $\fNL\sqrt{\mathcal{P}_\zeta}$ for an abundance of PBHs $\beta=10^{-4}$. It can be seen that the skewness does not depend linearly on $\fNL\sqrt{\mathcal{P}_\zeta}$, unlike that predicted by equation \eqref{eq:skewness}. However, in the central region where the skewness is small, $|\skewR |<\mathcal{O}(0.1)$, the relation is approximately linear, and the skewness may be used to parameterize the abundance of PBHs and constraints on the power spectrum.}
\label{skewness_fixed_beta}
\end{figure}

Figure \ref{skewness_fixed_beta} shows how the skewness varies as a function of $\fNL\sqrt{\mathcal{P}_\zeta}$. The plot is generated from the simulated density maps for a fixed abundance of PBHs, $\beta=10^{-4}$. Confirming the above calculation, the skewness is seen to be the largest for local non-Gaussianity, and smallest for orthogonal non-Gaussianity. The relation is also strongly non-linear as the skewness becomes large - which indicates the region where skewness can no longer be used to parameterize the abundance of PBHs. The skewness saturates relatively quickly as $\fNL$ increases - representing the fact that the distribution has become dominated by the non-Gaussian components. The fact that the skewness reaches some constant value as $\fNL$ becomes larger also corresponds to the fact that the constraints asymptote to a constant level as $\fNL$ becomes larger.

\begin{figure}[t]
\centering
\subfloat[]{%
      \includegraphics[width=0.49\textwidth]{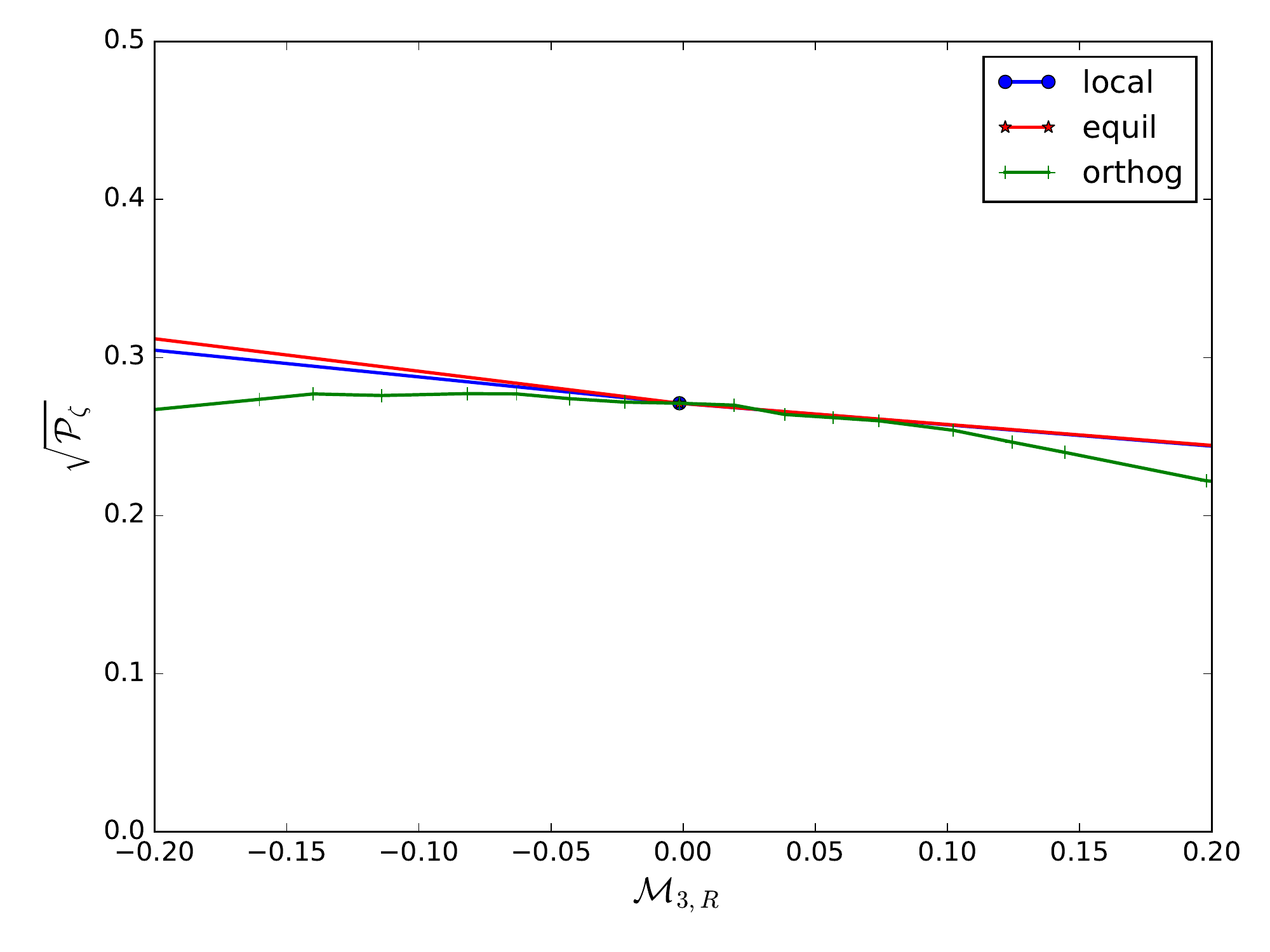}
    }
    \hfill
    \subfloat[]{%
      \includegraphics[width=0.49\textwidth]{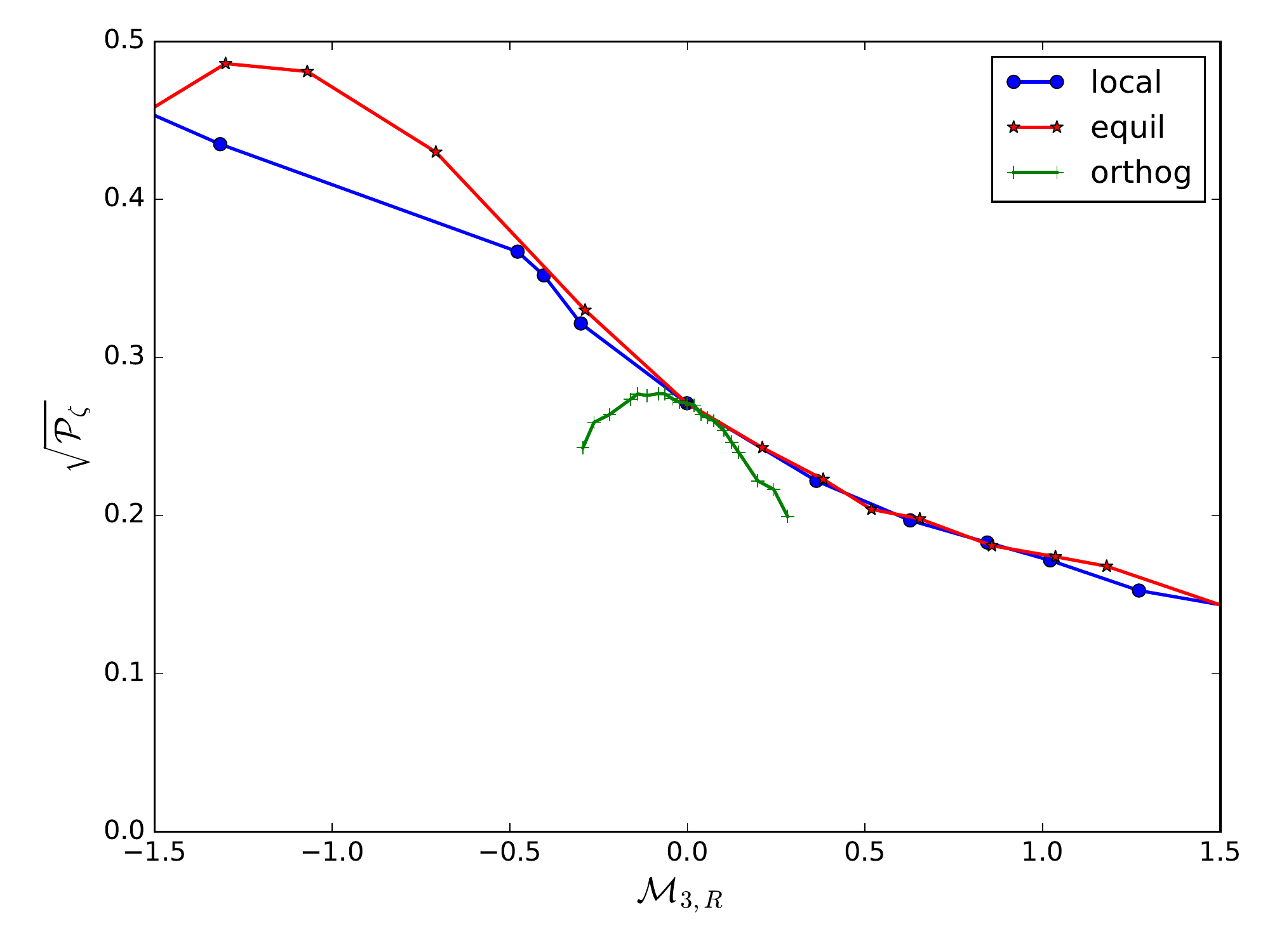}
    }
\caption[Constraints on the power spectrum as a function of the skewness]{The constraint on the power spectrum corresponding to a constraint on the abundance of PBHs, $\beta<10^{-4}$, as a function of the skewness is plotted. As can be seen in the left plot, for the three bispectrum shapes considered, the constraints on the power spectrum show good agreement whilst the skewness is small, $\skewR<\mathcal{O}(0.1)$. The skewness of the distribution is therefore the most important consideration, rather than the shape of the bispectrum. The right plot shows the behaviour as the skewness becomes large - the shape of the bispectrum has a large impact on the derived constraints and this must therefore be taken into account.}
\label{skew_vs_power}
\end{figure}

Figure \ref{skew_vs_power} plots the upper bound on the power spectrum corresponding to a constraint on the abundance of PBHs, $\beta<10^{-4}$, as a function of the skewness. Whilst the skewness is small, $\skewR<\mathcal{O}(0.1)$, the skewness of the distribution is the most important consideration, rather than the shape of the bispectrum. This can be seen in the left plot of figure \ref{skew_vs_power}. However, as the non-Gaussianity, $\fNL$, and the skewness, $\skewR$, become larger, this is no longer the case - and a large discrepancy between the different bispectrum configurations can be seen the right plot.

\section{Summary}\label{sec:conclusions}
The lack of observation of PBHs allows tight constraints to be placed on the mass fraction of the universe collapsing into PBHs at the time of formation, $\beta$. This, in turn, allows unique bounds to be placed on the small-scale primordial curvature perturbation power spectrum, $\mathcal{P}_\zeta$, at scales which are otherwise unobservable - although these bounds are orders of magnitude weaker than constraints from sources such as the CMB. Non-Gaussian density maps were generated and used to predict the abundance of PBHs for different shapes of bispectrum, in the local, equilateral and orthogonal configurations. These predictions were then used to place constraints on $\mathcal{P}_\zeta$ as a function of the amplitude and shape of the bispectrum.

As an improvement on previous work, this method allows the consideration of bispectra of arbitrary shape and amplitude. We confirmed the previous findings using analytic methods of the effects of local-type non-Gaussianity \cite{Byrnes:2012yx,Young:2013oia,Young:2014oea} - non-Gaussianity can have a strong effect on constraints on the power spectrum, typically becoming stronger (weaker) for positive (negative) values of $\fNL$. The effect of the skewness was also considered, confirming results seen in \cite{Shandera:2012ke} (note that \cite{Clark:2015tha} also recently used a similar technique to calculate constraints arising from ultra-compact mini-haloes) - but demonstrate that using the skewness to parameterize the abundance of PBHs is only valid for small amounts of non-Gaussianity. As seen in figure \ref{skew_vs_power}, for small amounts of skew the shape of the bispectrum has little effect on the constraints - and the skewness of the distribution can be considered the most important factor (but note that the constant of proportionality relating $\fNL$ to the skewness does strongly depend on the non-Gaussian template). However, for large non-Gaussianity, constraints on the power spectrum become strongly dependent on the shape of the bispectrum. \

For the local and equilateral shapes the constraints become tighter for positive $\fNL$ but dramatically weaker for small negative $\fNL$. For orthogonal-type non-Gaussianity, the effects are qualitatively similar, but much less dramatic - due to the relatively small skewness generated by this bispectral shape. Previous findings that the effect of modal coupling and positive skew is to increase PBH formation, whilst negative skew decreases PBH formation, are also confirmed.

\section*{Acknowledgements} SY is supported by an STFC studentship, and CB is supported by a Royal Society University
Research Fellowship. DR acknowledges support from the European Research Council under the European Union's Seventh Framework Programme (FP/2007-2013) / ERC Grant Agreement No. [308082].


\newpage

%

\chapter{Conclusion}
\markboth{Conclusion}{Conclusion}
\label{chap:conc}
\section{Summary of this thesis}

Primordial black holes (PBHs) may have formed very early on in the history of the Universe, and represent a unique probe to study the small scales in the early Universe - providing unique constraints on the power spectrum over a broad range of scales. However, these constraints only provide an upper bound on the power spectrum and are much weaker than constraints from the cosmic microwave background or large scale structure. Constraints on the abundance of PBHs of different masses vary from $\beta<10^{-27}$ to $\beta<10^{-5}$, where $\beta$ is the energy fraction of the Universe contained within PBHs at the time of their formation. Assuming a perfectly Gaussian distribution, there is a unique relation between the amplitude of the power spectrum $\mathcal{P}_\zeta$ at a given scale and the abundance of PBHs $\beta$ at the corresponding mass scale. Therefore, a constraint on the abundance of PBHs at a given mass can give a constraint on the power spectrum at a corresponding scale. However, whilst the constraint on $\beta$ varies by many orders of magnitude, the corresponding constraints on the power spectrum only vary by a factor of $4-5$, with $\mathcal{P}_\zeta<\mathcal{O}(10^{-2})$. 

Historically, the abundance of PBHs from has been calculated using the density contrast $\Delta$, and typically assumed a Gaussian distribution. The variance of the perturbations $\langle\Delta^2\rangle$ is calculated by integrating the power spectrum $\mathcal{P}_\Delta(k,t)$,
\begin{equation}
\langle\Delta^2\rangle=\int\limits_0^\infty\frac{\mathrm{d}k}{k}\mathcal{P}_\Delta(k,t)W^2(k,R),
\end{equation}
where $W(k,R)$ is the smoothing function and $R$ is the smoothing scale. The abundance of PBHs is then calculated by integrating the probability density function (PDF) of the perturbations over the range which would form a PBH, from the minimum value $\Delta_c$ to infinity.
\begin{equation}
\beta=2\int\limits_{\Delta_c}^\infty\mathrm{d}\Delta P(\Delta)=2\int\limits_{\Delta_c}^\infty\mathrm{d}\Delta \frac{1}{\sqrt{2\pi\langle\Delta^2\rangle}}\exp{-\frac{\Delta^2}{2\langle\Delta^2\rangle}}.
\end{equation}
However, \cite{Green:2004wb} introduced a calculation to derive $\beta$ from the primordial curvature perturbation power spectrum. The approach used the theory of peaks, as described in \cite{Bardeen:1985tr}, and used a critical value in terms of the curvature perturbation, $\zeta_c$. It was erroneously found that the two calculations were in good agreement - and this calculation became the standard approach for calculating PBH abundance when studying inflationary models.

In chapter \ref{chap:paper1} the calculation of the PBH abundance is reviewed, finding that an accurate calculation of $\beta$ from the primordial curvature perturbation power spectrum first requires a calculation of the density power spectrum, and an integral over the PDF of the density. Using the peaks theory approach of \cite{Green:2004wb} yields an error of many orders of magnitude due to the effect of super-horizon modes, which should not affect the (short-term) evolution of a region of the Universe, and so should not affect PBH formation - as shown in figure \ref{dontusezeta}. A threshold value for PBH formation should therefore not be stated in terms of the curvature perturbation $\zeta$ unless care is taken to neglect super-horizon modes. This effect was later seen and quantified for an analytic model involving spherically-symmetric top-hat perturbations by \cite{Harada:2015yda}, and had important implications for future work in this area of research.

This chapter also considers a fast approximation using the curvature perturbation (first used in \cite{Byrnes:2012yx}), and places the calculation on a much firmer theoretical platform. Essentially, the calculation states that the amplitude of density perturbations at the time of horizon re-entry is approximate to the amplitude of the curvature perturbation power spectrum at that scale. This approach is then used to calculate the PBH abundance throughout the remainder of the thesis.

The effect of non-Gaussianity on the number of PBHs forming in the early Universe has been the subject of extensive research over the years (i.e. \cite{Bullock:1996at,Ivanov:1997ia,Hidalgo:2007vk,Byrnes:2012yx,Shandera:2012ke}) - finding that non-Gaussianity can have a strong effect on the abundance of PBHs formed. Typically, only quadratic type non-Gaussianity had been considered - either $\zeta=\pm(\zeta_G^2-\sigma^2)$ or local-type non-Gaussianity, $\zeta=\zeta_G+\frac{3}{5}f_{NL}(\zeta_G^2-\sigma^2)$. \cite{Byrnes:2012yx} considered local-type non-Gaussianity up to third order and calculated the effect on the abundance of PBHs and derived constraints on the power spectrum. It was shown that, for even small non-Gaussianity parameters, the constraints on the power spectrum could change by an order of magnitude. However, this work considered second- or third-order terms separately, and did not consider the effects of simultaneously having a non-zero quadratic and cubic term. Shortly after this paper came the work by \cite{Shandera:2012ke}, which considered the effect of skewness arising from non-Gaussianity. The PDF of the distribution can be calculated by an expansion in terms of the skewness parameter $\mathcal{M}_3$ and then the number of PBHs forming can be calculated. However, the expansion is only valid while the distribution is close to Gaussian - and so cannot be trusted when the non-Gaussianity becomes significant.  The conclusions drawn in this paper were in agreement with those by \cite{Byrnes:2012yx}.

In chapter \ref{chap:paper2}, the calculation performed in \cite{Byrnes:2012yx} is extended to consider simulataneously local-type non-Gaussianity parameters up to $5^{th}$ order,
\begin{equation}
\zeta=\zeta_{g}+\frac{3}{5}f_{NL}\left(\zeta_{g}^{2}-\sigma^{2}\right)+\frac{9}{25}g_{NL}\zeta_{g}^3+\frac{27}{125}h_{NL}\left(\zeta_{g}^{4}-3\sigma^{4}\right)+\frac{81}{625}i_{NL}\zeta_{g}^5+\cdots .
\end{equation}
The results from previous papers are verified, and for certain simple models relating the non-Gaussianity parameters, it is shown that the results from such an expansion converge - although, notably, for negative terms it is shown to be important whether the series is truncated at odd or even terms. It was also shown that this may not always be the case, as in the curvaton model - where results diverge significantly depending on the order at which the above series is truncated. Another significant finding was that the even order terms (i.e. the $f_{NL}$ and $h_{NL}$ terms) all have a similar effect on the PDF and derived constraints - a statement which is also true for odd order terms.

It has been shown that non-Gaussianity has a strong effect on the abundance of PBHs, affecting their abundance by many orders of magnitude, and the constraints on $\mathcal{P}_\zeta$ depends strongly on the amount of non-Gaussianity. It is possible for the constraint on the power spectrum to vary by 4 orders of magnitude, from $\mathcal{P}_\zeta<\mathcal{O}(10^{-4})$ to $\mathcal{P}_\zeta<\mathcal{O}(1)$. For the majority of this thesis, only local-type non-Gaussianity was considered as this allows an analytic calculation of the constraints, although these results were verified and other types of non-Gaussianity considered with numerical simulations in chapter \ref{chap:paper5}.

There are 2 different ways in which non-Gaussianity can affect the abundance of PBHs (and the constraints on $\mathcal{P}_\zeta$):
\begin{enumerate}
\item As discussed in chapter \ref{chap:paper2}, the first is by affecting the skewness and kurtosis of the probability distribution function (PDF) - with even order non-Gaussianity ($\fNL$, $h_{\mathrm{NL}}$, etc) affecting the skewness, and odd order non-Gaussianity ($g_{\mathrm{NL}}$, $i_{\mathrm{NL}}$, etc) affecting the kurtosis. Positive (negative) skewness enhances (diminishes) the positive tail of the PDF where PBHs form, leading to an increase (decrease) in their abundance, and a tightening (weakening) of the constraints. Kurtosis typically serves to amplify both the positive and negative tails of the distribution, enhancing PBH formation and tightening the constraints. However, for small negative values, the tail of the PDF where PBHs form is reduced, and constraints weaken sharply for such values. 

\item The second effect, as discussed in chapter \ref{chap:paper3}, is modal coupling arising from the non-Gaussian component of the distribution. Whilst super-horizon perturbations do not directly affect whether or not a PBH forms, they have an indirect effect on the abundance of PBHs by increasing or decreasing the amplitude of the power spectrum in different regions of the Universe. Whilst this means that some areas of the Universe form more PBHs and some form less, the overall effect is to increase the number of PBHs forming. This is due to the exponential dependance of the abundance of PBHs on the power spectrum, and can be seen by considering a region which produces twice as many PBHs as expected due to coupling to a super-horizon mode, and a second region which produces half as many PBHs - resulting in a higher number of PBHs in total. The importance of modal coupling depends on the shape and amplitude of the bispectrum as well as the form of the power spectrum.
\end{enumerate}

The effects of skewness were studied by \cite{Shandera:2012ke}, who used the skewness parameter $\mathcal{M}_3$ to parameterise the primordial non-Gaussianity - which is valid for small amounts of non-Gaussianity. The skewness parameter is related to the non-Gaussianity parameter $\fNL$ for squeezed (local) and equilateral shape bispectra, $\mathcal{M}_3=3.13\fNLloc\mathcal{P}^{1/2}=1.22\fNLeq\mathcal{P}^{1/2}$. Higher order moments of the distribution, $\mathcal{M}_N$, are derived from the skewness parameter and this is used to reconstruct the probability density function for primordial perturbations, and this is then used to investigate the effect of skewness on the abundance of PBHs and constraints on the power spectrum. In agreement with the results found by \cite{Byrnes:2012yx} and with the results presented in this thesis, it was found that positive skewness increases the amount of PBHs forming, and negative skewness decreases it. One of the main drawbacks of this approach was that the expansions used are only valid for small non-Gaussianity - when the Gaussian component of the perturbations is still dominant. However, because the power spectrum must necessarily become large in order for a significant number of PBHs to form, even a small value for the non-Gaussianity parameters leads to the non-Gaussian components of the perturbations becoming large. 

In order to investigate this further in chapter \ref{chap:paper5}, a numerical approach was used to simulate non-Gaussian density maps of the early Universe. This allowed for an arbitrary shape bispectrum to be considered (local, equilateral and orthogonal shapes were used) and constraints on the power spectrum arising from PBHs to be calculated. It was found that, for a skewness parameter $|\mathcal{M}_3|<0.1$, the skewness alone can be used to place constraints on the power spectrum - which is in agreement with the work by \cite{Shandera:2012ke}. However, for larger skewness, it is necessary not only to know the full bispectrum shape, but also the form of the power spectrum at all scales in order to predict the abundance of PBHs and calculate constraints on the power spectrum.


PBHs are a viable cold dark matter (CDM) candidate, and there are many inflationary models which predict a significant amount of both PBHs and non-Gaussianity. In chapter \ref{chap:paper4} the effect of modal coupling between large-scale modes visible in the CMB, and small scale-scale PBH forming modes is considered in the scenario that dark matter is composed partially or entirely of PBHs. Modal coupling can have a strong effect on the abundance of PBHs in different regions of the Universe, and this results in (additional) perturbations in the PBH density - which would be seen as isocurvature perturbations. If CDM is composed of PBHs, either in whole or in part, this means that constraints on the amplitude of dark matter isocurvature perturbations from the Planck satellite can be used to place constraints on the non-Gaussianity parameters. Unless PBHs make up only a very small fraction of dark matter, less than around $0.1\%$, very tight constraints can be placed on the non-Gaussianity parameters at all orders - $|\fNL^{local},\gNL^{local}|<\mathcal{O}(10^{-3})$. This rules out nearly all inflationary models except single-scalar-field inflation as a method for producing PBH dark matter, and would provide great insight into the early Universe if PBHs, primordial non-Gaussianity, or isocurvature perturbations are detected in the future. A similiar study was also performed by \cite{Tada:2015noa} at the same time, and whilst their results were verified by the calculations in chapter \ref{chap:paper4}, their calculations did not account for the non-Gaussian distribution on small scales and only considered an $\fNL$ term.

\section{Directions for future study}

There are still many open questions and avenues of research to pursue. Possible future directions include:
\begin{itemize}
\item Ultra-compact mini-haloes provide similar constraints on the power spectrum as PBHs. The constraints are stronger, but cover a smaller range of scales. The constraints are not likely to be as strongly dependent on non-Gaussianity as for PBHs, but it is possible that they may be used to place a constraint on $\fNL$ that is competitive with constraints from the CMB if they are detected.

\item In terms of the density contrast, the critical value for PBH formation is $\delta_c\approx0.45$, although it is known that this depends on the shape of the perturbation considered, and there has been extensive research into this in the past. However, the effect of super- and sub-horizon perturbations has not been well studied, and could have a significant effect on the number of PBHs formed. There is currently ongoing research to investigate the effects of super-horizon modes - even small effects can have a significant effect on their abundance. Such an investigation can lead refinements in the calculation of the abundance of PBHs, as well as having important implications for their primordial clustering.

\item The recent detection of gravitational waves from merging black holes \cite{Abbott:2016blz} raises the interesting possibility of using future gravitational wave detectors as a tool to search for PBHs. Depending upon the sensitivity and scale of such detectors, it may be possible to detect PBH mergers - and there has been significant work in recent months \cite{Bird:2016dcv,Clesse:2016vqa,Sasaki:2016jop} discussing whether the observed merging black holes were primordial in origin. Factors which can have a significant effect on the marger rate, and thus the observability of such mergers, include the small-scale clustering of PBHs (both primordial clustering and as a result of there gravitational interactions since formations) and the mass function of PBHs (including effects from the time of formation and their subsequent merger history).
\end{itemize}

\bibliographystyle{JHEP}
\bibliography{bibfile}


\end{document}